\documentclass[a4paper,fleqn,usenatbib]{mnras}
 
\usepackage{newtxtext,newtxmath}
\usepackage[T1]{fontenc}
\usepackage{ae,aecompl}

\usepackage{graphicx}	% Including figure files
\usepackage{amsmath}	% Advanced maths commands
\usepackage{amssymb}	% Extra maths symbols
\usepackage{float}
\usepackage{hyperref}
\usepackage{color}

\usepackage{pdflscape}

\newcounter{species} 
\def\ion#1#2{\hbox{\setcounter{species}{#2}#1\,{\scriptsize\Roman{species}}\relax}}

\title[Attenuation in Star Forming Regions]{A Cautionary Tale of Attenuation in Star Forming Regions}

\author[M. Molina et al.]{
Mallory Molina$^{1,2}$,\thanks{E-mail: mallory.molina@montana.edu (MM)}
Nikhil Ajgaonkar$^{3}$,
Renbin Yan$^{3}$,
Robin Ciardullo$^{1}$,
Caryl Gronwall$^{1}$,
\newauthor 
Michael Eracleous$^{1}$,
M\'{e}d\'{e}ric Boquien$^{4}$ and
Donald P. Schneider$^{1}$
\\
$^{1}$Department of Astronomy and Astrophysics and Institute for Gravitation and the Cosmos, The Pennsylvania State University 525 Davey Lab,\\
University Park, PA 16802, USA\\
$^{2}$eXtreme Gravity Institute, Department of Physics, Montana State University, Bozeman, MT 59715, USA\\
$^{3}$Department of Physics and Astronomy, University of Kentucky, 505 Rose St., Lexington, KY 40506-0057, USA\\
$^{4}$Centro de Astronom\'{i}a (CITEVA), Universidad de Antofagasta, Avenida Angamos 601, Antofagasta 1270300, Chile
}

\date{Accepted XXX. Received YYY; in original form ZZZ}

\pubyear{2019}

\hypersetup{draft}
\begin{document}
\label{firstpage}
\pagerange{\pageref{firstpage}--\pageref{lastpage}}
\maketitle

% Abstract of the paper
\begin{abstract}
The attenuation of light from star forming galaxies is correlated with a multitude of physical parameters including star formation rate, metallicity and total dust content. This variation in attenuation is even more evident on kiloparsec scales, which is the relevant size for many current spectroscopic integral field unit surveys. To understand the cause of this variation, we present and analyse \textit{Swift}/UVOT near-UV (NUV) images and SDSS/MaNGA emission-line maps of 29 nearby ($z<0.084$) star forming galaxies. We resolve kiloparsec-sized star forming regions within the galaxies and compare their optical nebular attenuation (i.e., the Balmer emission line optical depth, $\tau^l_B\equiv\tau_{\textrm{H}\beta}-\tau_{\textrm{H}\alpha}$) and NUV stellar continuum attenuation (via the NUV power-law index, $\beta$) to the attenuation law described by Battisti et al. We show the data agree with that model, albeit with significant scatter. We explore the dependence of the scatter of the $\beta$--$\tau^l_B$ measurements from the star forming regions on different physical parameters, including distance from the nucleus, star formation rate and total dust content. Finally, we compare the measured $\tau^l_B$ and $\beta$ values for the individual star forming regions with those of the integrated galaxy light. We find a strong variation in $\beta$ between the kiloparsec scale and the larger galaxy scale that is not seen in $\tau^l_B$. We conclude that the sight-line dependence of UV attenuation and the reddening of $\beta$ due to the light from older stellar populations could contribute to the scatter in the $\beta$--$\tau^l_B$ relation.
\end{abstract}
% Select between one and six entries from the list of approved keywords.
% Don't make up new ones.

%INTRODUCTION
%%%%%%%%%%%%%%%%%%%%%%%%%%%%%%%%%%%%%%%%%%%%%%%%%%%%%%%%%%%%%%%%%%%%%%%%%%%%%%%%%
\section{Introduction}
\label{sec:intro}

Attenuation relations describe how the light from both the stellar continuum and nebular emission are affected by dust intrinsic to the source. They differ from extinction laws as they account for the dust, gas, and stars commingling within the observed aperture, which is appropriate for most extragalactic observations. \citet[][]{Calzetti1994} found a relation between the difference in attenuation experienced by the stellar continuum and that affecting the nebular emission in the integrated galaxy light from local ultra-violet (UV)-bright starburst galaxies. The relation was quantified by \citet[][]{Calzetti2000} as $E(B - V)_{\textrm{star}}=0.44E(B - V)_{\textrm{gas}}$, which was used to derive attenuation as a function of wavelength in the UV band. In response to this empirical result, \citet[][]{Charlot2000} described a physical model where the line-emitting gas is concentrated near star forming regions which are enshrouded in dust, while older stars are more evenly distributed among the diffuse interstellar medium; therefore the observed stellar continuum is not as strongly affected by attenuation as nebular emission. The Calzetti law was updated by \citet[][hereafter B16]{Battisti2016} for application to the unresolved nuclear regions of local ($z\lesssim0.1$) Sloan Digital Sky Survey (SDSS) galaxies, and systems with much lower star formation rates (SFRs; $\log[\textrm{SFR}/(M_\odot~\textrm{yr}^{-1})]<1.6$).

In general, attenuation laws have three distinct components: the general shape, which approaches a power law in the UV band, the relation between the stellar and nebular attenuation, and the presence of excess attenuation at 2175\,\AA\ (commonly referred to as the 2175\,\AA\ bump). All three components can individually correlate with the intrinsic properties of the galaxy, which include SFR, mass, optical opacity, and total attenuation \citep[][]{Charlot2000,Calzetti2000,Calzetti2001,Wild2011,Xiao2012,Salim2018}. The assumed attenuation law will strongly impact any estimate of SFRs determined from the UV stellar continuum, which traces recent star formation \citep[$\textrm{Age}\lesssim100$--200~Myr, see][for a complete review]{Kennicutt1998,Kennicutt2012,Calzetti2013}. Adding to that complication, the UV band is where most of the differences in the attenuation laws lie \citep[see][for a review and update]{Calzetti2000,Salim2018}. Understanding UV attenuation is especially important at higher redshifts, where the rest-frame UV is often the only accessible band for exploring star formation and other activity in galaxies. Therefore the choice of attenuation law is critical for accurately measuring recent star formation, both in nearby and more distant galaxies.

Unfortunately, the physical properties that correlate with the overall shape of the attenuation law, such as the dust composition and distribution relative to the stars and gas, all change across the face of a galaxy \citep[][]{Charlot2000,Calzetti2000,Wild2011,Xiao2012}. There is ample evidence of variations in attenuation curves, not only from galaxy to galaxy, but also within galaxies themselves \citep[][]{Buat2002,Buat2005,Gordon2004,Kong2004,Calzetti2005, Boissier2007, Dale2009, Boquien2009, Boquien2012,Decleir2019}. With the advent of IFU surveys such as the Spectrographic Area Unit for Research on Optical Nebulae \citep[SAURON;][]{Bacon2001}, ATLAS$^{3\textrm{D}}$ \citep{Cappellari2011}, DiskMass \citep{Bershady2010}, the Calar Alto Legacy Integral Field Area Survey \citep[CALIFA;][]{Sanchez2012,Sanchez2016,Walcher2014}, the Sydney-Australian Astronomical Observatory Multi-Object Integral Field Spectrograph \citep[SAMI;][]{Croom2012} and the Mapping Nearby Galaxies at Apache Point Observatory \citep[MaNGA;][]{Bundy15, Yan2016,Blanton2017}, we are no longer limited by data, but by our understanding of local attenuation and SFR estimates. 

In order to understand the appropriate attenuation law for the spatially resolved emission prevalent in current IFU surveys, we set out to test the validity of the B16 model for \textit{spatially resolved}, kpc-sized star forming regions across the faces of local star forming galaxies. The work of B16 utilized a set of galaxies with lower star formation rates than \citet{Calzetti1994,Calzetti2000}, and their results are therefore more representative of the average local star forming galaxy population (i.e., the sample from B16 has $-3.66\leq\log[\textrm{SFR}/(M_\odot~\textrm{yr}^{-1})]\leq1.6$, while that of \citet{Calzetti1994,Calzetti2000} has $0\lesssim\log[\textrm{SFR}/(M_\odot~\textrm{yr}^{-1})]\lesssim1.7$). We expand on their work by not only exploring spatially-resolved star forming regions across the faces of galaxies, but also by comparing those individual regions to the integrated galaxy light. Therefore this work explores the effects of aperture size on the attenuation of UV stellar continua and optical nebular emission in star forming galaxies. To effectively resolve regions within a galaxy, we combine data from the \textit{Swift} Ultraviolet Optical Telescope \citep[UVOT;][]{Roming2005} and the SDSS/MaNGA survey \citep[][]{Bundy15,Yan2016}. While the \textit{Galaxy Evolution Explorer} \citep[GALEX; ][]{galex} could give us a wider UV spectral range and is often used in attenuation studies, our emphasis on spatial resolution demands better image quality than the 5$\arcsec$ produced by GALEX.

Section~\ref{sec:obsdata} introduces the sample. Our goal is to measure the Balmer decrement and the UV spectral slope in resolved, kpc-sized star forming regions. In Section~\ref{sec:data} we discuss the basic data reduction; in Section~\ref{sec:stars} we list the requirements for a galaxy to be classified as star forming and compare GALEX to UVOT UV slope measurements, and give the criteria for defining star forming regions within the galaxy in Section \ref{sec:ddresult}. In Section~\ref{sec:analysis}, we compare the results to the attenuation relation for the unresolved nuclear regions of star forming galaxies as described by B16. After assessing the effectiveness of the B16 law for individual regions, we explore the effect of physical properties, such as metallicity and SFR, on the observed scatter between the measurements and the B16 law. In Section~\ref{sec:all} we study the dependence of the observed scatter on galaxy properties, and quantify the change in total attenuation for both the UV stellar continuum and the optical nebular emission between the individual, kpc-sized star forming regions and the integrated galaxy light for the star forming galaxies in our sample. We summarize our findings and present our conclusions in Section~\ref{sec:conclusion}. We assume a $\Lambda$CDM cosmology when quoting masses, distances, and luminosities, with $\Omega_{\textrm{m}}=0.3$, $\Omega_\Lambda = 0.7$ and $\textrm{H}_{0} = 70$~km~s\textsuperscript{$-1$}~Mpc\textsuperscript{$-1$}.

%Experimental Design and Target Selection
%%%%%%%%%%%%%%%%%%%%%%%%%%%%%%%%%%%%%%%%%%%%%%%%%%%%%%%%%%%%%%%%%%%%%%%%%%%%%%%%%
\section{The Sample}
\label{sec:obsdata}
This paper explores a subset of a larger sample that was chosen to study star formation and its quenching in the local universe. Our approach is to simultaneously analyse the near-UV (NUV) stellar continuum and optical nebular emission lines across the faces of galaxies, using data that are well matched in depth and physical scale. While this paper focuses on star forming galaxies as defined in Section \ref{sec:stars}, we present the basic properties of the total sample, including criteria for selection and basic data reduction. 
\subsection{SDSS-IV/MaNGA and \textit{Swift}/UVOT}
\label{ssec:sveys}
Our sample is drawn from the larger MaNGA survey \citep[][]{Bundy15, Yan2016}, which is one of the three major surveys in the fourth generation of SDSS \citep[SDSS-IV;][]{Blanton2017}. Target selection for the MaNGA survey is built on four basic requirements: (1) a sample size of $\sim10,000$ galaxies, (2) a uniform distribution in stellar mass above $M_{*} > 10^{9}~M_{\odot}$, as approximated by the SDSS \textit{i}-band absolute magnitude, (3) uniform spatial coverage in units of half-light radius, and (4) maximized signal-to-noise ratio and spatial resolution \citep{Wake2017}. The MaNGA data were taken with hexagonal IFU fiber bundles that are mounted on the Sloan 2.5-m telescope \citep[][]{Gunn2006}, observing up to 17 objects simultaneously. The IFU bundles can contain anywhere from 19 to 127 $2^{\prime\prime}$ fibers, and feed into the dual-channel Baryon Oscillation Spectroscopic  Survey (BOSS) spectrographs \citep[][]{Smee2013,Drory2015}. The resulting spectra cover the wavelength range of 3,622--10,354\,\AA\ at a resolving power of $R\sim2000$. The telescope is dithered to achieve near-critical sampling of the point-spread function (PSF), which has a full width at half maximum $\textrm{FWHM}\approx2.\!\!^{\prime\prime}$5. Meanwhile the data cubes have a spatial sampling of $0.\!\!^{\prime\prime}5$. These spectra provide the necessary coverage for the use of most nebular diagnostics, including BPT diagrams \citep{BPT81}. For this work, the most relevant emission lines are H$\alpha$ and H$\beta$.
 
\begin{figure}
\includegraphics[width=0.45\textwidth]{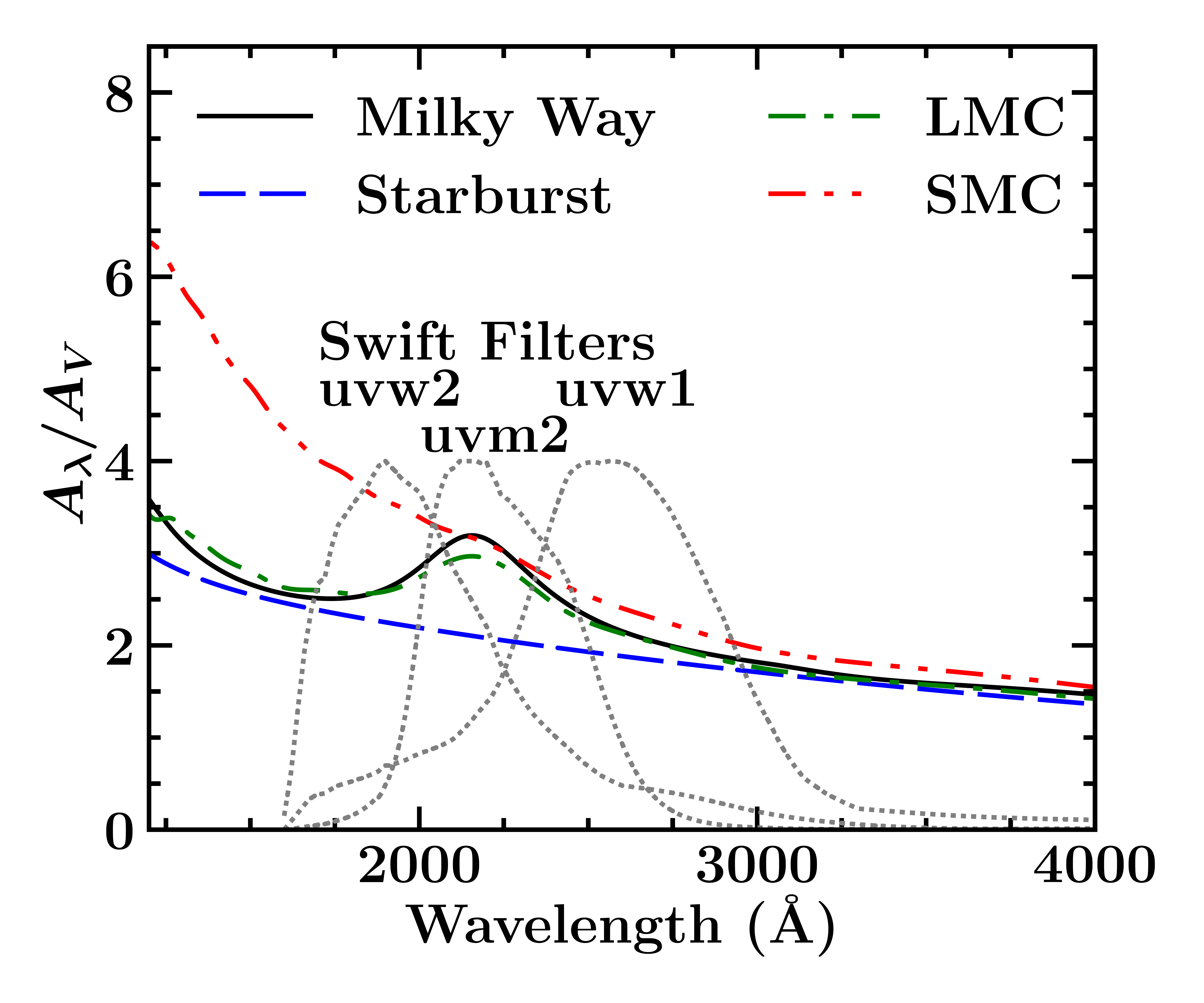}\vspace{-4mm}
\caption{The \textit{Swift}/UVOT uvw2, uvm2 and uvw1 NUV filter relative transmission curves (gray dotted lines) overlaid on the \citet{Cardelli1989} Milky Way extinction curve shown in a black solid line, the Large and Small Magellanic Cloud extinction curves from \citet{Gordon2003} in the green dash-dot line and red dash-dot-dot lines respectively, and the \citet{Calzetti2000} local starburst attenuation curve as the blue dashed line. The uvw2 and uvw1 filters can easily measure the UV slope of the observed attenuation curve, while uvm2 can constrain the UV bump strength.}
\label{fig:uvot_filt}\vspace{-2mm}
\end{figure}

In order to simultaneously study NUV and optical properties, we cross-referenced the MaNGA Product Launch 7 (MPL-7), corresponding to the SDSS Data Release 15, or DR15 \citep{Aguado2019}, with the \textit{Swift}/UVOT NUV archive as of April 26, 2018. UVOT is a 30-cm telescope that has a field of view (FOV) of $17^{\prime}\times17^{\prime}$ with an effective plate scale of $1^{\prime\prime}$~pixel\textsuperscript{$-1$}. The FWHM of the PSF varies across the NUV filters as reported in Table~\ref{table:sw_spec}, but is around $2.\!\!^{\prime\prime}5$, similar to the angular resolution of the MaNGA data. The NUV UVOT filter transmission curves are displayed over representative attenuation and extinction curves in Figure~\ref{fig:uvot_filt}. While both the uvw1 and uvw2 filters have small red leaks, this issue is unimportant here as we are targeting galaxies that exhibit strong star formation. The central wavelengths, bandpass FWHM, PSF FWHM, and the median and minimum exposure times for the observations of the galaxies in our sample are listed in Table~\ref{table:sw_spec}. These exposure time statistics apply only to the star forming galaxies studied in this paper. While there is a significant range in total exposure times, all of these galaxies are bright and well detected, with a minimum S/N of 15 in both filters for the integrated galaxy light and 5 for that of the individual regions.

The UVOT detector is a microchannel plate intensified CCD that operates in a photon counting mode. As a result, bright sources can suffer from coincidence loss \citep[as described in][]{Poole2008,Breeveld2010}. The objects in our sample are too faint for this to become a significant issue, as discussed in Section \ref{ssec:swft_dr}.

\begin{table}
\begin{minipage}{\linewidth}
\renewcommand{\thefootnote}{\textrm{\alph{footnote}}}
	\centering
	\setlength{\tabcolsep}{4pt}
	\caption{\textit{Swift}/UVOT NUV Observation Properties\label{table:sw_spec}}
	\begin{tabular}{lccccc} % four columns, alignment for each
		\hline
		\hline
		{} & {Central} & {} & {PSF} & {Median} & {Minimum}\vspace{-1mm}\\
		{ } &{Wavelength\footnotemark[1]$^,$\footnotemark[2]} & {FWHM\footnotemark[1]} & {FWHM\footnotemark[1]} & {Exposure\footnotemark[3]} & {Exposure\footnotemark[3]}\\
		{Filter} & {(\AA)} & {(\AA)} & {(arcsec)} & {(s)} & {(s)}\\
		\hline
		{uvw2} & {1928} & {657} & {$2.92$} &{2375} & {277}\\
		{uvm2\footnotemark[4]} &  {2246} & {498} & {$2.45$} &{2548} & {166}\\
		{uvw1} & {2600} & {693} & {$2.37$} &{2296} & {279}\\
		\hline \vspace{-6mm}
	\end{tabular}
\footnotetext[1]{The quoted filter properties are from \citet[][]{Breeveld2010}}
\footnotetext[2]{The central wavelength assumes a flat spectrum in $f_{\nu}$.}\vspace{-2mm}
\footnotetext[3]{The exposure time statistics are based on the subset of ``star forming'' galaxies defined by the criteria in Section~\ref{sec:stars}.}\vspace{-2mm}
\footnotetext[4]{The uvm2 exposure time statistics are calculated for the 25 star forming galaxies that have uvm2 images.}
\end{minipage}
\end{table}

Since this project aims to study the NUV slope, we required that all objects in our sample have UVOT images in both the uvw1 and uvw2 filters. We place a secondary constraint on the X-ray emission, $L_{X}(2$--10) keV $\lesssim10^{41}$~erg~s\textsuperscript{$-1$}, to exclude luminous active galactic nuclei (AGN), using the Swift X-ray Telescope (XRT) point source catalog \citep{Evans_2013}. These requirements produce a sample of 139 galaxies. The median redshift of these objects is 0.03, which corresponds to a luminosity distance of $D_{L}=130$~Mpc and a spatial scale of 600 pc$/^{\prime\prime}$; this redshift is similar to the average for MaNGA galaxies ($\langle z\rangle\approx0.037$). The stellar masses of the galaxies in the sample, obtained via spectral energy distribution (SED) fits from the NASA-Sloan Atlas \citep[][]{nsatlas}, span the range $8.73\leq\log(M/M_\odot)\leq11.11$. A full description of the total sample, which does not include the secondary constraint on the X-ray emission, is presented in Molina et al. (in preparation). From the sample of 139 galaxies, we select 29 galaxies with SFRs appropriate for our study. These criteria are described in detail in Section~\ref{subsec:int_sf}.

%Observations and Basic Data Reduction
%%%%%%%%%%%%%%%%%%%%%%%%%%%%%%%%%%%%%%%%%%%%%%%%%%%%%%%%%%%%%%%%%%%%%%%%%%%%%%%%%
\section{Data Reduction}
\label{sec:data}
\subsection{MaNGA Data Reduction}
\label{ssec:basic_dr}
The MaNGA spectra are reduced using the MaNGA Data Reduction Pipeline \citep[DRP; ][]{Law2016} which carries out the spectral extraction and calibration to produce datacubes. The datacubes are then processed through the MaNGA Data Analysis Pipeline \citep[DAP;][]{Westfall2019, Belfiore2019}, which creates 2--D maps of measured emission line strengths and spectral indices, as well as the best-fitting model spectra for all pixels that were successfully fit. We use the reduction version from MPL-6, which is nearly identical to SDSS DR15. DAP fits the stellar continuum using a list of 42 average stellar spectral templates constructed from the MILES library \citep[][]{Sanchez2006} using a hierarchical clustering algorithm \citep[see][for details]{Westfall2019}. The fitting employs the Penalized Pixel-Fitting code \citep[][]{Cappellari2004}. For this work, we employ the ``hybrid'' binning scheme from the DAP, which is optimal for emission line measurements. This scheme involves a two-step fitting process. First the stellar continua of the Voronoi-binned spectra \citep{Cappellari2003} are fitted, with the resulting fit saved as a template. Then the emission lines in the individual pixels are fitted simultaneously with the stellar continuum, using the fixed template from the previous step. This process is carried out for each pixel to create 2--D maps of emission line flux across the face of the galaxy. Spectral indices such as the Balmer break, or D$_n$(4000), are measured in the Voronoi-binned spectra from the first step. As this project relies on spatial resolution, we recalculate the D$_n$(4000) maps without Voronoi binning, using the continuum windows of $\lambda_{\textrm{blue}}=3850$--$3950$\,\AA\ and $\lambda_{\textrm{red}}=4000$--4100\,\AA. The reported errors include the correction factors derived from repeat observations as described in \cite{Westfall2019,Belfiore2019}: the flux errors are multiplied by 1.25, and the D$_n$(4000) errors by 1.9. These correction factors reliably account for the random and systematic errors associated with varying conditions during the observations and the methods used to make the measurements. The stellar mass and SFR within 1 effective radius (1~R$_\textrm{e}$) quoted in Table \ref{table:tgt_prop} are drawn from the NASA-Sloan Atlas \citep{nsatlas} and MaNGA DAP, respectively.

\subsection{Swift/UVOT Data Processing}
\label{ssec:swft_dr}
The UVOT data are obtained from the High Energy Astrophysics Science Archive Research Center (HEASARC), and are processed using the \textit{Swift} UVOT Pipeline\footnote{\url{github.com/malmolina/Swift-UVOT-Pipeline}}, an updated and automated version of \texttt{uvot\_deep.py}, which is a subroutine in UVOT Mosaic\footnote{\url{github.com/lea-hagen/uvot-mosaic}}. These routines complete the basic data processing as described in the UVOT Software Guide\footnote{\url{heasarc.gsfc.nasa.gov/docs/swift/analysis}}. Both the counts and exposure maps are aspect corrected, which reduces any uncertainty in the defined world coordinate system to $\sim0.\!\!^{\prime\prime}5$. In order to ensure reliable final images, we require that all individual frames are $2\times2$ binned, which yields a plate scale of $1^{\prime\prime}$ pixel\textsuperscript{$-1$}. UVOT images are composed of single frames, with very short exposure times, which are stacked together to create an image with the desired total exposure time. While different UVOT frame exposure times are possible, the UVOT software does not allow for the combination of different frame times as it complicates the image analysis. Therefore, we use the standard full frame time of 11.0322 ms.

All individual events in each UVOT frame are identified and the centroid of the event location is saved. The location of each event is placed into the final UVOT image, with each event stored as a count. A cosmic ray hitting the detector will register at most a few counts at a single location on the UVOT image rather than the thousands of counts expected from a relatively bright stationary source. Therefore, even though cosmic rays are non-uniform, they affect very few frames, and thus are incorporated into the background sky counts. 

The resulting UVOT images were corrected for the dead time and the degradation of the detector. Approximately 2\% of the full frame time is dedicated to transferring the charge out of the detector, which must be corrected by increasing the count rate \citep[][]{Poole2008}. The decline in count rate measured by the UVOT NUV filters due to the degradation of the detector is well characterized and described in the UVOT calibration documents\footnote{\url{heasarc.gsfc.nasa.gov/docs/heasarc/caldb/swift/docs/uvot}}, resulting in a 2.5\% correction for the most recent observations. 

As UVOT operates in photon-counting mode, coincidence loss can be a significant problem for bright sources. Coincidence loss occurs when two or more photons arrive at a similar location within the same frame, and their pile-up can affect the measured count rate. While \citet[][]{Breeveld2010} describe coincidence loss for the case when a point source is superposed on some diffuse background (e.g., knots of star formation on top of diffuse galactic background), the spatial resolution of our data ($\sim2$~kpc) does not allow us to resolve individual \ion{H}{2} regions. Therefore, that method is not appropriate; instead, we approached this problem using the prescription in \citet{Poole2008}. Their formulated coincidence loss corrections are only valid for point sources, but the effect is only significant when the count rate is above 10~counts~s\textsuperscript{$-1$}~pixel$^{-1}$. The UVOT observations of our sample have a maximum count rate of $< 0.9$ counts s\textsuperscript{$-1$} pixel\textsuperscript{$-1$}, translating to a correction of $\sim0.6$\%, significantly smaller than the dead time correction of 2\%. Therefore the effects of coincidence loss are negligible and we ignore them in this work. A full description of the \textit{Swift}/UVOT data reduction and the \textit{Swift} UVOT Pipeline is given in Molina et al. (in preparation).

\subsection{\textit{Swift}/UVOT Sky Subtraction}
\label{ssec:swft_sky}
The final reduction step for the UVOT images was to subtract the local background sky emission. In order to identify the local background, we quantified the extent of each galaxy by calculating its Petrosian radius (\citeyear{Petrosian1976}) in the uvw2 filter, i.e., where the local intensity equaled 20\% of the average flux contained within that aperture. The average Petrosian radius for our sample was 11.\!\!$^{\prime\prime}$9, much smaller than the $17^\prime\times17^\prime$ UVOT FOV. The sky annulus was defined with an inner radius of twice the calculated Petrosian radius and a set area of 1000 square arcseconds. We used the biweight estimator of the central location to estimate the sky counts. This estimation method was used to avoid the effect of any objects that fall within the sky annulus. The resulting sky counts were minimal as expected, and yield integrated magnitudes consistent with those from the \textit{Galaxy Evolution Explorer} \citep[GALEX; ][see Section~\ref{ssec:bcalc} for details]{galex}. We then calculated the background sky emission in the uvw1 filter via the same procedure; as there was at most a 1--2 pixel difference between the Petrosian radii calculated for uvw1 and uvw2, we simply adopted the uvw2 radius for this measurement.

\subsection{Spatial Sampling and Resolution Matching of SDSS-IV/MaNGA Maps and \textit{Swift} Images}
\label{ssec:matching}
While the NUV images from UVOT and the optical spectra from MaNGA have approximately the same angular resolution, the sampling is not the same. In order to directly compare the two data sets, the MaNGA 2--D maps must be re-sampled to the UVOT pixel size ($1^{\prime\prime}$~pixel\textsuperscript{$-1$}, double that of MaNGA), and angular resolution. To do this, we adopt the resolution of the UVOT uvw2, which has the poorest spatial resolution (see Table~\ref{table:sw_spec}). We first re-sample the 2--D MaNGA maps with a $2\times2$ top hat convolution kernel to match the UVOT pixel scale of 1$^{\prime\prime}$. After re-sampling the maps, we change the angular resolution of the MaNGA data cube to match that of uvw2 by employing a Gaussian convolution kernel of standard deviation $\sigma_M$ such that $\sigma_{MaNGA}^2 + \sigma_M^2 = \sigma_{Swift}^2$. This process was completed for each 2--D MaNGA map and associated error image of interest. A full description of this process is presented in Molina et al. (in preparation).

%identifying star-forming galaxies (integrated)
%%%%%%%%%%%%%%%%%%%%%%%%%%%%%%%%%%%%%%%%%%%%%%%%%%%%%%%%%%%%%%%%%%%%%%%%%%%%%%%%%
\section{Properties of Star Forming Galaxies}
\label{sec:stars}
\subsection{Identifying Star Forming Galaxies}
\label{subsec:int_sf}
While most of our galaxies have single-fiber SDSS spectra, about 45\% of the objects in our sample do not have detailed classifications beyond ``galaxy'' in the SDSS DR15 release \citep[see][for a description of the classification routine]{Bolton2012}. Additionally, as this study focuses on understanding attenuation in galaxies, we wish to identify star forming galaxies in a reddening independent manner. To this end, we used the MaNGA integrated emission line flux maps to create masks based on BPT diagrams to determine the locations of star formation. We used the starburst models described in \citet[][]{Kewley2006}, which include the theoretical extreme starburst line described in \citet[][]{Kewley01}, as well as the empirical line defining composite objects described in \citet[][]{Kauffmann03}. We then used the MaNGA data to create maps of the reddening-insensitive diagnostic ratios $[\textrm{\ion{O}{3}}]/\textrm{H}\beta$, $[\textrm{\ion{N}{2}}]/\textrm{H}\alpha$, $[\textrm{\ion{S}{2}}]/\textrm{H}\alpha$ and $[\textrm{\ion{O}{1}}]/\textrm{H}\alpha$. These emission line ratio maps were used to create masks that identified pixels that were classified as star forming in all three of the $[\textrm{\ion{O}{3}}]/\textrm{H}\beta$ vs.~$[\textrm{\ion{N}{2}}]/\textrm{H}\alpha$, $[\textrm{\ion{S}{2}}]/\textrm{H}\alpha$ and $[\textrm{\ion{O}{1}}]/\textrm{H}\alpha$ diagrams. For the $[\textrm{\ion{O}{3}}]/\textrm{H}\beta$ vs.~$[\textrm{\ion{N}{2}}]/\textrm{H}\alpha$ diagram, we additionally required that the measured emission line ratios were not in the composite object region defined by \citet[][]{Kauffmann03}.

\begin{figure}
\includegraphics[width=0.45\textwidth]{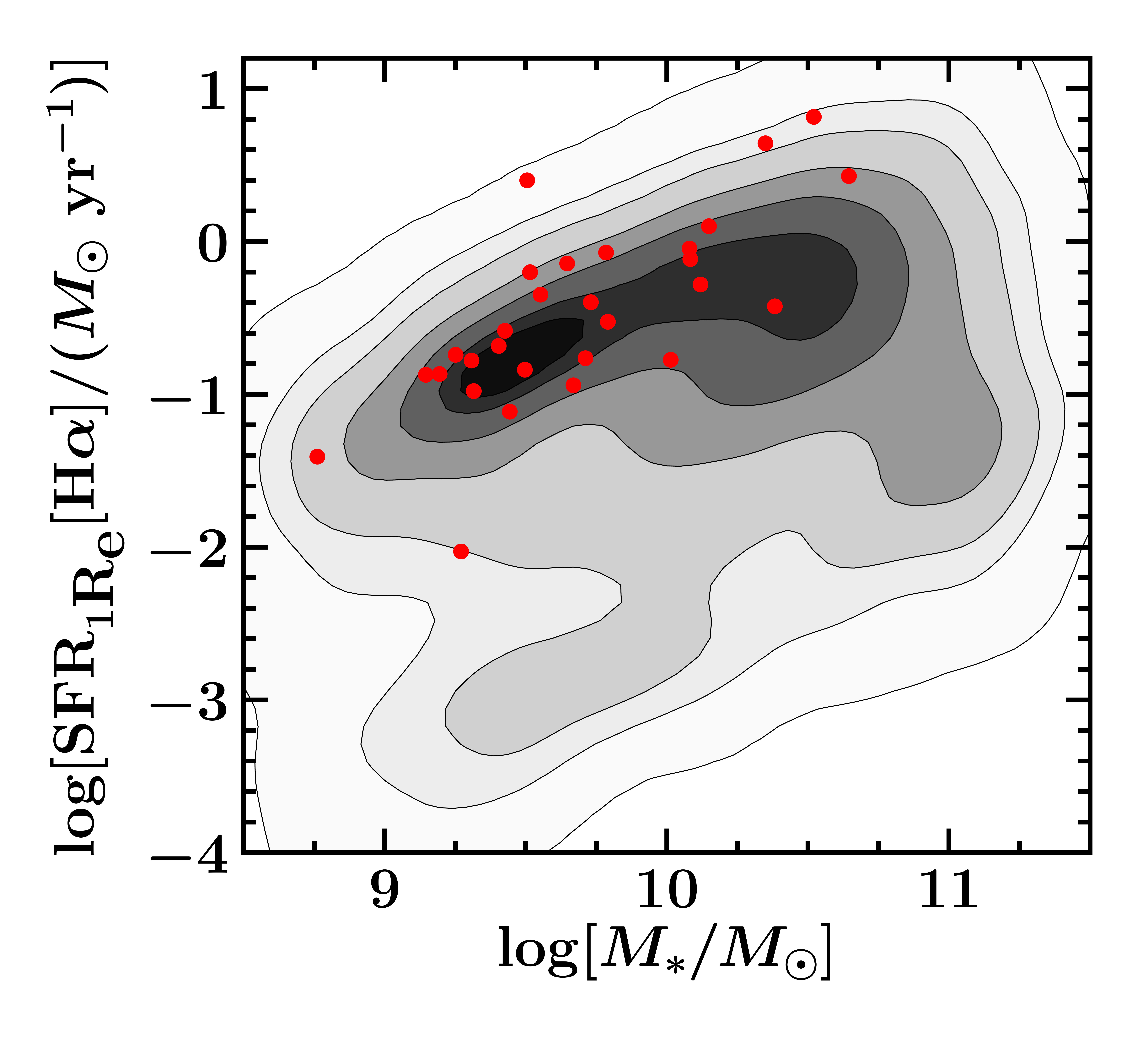}\vspace{-4mm}
\caption{The SFR within 1~R$_e$ vs.~stellar mass for the 29 star forming galaxies in the sample (red filled squares), overlaid on the full MaNGA sample (black and grey contours). The contours represent fractional values of the maximum number density, with lines of constant number density drawn at the 90\%, 80\%, 60\%, 40\%, 20\%, 10\% and 1\% levels. The galaxies studied in this work primarily fall on or above the star forming cloud. The stellar mass measurements are from the NASA-Sloan Atlas \citep{nsatlas}, and the SFR is calculated within 1 effective radius using the H$\alpha$ map from the MPL-6 version of the MaNGA DAP \citep{Westfall2019}, after applying an internal reddening correction using the \citet{Odonnell1994} Milky Way extinction curve.\vspace{-2mm}}
\label{fig:int_ms}
\end{figure}

The masks were used to identify galaxies where a large majority of the pixels, especially in the nuclear and circumnuclear regions, are star forming according to all three BPT diagrams. We identified 29 galaxies that satisfied these criteria and verified that none of these objects have an SDSS classification of AGN. We then confirmed that the 29 objects selected were indeed star forming galaxies by plotting their locations in the SFR--stellar mass plane against the entirety of the MaNGA sample as shown in Figure~\ref{fig:int_ms}. Almost all of the galaxies in our sample fall on or above the cloud of star forming galaxies. Their SFRs within $1~R_e$, SFR$_{1Re}$(H$\alpha$)$\lesssim$ 10~M$_{\odot}$ yr$^{-1}$, are well under the maximal value of those studied by B16 (SFR[H$\alpha$]$\lesssim40$~M$_{\odot}$ yr$^{-1}$), making their star forming galaxy attenuation law more appropriate than the \citet{Calzetti2000} starburst law. We therefore conclude that these galaxies are star forming, not subject to AGN contamination, and are appropriate for the proposed analysis. This final sample of galaxies has similar properties to that of B16: their stellar masses are in the range $8.76\leq~\!\log(M_{*}/M_\odot)~\!\leq10.7$ with a median of 9.65, and their SFRs are in the range $-1.52\leq~\!\log[\textrm{SFR}_{1R_e}\textrm{[H}\alpha\textrm{]}/(M_\odot~\textrm{yr}^{-1})]\leq~\!0.94$ with a median of $-0.45$. However, the sample is dominated by objects with lower stellar masses and SFRs; 69\% of the galaxies have $\log(M_{*}/M_\odot)\leq10.0$, while 76\% have SFR$_{1 R_e}$(H$\alpha$)$\leq1$ M$_\odot$ yr$^{-1}$. The stellar mass, SFRs, and other basic properties for the star forming galaxies in our sample are listed in Table \ref{table:tgt_prop}.

\subsection{Attenuation in Integrated Light of Star forming and Starburst Galaxies}
\label{ssec:bcalc}
To employ the B16 approach to study attenuation, we must assume that the UV stellar continuum is well described by a power law of the form $F_{\lambda}(\lambda)\propto\lambda^{\beta}$\null. Here $F_{\lambda}(\lambda)$ is the observed flux density per unit wavelength, and the power-law index (or slope in log space) is $\beta$. We are interested in vigorously star forming galaxies as we expect their integrated light will generally follow the relation between optical nebular and UV stellar continuum attenuation described by either \citet[][]{Calzetti1994} or B16. We confirmed this behavior by measuring the $\tau^l_B$ and $\beta$ of the integrated NUV stellar continua and optical nebular light for all galaxies in the sample, as described in Sections~\ref{ssec:tcalc} and \ref{ssec:bcalc}. For consistency with previous analysis, we then compare the values of $\beta$ calculated using the \textit{Swift} NUV filters to those derived from the GALEX FUV and NUV filters, as described in Section~\ref{ssec:bgal}. In cases where the galaxy was face-on, the circular uvw2 Petrosian radius (\citeyear[][]{Petrosian1976}), described in Section~\ref{ssec:swft_sky}, is adopted. If the galaxy was edge-on, or very elongated, an elliptical aperture that traced the majority of the measured H$\alpha$ flux was used. We compared the measured apparent magnitude of the integrated light in the uvm2 filter, when available, to the apparent NUV magnitude from GALEX measurements. All the uvm2 apparent magnitudes agreed with those from the GALEX NUV filters within the error, i.e., $|m_{\textrm{uvm2}} - m_{\textrm{GALEX}}|\lesssim 0.1$, and $\langle|m_{\textrm{uvm2}} - m_{\textrm{GALEX}}|\rangle=0.03$. The typical errors for the apparent uvm2 and GALEX magnitudes are 0.06 and 0.08, respectively.

\subsubsection{Calculating the Integrated $\tau^{l}_{B}$}
\label{ssec:tcalc}
\citet[][]{Calzetti1994} and B16 follow the same general procedure to develop their attenuation laws: they measure the relation between the UV spectral slope and Balmer optical depth for the integrated light of galaxies and plot them against each other to quantify the difference in colour excess between the UV stellar and optical nebular emission. The calculation of $\tau^{l}_{B}$ is carried out via
\begin{equation}
\tau^{l}_{B}=\tau_{\beta}-\tau_{\alpha}=\ln\!\bigg(\frac{f(\textrm{H}\alpha)/f(\textrm{H}\beta)}{2.86}\bigg),
\end{equation}
where 2.86 is the theoretical value for Case B recombination in a low density, $T=10^{4}$~K plasma \citep[][Section 4]{osterbrock2006}. $\tau^l_B$ is defined to be independent of the assumed extinction law. When $\tau^l_B = 0.0$ the measured Balmer decrement is equal to the intrinsic value and there is assumed to be no dust in the system. Therefore, if the assumed star formation history (SFH) is correct and the sightline coverage of both the nebular emission and UV stellar continuum by the attenuating dust are identical, the value of $\beta$ at $\tau^l_B = 0.0$ is also the intrinsic UV spectral slope of the stellar continuum. Any offset in $\beta$ at $\tau^l_B = 0.0$ in the B16 relation would then be a result of the SFH and/or the geometry of the attenuating dust, the stars and the gas \citep[see B16; ][for discussions on this topic]{Calzetti2013,Calzetti2015}.

We calculated the integrated $\tau^{l}_{B}$ by using the observed H$\alpha$ and H$\beta$ MaNGA flux maps. The integrated emission was defined as all valid pixels that fall within both the MaNGA IFU footprint and the apertures used to compare uvm2 measurements to GALEX measurements, regardless of whether the emission was dominated by star formation according to BPT diagrams. We define valid pixels as all pixels that are not flagged as untrustworthy by the MANGA DAP \citep[see ][for details]{Westfall2019}. The same apertures and masks were used for the integrated UV spectral slope ($\beta$) measurement, for consistency. The larger aperture will introduce scatter between the integrated light and individual regions, which will be explored in Section~\ref{sec:all}. All maps from MaNGA are corrected for foreground extinction by the MaNGA DAP, using the values from \citet{Schlegel1998} listed in Table \ref{table:tgt_prop}, and assuming the \citet[][]{Odonnell1994} Milky Way dust extinction curve with $\textrm{R}_{V}=3.1$. 

\subsubsection{UVOT Foreground Extinction Correction}
The NUV foreground extinction corrections were calculated using a multistep process. We began by combining the \citet{fitzpatrick1999} Milky Way extinction curve assuming $\textrm{R}_V=3.1$ and a range of $E(B-V)$ values, with each of the 13 solar metallicity, continuous star forming SEDs from Starburst99 \citep{s99}. The resulting SEDs were multiplied by the appropriate filter curve and integrated over all wavelengths. We calculated an average total-to-selective extinction, $k(\lambda)\equiv A_{\lambda}/E(B-V)$, for the two filters: $k_{\textrm{uvw1}}=6.269$ and $k_{\textrm{uvw2}}=8.069$. The stellar age of the SED did not significantly increase the scatter in the $k(\lambda)$ calculation; the choice in extinction law \citep[i.e.,][]{Cardelli1989,fitzpatrick1999,Odonnell1994} was the dominant source of uncertainty. We employed the Fitzpatrick extinction law for two reasons: (1) this extinction law is the same used in B16, and (2) this law has a thorough, careful treatment of the UV light. The difference in the total-to-selective extinction between the different extinction laws is $k_{\textrm{uvw1,Fitzpatrick}}= 0.9k_{\textrm{uvw1,O'Donnell}}$ and $k_{\textrm{uvw2, Fitzpatrick}} =0.98k_{\textrm{uvw2, O'Donnell}}$. When assuming the maximal $E(B-V)$ value for the objects in our sample, the foreground extinction-corrected colour $\textrm{uvw2}-\textrm{uvw1}$ when assuming the Fitzpatrick law is 0.03~mag larger than when assuming the O'Donnell law. We note this systematic offset but do not include it in our calculated errors, especially as it is negligible compared to the systematic uncertainty associated with the transformation of $\beta$ as described in Section~\ref{ssec:bgal}. Finally, as we are only interested in the power-law slope of the UV stellar continuum using the $\textrm{uvw2}-\textrm{uvw1}$ colour, there is no need to perform K-corrections on the flux density.

\subsubsection{Calculating the Integrated $\beta_{\rm Swift}$}
\label{ssec:bcalc}
Calculating $\beta$ is a more complex, object-specific process than that for $\tau^l_B$, and is complicated by the fact that we do not know the shape of the intrinsic spectrum. The B16 relation relied on the work first done by \citet[][]{Calzetti1994}, which used spectra covering a wavelength range of $1200$--$3200$~\AA. \citet[][]{Calzetti1994} used 10 continuum windows in the UV spectrum to calculate $\beta$, while B16 and this work use colours from GALEX and UVOT, respectively. Fortunately, our sample has very similar properties to that of B16 (see Table 2 in B16), so we calculated $\beta_{\rm Swift}$ in a similar manner, as described below. 

We started with the 100~Myr continuous star formation stellar continuum model generated by \texttt{Starburst99} \citep[][]{s99}, which assumes a constant SFR of 1~M$_\odot$~yr$^{-1}$. This model was chosen for consistency with B16, who note that the exact age of the reference spectrum is less important than the assumption of a continuous SFR. Thus we assume a nominal intrinsic, unreddened $\beta_0=-2.38$ \citep{Calzetti2001}. This choice will have consequences on our $\beta$ calculations, which will be discussed in detail in Section~\ref{ssec:sfh}.

The 100~Myr continuous star formation SED was redshifted according to the recessional velocity of each object. We then reddened the resulting model using the B16 attenuation law for a range of $E(B-V)$ values. For each reddened SED, we measured the UV spectral slope via the same continuum windows used by \citet[][]{Calzetti1994}, and also calculated the UVOT uvw2$-$uvw1 colour. We then used the measured $\beta_{\rm Swift}$ values as a function of UVOT colour to calculate the integrated galaxy $\beta_{\rm Swift}$ measurements for the objects in our sample. Finally, we compared the $\beta_{\rm Swift}$ values to GALEX $\beta$ measurements using the process described in Section~\ref{ssec:bgal}. The integrated $\beta$ and $\tau^{l}_{B}$ for the sample are given in Table \ref{table:tgt_prop}, with the exception of NGC~3191, where the MaNGA observations do not cover a significant portion of the galaxy disc. The difference in the uvw1 and uvw2 Petrosian radii is unimportant for this study as we are limited by the MaNGA FOV and pixel masks, as described in Section~\ref{ssec:tcalc}.

\subsection{Conversion of $\beta$ from UVOT to GALEX}\label{ssec:bgal}
While the calculations described in Section 4.2.3 are consistent with the methodology described in B16, our baseline is restricted to the wavelength range between $\sim 2000$~\AA\ and $\sim 2600$~\AA, i.e., the NUV, while B16 used the GALEX NUV and FUV filters. Thus, there could be a systematic offset between $\beta_{\rm Swift}$ and $\beta$ values obtained using entire UV wavelength range \citep[see Appendix B in][for a complete discussion]{Calzetti2001}. In short the ``iron curtain'' at longer UV wavelengths can depress the continua in the NUV, giving UV slopes with shorter baselines more negative values. They found that the relationship between the two $\beta$ values in UV-selected starbursts is linear, with a slope of order unity. Meanwhile, \cite{Meurer1999} found a constant, negative offset between UV slopes calculated with longer and shorter wavelengths, such that the short-wavelength slope is bluer.

In order to provide a more faithful comparison to B16, we compare our $\beta_{\rm Swift}$ values with UV slopes calculated with GALEX.  To do this, we measured integrated, aperture-matched magnitudes from both the NASA-Sloan Atlas and our Swift archival data (full details described in Molina et al. in preparation) using the $r$--band elliptical aperture from the NASA-Sloan Atlas, corrected for inconsistencies in the PSFs of the instruments.

\begin{figure}
\includegraphics[width=0.45\textwidth]{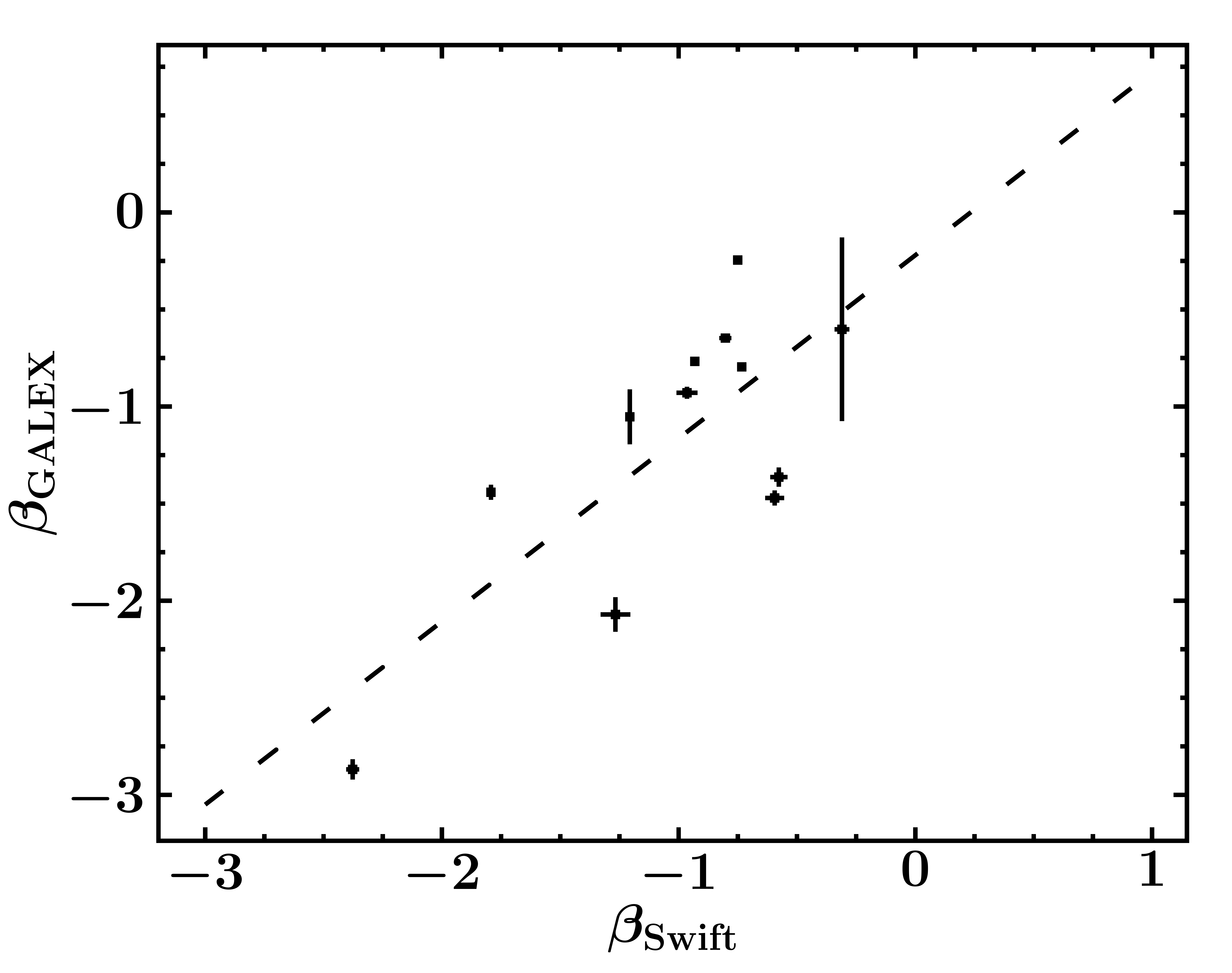}
\caption{Comparison of the UV spectral slope ($\beta$) as computed with GALEX and \textit{Swift} data. We only include objects with uncertainties in both GALEX and \textit{Swift} filters less than 0.5~mag, an axial ratio $b/a > 0.42$, and ${\rm D}_n(4000)\lesssim1.4$. The resulting fit is given by Equation~\ref{eq1}, but the uncertainty in this relation is driven by the dispersion, $\sigma_{\rm disp}=0.2$~mag. To account for this, we adopt an additional uncertainty of $\sigma=0.2$ in our calculation of $\beta$.\vspace{-4mm}}
\label{fig:gal_swft}
\end{figure}

$\beta_{\rm GALEX}$ is calculated via the method used to calculate $\beta_{\rm IUE}$ described in B16, while the $\beta_{\rm Swift}$ is calculated as described in Section~\ref{ssec:bcalc}. We only include objects that: (1) have GALEX detections in both bands with $\sigma_{\rm m, GALEX} < 0.5$~mag, (2) have an axial ratio $b/a > 0.42$ \citep[this will minimize contamination by the 2175\,\AA\ bump as described in][]{Battisti2017}, and (3) have ${\rm D}_n(4000)\lesssim1.4$ to create a sample similar to that in B16. The resulting relation between $\beta_{\rm Swift}$ and $\beta_{\rm GALEX}$ is shown in Figure~\ref{fig:gal_swft}, and is well-fit with a linear model of the form
\begin{equation}\label{eq1}
\beta_{\rm GALEX} = (0.9\pm0.3)\beta_{\rm Swift} - (0.2\pm0.3),
\end{equation}
with a scatter about the relation of $\sigma_{\rm disp}=0.2$. Therefore, while we find a similar result to that of \cite{Calzetti2001} and \cite{Meurer1999}, the large dispersion drives the overall observed uncertainty in this relation. Furthermore, we have no robust way to differentiate between the systematic and random components associated with that dispersion. As the two $\beta$ values are statistically equivalent given the errors on the fitted parameters, we do not transform $\beta_{\rm Swift}$. However, to account for this observed dispersion between the GALEX and Swift-calculated $\beta$ values, we adopt an additional uncertainty of $\sigma = 0.2$ into our error bar. This transformation is the largest source of uncertainty in the $\beta$ error budget.

\subsection{Continuous vs.~instantaneous star formation}\label{ssec:sfh}
Our calculation of $\beta$ assumes a continuous SFH instead of an instantaneous burst; this is significant, as $\beta$ is highly dependent on the intrinsic slope and thus on the assumed SFH. To quantify the difference between the two families of models, we measured $\beta$ in the same manner described in Section~\ref{ssec:bcalc} and compared the UVOT colour and intrinsic $\beta$ for instantaneous burst and continuous star formation stellar continua generated by \texttt{Starburst99} \citep{s99}. The two families of models agree for $\beta\lesssim-1.8$, which translates to an age of $\sim50$--$100$~Myr \citep{Calzetti2001}. 

However for more realistic SFHs, rapid changes in the SFH could easily affect the measured $\beta$. For example, a SFH with an exponential decay and a short e-folding time would produce a sharp increase in $\beta$ shortly after the initial burst. Thus, unless we caught the burst very close to its onset, the calculated $\beta$ values under this assumption would be systematically too blue for such a system.

We have tried to minimize this effect by selecting galaxies that exhibit strong H$\alpha$ emission ($\langle\Sigma_{{\rm H}\alpha}\rangle\gtrsim10^{38}$~erg~s$^{-1}$~kpc$^{-2}$), which implies that they have undergone vigorous star formation within the last 10~Myr \citep{Kennicutt2012}. In addition, we confirm that most of the galaxy has nebular emission consistent with star formation; this avoids galaxies undergoing quenching. Finally, while our presented $\beta$ values for the individual galaxies and kpc-sized regions may be systematically offset from their true value, we emphasize that the galaxies have similar mass, and both the galaxies and regions identified are currently forming stars at a similar rate to those presented in B16. Therefore, as with B16, \textit{the distribution} of the sample about the assumed intrinsic $\beta$ should be reliable.

\begin{figure}
\includegraphics[width=0.45\textwidth]{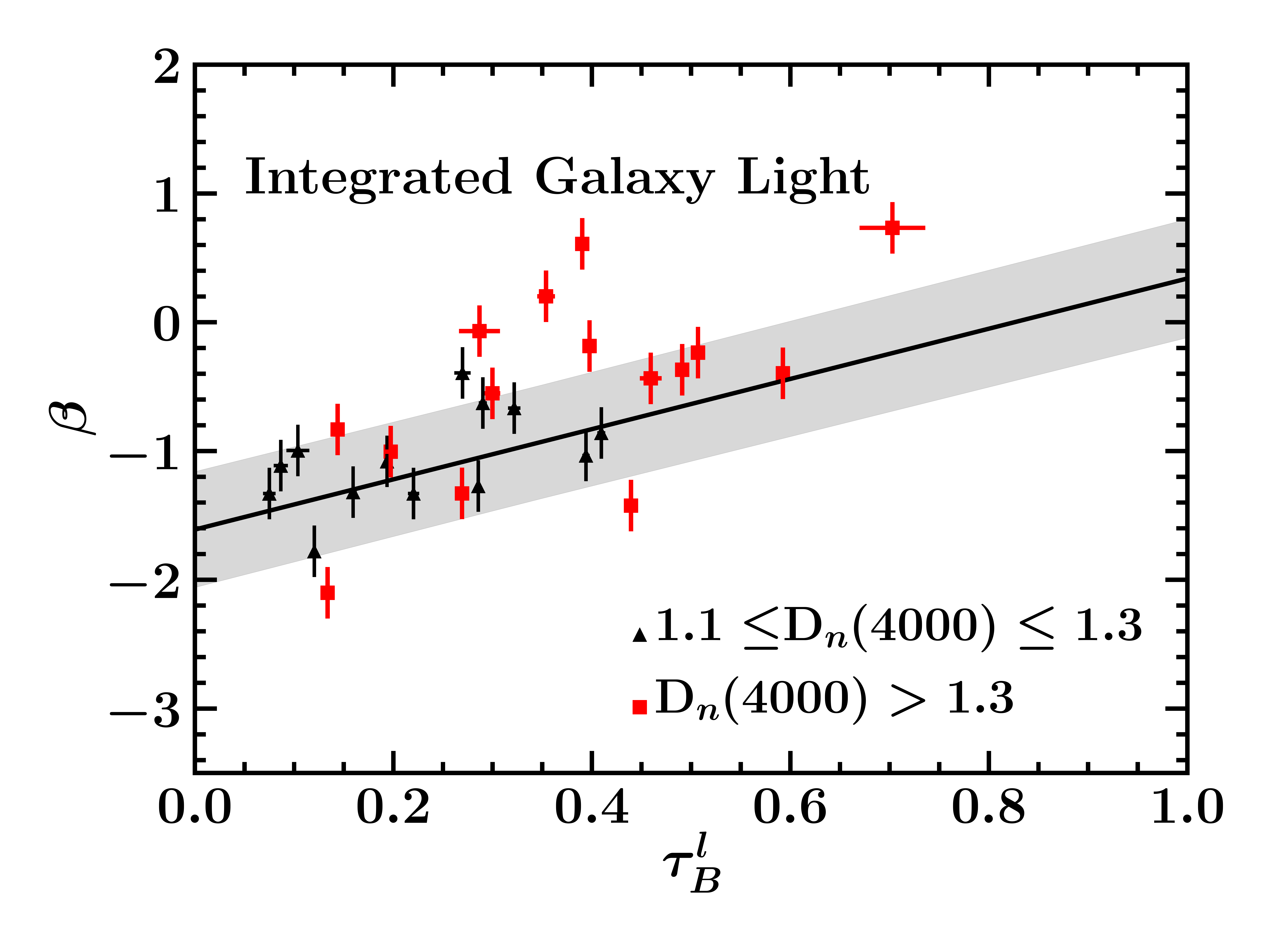}\vspace{-4mm}
\caption{The UV spectral slope ($\beta$) vs.~the Balmer optical depth ($\tau^{l}_{B}$) of the integrated galaxy light from our 29 galaxies, compared to the relation reported in B16 (solid black line). The grey shaded region represents the sample dispersion of the B16 relation, which is larger than the error in the fitted model. B16 only considered galaxies with $1.1\leq\textrm{D}_n(4000)\leq1.3$ when deriving the relation between $\beta$ and $\tau^{l}_{B}$. The symbols represent galaxies with different D$_n(4000)$ values: the black triangles denote galaxies with $1.1\leq\textrm{D}_n(4000)\leq1.3$, and the red squares represent redder systems with D$_n(4000)>1.3$. Galaxies with larger D$_n(4000)$ lie preferentially above the fitted relation, consistent with the trends seen in B16.}\vspace{-10mm}
\label{fig:int_btau}
\end{figure}

\subsection{Effects of Stellar Age, the 2175\,\AA\ bump, and S/N on $\beta$}
\label{ssec:bump}
Figure~\ref{fig:int_btau} compares the relation from B16 to the integrated $\beta$ and $\tau^l_B$ from the 29 galaxies in our sample. The gray shaded bar represents the sample dispersion from B16, which is larger than the error associated with the fit. Both B16 and this work probe stellar age using as a proxy the D$_n(4000)$ index, which samples the 4000\,\AA\ break \citep{Balogh1999}. This feature is produced by the absorption of light by metals in the atmospheres of cool stars, and is thus stronger in older stellar populations; younger systems have stronger UV fluxes, which dilute this break. For our purposes, we consider D$_n(4000)$ as an indication of the relative contribution of older stars compared to younger stars. The D$_n$(4000) values presented in this work are measured using the MaNGA maps, and represent the integrated value within the aperture. 

B16 focused on the unresolved nuclear regions of galaxies, with scales similar to the individual regions we study in detail starting in Section~\ref{sec:ddresult}. We find qualitatively that the integrated light in star forming galaxies follows the same relation as their nuclear regions. We will explore the change in attenuation across the galaxy and the effect of aperture size in Section~\ref{sec:all}. 

The integrated light for all of the galaxies in our sample have D$_n$(4000) measurements smaller than the upper limit in the full B16 sample. However, B16 based their fitted relation on galaxies with $1.1\leq\textrm{D}_n(4000)\leq1.3$. They do note that the values of $\beta$ and $\tau^{l}_{B}$ tend to increase with D$_n(4000)$, which is consistent with our findings. As clearly demonstrated, the integrated light from our sample, on average, has a higher D$_n$(4000) value than the galaxies used to define the fitted B16 relation. However, the UV light from the individual star forming regions should be dominated by massive, young stars; therefore, we will consider the entirety of the sample.

For this project, we assume that our galaxies are similar to those studied by B16, which themselves were further explored by \cite{Battisti2017}. They found that galaxies with $b/a < 0.42$ show evidence of a 2175\,\AA\ bump in their attenuation curves. We only have 3 such objects in our sample, and do not see any systematic trends with inclination in the integrated light. In addition to these physical effects, we explored the effect of the large range in the signal-to-noise (S/N) ratio on our measurement of $\beta$. We do not see any strong systematic trends with the measured S/N in either filter with respect to the measured $\beta$ in the integrated aperture.\vspace{-4mm}

%identifying star-forming REGIONS
%%%%%%%%%%%%%%%%%%%%%%%%%%%%%%%%%%%%%%%%%%%%%%%%%%%%%%%%%%%%%%%%%%%%%%%%%%%%%%%%%
\section{kiloparsec-scale Star Forming Regions}
\label{sec:ddresult}
\subsection{Isolating Individual Regions}\label{ssec:reg_calc}
The combination of the kpc-scale spatial resolution of our data and the diffuse galaxy light severely restricts the effectiveness of most commonly used methods for the identification of star forming regions \citep[e.g., Source Extractor;][]{sextractor}. Instead, we used the dereddened H$\alpha$ surface brightness maps ($\Sigma_{\textrm{H}\alpha}$) in combination with BPT diagrams and the H$\alpha$ equivalent width, EW(H$\alpha$), to define the kpc-sized regions of star formation. While $\Sigma_{\textrm{H}\alpha}$ measurements do require a correction for internal attenuation, all attenuation curves yield approximately the same correction at H$\alpha$, making it the most robust tracer of recent star formation. To create our sample of star forming regions, we began by selecting an initial minimum H$\alpha$ surface brightness of $\log[\Sigma_{\textrm{H}\alpha}$/(erg~s$^{-1}$~kpc$^{-2})]=39$; this threshold prevented excessive contamination from diffuse ionized gas (DIG) in the defined region. Such contamination is ubiquitous on MaNGA spatial scales as described in \citet{mangadig}, but they recommend a limit of  $\log[\Sigma_{\textrm{H}\alpha}$/(erg~s$^{-1}$~kpc$^{-2})]=39$ to identify \textit{pixels} dominated by light from the \ion{H}{2} regions; in contrast pixels with $\log[\Sigma_{\textrm{H}\alpha}$/(erg~s$^{-1}$~kpc$^{-2})]<38$ are dominated by DIG. This means that all of the regions we identify in the first pass meet this strong constraint. 

While subtraction of the diffuse galaxy component would allow us to exclude light from older stellar populations, the observed spatial scale does not allow us to cleanly identify individual \ion{H}{2} regions. Therefore, any ``diffuse component'' we measure could have low levels of star formation, which could impact the measured $\beta$. We thus decided not to subtract out the diffuse component from our measurements. To mitigate the presence of a large diffuse component, we also required that the mass-specific SFR proxy EW(H$\alpha) > 15$\AA. This step is done after defining our individual regions to identify those that are ``DIG-dominated''. Finally, to avoid any potential uncertainties due to the difference in the PSF between uvw1 ($2.\!\!^{\prime\prime}37$) and uvw2 ($2.\!\!^{\prime\prime}92$), we adopt a minimum aperture diameter of $5^{\prime\prime}$.

Using a combination of BPT and $\Sigma_{\textrm{H}\alpha}$ diagnostics, we iteratively identified the kpc-size star forming regions using the following four step process:

\begin{enumerate}
\item Identify local peaks where the observed, reddening-corrected H$\alpha$ surface brightness is $\log[\Sigma_{\textrm{H}\alpha}$/(erg~s$^{-1}$~kpc$^{-2})]\geq39$, and confirm they are separated by at least $5^{\prime\prime}$ (corresponding to the diameter of two resolution elements). Both foreground and internal reddening corrections assume the \citet[][]{Odonnell1994} Milky Way dust extinction curve, and $\textrm{R}_{V}=3.1$. The choice of attenuation law did not affect the value of $\Sigma_{\textrm{H}\alpha}$ within our measurement errors.
\item Define the largest isophotal contour that corresponds to $\log[\Sigma_{\textrm{H}\alpha}$/(erg~s$^{-1}$~kpc$^{-2})]\geq39$ and separates the local $\Sigma_{\textrm{H}\alpha}$ peaks. If the contour has a diameter less than $5^{\prime\prime}$, adopt a circular aperture of that diameter. The contour or assumed circular aperture is then defined as a star forming region. 

\clearpage
\thispagestyle{plain}
\begin{landscape}
\begin{table}
\begin{minipage}{\linewidth}
\hspace*{-6mm}
\renewcommand{\thefootnote}{\textrm{\alph{footnote}}}
	\setlength{\tabcolsep}{1.75pt}
	\caption{Basic Properties of Star Forming Galaxies and UVOT images}\label{table:tgt_prop}
	\begin{tabular}{cllcccccccccccccccc} 
		\hline
		\hline
{Object} & {Object} & {SDSS} & {} & {}& {} & {$R_{pet}$\footnotemark[4]}& {$\log(M_{*}$)\footnotemark[5]} & {$\log$(SFR)\footnotemark[6]} & {$t_{uvw2}$} & {$t_{uvw1}$} &{$f_{lim}$\footnotemark[7]}& {$f_{lim}$\footnotemark[7]} & {} & {} &{uvw2\footnotemark[9]} & {uvw1\footnotemark[9]} &{} & {}\\
{I.D.} &{Name} & {Class.\footnotemark[1]} & {$z$\footnotemark[2]} & {E(B$-$V$)_{G}$\footnotemark[3]} & {$b/a$\footnotemark[2]}  & {(arcsec)}& {($M_{\odot}$)} & {($M_{\odot}$ yr$^{-1}$)} & {(s)} & {(s)} & {{(uvw2)}}  & {{(uvw1)}} & {{$\Sigma_{\rm uvw2}$}\footnotemark[8]} & {{$\Sigma_{\rm uvw1}$}\footnotemark[8]} & {(mag)} & {(mag)} & {$\tau^l_{B,int}$\footnotemark[10]$^{,}$\footnotemark[11]} & {$\beta_{int}$\footnotemark[10]$^{,}$\footnotemark[12]}\\		
\hline
{1} & {UGC 11696 NOTES01} & {SF} & {0.017} & {0.101}  & {0.52}  & {16} & {8.76} & {$-1.23$} & {1439} & {816} & {5.7} & {7.0} & {2.3$\pm$0.1} & {1.8$\pm$0.1} & {18.31$\pm$0.03} & {18.15$\pm$0.07} & {0.075$\pm$0.006} & {$-1.3\pm$0.2}\\
{2} &{SDSS J170653.67+321010.1} & {SF} &{0.036}&{0.039}  & {0.82} & {15} & {9.15}  & {$-0.68$} & {1420} & {6283} & {5.2} & {1.3} & {0.84$\pm$0.03} & {0.63$\pm$0.01} &{18.81$\pm$0.04} & {18.59$\pm$0.02} & {0.086$\pm$0.008} & {$-1.1\pm$0.2}\\
{3} & {WISE J162856.63+393634.1} & {SF} & {0.035} &{0.010}& {0.48}  & {9} & {9.19} & {$-0.62$} & {8412} & {5674} & {1.2} & {1.3} & {1.91$\pm$0.04} & {1.49$\pm$0.03} & {19.85$\pm$0.02} & {19.44$\pm$0.02} & {0.270$\pm$0.009} &{$-0.4\pm$0.2}\\
{4} &{2MASS J04072365-0641117} & {SF} & {0.038} &{0.098}& {0.60}   & {18} & {9.25}   & {$-0.71$} & {788} & {1785} & {8.9} & {3.8} & {1.2$\pm$0.1} & {0.72$\pm$0.03}& {18.68$\pm$0.06} & {18.71$\pm$0.04} & {0.134$\pm$0.005} &{$-2.1\pm$0.2}\\
{5} &{2MASX J11044509+4509238} & {SF} & {0.022} &{0.008}& {0.88}    & {6} & {9.27}   & {$-1.52$} & {2265} & {13689} & {3.8} & {0.8} & {2.5$\pm$0.1} & {2.17$\pm$0.04}& {20.81$\pm$0.06} & {20.31$\pm$0.02} & {0.29$\pm$0.03} & {$-0.1\pm$0.2}\\
{6} &{2MASS J14135887+4353350} & {SB} & {0.040} & {0.013} & {0.71}    & {7} & {9.31}  & {$-0.61$} & {2066} & {1048} & {3.5} & {4.8} & {5.1$\pm$0.2} & {3.5$\pm$0.1}& {19.42$\pm$0.03} & {19.13$\pm$0.04} & {0.144$\pm$0.005} & {$-0.8\pm$0.2}\\
{7} &{KUG 1016+468} & {SF} & {0.024} & {0.013}       & {0.45}        & {13} & {9.32} & {$-0.87$} & {10027}  & {2296} & {1.4} & {2.8} & {3.72$\pm$0.04} & {2.54$\pm$0.05}& {18.66$\pm$0.01} & {18.41$\pm$0.02} & {0.197$\pm$0.006} & {$-1.0\pm$0.2}\\
{8} &{SDSS J030659.79-004841.5} & {SF} & {0.038} & {0.070}& {0.78}  & {8} & {9.40}  & {$-0.61$} & {4084} &{3567} & {2.5} & {2.4} & {3.66$\pm$0.08} & {2.56$\pm$0.07}& {18.93$\pm$0.03} & {18.76$\pm$0.03} & {0.220$\pm$0.006} & {$-1.3\pm$0.2}\\
{9} &{2MASS J14132780+4354501} & {SF} & {0.040} & {0.013}& {0.40}  & {13} & {9.43} & {$-0.56$} & {2066} & {1048} & {3.5} & {4.8} & {2.81$\pm$0.06} & {1.76$\pm$0.06}& {18.82$\pm$0.02} & {18.65$\pm$0.03} & {0.159$\pm$0.004} & {$-1.3\pm$0.2}\\
{10} &{SDSS J024112.93-005236.9} & {SF} & {0.038} & {0.034} & {0.56} & {10} & {9.44} & {$-1.07$} & {20519} & {3474} & {0.7} & {2.2} & {1.29$\pm$0.02} & {0.97$\pm$0.03}& {19.18$\pm$0.02} & {18.93$\pm$0.03} & {0.10$\pm$0.01} & {$-1.0\pm$0.2}\\
{11} &{2MASS J15004786+4836270} & {SF} & {0.037} & {0.020}& {0.54} &{7} & {9.50}  & {$-0.54$}  &{2108} &{1579} & {4.0} & {3.8} & {2.3$\pm$0.1} & {2.0$\pm$0.1}& {20.60$\pm$0.07} & {20.04$\pm$0.06} & {0.354$\pm$0.009} & {0.2$\pm$0.2}\\
{12} & {2MASXi J0740537+400411} & {N/A} & {0.042} & {0.052} & {0.59} & {9} & {9.50} & {\phantom{1}\,\,0.47} &{4558}&{3904} & {2.3} & {2.2} & {17.7$\pm$0.1\phantom{1}} & {10.9$\pm$0.1\phantom{1}}& {16.83$\pm$0.03} & {16.78$\pm$0.01} & {0.120$\pm$0.009} & {$-1.8\pm$0.2}\\
{13} & {KUG 0757+468} & {SF} & {0.019} & {0.065}    & {0.53}            & {14} & {9.52} & {$-0.03$} &{559}&{279} & {12.0} & {18.0} & {5.7$\pm$0.1} & {4.2$\pm$0.1}& {16.78$\pm$0.02} & {16.55$\pm$0.04} & {0.194$\pm$0.001} & {$-1.1\pm$0.2}\\
{14} &{2MASX J14403849+5328414} & {SB} & {0.038} & {0.011}& {0.80}  & {8} & {9.55}  & {$-0.12$}   & {7621} & {1314} & {1.4} & {4.2} & {1.18$\pm$0.04} & {1.21$\pm$0.08}& {20.65$\pm$0.03} & {19.98$\pm$0.07} & {0.390$\pm$0.004} & {\,\,\,\,$0.6\pm$0.2}\\
{15} &{KUG 0254-004} & {SF} & {0.029} & {0.066}     & {0.80}         & {10} & {9.65}  & {\phantom{1}\,\,0.00} & {2375}&{2307} & {3.6} & {2.7} & {8.2$\pm$0.1} & {5.77$\pm$0.07}& {17.29$\pm$0.01} & {17.10$\pm$0.01} & {0.286$\pm$0.001} & {$-1.3\pm$0.2}\\
{16} &{2MASX J07522873+4950192} & {SF} & {0.022} & {0.059}  & {0.50}  & {30} & {9.67} & {$-0.86$} & {946} & {473} & {7.0} & {10.0} & {0.95$\pm$0.05} & {0.87$\pm$0.05}& {18.60$\pm$0.05} & {18.23$\pm$0.07} & {0.300$\pm$0.008} & {$-0.6\pm$0.2}\\
{17} &{KUG 0751+485} & {SB} & {0.022} & {0.035}          & {0.71}          & {5} & {9.71}  & {$-0.46$} & {726} & {2669} & {9.0} & {2.8} & {1.06$\pm$0.06} & {0.90$\pm$0.03}& {19.37$\pm$0.06} & {18.91$\pm$0.03} & {0.398$\pm$0.001} & {$-0.2\pm$0.2}\\
{18} & {2MFGC 8582} & {SF} & {0.025} & {0.014}       & {0.33}        & {22} & {9.73} & {$-0.23$} & {4106} & {1017} & {2.0} & {4.6} & {1.81$\pm$0.05} & {1.65$\pm$0.07}& {19.43$\pm$0.03} & {19.02$\pm$0.05} & {0.593$\pm$0.005} & {$-0.4\pm$0.2}\\
{19} &{2MASX J07595878+3153360} & {SB} & {0.045} & {0.048}  & {0.57}   & {8} & {9.78}  & {\phantom{1}\,\,0.08} &{1402} & {1544} & {5.7} & {4.7} & {4.4$\pm$0.2} & {3.1$\pm$0.1}& {19.00$\pm$0.04} & {18.75$\pm$0.04} & {0.394$\pm$0.004} & {$-1.0\pm$0.2}\\
{20} &{KUG 1121+239} & {SF} & {0.028} & {0.019}       & {0.80}       & {14} & {9.79}& {$-0.45$} & {277} & {346} & {22.0} & {16.0} & {3.7$\pm$0.2} & {2.5$\pm$0.1}& {17.82$\pm$0.03} & {17.65$\pm$0.06} & {0.269$\pm$0.004} & {$-1.3\pm$0.2}\\
{21} &{WISE J163226.31+393104.0} & {SF} &{0.029}&{0.009} & {0.87}   & {10} & {10.0} & {$-0.53$} & {994} & {599} & {7.1} & {7.9} & {1.22$\pm$0.06} & {0.97$\pm$0.05}& {19.15$\pm$0.05} & {18.75$\pm$0.05} & {0.46$\pm$0.01} & {$-0.4\pm$0.2}\\
{22} &{2MASX J09115605+2753575} & {SB} & {0.047} & {0.028}  & {0.48}& {6} & {10.1}  & {\phantom{1}\,\,0.17}& {495}&{644} & {12.0} & {8.4} & {4.3$\pm$0.2} & {3.2$\pm$0.1}& {18.64$\pm$0.05} & {18.35$\pm$0.04} & {0.410$\pm$0.004} & {$-0.9\pm$0.2}\\
{23} &{KUG 1343+270} & {SF} & {0.030} & {0.017}    & {0.54}  & {21} & {10.1}  & {$-0.22$}  & {2998}     &{6144} & {2.7} & {1.3} & {4.33$\pm$0.07} & {3.27$\pm$0.04}& {18.36$\pm$0.02} & {18.02$\pm$0.01} & {0.290$\pm$0.004} & {$-0.6\pm$0.2}\\
{24} &{2MASX J11532094+5220438} & {SF} & {0.049}&{0.029}&{0.87}&{10}&{10.1}&{\phantom{1}\,\,0.00}&{2481}&{7471} & {2.9} & {1.1} & {3.10$\pm$0.06} & {2.35$\pm$0.03}& {18.95$\pm$0.02} & {18.61$\pm$0.01} & {0.327$\pm$0.006} & {$-0.7\pm$0.2}\\
{25} &{KUG 1626+402} & {SF} & {0.026} & {0.009}  & {0.70}& {5} & {10.2}& {\phantom{1}\,\,0.15} & {13227}&{8908} & {1.1} & {1.1} & {4.71$\pm$0.02} & {3.38$\pm$0.01}& {18.80$\pm$0.01} & {18.38$\pm$0.01} & {0.491$\pm$0.004} & {$-0.4\pm$0.2}\\
{26} &{NGC 3191\footnotemark[13]} & {SF} & {0.031} & {0.011}& {0.86}&{17}&{10.3}&{\phantom{1}\,\,0.73}& {11850}&{2296} & {1.3} & {2.9} & {24.06$\pm$0.04\phantom{1}} & {10.10$\pm$0.05\phantom{1}}& {\ldots} & {\ldots} & {\ldots} & {\ldots}\\
{27} &{LCSB S1611P} & {Galaxy} & {0.056} & {0.010}& {0.40}  & {4}  &  {10.4}  & {$-0.13$}  & {12190} &             {7209} & {1.0} & {1.2} & {0.50$\pm$0.02} & {0.53$\pm$0.02}& {20.90$\pm$0.04} & {20.20$\pm$0.04} & {0.70$\pm$0.04} & {\,\,\,\,$0.7\pm$0.2}\\
{28} &{2MASX J07333599+4556364}&{SF}&{0.077}&{0.091}& {0.89}  & {11} & {10.5}  & {\phantom{1}\,\,0.94}&  {3453}&{2048} & {2.5} & {3.4} & {3.40$\pm$0.07} & {3.21$\pm$0.08}& {18.62$\pm$0.02} & {18.16$\pm$0.03} & {0.507$\pm$0.003} & {$-0.2\pm$0.2}\\
{29} &{2MASX J03065213-0053469}&{SF}&{0.084}&{0.070}&{0.88}&{11} & {10.7} & {\phantom{1}\,\,0.56}&{4084}&{3567} & {2.4} & {2.3} & {2.97$\pm$0.05} & {2.06$\pm$0.04}& {18.30$\pm$0.02} & {18.15$\pm$0.02} & {0.439$\pm$0.006} & {$-1.4\pm$0.2}\\
\hline \vspace{-1cm}
	\end{tabular}
\footnotetext[1]{The SDSS Classification of the galaxy \citep{Bolton2012}. SB stands for starburst, while SF stands for star forming galaxy.}
\footnotetext[2]{The redshift {and $b/a$ values} are taken from the NASA-Sloan Atlas \citep[\url{http://nsatlas.org}; ][]{nsatlas}.}
\footnotetext[3]{ The foreground extinction measurements are from \citet{Schlegel1998}.}
\footnotetext[4]{The Petrosian Radii are calculated using the uvw2 filter as described in Section~\ref{ssec:swft_sky}.}
\footnotetext[5]{Stellar masses are taken from the SED fits from the MaNGA DRP \citep[][]{Law2016}.}
\footnotetext[6]{The SFRs within $1~R_e$ using the H$\alpha$ luminosity are drawn from the MaNGA Data Analysis Pipeline \citep{Westfall2019}.}
\footnotetext[7]{$f_{min}$ represents the minimum flux density detections, in units of $10^{-18}$~erg~s$^{-1}$~cm$^{-2}$~\AA$^{-1}$, for a point source with a diameter of $2\farcs5$ at the redshift of the galaxy with a $S/N=3$}
\footnotetext[8]{The UV surface densities are given in units of $10^{-18}$~erg~s$^{-1}$~cm$^{-2}$~\AA$^{-1}$~arcsec$^{-2}$, and are calculated by dividing the integrated, observed fluxes by the area of the integrated aperture.}
\footnotetext[9]{The uvw2 and uvw1 magnitudes are presented in the AB magnitude system without K-corrections.}
\footnotetext[10]{The Balmer optical depth ($\tau^l_B$) and UV spectral slope ($\beta$) are for the integrated light in the galaxy, See Section~\ref{ssec:bcalc} for details.}
\footnotetext[11]{The systematic uncertainty in the emission line flux as described by \citet{Belfiore2019} is included in the reported error. See Section~\ref{ssec:basic_dr} for details.}
\footnotetext[12]{The systematic uncertainty of 0.03~mag due to choice in foreground attenuation curves is not included in these 1$\sigma$ errors, as the $\beta_{Swift}$--$\beta$ ``conversion'' dominates the error budget. See Section~\ref{ssec:bgal} for details.}
\footnotetext[13]{The MaNGA IFU does not cover a majority of the galaxy disc, so no $\tau_{B, int}^{l}$ and $\beta_{int}$ are reported.}
\end{minipage}
\end{table}
\end{landscape}

\clearpage

\item Confirm the nebular emission line ratios from the defined region fall within the star forming locus in all three BPT diagrams described in Section~\ref{subsec:int_sf}.
\item Repeat steps 1--3, stopping at a threshold of  $\log[\Sigma_{\textrm{H}\alpha}$/(erg~s$^{-1}$~kpc$^{-2})]\geq 38$ in increments of 0.2 in $\log(\Sigma_{\textrm{H}\alpha})$, while ensuring no region is identified more than once.
\end{enumerate}

\begin{figure}
\includegraphics[width=0.45\textwidth]{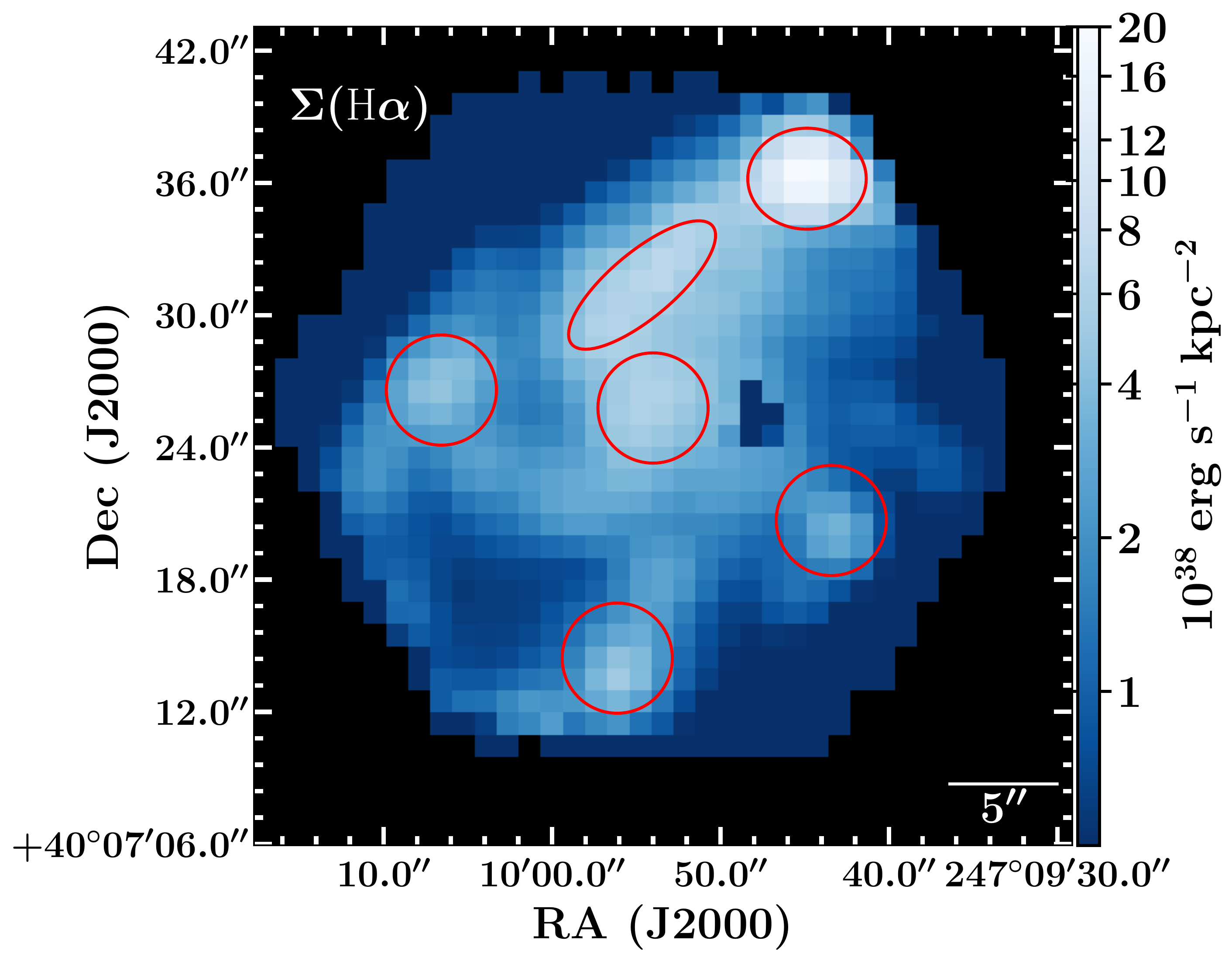}
\vspace{-3mm}
\caption{An example of the identified star forming regions in KUG 1626+402. The map displays the H$\alpha$ surface brightness across the face of the galaxy within the MaNGA FOV, and increases in intensity from dark blue to white as indicated by the colour bar. The angular resolution of the image is $\approx2.\!\!^{\prime\prime}5$. A scale bar of $5^{\prime\prime}$ is shown on the lower right, corresponding to a physical scale of 2.61~kpc. The identified star forming regions are represented by the red circles and ellipses. Six regions were identified in this galaxy. See Appendix~\ref{app:a}, which is available online, for more information.}
\label{fig:sfreg}
\end{figure}
\begin{figure}
\includegraphics[width=0.45\textwidth]{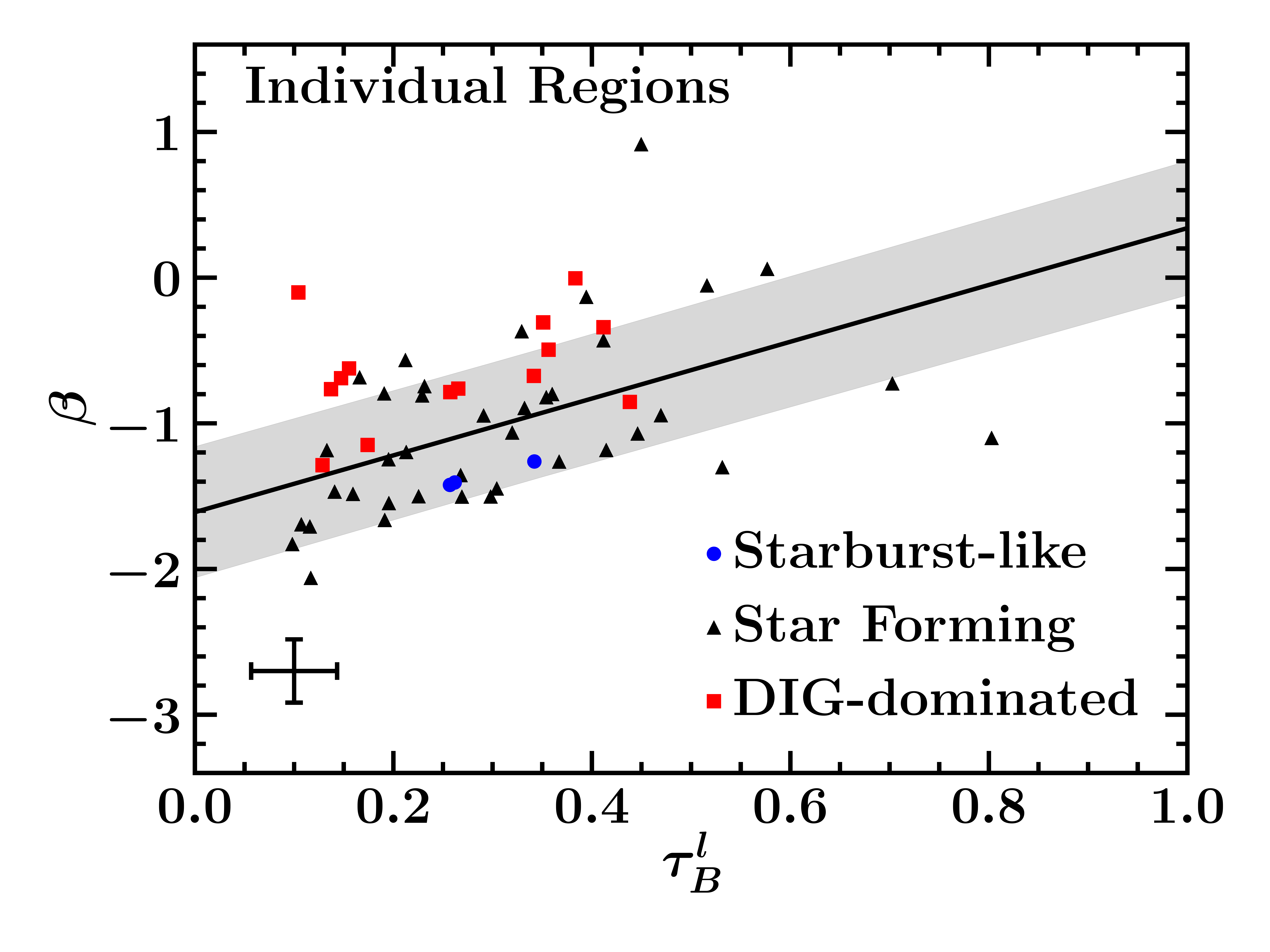}\vspace{-4mm}
\caption{The UV spectral slope ($\beta$) vs.~the Balmer optical depth ($\tau^{l}_{B}$) of the individual star forming regions in our sample. The blue circles are starburst-like regions ($\log[\Sigma_{\textrm{H}\alpha}$/(erg~s$^{-1}$~kpc$^{-2})]\geq39.4$) and the black triangles are normal star forming regions ($38.0 < \log[\Sigma_{\textrm{H}\alpha}$/(erg~s$^{-1}$~kpc$^{-2})]\leq 39.4$ and EW(H$\alpha) > 15$\AA), as defined in Section~\ref{ssec:reg_calc}. The red squares are regions likely dominated by DIG ($\log[\Sigma_{\textrm{H}\alpha}$/(erg~s$^{-1}$~kpc$^{-2})]< 38.0$ or  EW(H$\alpha) < 15$\AA; see Section~\ref{ssec:reg_calc} for discussion). The error bar on the lower left represents the characteristic uncertainties in $\beta$ and $\tau^l_B$. The relation between $\beta$ and $\tau^{l}_{B}$ from B16 is indicated by the solid black line, while the intrinsic dispersion of their data is shown by the grey shaded region.}
\label{fig:reg_btau}
\end{figure}
\begin{figure}
\hspace{-2mm}\includegraphics[width=0.5\textwidth]{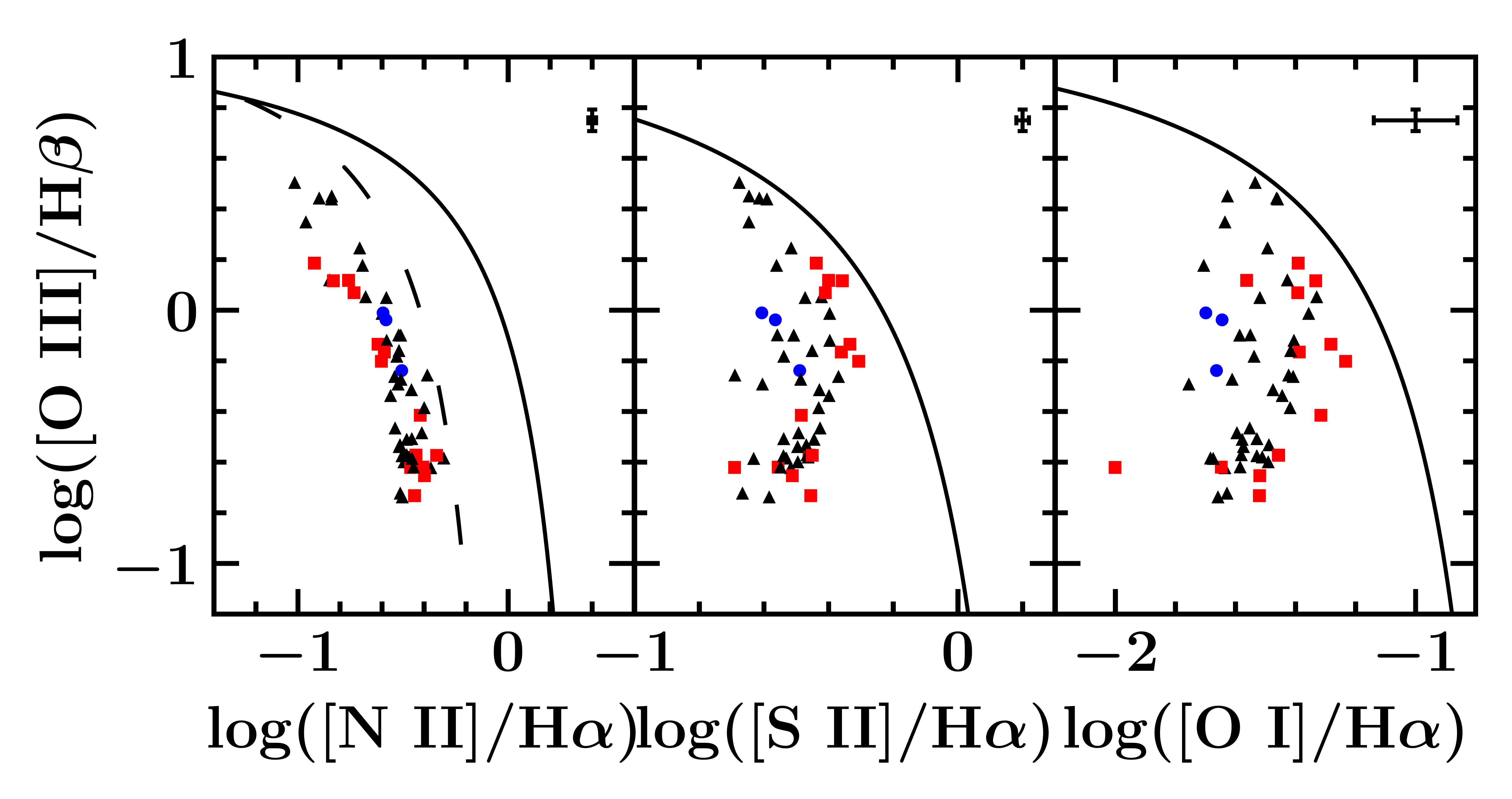}\vspace{-3mm}
\caption{BPT diagrams for the individual star forming regions in our sample. The colour coding for the data is the same as in Figure~\ref{fig:reg_btau}. The solid black line in each panel is the extreme starburst limit from \citet[][]{Kewley2006}; the dashed line is the composite object locus from \citet{Kauffmann03}. All regions fall inside the star forming region of all three diagrams. The errors bars on the top right of each panel show the characteristic uncertainty in the emission line ratios.}
\label{fig:reg_bpt}
\end{figure}

By iterating this process, we ensure that all peaks with $\log[\Sigma_{\textrm{H}\alpha}$/(erg~s$^{-1}$~kpc$^{-2})]>38$ are identified. The $\log(\Sigma_{\textrm{H}\alpha})$ limits exclude excessive amounts of DIG from the aperture, as the kpc-scales probed here do not allow for its clean subtraction. An example of the identified star forming regions for KUG 1626+402 is presented in Figure~\ref{fig:sfreg}. We present the individual region positions and a full suite of images in Appendix~\ref{app:a}, which is available online.

Since we are interested in comparing ``starburst''-like regions to normal star forming regions and are resolving kpc-sized regions instead of relying on pixels, we classify our regions using a combination of the criteria from \citet{Oey2007}, \citet{mangadig}, \citet{Sanchez2014} and \citet{Wang2018}. We define a starburst region as one with $\log[\Sigma_{\textrm{H}\alpha}$/(erg~s$^{-1}$~kpc$^{-2})]\geq39.4$, while a normal star forming region has $38.0 < \log[\Sigma_{\textrm{H}\alpha}$/(erg~s$^{-1}$~kpc$^{-2})]\leq 39.4$ and EW(H$\alpha) > 15$~\AA. This is similar to the criterion for ``star forming'' regions defined by \citet{Oey2007}, but combines the slightly lower $\Sigma_{\textrm{H}\alpha}$ cut presented in \citet{mangadig}, and requires more than double the EW(H$\alpha) > 6$~\AA\ cut used by \citet{Sanchez2014} and \citet{Wang2018} to define star forming regions. Thus our regions appear star forming to both our area- and mass-specific SFR proxies. For regions that do not meet both of these criteria, the UV continuum may not be dominated by the light from massive stars; hence the assumptions behind the calculation of $\beta$ may not hold. All such regions which were selected to meet the BPT constraints for a star forming region, but have either $\log[\Sigma_{\textrm{H}\alpha}$/(erg~s$^{-1}$~kpc$^{-2})] \leq 38.0$ or EW(H$\alpha) < 15$~\AA, are denoted as DIG-dominated. Figure~\ref{fig:reg_btau} displays our identified kpc-sized star forming regions on the B16 relation including the intrinsic dispersion in their data set, while Figure~\ref{fig:reg_bpt} shows the nebular emission line ratios compared to the \citet[][]{Kewley2006} and \citet[][]{Kauffmann03} models in the standard BPT diagrams. The starburst-like, normal star forming, and DIG dominated regions are plotted in both Figures~\ref{fig:reg_btau} and \ref{fig:reg_bpt} for completeness. We note that there is no systematic trend with inclination, and almost all of our regions meet the $b/a>0.42$ requirement described in \citet{Battisti2017}. The measured flux densities in the \textit{Swift} NUV filters and the emission line flux from MaNGA are listed in Table~\ref{table:reg_flx}, and the calculated properties for the 56 identified regions are listed in Table~\ref{table:reg_val}. 

We tested the effects of aperture positioning and pixelation on the photometry by comparing two different error measurement methods: (1) we only include pixels whose center lies within the aperture, and (2) we calculate the exact boundary of the aperture within the pixel. We find the difference between these two methods was $\sim2$\% of the total error calculated for method (2), which is the most commonly adopted error calculation technique. Thus we find that the effects of aperture positioning and pixelation on the photometry is minimal, and we adopt method (2). Additionally, we find no systematic trends with $\beta$ and signal-to-noise in either uvw1 or uvw2.

\begin{table*}
\begin{minipage}{\linewidth}
\renewcommand{\thefootnote}{\textrm{\alph{footnote}}}
	\centering
	\caption{Flux Measurements of Star Forming Regions\label{table:reg_flx}}
	\begin{tabular}{ccccccccccc} % four columns, alignment for each
		\hline
		\hline
		{} &{}&\multicolumn{2}{c}{Flux Densities\footnotemark[1]}& \multicolumn{7}{c}{Flux Measurements\footnotemark[2]$^{,}$\footnotemark[3]}\\
		{Object} & {Region}& \multicolumn{2}{c}{\hrulefill}& \multicolumn{7}{c}{\hrulefill}\\
		 {I.D.}  &{I.D.} & {uvw2} & {uvw1} & {H$\beta\lambda4861$} & {[\ion{O}{3}]$\lambda5007$} & {[\ion{O}{1}]$\lambda6300$} & {H$\alpha\lambda6563$}  & {[\ion{N}{2}]$\lambda6583$}  & {[\ion{S}{2}]$\lambda$6716} & {[\ion{S}{2}]$\lambda$6731}\\
		\hline
{1}  & {1.1}    & {16$\pm$2} & {11$\pm$2}& {$27\pm1$} &{$35\pm1$} & {$3\pm1$} & {$86\pm1$} & {$12\pm1$} & {$20\pm1$} & {$14\pm1$}\\
{1}  & {1.2}    & {\phantom{1}9$\pm$1} & {\phantom{1}8$\pm$1}& {$16\pm1$} &{$21\pm1$} & {$2\pm1$} & {$54\pm1$} & {\phantom{1}$8\pm1$} & {$14\pm1$} & {\phantom{1}$9\pm1$}\\
{1}  & {1.3}    & {\phantom{1}9$\pm$1} & {\phantom{1}9$\pm$1}& {$14\pm1$} &{$22\pm1$} & {$2\pm1$} & {$45\pm1$} & {\phantom{1}$5\pm1$} & {$10\pm1$} & {\phantom{1}$7\pm1$}\\
{2}  & {2.1}    & {\phantom{1}7.2$\pm$0.8} & {\phantom{1}4.5$\pm$0.3}& {$19\pm1$} &{4$3\pm1$} & {$1\pm1$} & {$63\pm1$} & {\phantom{1}$7\pm1$} & {\phantom{1}$8\pm1$} & {\phantom{1}$6\pm1$}\\
{3}  & {3.1}    & {14.1$\pm$0.4} & {13.2$\pm$0.4}& {$43\pm1$} &{$15\pm1$} & {$5\pm1$} & {$170\pm1$\phantom{1}} & {$49\pm1$} & {$37\pm1$} & {$26\pm1$}\\
{4}  & {4.1}    & {17.3$\pm$2.0} & {12$\pm$1}& {$24\pm1$} &{$23\pm1$} & {$4\pm1$} & {$83\pm1$} & {$21\pm1$} & {$20\pm1$} & {$14\pm1$}\\
{5}  & {5.1}    & {\phantom{1}7.4$\pm$0.5} & {\phantom{1}7.0$\pm$0.2}& {$11\pm1$} &{\phantom{1}$3\pm1$} & {$2\pm1$} & {$43\pm1$} & {$16\pm1$} & {\phantom{1}$9\pm1$} & {\phantom{1}$6\pm1$}\\
{6}  & {6.1}   & {20$\pm$1} & {17$\pm$1}& {$68\pm1$} &{$119\pm1$\phantom{1}} & {$7\pm1$} & {$229\pm1$\phantom{1}} & {$45\pm1$} & {$40\pm1$} & {$30\pm1$}\\
{7}  & {7.1}    & {18.6$\pm$0.5} & {15.7$\pm$0.7}& {$31\pm1$} &{$17\pm1$} & {$4\pm1$} & {$113\pm1$\phantom{1}} & {$33\pm1$} & {$28\pm1$} & {$20\pm1$}\\
{7}  & {7.2}    & {11.0$\pm$0.4} & {\phantom{1}9.4$\pm$0.6}& {$12\pm1$} &{\phantom{1}$8\pm1$} & {$2\pm1$} & {$43\pm1$} & {$10\pm1$} & {$12\pm1$} & {\phantom{1}$8\pm1$}\\
{7}  & {7.3}    & {11.9$\pm$0.4} & {\phantom{1}9.3$\pm$0.6}& {$14\pm1$} &{\phantom{1}$9\pm1$} & {$3\pm1$} & {$48\pm1$} & {$12\pm1$} & {$15\pm1$} & {\phantom{1}$9\pm1$}\\
{8}  & {8.1}    & {27$\pm$1} & {20.5$\pm$0.8}& {$34\pm1$} &{$16\pm1$} & {$4\pm1$} & {$122\pm1$\phantom{1}} & {$34\pm1$} & {$29\pm1$} & {$19\pm1$}\\
{9}  & {9.1}    & {18.4$\pm$0.8} & {14.1$\pm$0.9}& {$40\pm1$} &{$30\pm1$} & {$5\pm1$} & {$139\pm1$\phantom{1}} & {$37\pm1$} & {$33\pm1$} & {$23\pm1$}\\
{9}  & {9.2}    & {12.8$\pm$0.7} & {\phantom{1}9.3$\pm$0.7}& {$24\pm1$} &{$27\pm1$} & {$4\pm1$} & {$80\pm1$} & {$17\pm1$} & {$18\pm1$} & {$12\pm1$}\\
{9}  & {9.3}    & {\phantom{1}9.3$\pm$0.6} & {\phantom{1}7.0$\pm$0.6}& {$18\pm1$} &{$21\pm1$} & {$2\pm1$} & {$58\pm1$} & {$11\pm1$} & {$13\pm1$} & {\phantom{1}$9\pm1$}\\
{10\phantom{1}} & {10.1\phantom{1}}    & {\phantom{1}8.6$\pm$0.3} & {\phantom{1}7.5$\pm$0.5}& {$10\pm1$} &{\phantom{1}$7\pm1$} & {$1\pm1$} & {$33\pm1$} & {\phantom{1}$8\pm1$} & {\phantom{1}$9\pm1$} & {\phantom{1}$6\pm1$}\\
{10\phantom{1}} & {10.2\phantom{1}}    & {\phantom{1}7.9$\pm$0.3} & {\phantom{1}6.8$\pm$0.5}& {$\phantom{1}6\pm1$} &{\phantom{1}$8\pm1$} & {$1\pm1$} & {$21\pm1$} & {\phantom{1}$4\pm1$} & {\phantom{1}$5\pm1$} & {\phantom{1}$4\pm1$}\\
{11\phantom{1}} & {11.1\phantom{1}}    & {10.0$\pm$0.7} & {10.0$\pm$0.7}& {$33\pm1$} &{$10\pm1$} & {$4\pm1$} & {$140\pm1$\phantom{1}} & {$46\pm1$} & {$29\pm1$} & {$21\pm1$}\\
{12\phantom{1}} & {12.1\phantom{1}}   & {73$\pm$3} & {48$\pm$1} & {$190\pm1$\phantom{1}} &{$527\pm2$\phantom{1}} & {$21\pm1$\phantom{1}} & {$600\pm2$\phantom{1}} & {$76\pm1$} & {$85\pm1$} & {$61\pm1$}\\
{12\phantom{1}} & {12.2\phantom{1}}   & {55$\pm$1} & {38$\pm$1}& {$157\pm1$\phantom{1}} &{$501\pm2$\phantom{1}} & {$15\pm1$\phantom{1}} & {$500\pm2$\phantom{1}} & {$48\pm1$} & {$61\pm1$} & {$44\pm1$}\\
{12\phantom{1}} & {12.3\phantom{1}}   & {99$\pm$2} & {72$\pm$1}& {$320\pm2$\phantom{1}} &{$880\pm2$\phantom{1}} & {$37\pm1$\phantom{1}} & {$1054\pm2$\phantom{1}\,\,} & {$152\pm1$\phantom{1}} & {$156\pm1$\phantom{1}} & {$114\pm1$\phantom{1}}\\
{13\phantom{1}} & {13.1\phantom{1}}   & {79$\pm$5} & {67$\pm$5}& {$275\pm2$\phantom{1}} &{$413\pm2$\phantom{1}} & {$19\pm1$\phantom{1}} & {$952\pm2$\phantom{1}} & {$193\pm1$\phantom{1}} & {$153\pm1$\phantom{1}} & {$109\pm1$\phantom{1}}\\
{13\phantom{1}} & {13.2\phantom{1}}   & {63$\pm$4} & {43$\pm$4}& {$129\pm1$\phantom{1}} &{$145\pm1$\phantom{1}} & {$14\pm1$\phantom{1}} & {$447\pm2$\phantom{1}} & {$118\pm1$\phantom{1}} & {$88\pm1$} & {$63\pm1$}\\
{13\phantom{1}} & {13.3\phantom{1}}   & {47$\pm$4} & {40$\pm$4}& {$261\pm2$\phantom{1}} &{$737\pm2$\phantom{1}} & {$22\pm1$\phantom{1}} & {$941\pm2$\phantom{1}} & {$136\pm1$\phantom{1}} & {$123\pm1$\phantom{1}} & {$89\pm1$}\\
{14\phantom{1}} & {14.1\phantom{1}}   & {\phantom{1}7.9$\pm$0.3} & {10.1$\pm$0.7}& {$93\pm1$} &{$74\pm1$} & {$12\pm1$\phantom{1}} & {$418\pm2$\phantom{1}} & {$126\pm1$\phantom{1}} & {$62\pm1$} & {$54\pm1$}\\
{15\phantom{1}} & {15.1\phantom{1}}   & {122$\pm$3\phantom{1}} & {93$\pm$2}& {$227\pm2$\phantom{1}} &{$131\pm2$\phantom{1}} & {$20\pm1$\phantom{1}} & {$913\pm2$\phantom{1}} & {$285\pm2$\phantom{1}} & {$174\pm2$\phantom{1}} & {$122\pm2$\phantom{1}}\\
{16\phantom{1}} & {16.1\phantom{1}}    & {\phantom{1}9$\pm$1} & {\phantom{1}8$\pm$1}& {$14\pm1$} &{\phantom{1}$5\pm1$} & {$3\pm1$} & {$56\pm1$} & {$21\pm1$} & {$11\pm1$} & {\phantom{1}$8\pm1$}\\
{16\phantom{1}} & {16.2\phantom{1}}    & {10$\pm$2} & {\phantom{1}7$\pm$1}& {$17\pm1$} &{\phantom{1}$5\pm1$} & {$2\pm1$} & {$65\pm1$} & {$22\pm1$} & {$13\pm1$} & {$10\pm1$}\\
{17\phantom{1}} & {17.1\phantom{1}}   & {29$\pm$2} & {27$\pm$1}& {$110\pm1$\phantom{1}} &{$26\pm1$} & {$11\pm1$\phantom{1}} & {$476\pm2$\phantom{1}} & {$204\pm1$\phantom{1}} & {$80\pm1$} & {$65\pm1$}\\
{18\phantom{1}} & {18.1\phantom{1}}    & {\phantom{1}6.8$\pm$0.5} & {\phantom{1}5.9$\pm$0.7}& {$26\pm1$} &{$11\pm1$} & {$6\pm1$} & {$151\pm1$\phantom{1}} & {$60\pm1$} & {$33\pm1$} & {$23\pm1$}\\
{19\phantom{1}} & {19.1\phantom{1}}   & {26$\pm$1} & {21$\pm$1}& {$97\pm1$} & {$25\pm1$} & {$9\pm1$} & {$433\pm2$\phantom{1}} & {$214\pm1$\phantom{1}} & {$69\pm1$} & {$58\pm1$}\\
{20\phantom{1}} & {20.1\phantom{1}}    & {21$\pm$3} & {15$\pm$2}& {$26\pm1$} &{\phantom{1}$7\pm1$} & {$3\pm1$} & {$103\pm1$\phantom{1}} & {$33\pm1$} & {$20\pm1$} & {$13\pm1$}\\
{20\phantom{1}} & {20.2\phantom{1}}    & {16$\pm$3} & {14$\pm$2}& {$23\pm1$} &{\phantom{1}$4\pm1$} & {$3\pm1$} & {$85\pm1$} & {$30\pm1$} & {$18\pm1$} & {$12\pm1$}\\
{20\phantom{1}} & {20.3\phantom{1}}    & {11$\pm$2} & {\phantom{1}9$\pm$2}& {$18\pm1$} &{\phantom{1}$5\pm1$} & {$2\pm1$} & {$70\pm1$} & {$24\pm1$} & {$13\pm1$} & {$11\pm1$}\\
{21\phantom{1}} & {21.1\phantom{1}}    & {16$\pm$2} & {13$\pm$1}& {$18\pm1$} &{\phantom{1}$4\pm1$} & {$1\pm1$} & {$80\pm1$} & {$27\pm1$} & {\phantom{1}$9\pm1$} & {\phantom{1}$7\pm1$}\\
{22\phantom{1}} & {22.1\phantom{1}}   & {48$\pm$3} & {37$\pm$2}& {$95\pm1$} &{$31\pm1$} & {$10\pm1$\phantom{1}} & {$412\pm2$\phantom{1}} & {$160\pm1$\phantom{1}} & {$76\pm1$} & {$56\pm1$}\\
{23\phantom{1}} & {23.1\phantom{1}}    & {13.0$\pm$0.7} & {13.3$\pm$0.4}& {$23\pm1$} &{\phantom{1}$6\pm1$} & {$2\pm1$} & {$98\pm1$} & {$38\pm1$} & {$16\pm1$} & {$11\pm1$}\\
{23\phantom{1}} & {23.2\phantom{1}}    & {13.6$\pm$0.7} & {11.9$\pm$0.4}& {$17\pm1$} &{\phantom{1}$4\pm1$} & {$2\pm1$} & {$70\pm1$} & {$28\pm1$} & {$13\pm1$} & {\phantom{1}$9\pm1$}\\
{23\phantom{1}} & {23.3\phantom{1}}    & {10.0$\pm$0.6} & {\phantom{1}8.9$\pm$0.3}& {$20\pm1$} &{\phantom{1}$9\pm1$} & {$2\pm1$} & {$69\pm1$} & {$24\pm1$} & {$15\pm1$} & {$10\pm1$}\\
{23\phantom{1}} & {23.4\phantom{1}}    & {29$\pm$1} & {22.5$\pm$0.5}& {$52\pm1$} &{$41\pm1$} & {$4\pm1$} & {$168\pm1$} & {$52\pm1$} & {$31\pm1$} & {$22\pm1$}\\
{24\phantom{1}} & {24.1\phantom{1}}    & {\phantom{1}8.2$\pm$0.5} & {\phantom{1}7.0$\pm$0.3}& {$20\pm1$} &{\phantom{1}$6\pm1$} & {$2\pm1$} & {$80\pm1$} & {$28\pm1$} & {$13\pm1$} & {$10\pm1$}\\
{24\phantom{1}} & {24.2\phantom{1}}    & {10.3$\pm$0.6} & {\phantom{1}8.7$\pm$0.3}& {$22\pm1$} &{\phantom{1}$5\pm1$} & {$2\pm1$} & {$88\pm1$} & {$31\pm1$} & {$15\pm1$} & {$10\pm1$}\\
{25\phantom{1}} & {25.1\phantom{1}}    & {11.9$\pm$0.3} & {\phantom{1}9.0$\pm$0.3}& {$18\pm1$} &{\phantom{1}$5\pm1$} & {$2\pm1$} & {$74\pm1$} & {$22\pm1$} & {$15\pm1$} & {$10\pm1$}\\
{25\phantom{1}} & {25.2\phantom{1}}   & {20.3$\pm$0.4} & {16.0$\pm$0.4}& {$61\pm2$}  &{$31\pm1$} & {$7\pm1$} & {$390\pm2$\phantom{1}} & {$117\pm1$\phantom{1}} & {$54\pm1$} & {$43\pm1$}\\
{25\phantom{1}} & {25.3\phantom{1}}    & {19.1$\pm$0.4} & {15.6$\pm$0.4}& {$50\pm1$} &{\phantom{1}$9\pm1$} & {$5\pm1$} & {$228\pm1$\phantom{1}} & {$71\pm1$} & {$35\pm1$} & {$24\pm1$}\\
{25\phantom{1}} & {25.4\phantom{1}}    & {21.8$\pm$0.4} & {22.1$\pm$0.4}& {$41\pm1$} &{\phantom{1}$8\pm1$} & {$5\pm1$} & {$198\pm1$\phantom{1}} & {$61\pm1$} & {$25\pm1$} & {$18\pm1$}\\
{25\phantom{1}} & {25.5\phantom{1}}    & {\phantom{1}9.6$\pm$0.3} & {\phantom{1}8.0$\pm$0.3}& {$30\pm1$} &{\phantom{1}$9\pm1$} & {$3\pm1$} & {$120\pm1$\phantom{1}} & {$36\pm1$} & {$22\pm1$} & {$16\pm1$}\\
{25\phantom{1}} & {25.6\phantom{1}}    & {16.1$\pm$0.4} & {12.8$\pm$0.3}& {$27\pm1$} &{\phantom{1}$7\pm1$} & {$3\pm1$} & {$108\pm1$\phantom{1}} & {$34\pm1$} & {$17\pm1$} & {$14\pm1$}\\
{26\phantom{1}} & {26.1\phantom{1}}    & {32.7$\pm$0.5} & {23$\pm$1}& {$66\pm3$} &{$46\pm1$} & {$9\pm1$} & {$236\pm1$\phantom{1}} & {$72\pm1$} & {$48\pm1$} & {$35\pm1$}\\
{26\phantom{1}} & {26.2\phantom{1}}    & {133$\pm$1\phantom{1}} & {97$\pm$2}& {$509\pm2$\phantom{1}} &{$497\pm2$\phantom{1}} & {$38\pm1$\phantom{1}} & {$1883\pm2$\phantom{1}\,\,} & {$478\pm2$\phantom{1}} & {$268\pm1$\phantom{1}} & {$198\pm1$\phantom{1}}\\
{26\phantom{1}} & {26.3\phantom{1}}    & {78.9$\pm$0.8} & {58$\pm$1}& {$266\pm2$\phantom{1}} &{$244\pm1$\phantom{1}} & {$22\pm1$\phantom{1}} & {$989\pm2$\phantom{1}} & {$259\pm1$\phantom{1}} & {$159\pm1$\phantom{1}} & {$110\pm1$\phantom{1}}\\
{26\phantom{1}} & {26.4\phantom{1}}    & {90$\pm$1} & {65$\pm$1}& {$115\pm1$\phantom{1}} &{$61\pm1$} & {$10\pm1$\phantom{1}} & {$429\pm2$\phantom{1}} & {$133\pm1$\phantom{1}} & {$83\pm1$} & {$57\pm1$}\\
{26\phantom{1}} & {26.5\phantom{1}}    & {27.2$\pm$0.5} & {19.5$\pm$0.8}& {$81\pm1$} &{$53\pm1$} & {$9\pm1$} & {$310\pm2$\phantom{1}} & {$92\pm1$} & {$53\pm1$} & {$37\pm1$}\\
{27\phantom{1}} & {27.1\phantom{1}}    & {\phantom{1}3.0$\pm$0.2} & {\phantom{1}2.8$\pm$0.2}& {\phantom{1}$6\pm1$} &{\phantom{1}$1\pm1$} & {$1\pm1$} & {$24\pm1$} & {$11\pm1$} & {\phantom{1}$5\pm1$} & {\phantom{1}$4\pm1$}\\
{28\phantom{1}} & {28.1\phantom{1}}    & {40$\pm$1} & {42$\pm$2}& {$148\pm2$\phantom{1}} &{$39\pm2$} & {$16\pm1$\phantom{1}} & {$755\pm2$\phantom{1}} & {$263\pm2$\phantom{1}} & {$105\pm2$\phantom{1}} & {$72\pm2$}\\
{29\phantom{1}} & {29.1\phantom{1}}    & {15.8$\pm$0.7} & {12.0$\pm$0.6}& {$13\pm1$} &{\phantom{1}$7\pm1$} & {$2\pm1$} & {$63\pm1$} & {$26\pm1$} & {\phantom{1}$7\pm1$} & {\phantom{1}$6\pm1$}\\
		\hline \vspace{-8mm}
	\end{tabular}
\footnotetext[1]{Flux densities for the Swift uvw1 and uvw2 filters are presented in units of $10^{-17}$~erg~s\textsuperscript{$-1$}~cm\textsuperscript{$-2$}~\AA$^{-1}$ and without a K-correction, but are corrected for foreground extinction. See Section~\ref{ssec:bcalc} for details.}\vspace{-2mm}
\footnotetext[2]{Fluxes for emission lines are presented in units of $10^{-17}$~erg~s\textsuperscript{$-1$}~cm\textsuperscript{$-2$} and are corrected for foreground extinction. See Section~\ref{ssec:tcalc} for details.}
\footnotetext[3]{The systematic uncertainty in emission line flux as described by \citet{Belfiore2019} is included in the reported error. See Section~\ref{ssec:basic_dr} for details.}
\end{minipage}
\end{table*}

\begin{table*}
\begin{minipage}{\linewidth}
\renewcommand{\thefootnote}{\textrm{\alph{footnote}}}
	\centering
		\setlength{\tabcolsep}{8pt}
	\caption{Derived Quantities of Star Forming Regions\label{table:reg_val}}
	\begin{tabular}{ccccccccc} % four columns, alignment for each
		\hline
		\hline
		{Region} & {Dist.\footnotemark[1]} & {{Area}} & {} & {} & {} & {} &{}&{EW(H$\alpha$)\footnotemark[2]} \\
 {I.D.} & {(kpc)} & {(kpc$^2$)}& {$\log(\Sigma_{\textrm{H}\alpha}$/[erg~s$^{-1}$~kpc$^{-2}$])\footnotemark[2]} & {{$\beta$}} & {$\tau^l_B$\footnotemark[2]} & {$12+\log$([O/H])\footnotemark[2]$^{,}$\footnotemark[3]} & {{D$_n$(4000)}\footnotemark[4]} & {(\AA)}\\
 		\hline
{1.1} & {2.6}   & {4.7} &   {38.024$\pm$0.006} & {$-$1.7$\pm$0.3} & {0.12$\pm$0.05} & {8.6$\pm$0.3}   & {1.26$\pm$0.06} & {32$\pm$2}\\
{1.2} & {0.1}   & {4.7} &   {37.968$\pm$0.008} & {$-$0.6$\pm$0.2} & {0.16$\pm$0.07} & {8.7$\pm$0.3}     & {1.31$\pm$0.04} & {23$\pm$1}\\
{1.3} & {1.9}   & {4.1} &   {37.89$\pm$0.01} & {$-$0.1$\pm$0.2} & {0.10$\pm$0.07} & {8.6$\pm$0.3}     & {1.23$\pm$0.06} & {40$\pm$3}\\
{2.1} & {4.3}   & {13.4\phantom{1}} &   {38.249$\pm$0.007} & {$-$2.1$\pm$0.3} & {0.12$\pm$0.05} & {8.5$\pm$0.3}   & {1.14$\pm$0.08} & {101$\pm$5}\\
{3.1} & {0.2}   & {21.5\phantom{1}} &   {38.454$\pm$0.003} & {$-$0.4$\pm$0.2} & {0.33$\pm$0.03} & {9.0$\pm$0.3} & {1.22$\pm$0.04} & {41$\pm$2}\\
{4.1} & {0.9}   & {20.2\phantom{1}} &   {38.247$\pm$0.006} & {$-$1.5$\pm$0.2} & {0.20$\pm$0.05} & {8.9$\pm$0.3}   & {1.33$\pm$0.04} & {21$\pm$1}\\
{5.1} & {0.5}   & {4.1} &   {38.12$\pm$0.01} & {$-$0.3$\pm$0.3} & {0.4$\pm$0.1} & {9.0$\pm$0.3}     & {1.44$\pm$0.04} & {8.4$\pm$0.8}\\
{6.1} & {1.3}   & {19.7\phantom{1}} &   {38.753$\pm$0.003} & {$-$0.7$\pm$0.2} & {0.17$\pm$0.02} & {8.7$\pm$0.3} & {1.26$\pm$0.02} & {33$\pm$1}\\
{7.1} & {0.3}   & {10.7\phantom{1}} &   {38.254$\pm$0.005} & {$-$0.8$\pm$0.2} & {0.23$\pm$0.04} & {8.9$\pm$0.3} & {1.33$\pm$0.04} & {19.0$\pm$0.8}\\
{7.2} & {3.9}   & {9.4} &   {37.83$\pm$0.01} & {$-$0.8$\pm$0.2} & {0.26$\pm$0.09} & {8.8$\pm$0.3}     & {1.26$\pm$0.04} & {20$\pm$1}\\
{7.3} & {4.1}   & {10.8\phantom{1}} &   {37.88$\pm$0.01} & {$-$1.1$\pm$0.2} & {0.17$\pm$0.08} & {8.9$\pm$0.3}     & {1.28$\pm$0.04} & {17$\pm$1}\\
{8.1} & {0.5}   & {15.7\phantom{1}} &   {38.517$\pm$0.004} & {$-$1.2$\pm$0.2} &{0.21$\pm$0.03}& {8.9$\pm$0.3} & {1.23$\pm$0.04} & {32$\pm$1}\\
{9.1} & {1.8}   & {33.8\phantom{1}} &   {38.286$\pm$0.004} & {$-$1.2$\pm$0.2} & {0.20$\pm$0.03} & {8.8$\pm$0.3} &   {1.25$\pm$0.02} & {28$\pm$1}\\
{9.2} & {5.6}   & {34.0\phantom{1}} &   {38.046$\pm$0.007} & {$-$1.5$\pm$0.2} & {0.16$\pm$0.04} & {8.7$\pm$0.3}   & {1.24$\pm$0.04} & {35$\pm$2}\\
{9.3} & {8.7}   & {34.0\phantom{1}} &   {37.903$\pm$0.008} & {$-$1.3$\pm$0.2} & {0.13$\pm$0.05} & {8.7$\pm$0.3}   & {1.21$\pm$0.04} & {37$\pm$2}\\
{10.1\phantom{1}}  & {0.9} & {21.8\phantom{1}} & {37.807$\pm$0.01} & {$-$0.7$\pm$0.2} & {0.1$\pm$0.1} & {8.8$\pm$0.3} & {1.29$\pm$0.04} & {11$\pm$1}\\
{10.2\phantom{1}}  & {3.3} & {17.5\phantom{1}} & {37.61$\pm$0.02} & {$-$0.8$\pm$0.2} & {0.1$\pm$0.1} & {8.7$\pm$0.3} & {1.2$\pm$0.1} & {15$\pm$2}\\
{11.1\phantom{1}}  & {0.2} & {22.0\phantom{1}} & {38.420$\pm$0.004} & {$-$0.1$\pm$0.2} & {0.39$\pm$0.04}&{9.0$\pm$0.3} & {1.32$\pm$0.04} & {28$\pm$1}\\
{12.1\phantom{1}}  & {2.2} & {25.3\phantom{1}} & {39.186$\pm$0.001} & {$-$1.8$\pm$0.2} & {$0.098\pm0.008$} & {8.5$\pm$0.3} & {1.00$\pm$0.02} & {185$\pm$2}\\
{12.2\phantom{1}}  & {4.8} & {25.3\phantom{1}} & {39.011$\pm$0.001} & {$-$1.7$\pm$0.2}  & {0.107$\pm$0.009} &{8.4$\pm$0.3} & {1.02$\pm$0.02} & {149$\pm$2}\\
{12.3\phantom{1}}  & {1.7} & {25.3\phantom{1}} & {39.335$\pm$0.001} & {$-$1.5$\pm$0.2} & {0.141$\pm$0.005} & {8.5$\pm$0.3} & {1.04$\pm$0.02} & {142$\pm$1}\\
{13.1\phantom{1}}  & {2.9} & {5.8} & {39.298$\pm$0.001} & {$-$0.8$\pm$0.2} & {0.191$\pm$0.006} & {8.7$\pm$0.3} & {1.08$\pm$0.02} & {155$\pm$2}\\
{13.2\phantom{1}}  & {0.3} & {5.8} & {38.971$\pm$0.002} & {$-$1.7$\pm$0.2} & {0.19$\pm$0.01} & {8.8$\pm$0.3}   &   {1.21$\pm$0.02} & {49$\pm$1}\\
{13.3\phantom{1}}  & {2.8} & {7.6} & {39.296$\pm$0.001} & {$-$0.7$\pm$0.2} & {0.231$\pm$0.007} & {8.5$\pm$0.3} & {1.02$\pm$0.02} & {236$\pm$3}\\
{14.1\phantom{1}}  & {0.6} & {15.3\phantom{1}} & {39.064$\pm$0.002} & {\,\,\phantom{1}0.9$\pm$0.2} & {0.45$\pm$0.02} & {8.8$\pm$0.3}    & {1.37$\pm$0.04} & {50$\pm$1}\\
{15.1\phantom{1}}  & {0.3} & {8.9} & {39.206$\pm$0.001} & {$-$1.3$\pm$0.2} & {0.342$\pm$0.008}&{8.9$\pm$0.3} & {1.18$\pm$0.02} & {47$\pm$1}\\
{16.1\phantom{1}}  & {1.1} & {13.0\phantom{1}} & {37.855$\pm$0.002} & {$-$0.5$\pm$0.2} & {0.36$\pm$0.09} & {9.0$\pm$0.3}     &         {1.50$\pm$0.06} & {8.1$\pm$0.8}\\
{16.2\phantom{1}}  & {4.6} & {8.2} & {37.99$\pm$0.01} & {$-$1.4$\pm$0.2} & {0.27$\pm$0.08} & {9.0$\pm$0.3}     &         {1.39$\pm$0.08} & {17$\pm$2}\\
{17.1\phantom{1}}  & {0.3} & {5.8} & {39.068$\pm$0.009} & {$-$0.4$\pm$0.2} & {0.41$\pm$0.01} & {9.1$\pm$0.3}   &         {1.32$\pm$0.02} & {35$\pm$1}\\
{18.1\phantom{1}}  & {0.9} & {16.2} & {38.222$\pm$0.002} & {$-$0.7$\pm$0.3} & {0.70$\pm$0.04} & {9.0$\pm$0.3}  &          {1.45$\pm$0.04} & {19.4$\pm$0.8}\\
{19.1\phantom{1}}  & {0.9} & {29.9\phantom{1}} &{38.939$\pm$0.004} & {$-$1.1$\pm$0.2} & {0.45$\pm$0.01} &{9.1$\pm$0.3} & {1.24$\pm$0.02} & {43$\pm$1}\\
{20.1\phantom{1}}  & {1.0} & {8.4} & {38.452$\pm$0.002} & {$-$1.4$\pm$0.2} & {0.30$\pm$0.04}&{9.0$\pm$0.3} &             {1.27$\pm$0.04} & {28$\pm$1}\\
{20.2\phantom{1}}  & {0.7} & {8.4} & {38.370$\pm$0.005} & {$-$0.8$\pm$0.2} & {0.27$\pm$0.05} & {9.1$\pm$0.3} &           {1.35$\pm$0.02} & {14.2$\pm$0.8}\\
{20.3\phantom{1}}  & {1.2} & {8.3} & {38.381$\pm$0.006} & {$-$0.9$\pm$0.2} & {0.29$\pm$0.06} & {9.0$\pm$0.3} &           {1.29$\pm$0.04} & {23$\pm$2}\\
{21.1\phantom{1}}  & {0.5} & {11.0\phantom{1}} & {38.840$\pm$0.007} & {$-$0.9$\pm$0.2} & {0.44$\pm$0.07} & {9.1$\pm$0.3}   &           {1.54$\pm$0.04} & {7.5$\pm$0.5}\\
{22.1\phantom{1}}  & {1.4} & {39.0\phantom{1}} & {38.260$\pm$0.002} & {$-$1.2$\pm$0.2} & {0.41$\pm$0.01} & {9.0$\pm$0.3}&{1.21$\pm$0.02} & {40$\pm$1}\\
{23.1\phantom{1}}  & {0.7} & {14.2\phantom{1}} & {38.260$\pm$0.006} & {\phantom{1}0.0$\pm$0.2} & {0.38$\pm$0.05} & {9.1$\pm$0.3}  &  {1.47$\pm$0.02} & {9.1$\pm$0.4}\\
{23.2\phantom{1}}  & {2.3} & {14.2\phantom{1}} & {38.115$\pm$0.007} & {$-$0.7$\pm$0.2} & {0.34$\pm$0.06} & {9.1$\pm$0.3} &           {1.41$\pm$0.04} & {11.1$\pm$0.6}\\
{23.3\phantom{1}}  & {5.1} & {14.2\phantom{1}} & {38.109$\pm$0.007} & {$-$0.6$\pm$0.2} & {0.21$\pm$0.05} & {8.9$\pm$0.3} &           {1.30$\pm$0.02} & {26$\pm$2}\\
{23.4\phantom{1}}  & {5.4} & {14.2\phantom{1}} & {38.496$\pm$0.003} & {$-$1.2$\pm$0.2} & {0.13$\pm$0.02} & {8.8$\pm$0.3}&            {1.15$\pm$0.02} & {44$\pm$2}\\
{24.1\phantom{1}}  & {5.5} & {23.1\phantom{1}} & {38.386$\pm$0.007} & {$-$0.8$\pm$0.2} & {0.36$\pm$0.05} & {9.0$\pm$0.3}&{1.27$\pm$0.06} & {26$\pm$2}\\
{24.2\phantom{1}}  & {3.7} & {23.3\phantom{1}} & {38.427$\pm$0.006} & {$-$0.8$\pm$0.2} & {0.35$\pm$0.05} & {9.1$\pm$0.3}&{1.27$\pm$0.06} & {27$\pm$2}\\
{25.1\phantom{1}}  & {4.4} & {8.2} & {38.253$\pm$0.007} & {$-$1.3$\pm$0.2} & {0.37$\pm$0.06} & {9.0$\pm$0.3} & {1.22$\pm$0.04} & {38$\pm$2}\\
{25.2\phantom{1}}  & {6.4} & {8.2} & {39.984$\pm$0.002} & {$-$1.1$\pm$0.2} & {0.80$\pm$0.03} & {8.9$\pm$0.3} & {1.21$\pm$0.04} & {113$\pm$2}\\
{25.3\phantom{1}}  & {3.4} & {11.9\phantom{1}} & {39.747$\pm$0.003} & {$-$0.9$\pm$0.2} & {0.47$\pm$0.03} & {9.1$\pm$0.3} & {1.27$\pm$0.02} & {36$\pm$1}\\
{25.4\phantom{1}}  & {0.7} & {8.2} & {39.683$\pm$0.003} & {$-$0.1$\pm$0.2}&{0.52$\pm$0.03} & {9.1$\pm$0.3} &   {1.37$\pm$0.02} & {16$\pm$1}\\
{25.5\phantom{1}}  & {5.8} & {8.1} & {38.464$\pm$0.004} & {$-$0.9$\pm$0.2} & {0.33$\pm$0.04} & {9.0$\pm$0.3} & {1.25$\pm$0.04} & {40$\pm$2}\\
{25.6\phantom{1}}  & {6.1} & {8.2} & {38.419$\pm$0.005} & {$-$1.1$\pm$0.2} & {0.32$\pm$0.04} & {9.0$\pm$0.3} & {1.21$\pm$0.04} & {39$\pm$2}\\
{26.1\phantom{1}}  & {6.4} & {9.8} & {38.847$\pm$0.003} & {$-$1.5$\pm$0.2} & {0.23$\pm$0.05} & {8.9$\pm$0.3}   & {1.19$\pm$0.02} & {46$\pm$2}\\
{26.2\phantom{1}}  & {4.9} & {9.5} & {39.748$\pm$0.001} & {$-$1.4$\pm$0.2} & {0.257$\pm$0.004} & {8.8$\pm$0.3} & {1.05$\pm$0.02} & {217$\pm$2}\\
{26.3\phantom{1}}  & {6.3} & {12.7\phantom{1}} & {39.469$\pm$0.001} & {$-$1.4$\pm$0.2} & {0.262$\pm$0.006} & {8.8$\pm$0.3} & {1.08$\pm$0.02} & {152$\pm$2}\\
{26.4\phantom{1}}  & {7.8} & {9.5} & {39.110$\pm$0.002} & {$-$1.5$\pm$0.2} & {0.27$\pm$0.01} & {8.9$\pm$0.3} & {1.12$\pm$0.02} & {54$\pm$1}\\
{26.5\phantom{1}}  & {8.0} & {9.5} & {38.965$\pm$0.002} & {$-$1.5$\pm$0.2} & {0.30$\pm$0.02} & {8.9$\pm$0.3}   &   {1.19$\pm$0.02} & {78$\pm$2}\\
{27.1\phantom{1}}  & {9.0} & {65.8\phantom{1}} & {37.53$\pm$0.02} & {$-$0.3$\pm$0.2} & {0.4$\pm$0.1} & {9.1$\pm$0.3}   &     {1.50$\pm$0.08} & {11$\pm$1}\\
{28.1\phantom{1}}  & {1.2} & {56.3\phantom{1}} & {39.243$\pm$0.001} & {\phantom{1}0.1$\pm$0.2} & {0.58$\pm$0.01} & {9.1$\pm$0.3}  & {1.28$\pm$0.02} & {37$\pm$1}\\
{29.1\phantom{1}}  & {1.3} & {67.1\phantom{1}} & {38.286$\pm$0.006} & {$-$1.3$\pm$0.2} & {0.53$\pm$0.06} & {9.0$\pm$0.3} & {1.38$\pm$0.02} & {15.4$\pm$0.4}\\
\hline
	\end{tabular}
	\vspace{-4mm}
\footnotetext[1]{Distance is the difference between the nuclear coordinate of the galaxy from Table~\ref{table:tgt_prop} and the centre of the region's aperture.}\vspace{-2mm}
\footnotetext[2]{The systematic uncertainty in emission line flux as described by \citet{Belfiore2019} is included in the reported error. See Section~\ref{ssec:basic_dr} for details.}
\footnotetext[3]{Metallicity is calculated using the ([\ion{O}{3}]/H$\beta$)/([\ion{N}{2}]/H$\alpha$) ratio as described in as described in \citet[][]{Pettini2004}, and converted to \citet{Tremonti04} via \citet{Kewley08}. The error bar includes the 0.3~dex uncertainty in the \citet{Pettini2004} relation, which is the largest uncertainty associated with the measurement. See Section~\ref{sec:phys} for details.}\vspace{-2mm}
\footnotetext[4]{The systematic uncertainty in D$_n$(4000) as described by \citet{Westfall2019} is included in the reported error. See Section~\ref{ssec:basic_dr} for details.}
\end{minipage}
\end{table*}

\subsection{DIG-dominated regions}\label{dig}
Of the 56 regions we identified, 14 are DIG-dominated according to the description above. For these regions, the $\langle\textrm{EW(H}\alpha)\rangle$ is about 17~\AA\ but they are ``star forming'' according to the BPT line ratios, and have a similar scatter in $\tau^l_B$ to the star forming and starburst-like regions. Kolmogorov-Smirnov (K-S) two-sample tests for each grouping of the three $\tau^l_B$ populations fail to show that the populations are statistically distinct, with p-values 0.59--0.79. However, the DIG-dominated regions have systematically higher values of $\beta$; K-S two-sample and Anderson-Darling tests \citep{anderson_darling} show distinct populations at the 95\% significance level. The discrepant behavior in $\beta$ and $\tau^{l}_{B}$ is likely because the UV light in the DIG-dominated regions is not dominated by massive stars, making $\beta$ an unreliable tracer of the stellar continuum colour excess. As such, we exclude the DIG-dominated regions from future analysis.

\subsection{Demographics of Star Forming Regions}
Of the 42 remaining regions, 95\% are defined as star forming and 5\% as starburst-like. These regions have, larger $\langle\textrm{EW(H}\alpha)\rangle$ than the DIG-dominated regions, $\sim$58~\AA\ for the star forming, and $\sim$130~\AA\ for the starburst-like, and are quite similar to those from B16 ($\textrm{EW}\sim40$~\AA) and \citet[][$\textrm{EW}\sim110$~\AA]{Calzetti1994}, respectively. These regions have physical areas that span a range of 4.7--67.1~kpc$^2$ with an average of 17.4~kpc$^2$ and a median of 13~kpc$^2$. Most of the regions are truly kpc-sized, as 80\% have areas less than 25~kpc$^2$. B16 adopted a 4.\!\!$^{\prime\prime}$5 aperture and probed a similar redshift range so their regions are approximately the same size as those in our study.  

Therefore this work builds on that of B16 by studying \textit{resolved} regions across the faces of galaxies. We will begin by establishing consistency between the regions studied by B16 and those in this work, and employ statistical tests to characterize any dependence on the physical properties of both the resolved regions and the integrated galaxy light. Finally we will explore the dependence on aperture size by comparing the resolved regions to the integrated galaxy light, which was a test not possible in the B16 study.

%B16 model comparison
%%%%%%%%%%%%%%%%%%%%%%%%%%%%%%%%%%%%%%%%%%%%%%%%%%%%%%%%%%%%%%%%%%%%%%%%%%%%%%%%

\section{Comparison to the B16 Model}
\label{sec:analysis}

\begin{table}
\begin{minipage}{\linewidth}
\renewcommand{\thefootnote}{\textrm{\alph{footnote}}}
	\centering
		\setlength{\tabcolsep}{10pt}
	\caption{Values for $\tau^l_B$ bins\label{table:scatter}}
	\begin{tabular}{ccccc} % four columns, alignment for each
		\hline
		\hline
		{Bin Size} & {N\footnotemark[1]}  & {$\langle\tau^l_B\rangle$\footnotemark[2]} & {$\langle\beta\rangle$\footnotemark[2]} & {$\sigma(\beta)$\footnotemark[3]}\\
		\hline
		{\hspace{10pt}$0.0\leq\tau^l_B<0.195$} & {10} & {$0.14$} & {$-1.5$} &{$0.4$} \vspace{0.3mm}\\
		{$0.195\leq\tau^l_B<0.27$} & {11} & {$0.23$} & {$-1.2$} &{$0.3$}\vspace{0.3mm} \\
		{$0.27\leq\tau^l_B<0.4$} & {11} & {$0.34$} & {$-1.0$} &{$0.4$}\vspace{0.3mm} \\
		{\hspace{8pt}$0.4\leq\tau^l_B<0.81$} & {10} & {$0.53$} & {$-0.6$} &{$0.7$}\vspace{0.3mm} \\
		\hline \vspace{-6mm}
	\end{tabular}
\footnotetext[1]{N is the number of regions within the $\tau^l_B$ bin.}
\footnotetext[2]{The reported values are the means of $\tau^l_B$ and $\beta$ in each bin.}\vspace{-2mm}
\footnotetext[3]{$\sigma(\beta)$ is the $1\sigma$ dispersion of the $\beta$ values in each bin.}\vspace{-2mm}
\end{minipage}
\end{table}

\begin{figure}
\hspace{0mm}\includegraphics[width=0.45\textwidth]{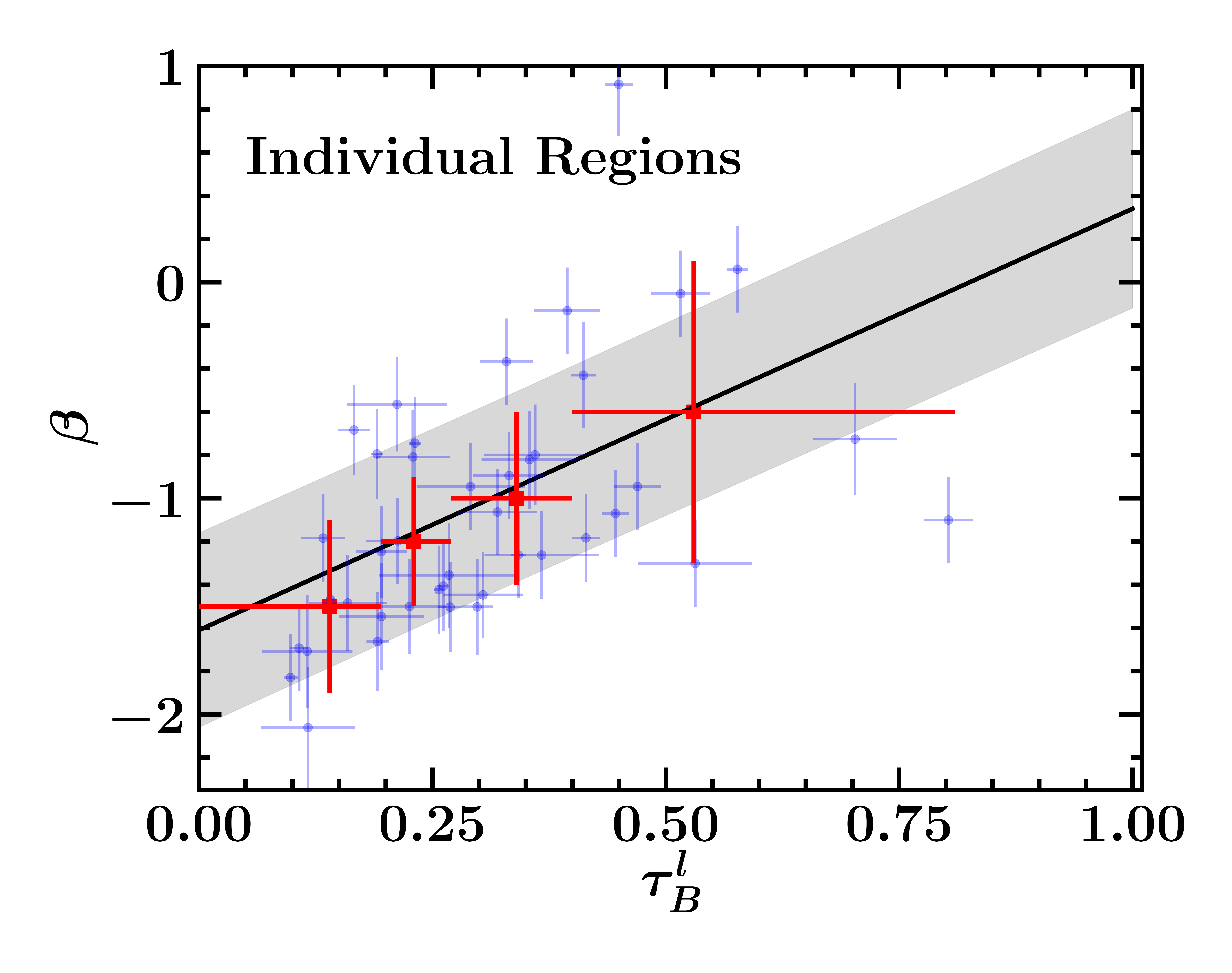}\vspace{-4mm}
\caption{Same as Figure~\ref{fig:reg_btau}, but without the DIG-dominated regions (see discussion in Section~\ref{dig}.) The light blue circles are the measurements for the individual regions, and the red squares are the mean and associated sample dispersion for the binned data. The error bars on $\tau^l_B$ represent the size of the bin, while those on $\beta$ represent the 1$\sigma$ dispersion within the bin. All bins are statistically consistent with the B16 relation.}
\label{fig:scatter}
\end{figure}

\subsection{Intrinsic Scatter for Individual Regions}
To examine the relationship between $\beta$ and $\tau^l_B$, we binned the data with respect to $\tau^l_B$, and calculated the unweighted mean for $\beta$ and $\tau^l_B$, and the unweighted $1\sigma$ sample dispersion in $\beta$ for each bin. The bins were defined to hold approximately equal numbers of data points and have a similar size to those used in B16. The results are reported in Table~\ref{table:scatter}. The mean values and associated sample dispersions are plotted against the B16 relation in Figure~\ref{fig:scatter}. All four bins are statistically consistent with the B16 relation. The third bin is dominated by regions with $1.1\leq\textrm{D}_n(4000)\leq1.3$, whose importance will be described in Section~\ref{ssec:d4000}. The bin with the highest $\tau^l_B$ values is sparsely populated. While more regions are needed to properly constrain the intrinsic scatter for $\tau^l_B>0.4$, they are not easily detectable due to the high levels of extinction. 

\subsection{Stellar Age}
\label{ssec:d4000}
\begin{figure}
\hspace{0mm}
\includegraphics[width=0.45\textwidth]{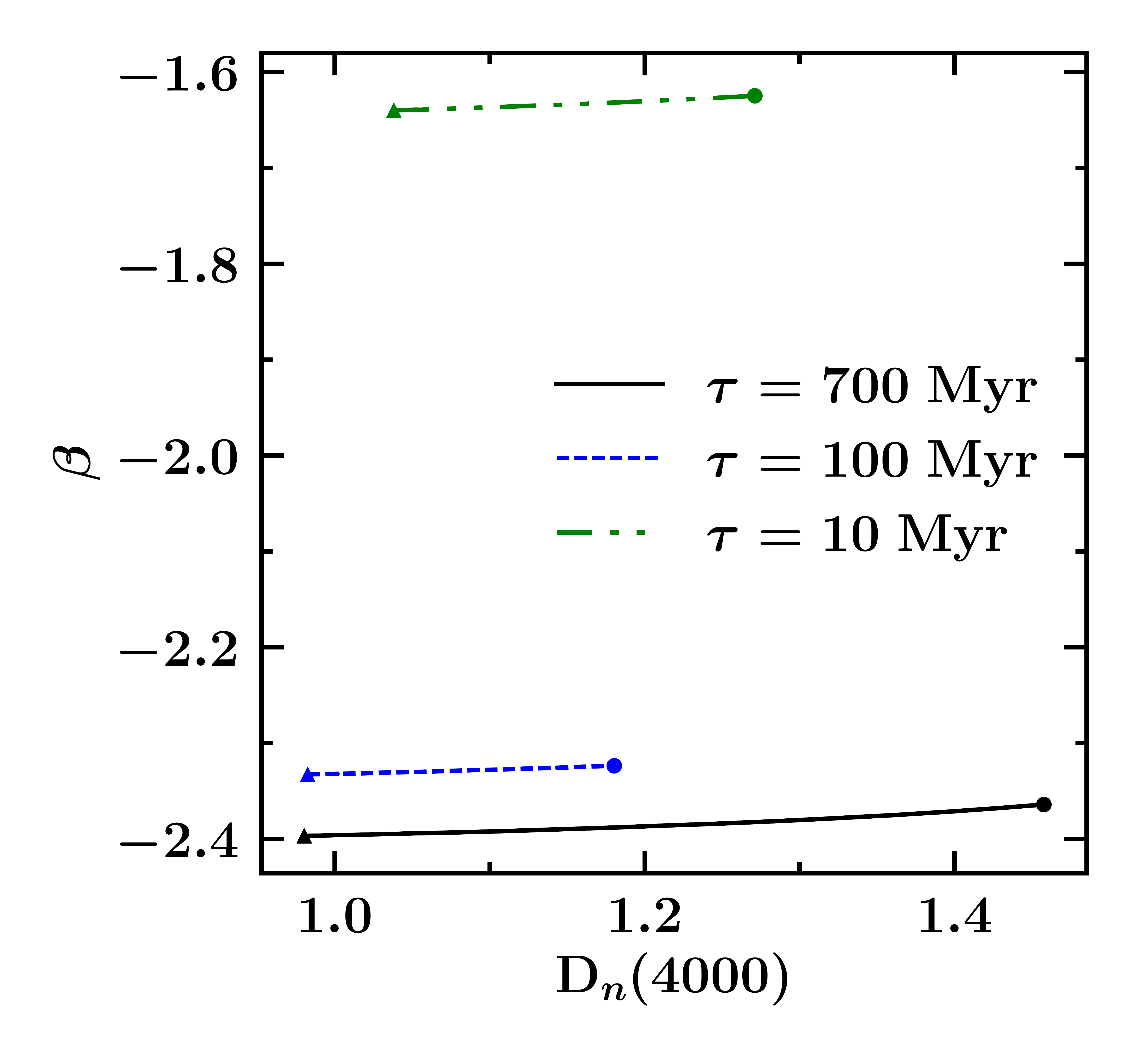}\vspace{-4mm}
\caption{$\beta$ vs.~D$_n$(4000) for the simulated multi-burst models described in Section~\ref{ssec:d4000}. In each case the old population is modeled with an exponentially decaying SFH and an e-folding time-scale of 700~Myr. This is combined with a younger burst with an exponentially decaying SFH and an e-folding time of either 700~Myr (black sold line), 100~Myr (blue dashed line) or 10~Myr (green dot-dashed line). The second burst is placed 10~Gyr after the first burst, and is computed with different mass ratios of the older to younger populations, ranging from 1:1 (filled triangle) to 100:1 (filled circle). We plot the values for $\beta$ and D$_n$(4000) 100~Myr after the second burst. The effect of the older stellar population on $\beta$ is minimal compared to the assumed SFH. Given the strong H$\alpha$ detections, implying recent star formation, we assume $\beta$ is a robust tracer of the stellar continuum attenuation.}
\label{fig:bd4000}
\end{figure}

As mentioned in Section~\ref{ssec:bump}, the value of D$_n(4000)$ is related to the relative contribution of the older and younger stellar populations, and is correlated with the measured slope between $\beta$ and $\tau^{l}_{B}$. Specifically, as the D$_n(4000)$ of the integrated light increases, so does the light from older stars. The result is a steeper slope as seen in Figure 20 of B16. We assessed this effect by using a combination of \citet[][BC03]{Bruzual2003} models. In order to simulate a star forming region with a strong contribution from an older stellar population, we created a multi-burst model by evolving the solar metallicity simple stellar population models. The first burst has an exponentially declining SFH with an e-folding time-scale of 700~Myr. The second burst also has an exponentially declining SFH, with a variable e-folding time-scale (700, 100 or 10~Myr). We separated the bursts by 10~Gyr and varied the relative mass ratio between the older and younger populations from 1:1 to 100:1. We show the results at $t=100$~Myr after the second burst for each of the three described SFHs in Figure~\ref{fig:bd4000}. The intrinsic $\beta$ in each case increases by $\sim0.03$ between these two older to younger mass ratios; this is much smaller than the 1$\sigma$ sample dispersion in $\beta$. However, the initial $\beta$ and D$_n$(4000) are strongly tied to the current SFR and SFH, as clearly demonstrated by Figure~\ref{fig:bd4000} (see Section~\ref{ssec:sfh} for a discussion of the assumed intrinsic $\beta$). This result corroborates the minimal effects of the older stellar population on the $\beta$--$\tau^l_B$ for D$_n$(4000)$\lesssim1.5$ presented in B16.
 
In addition to D$_n(4000)$, we measured the EW(H$\alpha$) for all the identified kpc-sized regions, a proxy for the specific SFR (sSFR), as well as an indicator of the contribution from the older stellar population. All of the identified regions have EW(H$\alpha) > 15$~\AA, a value 2.5 times larger than that used to classify kpc-sized regions as star forming \citep{Sanchez2014,Wang2018}. We will therefore assume for the rest of the paper that the measured $\beta$ values are not significantly affected by UV light from the older stellar population.

\begin{figure}
\hspace{-4mm}
\includegraphics[width=0.45\textwidth]{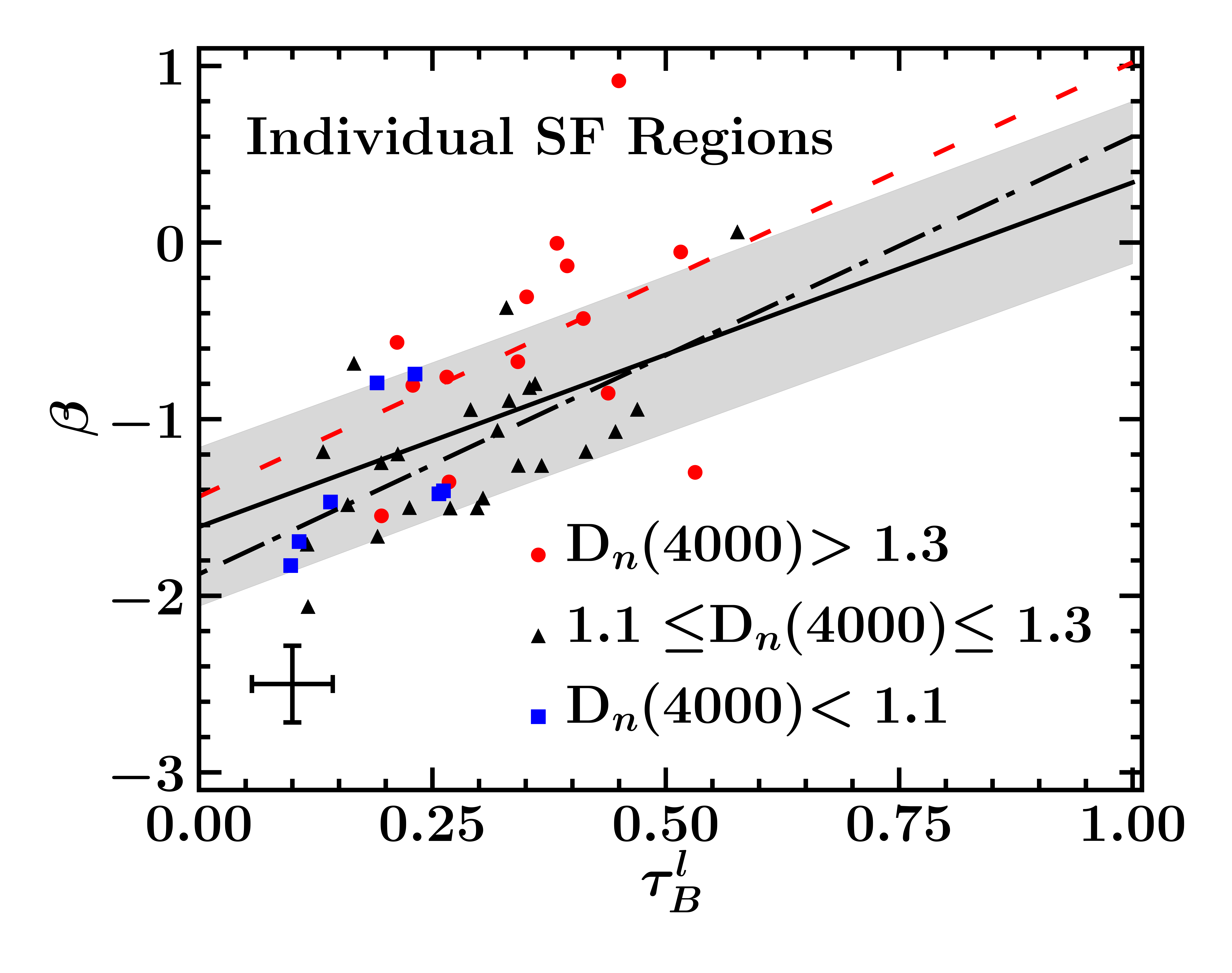}\vspace{-4mm}
\caption{Same as Figure~\ref{fig:scatter}, but colour coded according to the measured D$_n$(4000). The red circles represent regions with D$_n$(4000)$>1.3$, black triangles with $1.1\leq\textrm{D}_n(4000)\leq 1.3$, and blue squares with D$_n$(4000)$<1.1$. The B16 relation was fit only to galaxies with measured $1.1\leq\textrm{D}_n(4000)\leq 1.3$. The black dot-dashed line represents the fit to the regions in this work with the same D$_n$(4000) constraints as B16, and the fits are very similar. The red dashed line is the fit to the regions in this work with D$_n$(4000)$>1.3$. The fit to the older population has a higher intercept and steeper slope than that of B16. We do not attempt to fit the regions with D$_n$(4000)$<1.1$ because of the small number of such regions.}
\label{fig:reg_d4000}
\end{figure}

The individual regions exhibit a trend with stellar age similar to that found by B16, as shown in Figure~\ref{fig:reg_d4000}. The B16 relation is displayed, as well as the least squares linear fits to our data for the regions with D$_n(4000)>1.3$ and $1.1\leq\textrm{D}_n(4000)\leq1.3$. The latter corresponds to the values used for the B16 relation, and, within the errors, are well fitted by the slope and intercept derived by B16. 

Regions with larger D$_n(4000)$ measurements require a steeper slope and higher intercept than that of B16. A K-S two-sample test comparing $\beta$ for the two D$_n(4000)$ bins shows that the distributions are statistically distinct at the 99\% confidence interval. The physical environment may be different for young, dusty systems, but we lack sufficient data in the low D$_n(4000)$ bin to provide any quantitative explanation for their behavior. We conclude that the B16 law is only applicable to regions with a measured D$_n(4000)$ between 1.1 and 1.3. Finally, we see no significant trends with EW(H$\alpha$), confirming that our $\Sigma_{\textrm{H}\alpha}$ and EW(H$\alpha$) criteria are robust indicators of star forming regions.

\subsection{Dependence of $\beta$ and $\tau^l_B$ on Metallicity and SFR}\label{sec:phys}
 In addition to deriving the $\beta$--$\tau^l_B$ relation, B16 fitted second order polynomials to the $\beta$ and $\tau^l_B$ measurements with respect to different physical properties, including metallicity, SFR and $\Sigma_{\textrm{SFR}}$. We compare the measured properties for the regions in our sample to those fitted relations given by B16. In order to have a fair comparison, we calculated metallicities for our star forming regions using the ([\ion{O}{3}]/H$\beta$)/([\ion{N}{2}]/H$\alpha$) ratio as described in \citet[][]{Pettini2004}, and converted to the \citet{Tremonti04} system via the prescription of \citet{Kewley08}. The SFR and $\Sigma_{\textrm{SFR}}$ values were measured using the H$\alpha$ luminosity as described in \citet[][]{Kennicutt2012}. The SFR used in B16 came from \citet[][]{Brinchmann2004}, who also used H$\alpha$ to estimate a global SFR for SDSS galaxies. 
 
   \begin{figure}
\hspace{-3mm}
\includegraphics[width=0.45\textwidth]{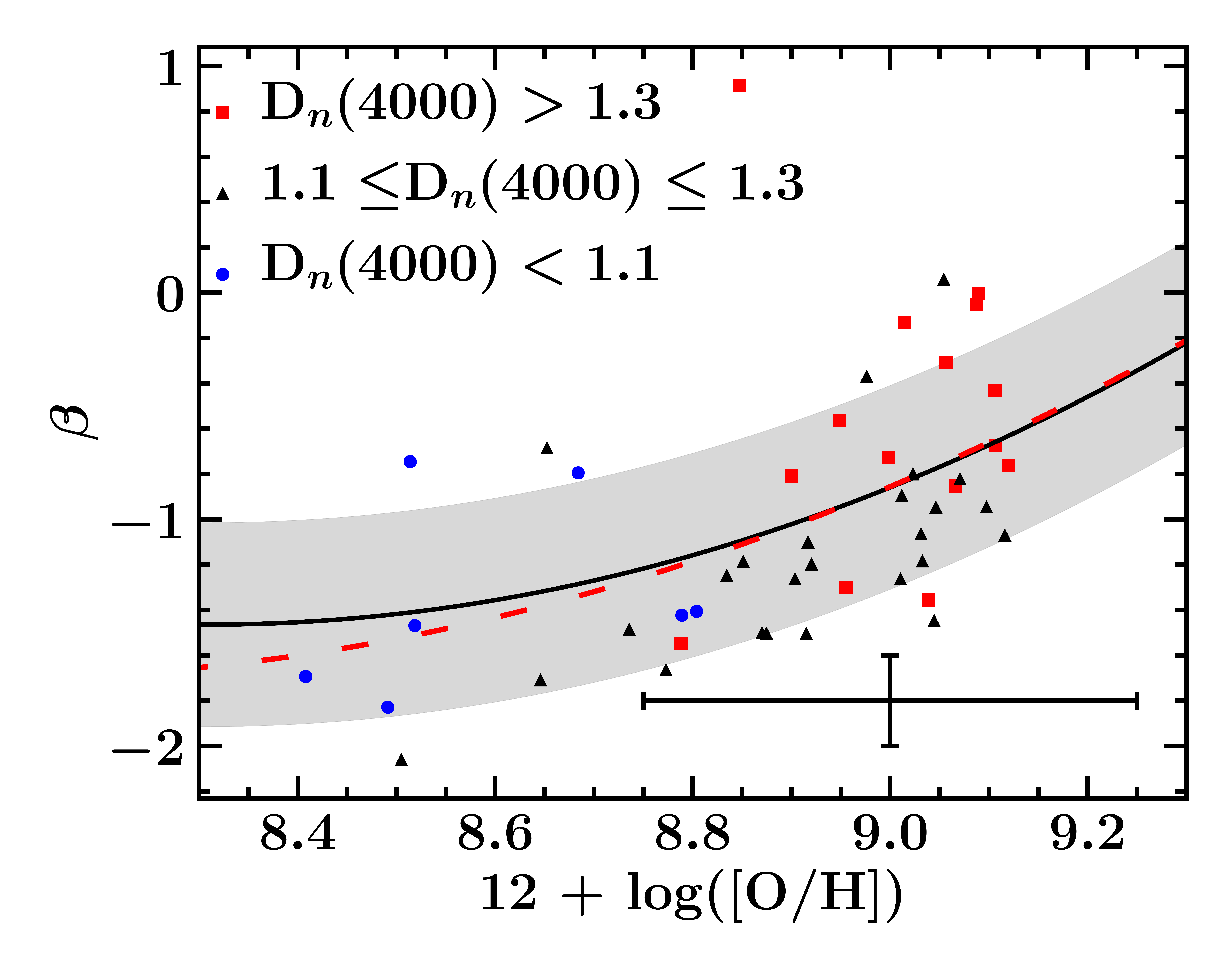}\vspace{-3mm}
\caption{The measured $\beta$ vs.~metallicity for individual star forming regions. The solid line and gray shaded region are the fitted model and scatter from B16. The red dashed line is the fit to the data from this work. The metallicity is converted from the \citet{Pettini2004} system to that of \citet{Tremonti04} for consistency. The uncertainty in the metallicity conversion and $\beta$ are indicated by the error bar in the bottom right. We approximately recover the B16 relation.}
\label{fig:reg_bmet}
\end{figure}
\begin{figure}
\hspace{-3mm}\includegraphics[width=0.45\textwidth]{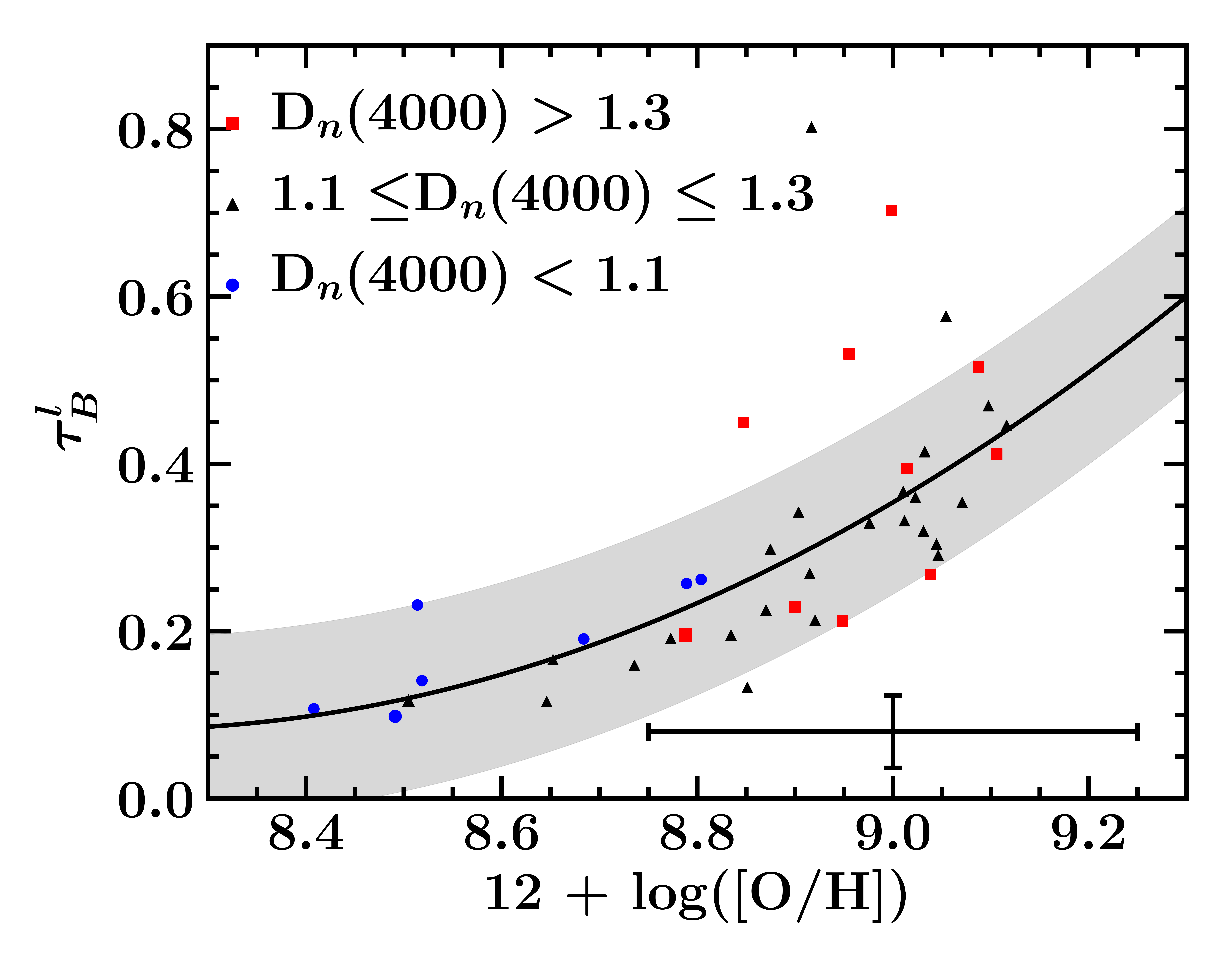}\vspace{-3mm}
\caption{Same as Figure~\ref{fig:reg_bmet}, but for $\tau^l_B$. The data agree well with the model for $\tau^l_B\lesssim0.5$.}
\label{fig:reg_tmet}
\end{figure}

Figures~\ref{fig:reg_bmet} and \ref{fig:reg_tmet} compare the measured metallicities to the fitted B16 relations for $\beta$ and $\tau^l_B$. We note  that unlike the $\beta$--$\tau^l_B$ relation, the metallicity, SFR, and $\Sigma_{\rm SFR}$ fits presented in B16 include {\it all} regions regardless of their D$_n$(4000) measurement. We fitted a second order polynomial to the $\beta$--metallicity relation and include all data, regardless of the D$_n$(4000) measurement, as shown in Figure~\ref{fig:reg_bmet}. The resulting fit for $\beta$ is almost identical to that of B16, albeit with a larger intrinsic scatter (0.51 in $\beta$). The regions with D$_n(4000)>1.3$ have higher metallicity and drive the overall fit. We conclude that, unlike the $\beta$--$\tau^l_B$ relation, the relation between nebular metallicity and $\beta$ seen in B16 is strongly influenced by the regions with D$_n(4000)>1.3$.
 
Our data are also consistent with the $\tau^l_B$--metallicity relation from B16 for $\tau^l_B\lesssim0.5$. Above that threshold the measured metallicity underestimates $\tau^l_B$. It is unclear whether this behavior is due to the inability of the B16 model to predict high values of $\tau^l_B$ or due to systematic errors when converting the metallicities. However, caution should be used when applying this model for high $\tau^l_B$ systems. Overall, metallicity, as estimated from the strong lines of [\ion{O}{3}] and [\ion{N}{2}], appears to be a valid indicator for both $\beta$ and $\tau^l_B$ in low dust-content star forming regions, regardless of the D$_n(4000)$ measurement. 

\begin{figure}
\hspace{-3mm}\includegraphics[width=0.45\textwidth]{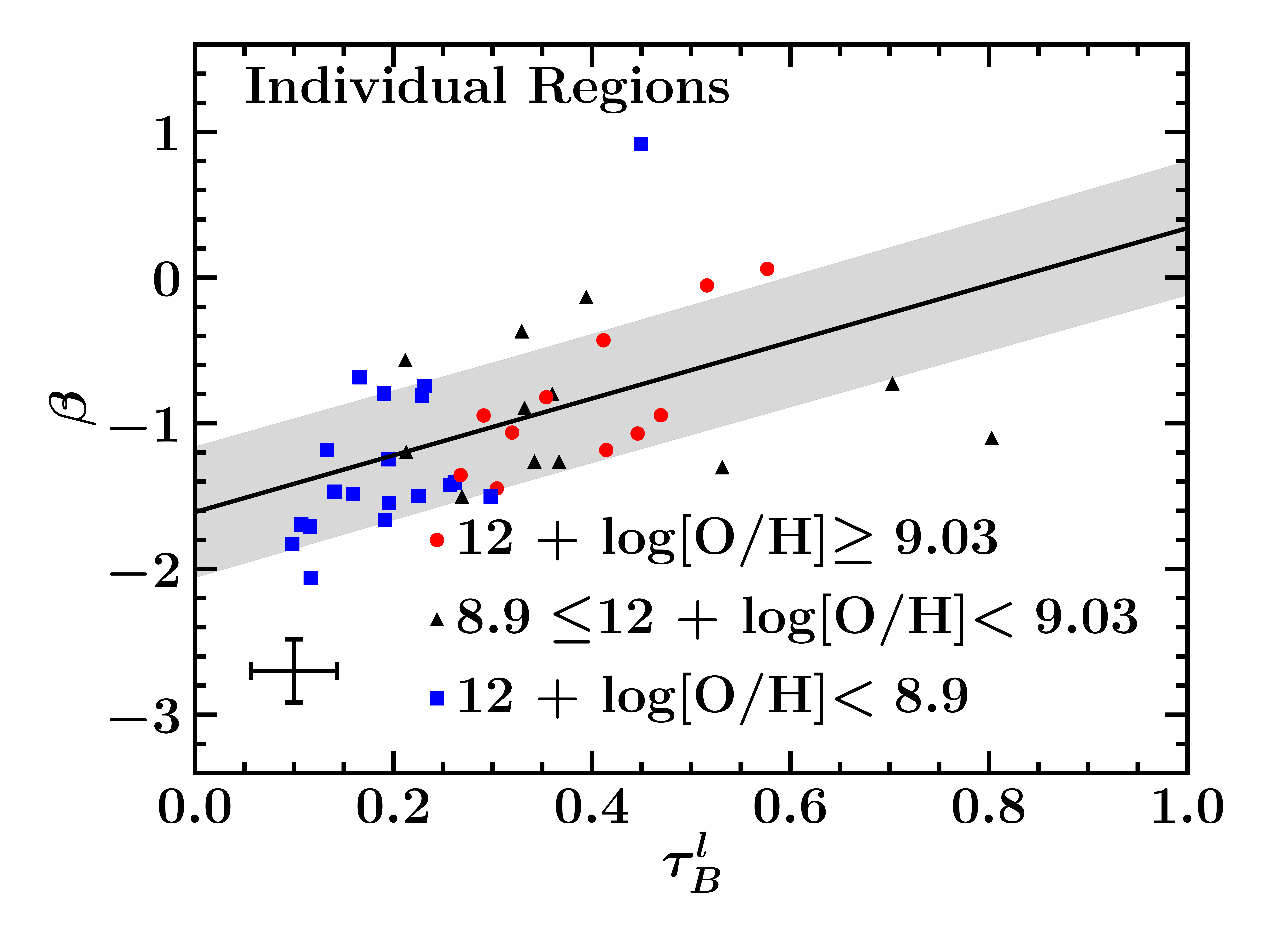}\vspace{-4mm}
\caption{Same as Figure~\ref{fig:reg_d4000}, but colour coded according to the metallicity for the individual regions. The metallicity is converted from the \citet{Pettini2004} system to that of \citet{Tremonti04} for consistency. The B16 relation describes the data regardless of metallicity.}
\label{fig:btau_met}
\end{figure}

Unfortunately, these relations between dust content and metallicity are subject to strong systematic effects, such as the viewing angle or total gas column density, which can vary independent of metallicity. Thus we also looked at the scatter of the individual regions about the $\beta$--$\tau^l_B$ relation with respect to metallicity, as shown in Figure~\ref{fig:btau_met}. The B16 law adequately describes the data regardless of metallicity. This lack of dependence on metallicity is consistent with the findings of \citet{Salim2018}.

B16 also found relations similar to those of Figures~\ref{fig:reg_bmet} and \ref{fig:reg_tmet}, but for SFR and $\Sigma_{\textrm{SFR}}$. We looked for the same relations in our star forming regions, using the SFR calculated from the H$\alpha$ emission line and the relation given in \citet[][]{Kennicutt2012}. We find very poor agreement between the properties of our individual star forming regions and the $\beta$--SFR, $\tau^l_B$--SFR, $\beta$--$\Sigma_{\textrm{SFR}}$ and $\tau^l_B$--$\Sigma_{\textrm{SFR}}$ relations described in B16. We conclude that these physical properties are not useful in constraining either $\beta$ or $\tau^l_B$.

\subsection{Distance from Galaxy Centre}
To test whether galactocentric radius plays a role in the $\beta$--$\tau^l_B$ relation, we calculated the distance between each galaxy's nucleus, as given by the MaNGA collaboration, and the centre of the apertures used to measure its star forming regions. The majority of the older regions, as probed by D$_n(4000)$, are near the nucleus as expected; however, we see no statistically significant change between the measured $\beta$, $\tau^l_B$ or metallicity between nuclear (within $\sim2$~kpc of galaxy centre) and off-nuclear regions at the 95\% confidence level. The marginal p-value for $\tau^l_B$, 0.09, is most likely due to the difference in total dust content, which falls exponentially with distance from the nucleus \citep{peletier1995,giovanelli1994}. Therefore, any change seen in the $\beta$--$\tau^l_B$ relation with distance from galaxy centre is most likely a direct consequence of the change in total dust content, and is not statistically significant. 
 
%Local/Global Properties comparison
%%%%%%%%%%%%%%%%%%%%%%%%%%%%%%%%%%%%%%%%%%%%%%%%%%%%%%%%%%%%%%%%%%%%%%%%%%%%%%%%
\section{Dependence of Local Attenuation on Galaxy-Scale and Local Properties}
\label{sec:all}
\subsection{Dependence on Large-Scale Properties}
\label{ssec:global}
The total dust content, metallicity, and SFR all scale with galaxy mass, and all are associated with changes in the intrinsic attenuation curve \citep[e.g.,][]{Tremonti04,Calzetti2000,Calzetti2001,Wild2011,Xiao2012,Salim2018}. Any scatter between the kpc-sized regions and the B16 model could therefore be the result of the different physical environments present within the galaxies in our sample. To test this hypothesis, we examined the effect of the \textit{integrated galaxy} stellar mass and SFR on the $\beta$--$\tau^l_B$ relation for the \textit{individual} star forming regions. Here we define the ``integrated'' emission as that within the IFU footprint on the galaxy, as described in Section~\ref{ssec:tcalc}, which may not perfectly represent the global galaxy value. We also compare the $\beta$ and $\tau^l_B$ measurements between the integrated galaxy light and individual star forming regions in Section~\ref{ssec:gl_lo}.
\subsubsection{Stellar Mass}

\begin{figure}
\hspace{-4mm}\includegraphics[width=0.45\textwidth]{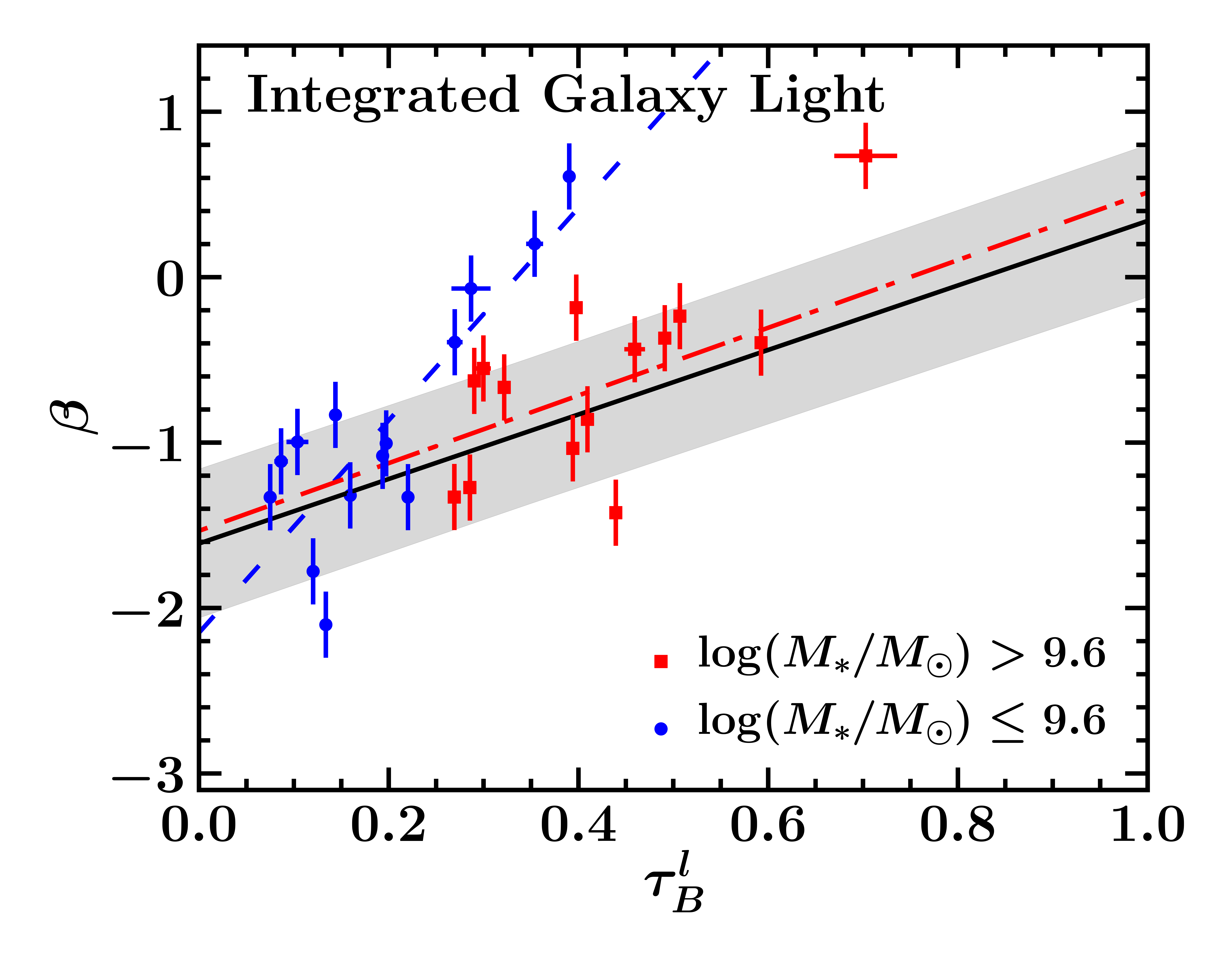}\vspace{-4mm}
\caption{The $\beta$ vs.~$\tau^l_B$ measurements from the integrated galaxy light of the objects in our sample, colour-coded according to the stellar mass of the galaxy. The solid black line and gray shaded region are the B16 relation and sample dispersion, respectively. The blue dashed line is the least squares linear fit to the galaxies to the lowest stellar mass bin [$\log(M_*/M_{\odot})\lesssim 9.6$] in this sample. The red dot-dashed line is the fit to the galaxies in the high stellar mass bin. The lowest mass bin has a significantly steeper slope than that measured for the high mass bin.}
\label{fig:int_mass}
\end{figure}
\begin{figure}
\hspace{-4mm}
\includegraphics[width=0.45\textwidth]{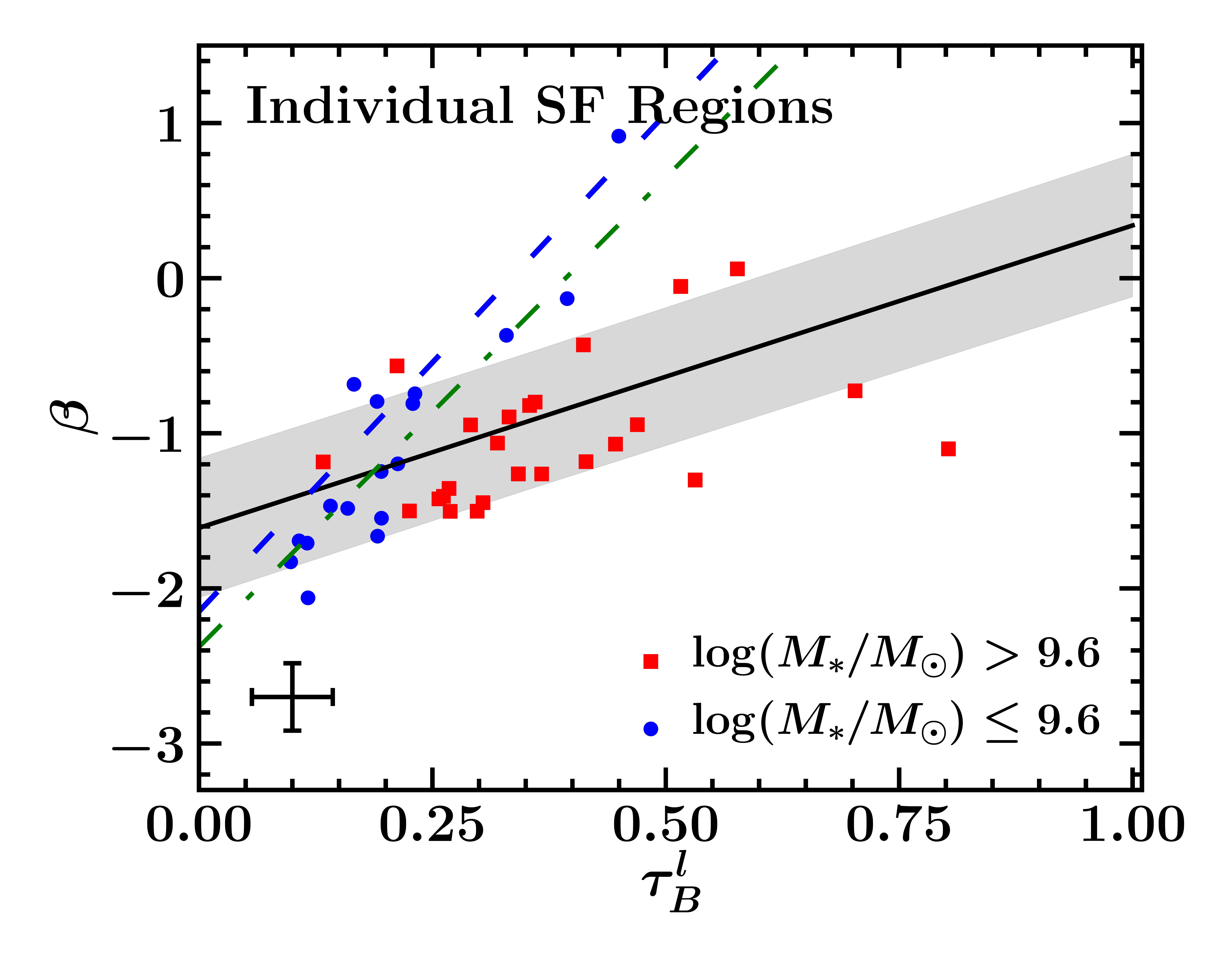}\vspace{-4mm}
\caption{$\beta$ vs.~$\tau^l_B$ measurements from the individual star forming regions colour-coded according to the stellar mass of the parent galaxy. The solid black line and gray shaded region are the B16 relation and sample dispersion, respectively. The blue dashed line is the fit to the integrated light of low-mass galaxies, as shown in Figure~\ref{fig:int_mass}. The green dot-dashed line is the least squares fit to the individual regions of low-mass galaxies \textit{excluding} the one outlier at $\beta\approx1$. The error bar on the lower left represents the characteristic uncertainty in $\beta$ and $\tau^l_B$.}
\label{fig:reg_mass}
\end{figure}

\citet[][]{Salim2018} concluded that the UV slope of the attenuation curve for the integrated light steepens in lower-mass galaxies, which they associate with a lower dust content. A similar trend with stellar mass \citep[or metallicity, as the two properties are related; see][]{Tremonti04} is present in our $\beta$--$\tau^l_B$ measurements for the integrated galaxy light (see Figure~\ref{fig:int_mass}). Given our small sample size, we separate our objects into two mass bins, those above and below $\log(M_*/M_\odot)=9.6$; this definition is roughly consistent with the commonly used division between dwarf and high mass galaxies ($\log[M_*/M_\odot]\sim9.5$). The observed trend in our data persists when galaxies with high inclination angles ($b/a < 0.42$) are excluded \citep[i.e., those where the UV bump might be present; see][for details]{Battisti2017}, and is not due to systematic changes in D$_n$(4000) or $\Sigma_{{\rm H}\alpha}$.

The individual regions from the galaxies sorted by the mass of the parent galaxy are compared to the $\beta$--$\tau^l_B$ relation from B16 in Figure~\ref{fig:reg_mass}. After excluding the region with $\beta\approx1$, the slope between $\beta$ and $\tau^l_B$ for the individual regions in low-mass galaxies is only slightly shallower than that for the parent galaxy. This result is likely driven by the fact that the individual regions in low-mass galaxies have preferentially lower $\tau^l_B$ values; the K\nobreakdash-S two sample test for $\tau^l_B$ shows evidence of distinctly different distributions at the 99\% confidence level, while that of $\beta$ does not. We additionally see no systematic trends with D$_n$(4000) or $\Sigma_{{\rm H}\alpha}$ between the two mass bins, similar to that of the integrated galaxy light. 

There are three possible methodological explanations for our result: 1) our $\Sigma_{\textrm{H}\alpha}$ and EW(H$\alpha$) cuts are not sufficient to eliminate all regions that are dominated by DIG, 2) our assumption that the 2175\,\AA\ bump is absent is not valid, and must be included in our estimate of the relation between $\beta$ and $\tau^l_B$, or 3) the D$_n(4000)$ or EW(H$\alpha$) are systematically different in low-mass galaxies hence the slope described by B16 is not appropriate. As shown in Table~\ref{table:reg_val}, the $\Sigma_{\textrm{H}\alpha}$ values of the individual regions from low-mass galaxies span the range $\log[\Sigma_{\textrm{H}\alpha}/(M_\odot~\textrm{yr}^{-1}~\textrm{kpc}^{-2})]=38.05$--$39.3$, while those from the high-mass galaxies extend to slightly higher surface brightnesses, $\log[\Sigma_{\textrm{H}\alpha}/(M_\odot~\textrm{yr}^{-1}~\textrm{kpc}^{-2})]=38.06$--$39.7$. Similarly, for low-mass galaxies, the median D$_n(4000) = 1.234$ is not dissimilar to the value from higher-mass galaxies, 1.247. While the average H$\alpha$ equivalent width is higher for the regions from low-mass galaxies (78~\AA) than that for high-mass galaxies (50~\AA), neither indicate a very low sSFR. Thus, it is unlikely that either the $\Sigma_{\textrm{H}\alpha}$ cut or dilution from the older stellar population are the main causes of the observed steeper slope.
 
 Therefore, the only methodological issue that could explain this discrepancy is the 2175\,\AA\ bump, which we are neglecting in this analysis. If the bump strength was similar to that of the Milky Way, the values of $\beta$ calculated in this work would be overestimated. Moreover, the 2175\,\AA\ bump is more prominent in lower-mass star forming galaxies and quiescent galaxies. However, \citet[][]{Battisti2017} found that the bump is only significant in star forming galaxies when $b/a < 0.42$. Also, the UVOT filters we are using in our analysis are broad: the uvw2 filter, which includes the 2175~\AA feature, is almost five times wider than the feature itself, thus diluting any effect this absorption feature might have. Finally, the trend is not dependent on any regions with galaxies with high inclination angles, implying that this is not caused by the 2175\,\AA\ bump.
 
Thus we cannot find any strong systematic issues that could cause this dependence of $\beta$--$\tau^l_B$ on stellar mass. This exact trend is also seen in \citet{Salim2018}, which they interpret to be a result of higher $A_V$ values in more massive galaxies. Future studies should include kiloparsec-sized regions both from starburst galaxies (which have a higher dust content due to the high SFR), and star forming galaxies spanning a wide range of stellar mass to disentangle this issue. Both of these requirements can easily be met by increasing the current overlap between \textit{Swift} and MaNGA galaxies.
 
\subsubsection{Star Formation Rate}
The total SFR of a galaxy affects the slope between the integrated values of $\beta$ and $\tau^l_B$. This connection is evident when comparing the results of \citet[][]{Calzetti1994} and B16. The individual regions show a bias to lower $\beta$ values when the \textit{integrated} SFR is above log[SFR$_{1R_e}$[H$\alpha]/(M_\odot~\textrm{yr}^{-1})]>0$. A K\nobreakdash-S 2 sample test for the $\beta$ measurements between the regions in the highest integrated SFR bin and the other two bins reveal that they are statistically distinct at the 90\% confidence level. However, all of the  regions in the high integrated SFR bin reside in a total of six galaxies (Table~\ref{table:tgt_prop}), all with log[SFR$_{1R_e}$[H$\alpha]/(M_\odot~\textrm{yr}^{-1})]<1$. In order to fully constrain this potential effect, we need a larger sample of high star formation rate galaxies.

\subsection{Comparing Local and Global Measures of $\beta$ and $\tau^l_B$}
\label{ssec:gl_lo}
\begin{figure}
\hspace{-4mm}\includegraphics[width=0.45\textwidth]{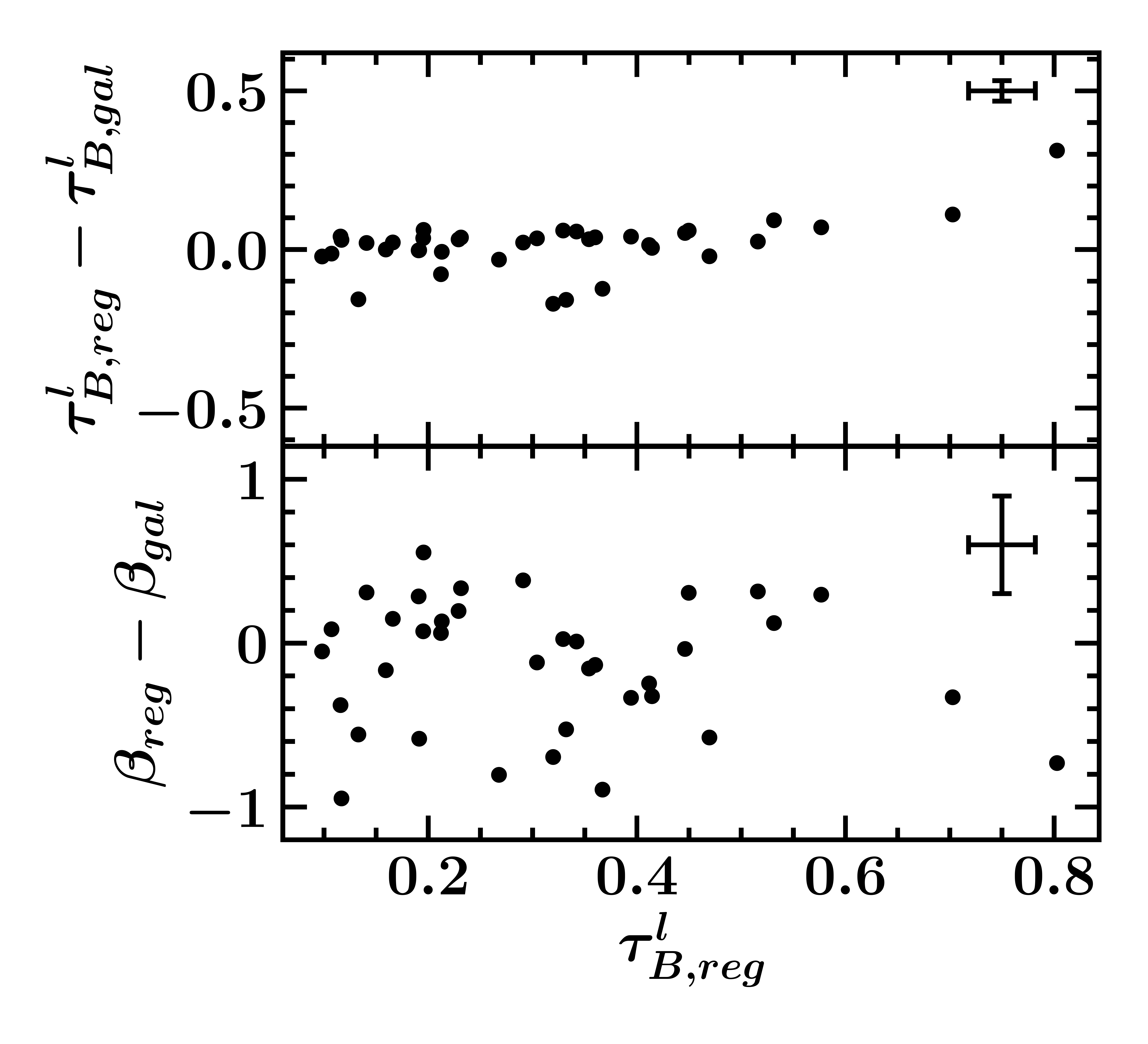}\vspace{-4mm}
\caption{The difference between the measured $\tau^l_B$ (top panel) and $\beta$ (lower panel) values from the individual region and those derived from the integrated galaxy light as a function of the measured $\tau^l_B$ from the individual regions. The ordinates span the same range of total attenuation, given the B16 relation of $\beta = 1.95\tau^l_B -1.61$. $\beta$ shows significant local variation, while $\tau^l_B$ is remarkably consistent between the two scales. The error bar in the upper right corner in each panel shows the characteristic uncertainty associated with these measurements.}
\label{fig:reg_gbtau}
\end{figure}

\begin{figure}
\hspace{-4mm}\includegraphics[width=0.45\textwidth]{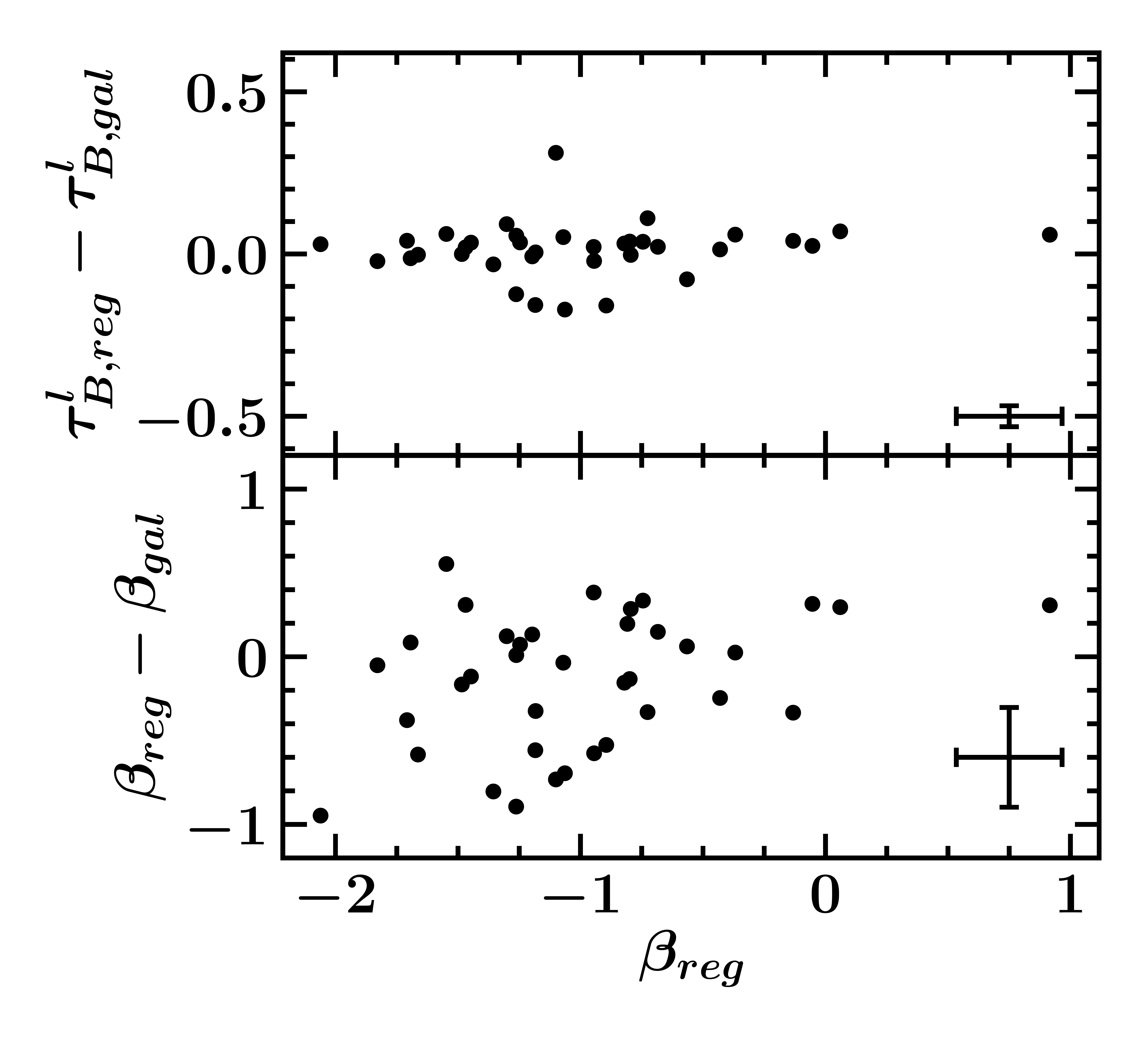}\vspace{-4mm}
\caption{Same as Figure~\ref{fig:reg_gbtau}, but as a function of $\beta$ from the individual regions. The error bar in the lower left corner in each panel shows the characteristic uncertainty for these measurements.}
\label{fig:reg_gbtaub}
\end{figure}

The integrated flux from galaxies includes emission from star forming regions, DIG, the underlying older stellar population, and all other sources of light. The kpc-sized star forming regions should have a relatively larger contribution from the younger stellar populations, albeit with some contamination from DIG (see Section~\ref{sec:ddresult}). As a result, $\beta$ and $\tau^l_B$ do not necessarily need to match between the two very different spatial scales. Figures~\ref{fig:reg_gbtau} and \ref{fig:reg_gbtaub} make this comparison using the measured $\tau^l_B$ and $\beta$ from the individual star forming regions and from the galaxy as a whole. The ordinates in the two panels have been scaled to represent the same amount of attenuation given the B16 relation. The dichotomy is striking: while $\tau^l_B$ for the individual regions largely agrees with that for the integrated galaxy light, $\beta$ values exhibit a large amount of scatter. We also note a trend at high levels of nebular attenuation where the values of $\beta$ from the individual regions are smaller than that for the integrated galaxy value and vice versa for $\tau^l_B$.
\begin{figure}
\hspace{-4mm}\includegraphics[width=0.45\textwidth]{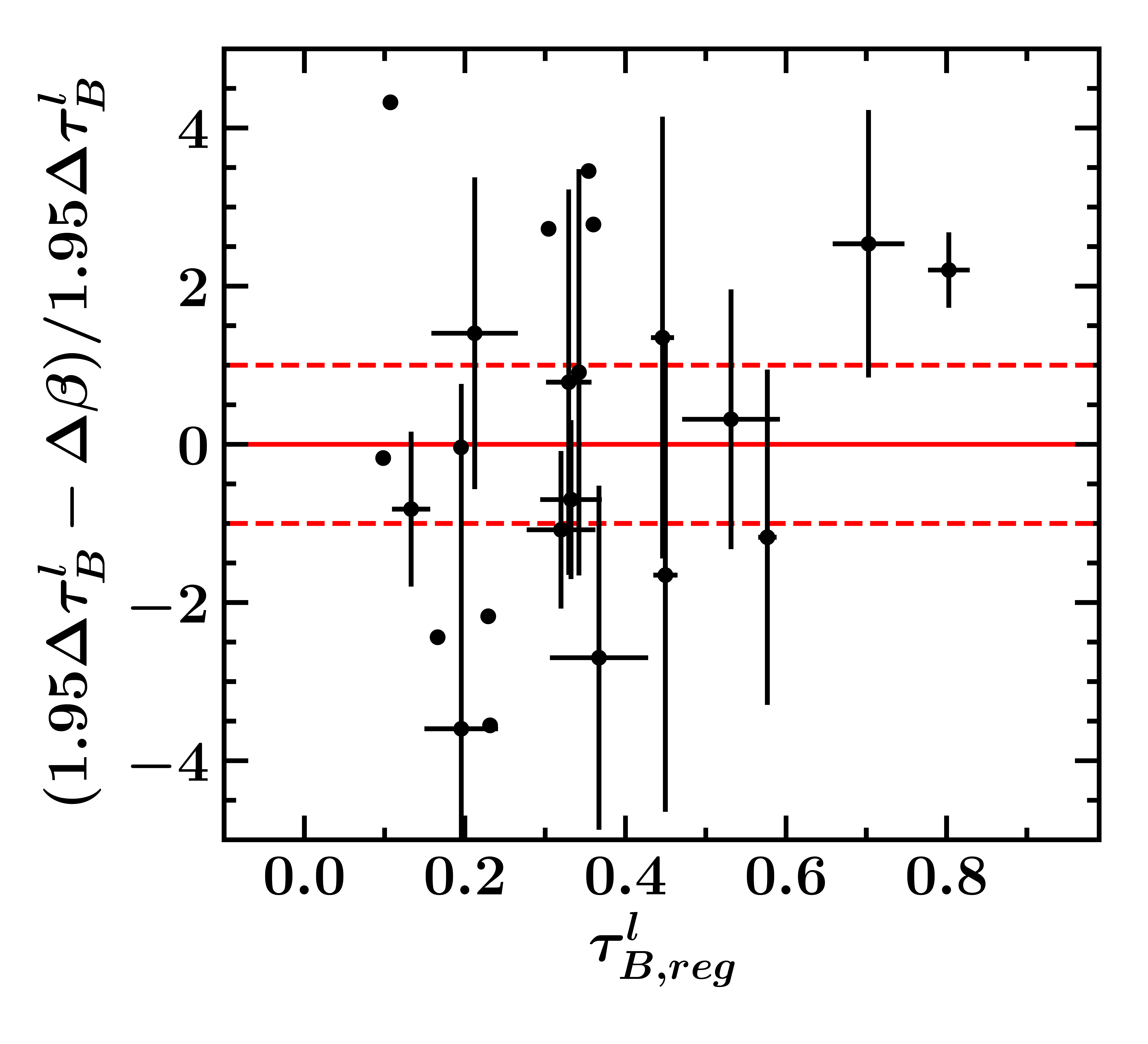}\vspace{-4mm}
\caption{The fractional difference between the measured difference in $\Delta\beta$ to that predicted by B16 from $\Delta\tau^l_B$ as described in Figure~\ref{fig:reg_gbtau}. The solid red line denotes perfect agreement with the model, while the dashed red lines represent a departure of 100\% from the expected value. The scatter between the optical and UV attenuation is significant even after including scaling differences. The uncertainties for measurements with $\Delta\tau^l_B < 0.05$ are not shown for clarity; while all of the values are well-measured, the small $\Delta\tau^l_B$ values create large errors when propagated.}
\label{fig:reg_fdif}
\end{figure}

Given the scaling effects between the optical nebular and UV stellar continuum attenuation, we present in Figure~\ref{fig:reg_fdif} the fractional difference $\Delta\beta$ between the observed values and the predicted value, using $\Delta\tau^l_B$ and the B16 relation, $\beta = 1.95\tau^l_B -1.61$. Clearly, $\beta$ varies drastically between the two aperture sizes while $\tau^l_B$ remains within $\sim50\%$ of the integrated galaxy value. The variation of $\beta$ due to aperture size is also larger than the intrinsic  scatter of $\sigma_{\textrm{int}}=0.44$ seen in B16. We attempted to quantify the observed trends in $\tau^l_B$ and $\beta$ by searching for any dependence on distance from the nucleus, D$_n$(4000), metallicity and other parameters, as described below. 

\subsubsection{Distance from Nucleus}
\begin{figure}
\hspace{-4mm}
\includegraphics[width=0.45\textwidth]{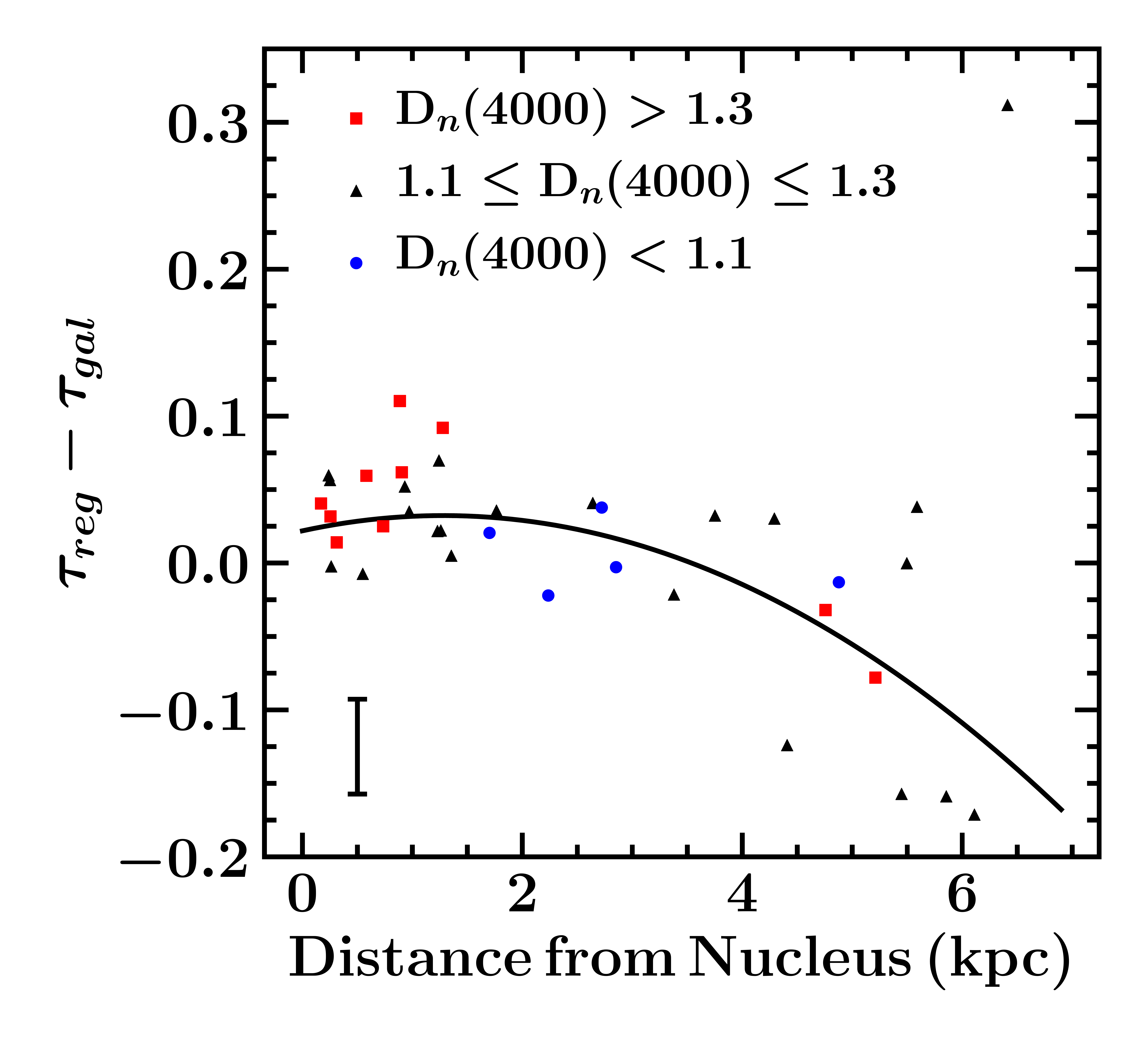}\vspace{-4mm}
\caption{The difference between the $\tau^l_B$ measurements for the individual regions and that for the integrated galaxy light vs.~distance from the nucleus, colour-coded by D$_n(4000)$. The second order polynomial distribution, shown by the black line, is a fit to the data. The discrepant point on the top right is also the region with the highest $\tau^l_B\approx0.8$, and is excluded from the fit. The error bar in the lower left corner represents the shows the characteristic uncertainty for these measurements.}
\label{fig:ctau_dist}
\end{figure}

The difference between $\tau^l_B$ for the individual regions and for the integrated galaxy value is negatively correlated with distance from the nucleus. This behavior is illustrated in Figure~\ref{fig:ctau_dist}. The Spearman and Kendall nonparametric correlation tests \citep{spearman,kendall} show the two parameters are correlated at the 99\% confidence level. We fitted the data with a second order polynomial, which produced the relation
\begin{equation}
\Delta\tau^l_B = -0.00637 d^2 + 0.01418 d+0.0328,
\end{equation}
where $d$ is the distance from the nucleus in kiloparsec; the intrinsic dispersion is $\sigma_{\textrm{int}}=0.08$. The one outlier seen at a distance of $\sim6$~kpc is the region with the highest $\tau^l_B\approx0.8$, is the only such measurement in that region of parameter space, and is excluded from the fit. We colour-code by D$_n(4000)$, but do not see any systematic trends associated with that parameter.

\begin{figure}
\hspace{-4mm}\includegraphics[width=0.46\textwidth]{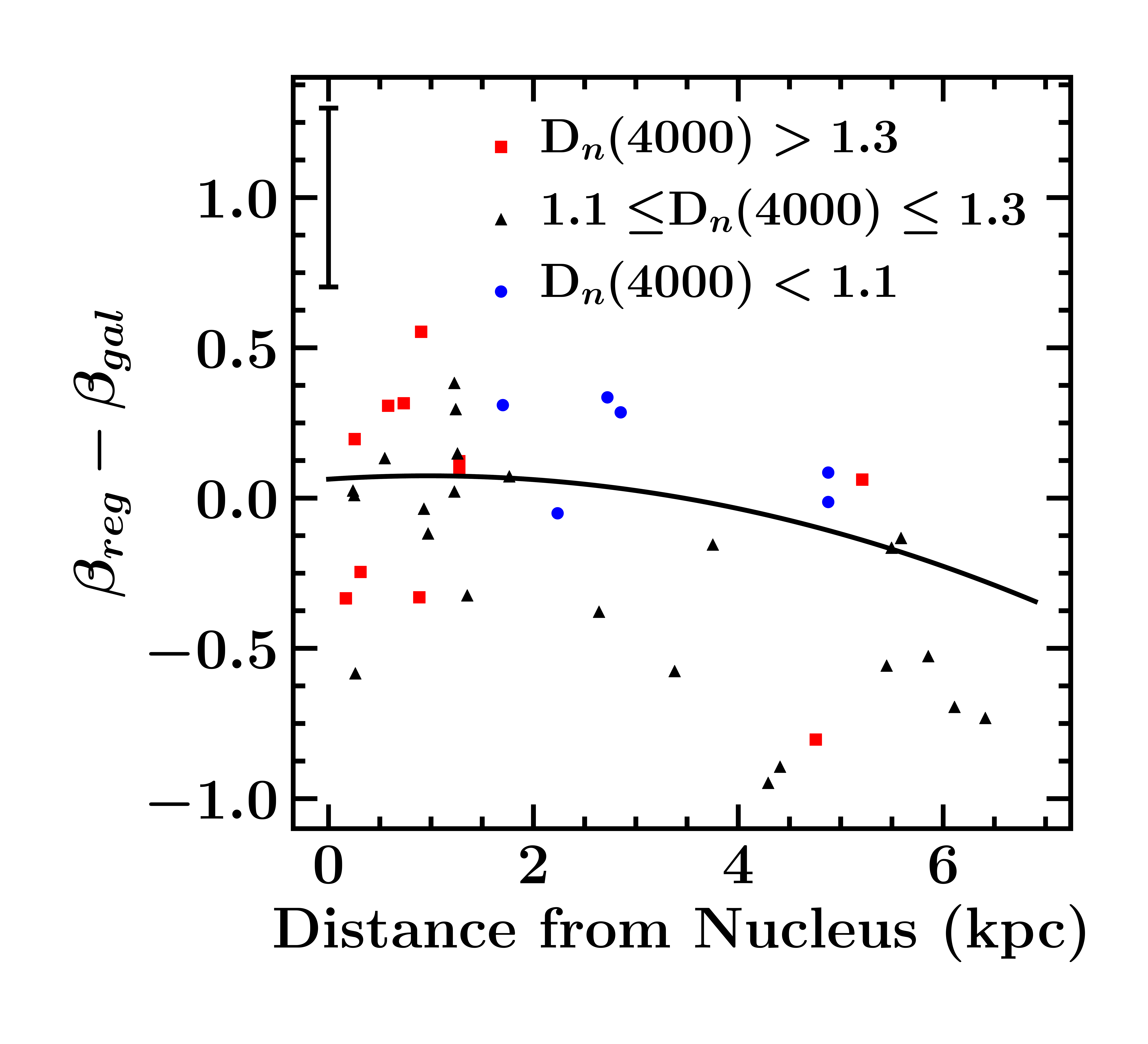}\vspace{-4mm}
\caption{The difference between the measured $\beta$ values for individual regions and galaxy light vs.~distance from the nucleus, colour-coded by D$_n(4000)$. The error bar in the top left corner represents the characteristic uncertainty of the measurements. The second order polynomial shown in Figure~\ref{fig:ctau_dist} is translated from $\Delta\tau^l_B$ to $\Delta\beta$ using the B16 relation (black line), as the large scatter prevents us from obtaining a robust fit.}
\label{fig:cbeta_dist}
\end{figure}

If we translate the fit from $\Delta\tau^l_B$ to $\Delta\beta$ using the B16 relation, the data qualitatively show a correlation that agrees with the fitted relation. However, the large intrinsic scatter undermines the utility of this diagnostic. Indeed, the average deviation of the measured vs.~expected $\Delta\beta$ is 0.277. The reason for the large scatter could be attributed to a number of causes, such as the reliability of the $\beta$ parameter given the SFH within the past few 100~Myr, the influence of the underlying older stellar population, the total dust content of the individual region, or the SFR seen in the observed region. To test for effects due to the underlying stellar population, we colour-code by D$_n$(4000) in Figures~\ref{fig:ctau_dist} and \ref{fig:cbeta_dist}. We see no clear correlation with D$_n(4000)$, and find this trend persists, even when excluding the regions with D$_n(4000)>1.3$. The scatter in $\beta$ is not simply due to changes in D$_n(4000)$, as evident by Figure~\ref{fig:cbeta_dist}.

\begin{figure}
\hspace{-4mm}
\includegraphics[width=0.46\textwidth]{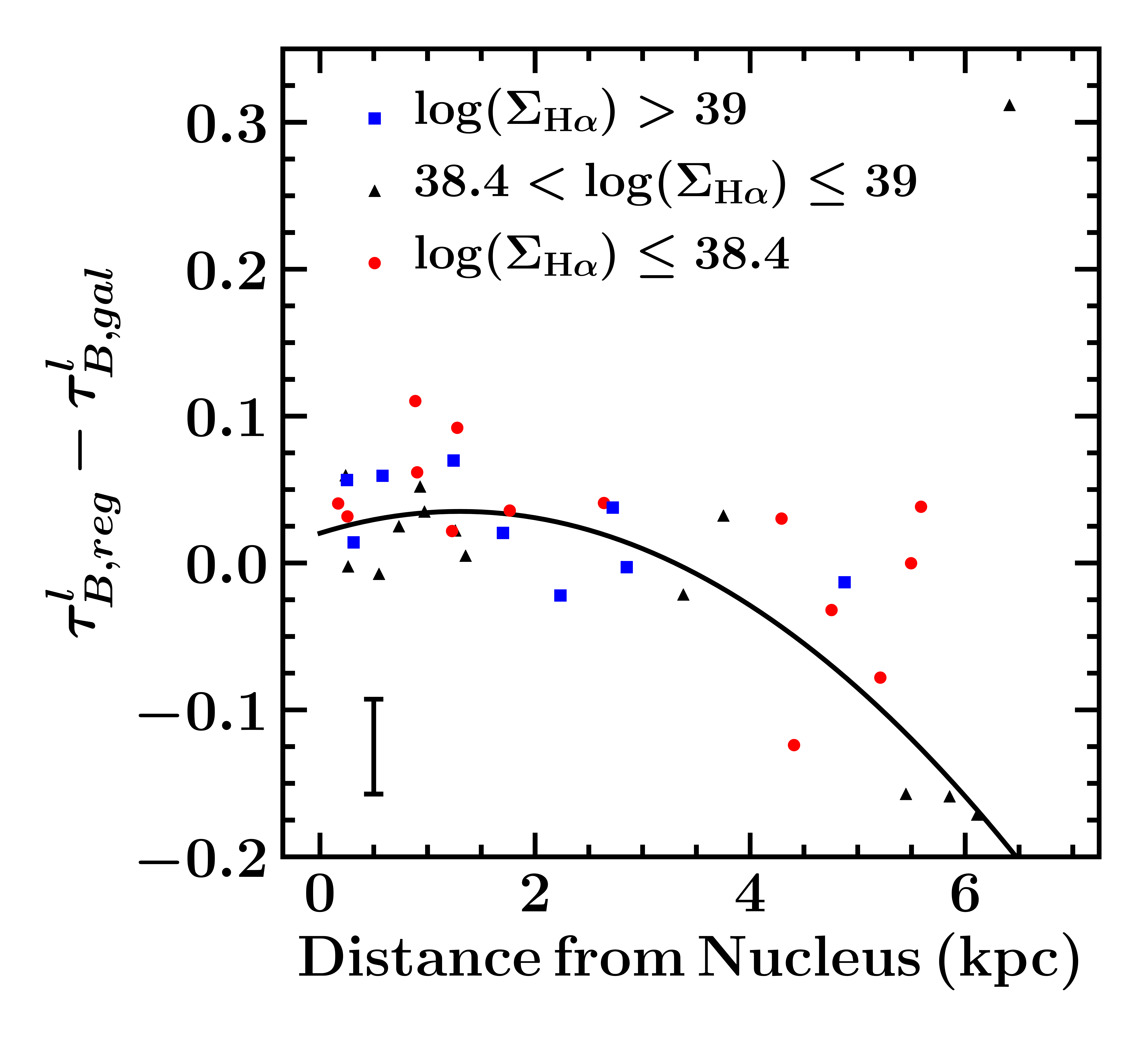}\vspace{-4mm}
\caption{Same as Figure~\ref{fig:ctau_dist}, except colour-coded according to $\Sigma_{\rm{H}\alpha}$, which is presented in units of erg~s$^{-1}$~kpc$^{-2}$. The regions with sparse star formation significantly contribute to the scatter seen in the relation.}
\label{fig:ctau_dist_sha}
\end{figure}

\begin{figure}
\hspace{-4mm}\includegraphics[width=0.46\textwidth]{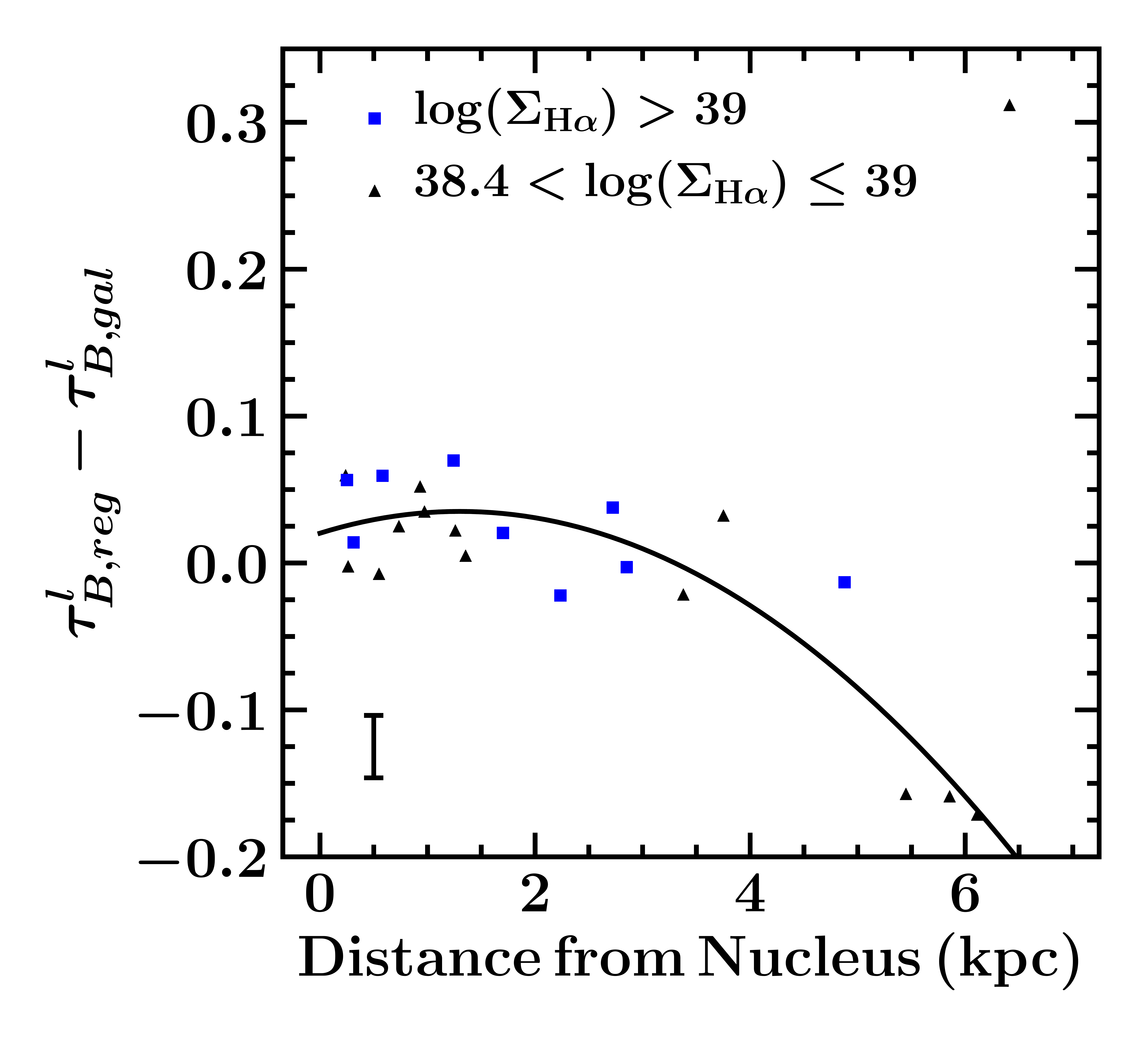}\vspace{-4mm}
\caption{Same as Figure~\ref{fig:ctau_dist_sha}, but excluding ``sparse'' star forming regions ($\Sigma_{\rm{H}\alpha}<10^{38.4}$~erg~s$^{-1}$~kpc$^{-2}$). The second order polynomial shown is from Equation~(\ref{eq4}).}
\label{fig:ctau_dist1_sha}
\end{figure}

\begin{figure}
\hspace{-4mm}\includegraphics[width=0.45\textwidth]{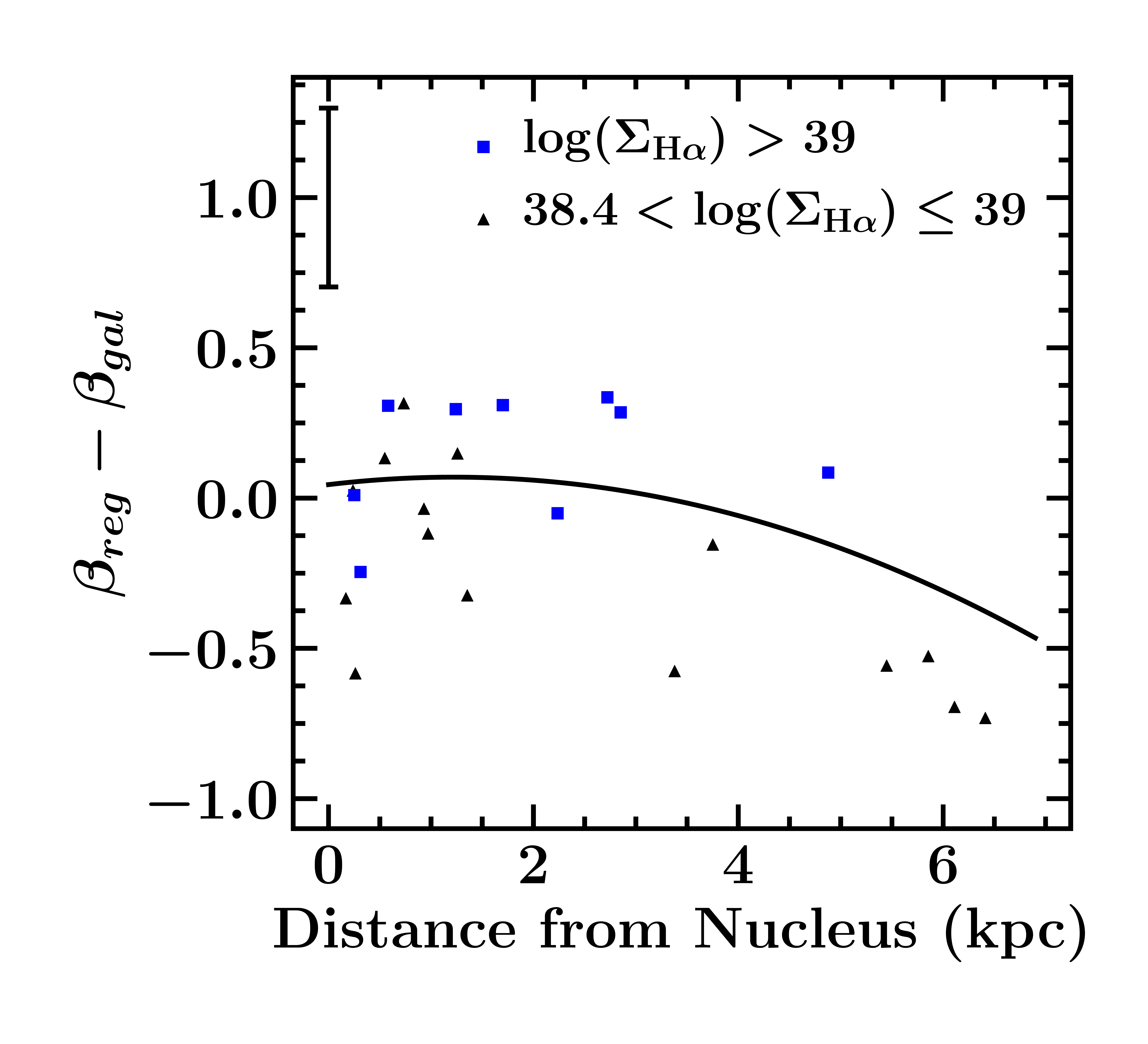}\vspace{-4mm}
\caption{The difference between the measured $\beta$ values for the individual regions and galaxy light vs.~distance from the nucleus, colour-coded by $\Sigma_{\rm{H}\alpha}$, which is presented in units of erg~s$^{-1}$~kpc$^{-2}$. The error bar in the top left corner represents the characteristic uncertainty. The second order polynomial shown in Figure~\ref{fig:ctau_dist} is translated from $\Delta\tau^l_B$ to $\Delta\beta$ using the B16 relation (black line), when the sparse star forming regions are excluded. The resulting dispersion associated with the fit is $\sigma_{\rm{disp}}=0.28$.}
\label{fig:cbeta_dist_sha}
\end{figure}

We similarly test for the effects caused by the DIG seen in the observed region by colour-coding our relation according to $\Sigma_{\rm{H}\alpha}$ in Figures~\ref{fig:ctau_dist_sha}--\ref{fig:cbeta_dist_sha}. We note that the regions that with sparse star formation ($\Sigma_{\rm{H}\alpha}\leq 10^{38.4}$~erg~s$^{-1}$~kpc$^{-2}$) introduce significant scatter in the $\Delta\tau^l_B$--distance relation. When these sparse regions are excluded, the resulting fit given by
\begin{equation}\label{eq4}
\Delta\tau^l_B = -0.00851 d^2 + 0.02082 d+0.02298.
\end{equation}
We show this fit in Figure~\ref{fig:ctau_dist1_sha}. If we translate this relation and exclude the sparse star forming regions, the $\Delta\beta$--distance relation is well-described by the model, as shown in Figure~\ref{fig:cbeta_dist_sha}. The best-fit model has an average deviation from the predicted value of 0.23, and a dispersion of $\sigma_{\rm{disp}}=0.28$. This indicates that the regions with $\Sigma_{\rm{H}\alpha}<10^{38.4}$~erg~s$^{-1}$~kpc$^{-2}$ contribute to the scatter seen in the $\Delta\beta$--distance relation. Finally, we do note a qualitative difference between those regions with high $\Sigma_{\rm{H}\alpha}$ (approximately starburst-like) and those that are ``star-forming''. A K-S 2-Sample test shows the two samples are distinct at the 95\% confidence level. This could be indicative of the change in dust structure as a function of sSFR. However, given the small number of regions in each category, we do not attempt to further quantify this behavior.

Another strong contributor to this deviation could be the shape of the attenuation curve. Numerous papers have been written on the variation between the ratio of infrared to UV luminosity (IRX) and the slope of the UV stellar continuum \citep[$\beta$; e.g., ][for a sample of reasonably star forming galaxies]{Salim2019}. However, a much larger sample would be needed to disentangle these different possibilities.


We conclude that the deviation from the integrated measure of nebular attenuation clearly scales as a function of distance from the nucleus, while that for the stellar continuum attenuation is dependent on both distance and $\Sigma_{\rm{H}\alpha}$. While the correlation between $\Delta\beta$ and distance is relatively strong for high $\Sigma_{\rm{H}\alpha}$ regions, the large uncertainty, small number of data points and significant scatter prevents us from obtaining a useful fit.

\subsubsection{D$_n$(4000), Metallicity, and Other Parameters}
We compared the deviation in $\beta$ and $\tau^l_B$ between local and global scales to the metallicity and three different D$_n$(4000) measurements: those from (1) the integrated galaxy light, (2) the individual star forming regions, and (3) the difference between (1) and (2). The scatter in both $\beta$ and $\tau^l_B$ increases with both the metallicity and the D$_n$(4000) index of the individual regions. We find a qualitative trend between D$_{reg}$(4000) and D$_{n,reg}$(4000)$-$D$_{n,gal}$(4000) with respect to $\beta_{reg}-\beta_{gal}$. Spearman and Kendall correlation tests both find a correlation which is significant at the 95\% confidence level. However, the large dispersion makes any quantitative description of the trend difficult. Additionally, for large deviations in either direction, the assumed relation does not hold well. Therefore, while the effect of the underlying older stellar population is definitely contributing, it cannot be the only factor causing the observed scatter between local and integrated measures of attenuation based on $\beta$. In addition to the physical properties described above, we also tested for the dependence of $\Delta\beta$ and $\Delta\tau^l_B$ on $\Sigma_{\textrm{H}\alpha}$, and the stellar mass and SFR of the parent galaxy, but we found no significant correlations. 

%Discussion and conclusions
%%%%%%%%%%%%%%%%%%%%%%%%%%%%%%%%%%%%%%%%%%%%%%%%%%%%%%%%%%%%%%%%%%%%%%%%%%%%%%%%%
\section{Summary and Discussion}
\label{sec:conclusion}
In this work, we introduced a sample of 29 nearby star forming galaxies that have both MaNGA and \textit{Swift}/UVOT data. This sample is a subset of a larger set of galaxies which will be used for a number of other investigations including the quenching of star formation (Molina et al. in preparation). We compared the measured UV spectral slope, $\beta$, and Balmer optical depth, $\tau^l_B$, of both the integrated galaxy light and individual kpc-sized star forming regions within the galaxies to the relation described in B16. Our goal was to test the validity of the B16 relation across the faces of galaxies and for the integrated galaxy light, as well as to study the dependence of $\beta$ and $\tau^l_B$ measurements on aperture size. While the results of this paper are limited by our sample size, they also illustrate the importance of comparing integrated to spatially-resolved measures of attenuation within a single galaxy.

A region's age as probed by D$_n$(4000), its SFR, metallicity, parent galaxy's mass and SFR all contribute to the observed scatter in the $\beta$--$\tau^l_B$ relation. Specifically, D$_n$(4000) appears to be a useful indicator of the observed slope of the $\beta$--$\tau^l_B$ relation, with older stellar populations having steeper slopes and higher intercepts. This effect cannot be attributed to the dilution of $\beta$ by the older stars alone, as shown in Figure~\ref{fig:bd4000}. We agree with B16 that taking stellar age into account via D$_n$(4000) is important when adopting an attenuation curve. However, we find large systematic offsets for star forming regions outside of the range $1.1\leq\textrm{D}_n(4000)\leq1.3$, which was used to define the B16 $\beta$--$\tau^l_B$ relation. We caution against using the B16 attenuation law in areas with more extreme D$_n$(4000) values.

The SFR of an individual region is a key factor in our study, as the interpretation of $\beta$ relies on the assumption that the UV light is dominated by the light of young stars. Regions with lower SFRs could be dominated by DIG, and represent a very different physical environment than those dominated by star formation. A demarcation at $\Sigma_{\textrm{H}\alpha}~>~10^{38}$~erg~s$^{-1}$~kpc$^{-2}$ in conjunction with EW(H$\alpha$) > 15\AA\ appears sufficient for discriminating between these two regimes.

The relation between metallicity and $\beta$ or $\tau^l_B$ described in B16 holds for individual kpc-sized regions; however there is significant scatter at high values of $\tau^l_B$. To agree with the B16 $\beta$ vs.~metallicity relation, one must use all star forming regions regardless of their measured D$_n$(4000), as per the method used by B16 in their original derivation. However, their fitted $\beta$--$\tau^l_B$ relation is limited to a specific range in D$_n$(4000). Therefore, caution must be used when utilizing the $\beta$--metallicity and $\beta$--$\tau^l_B$ relations concurrently.

Similarly, the mass of the parent galaxy affects the observed slope of not only the integrated light, similar to that seen in \citet[][and references therein]{Salim2018}, but also that of the kpc-sized regions. For both the integrated galaxy light and the individual star forming regions, lower-mass galaxies tend to have a steeper UV attenuation law than that of the higher-mass galaxies. This behavior could be attributed to both dust type and dust content as well as properties of the observed stars and gas within the region. The dust responsible for the attenuation can be different in low-mass galaxies, as total dust content is lower \citep{moustakas2006,zhu2009}, and the composition and geometry of the gas can also differ \citep{Calzetti2001,Kong2004,Gordon2004}. The former is related to the mass of the galaxy, while the latter is strongly tied to the SFR, stellar population, and the physical structure of the star forming regions. As these regions are independent from one another, their dust compositions and geometries, mean stellar age, metallicity and structure can be unique. However, the global dust composition and metallicity will be set by the environment, i.e., the host galaxy. Therefore the similar global conditions for low-mass galaxies may be the driver of the change in the observed UV attenuation law.

Finally, the measured $\tau^l_B$ and $\beta$ from the integrated galaxy light and kpc-sized star forming regions show drastically different behaviors. $\beta$ displays significant variation not present in $\tau^l_B$, even when accounting for the change in total attenuation between the two bands. The small change in $\tau^l_B$ with distance from the nucleus is well characterized by a second order polynomial fit. Conversely, the scatter in $\beta$ shows marginal dependence on both position in the galaxy and deviation in D$_n$(4000) from that of the integrated galaxy light. We therefore cannot statistically identify any one parameter as the main cause of the deviation in $\beta$.

Physically, the difference in the variation in $\beta$ and $\tau^l_B$ could be explained by either sightline dependence, the dilution of $\beta$, or a combination of the two effects. The nebular emission is not as strongly affected by the observed line of sight because of the large extent of the ionized gas and the fact that it involves isotropic radiation. Conversely the star forming regions are confined spatially within galaxies, forcing the UV attenuation of the continua of the young stars to be much more dependent on the small-scale dust distribution (i.e., granularity) within and between star-forming regions. Therefore, $\beta$ will have a much stronger dependence on both the sightline and aperture used than $\tau_B^l$. Additionally, the increased contribution from the older stellar populations in the larger aperture, as well as low area-specific SFRs in the individual regions, could be a contributing factor to the scatter in $\beta$ for some of the less vigorous star forming galaxies. All of these effects could combine to create the small variation in $\tau^l_B$ and the large variation in $\beta$ seen in Figures~\ref{fig:ctau_dist}--\ref{fig:cbeta_dist_sha}. 

We conclude that agreement with the B16 $\beta$--$\tau^l_B$ model for kpc-sized star forming regions is limited to the specific parameter range of $1.1 \leq\textrm{D}_n(4000)\leq 1.3$, $\Sigma_{\textrm{H}\alpha} > 10^{38}$~erg~s$^{-1}$~kpc$^{-2}$ and EW(H$\alpha) > 15$~\AA. While the overall qualitative trend between UV stellar continuum and optical nebular attenuation holds, adopting the B16 relation for specific objects will result in large uncertainties due to variation in physical properties, such as the SFR and the age of the underlying stellar population. The integrated light also qualitatively follows the B16 relation but with significant scatter. Finally, aperture size can greatly affect the measured $\beta$ and thus the adopted value of total attenuation. Therefore, while the B16 relation provides an accurate qualitative description for both the integrated galaxy light and kpc-sized star forming regions, the large scatter due to the dependence on a variety of physical parameters makes this a multi-parameter problem that requires a more comprehensive dataset to address properly.

%ACKNOWLEDGEMENTS
%%%%%%%%%%%%%%%%%%%%%%%%%%%%%%%%%%%%%%%%%%%%%%%%%%%%%%%%%%%%%%%%%%%%%%%%%%%%%%%%% 
\section*{Acknowledgements}
We thank the anonymous referee for insightful comments that helped us improve this paper. We thank Lea Hagen and Michael Siegel for their help with some of the technical aspects of this work. This work was supported by funding from the Alfred P. Sloan Foundation's Minority Ph.D. (MPHD) Program, awarded to MM in 2014--15. This research has made use of data and/or software provided by the High Energy Astrophysics Science Archive Research Center (HEASARC), which is a service of the Astrophysics Science Division at NASA/GSFC and the High Energy Astrophysics Division of the Smithsonian Astrophysical Observatory. This research made use of Astropy, a community-developed core Python package for Astronomy (Astropy Collaboration, 2018). This research has made use of the NASA/IPAC Extragalactic Database (NED),which is operated by the Jet Propulsion Laboratory, California Institute of Technology, under contract with the National Aeronautics and Space Administration. The Institute for Gravitation and the Cosmos is supported by the Eberly College of Science and the Office of the Senior Vice President for Research at the Pennsylvania State University. Funding for the Sloan Digital Sky Survey IV has been provided by the Alfred P. Sloan Foundation, the U.S. Department of Energy Office of Science, and the Participating Institutions. SDSS-IV acknowledges support and resources from the Center for High-Performance Computing at the University of Utah. The SDSS web site is www.sdss.org.

SDSS-IV is managed by the Astrophysical Research Consortium for the Participating Institutions of the SDSS Collaboration including the Brazilian Participation Group, the Carnegie Institution for Science, Carnegie Mellon University, the Chilean Participation Group, the French Participation Group, Harvard-Smithsonian Center for Astrophysics, Instituto de Astrof\'isica de Canarias, The Johns Hopkins University, Kavli Institute for the Physics and Mathematics of the Universe (IPMU) / University of Tokyo, the Korean Participation Group, Lawrence Berkeley National Laboratory, Leibniz Institut f\"ur Astrophysik Potsdam (AIP),  Max-Planck-Institut f\"ur Astronomie (MPIA Heidelberg), Max-Planck-Institut f\"ur Astrophysik (MPA Garching), Max-Planck-Institut f\"ur Extraterrestrische Physik (MPE), National Astronomical Observatories of China, New Mexico State University, New York University, University of Notre Dame, Observat\'ario Nacional / MCTI, The Ohio State University, Pennsylvania State University, Shanghai Astronomical Observatory, United Kingdom Participation Group,Universidad Nacional Aut\'onoma de M\'exico, University of Arizona, University of Colorado Boulder, University of Oxford, University of Portsmouth, University of Utah, University of Virginia, University of Washington, University of Wisconsin, Vanderbilt University, and Yale University.

%BIBLIOGRAPHY
%%%%%%%%%%%%%%%%%%%%%%%%%%%%%%%%%%%%%%%%%%%%%%%%%%%%%%%%%%%%%%%%%%%%%%%%%%%%%%%%%
\bibliographystyle{mnras}
\bibliography{all.041018.bib}
%%%%%%%%%%%%%%%%% APPENDICES %%%%%%%%%%%%%%%%%%%%%
\clearpage
\appendix
\onecolumn
\section{Identified Star Forming Regions}\label{app:a}
In this appendix, we present $\Sigma_{\rm{H}\alpha}$, uvw2 and uvw1 images of the 29 galaxies with the individual regions indicated by the black apertures. The objects are labelled by the object IDs listed in Table~2. We used a conservative approach to define star forming regions; each region must: (1) have a brightness peak with the observed $\Sigma_{\rm{H}\alpha} \geq10^{38}$~erg~s$^{-1}$~kpc$^{-2}$ and (2) appear star forming in all three BPT diagrams. Therefore, there may be bright regions in the $\Sigma_{\rm{H}\alpha}$ maps that are not identified as a star forming region. Finally, we also include a table with the positions, radii, $b/a$ and $\theta$ values for the regions in Table~\ref{table:aper}.
\vspace{3mm}

\begin{center}
\includegraphics[width=0.3\textwidth]{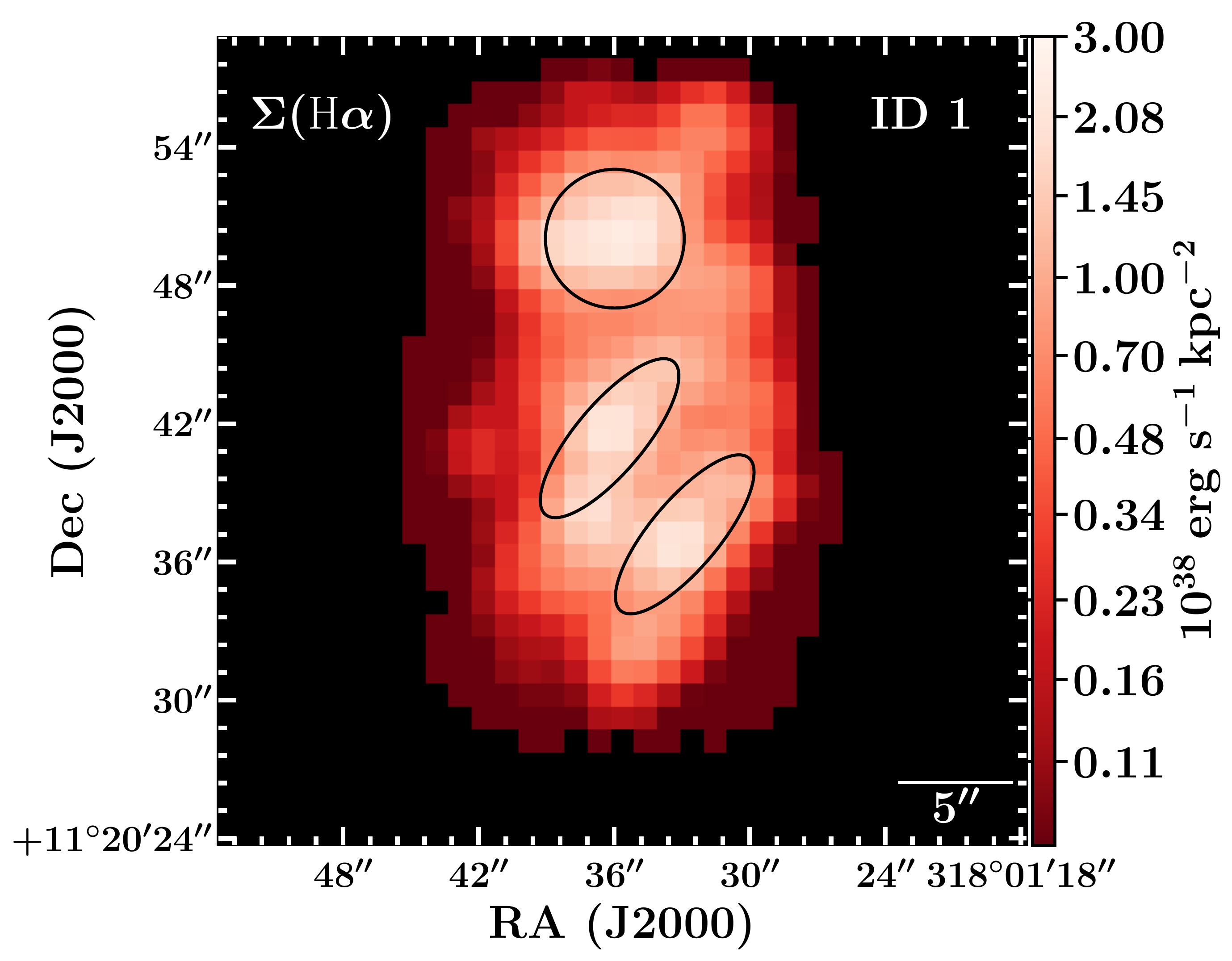}
\includegraphics[width=0.3\textwidth]{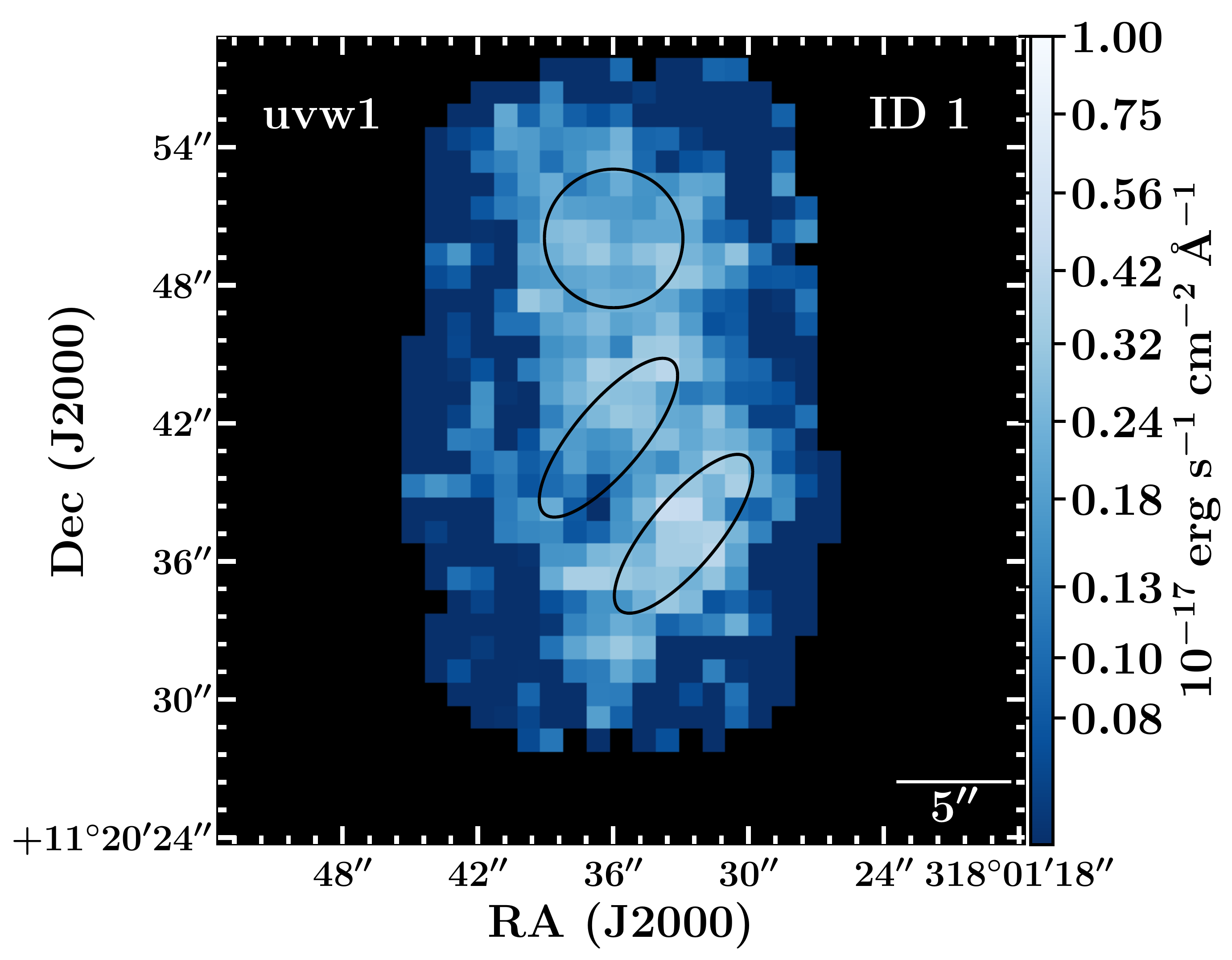}
\includegraphics[width=0.3\textwidth]{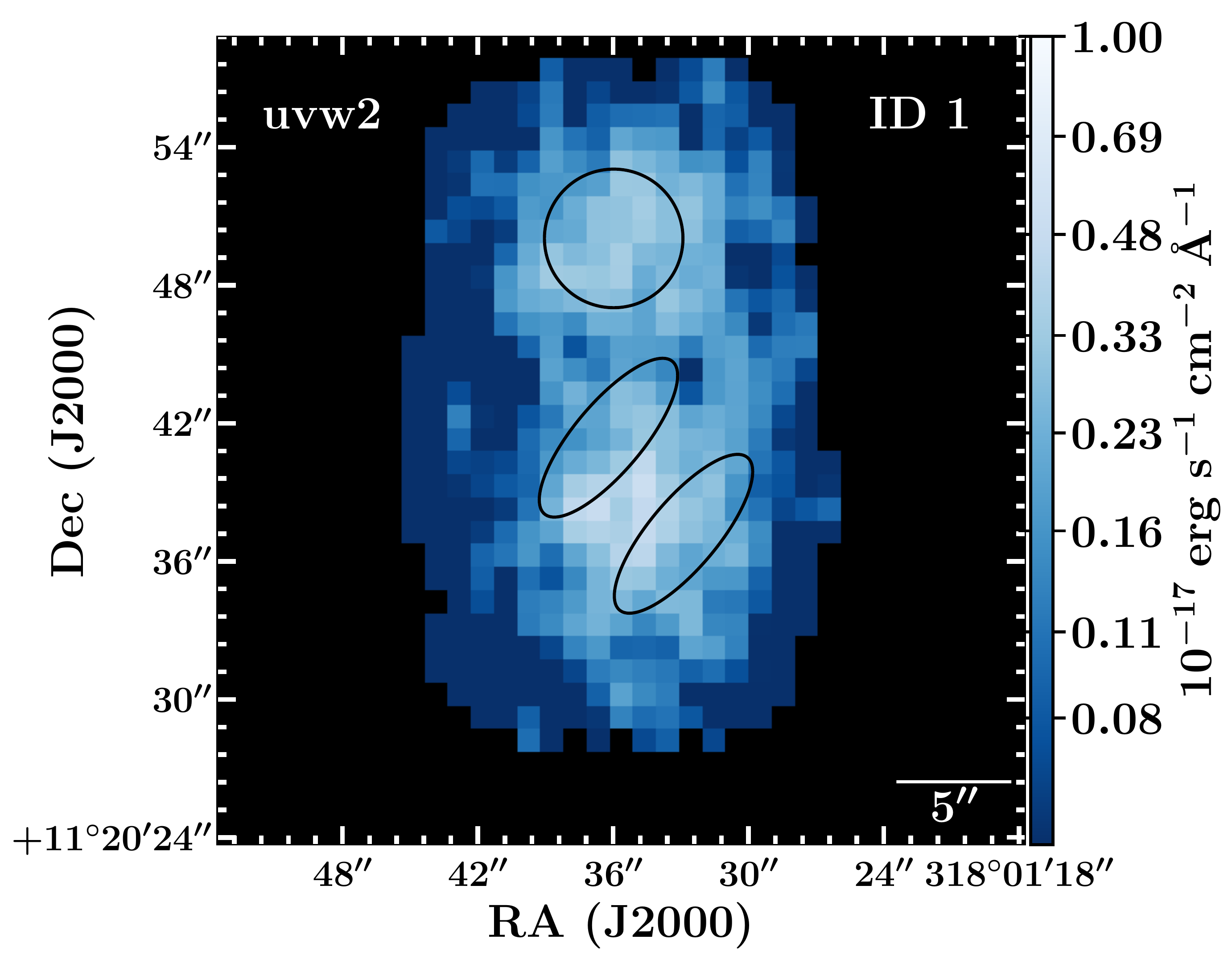}\\
\includegraphics[width=0.3\textwidth]{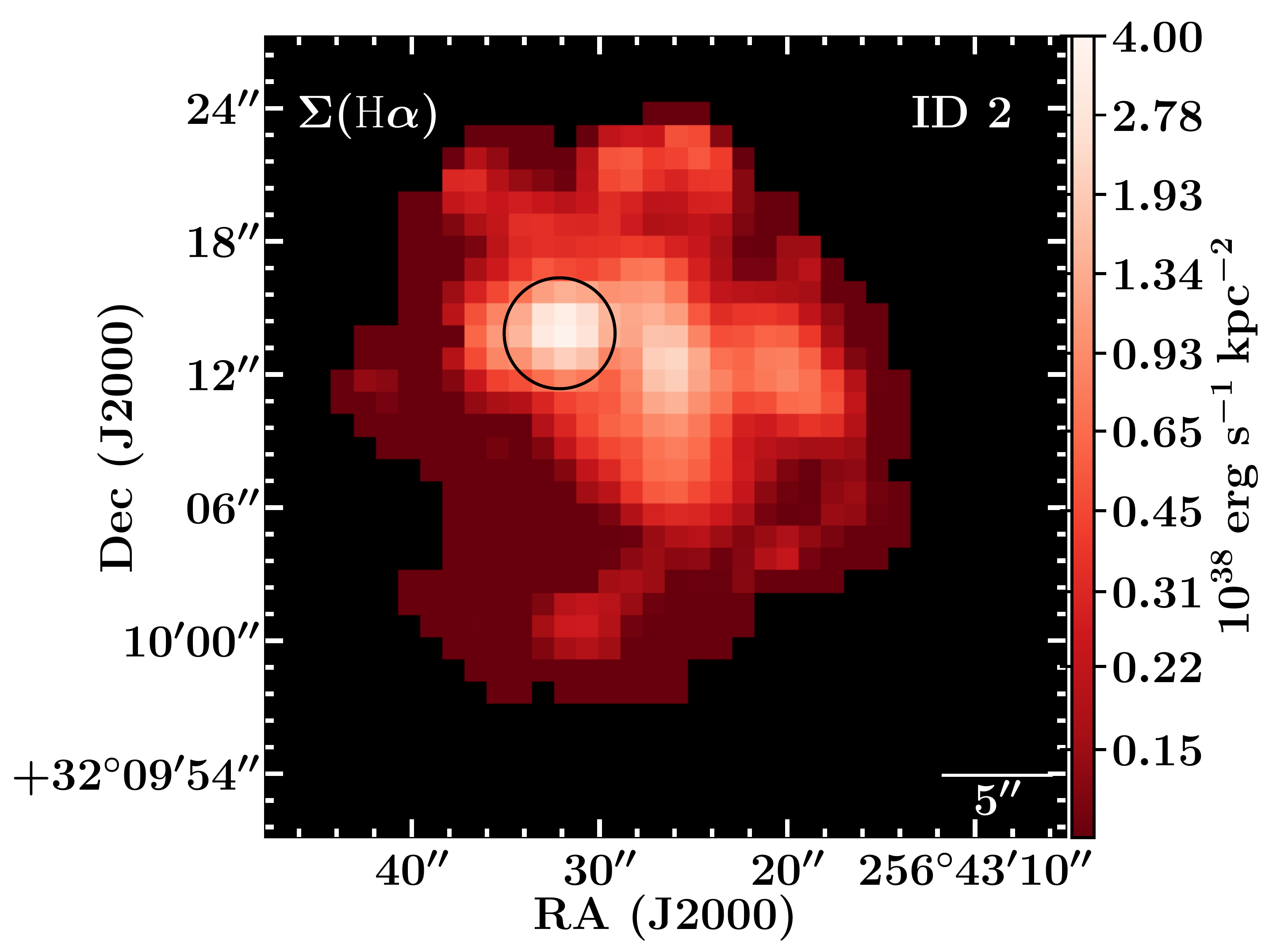}
\includegraphics[width=0.3\textwidth]{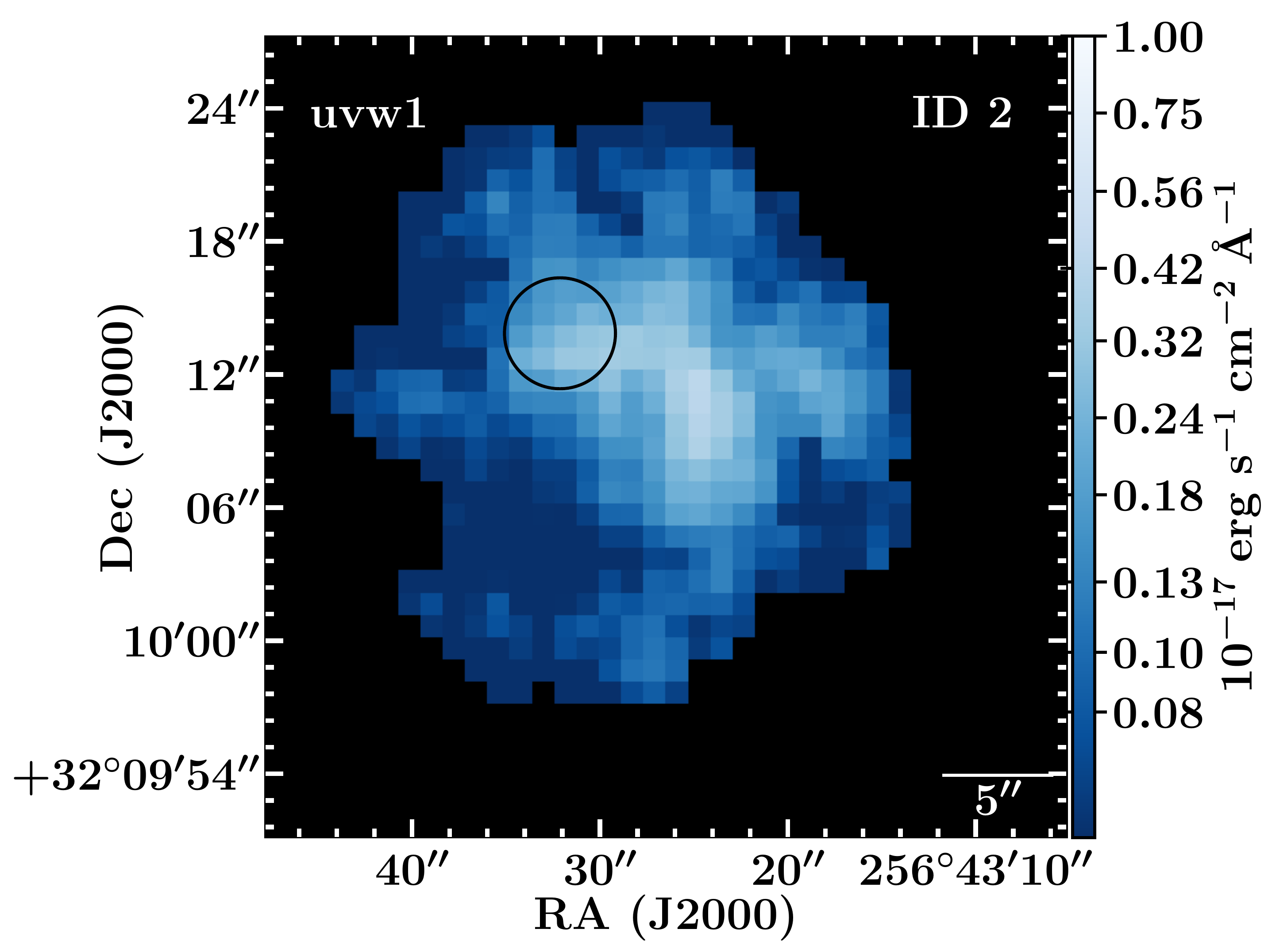}
\includegraphics[width=0.3\textwidth]{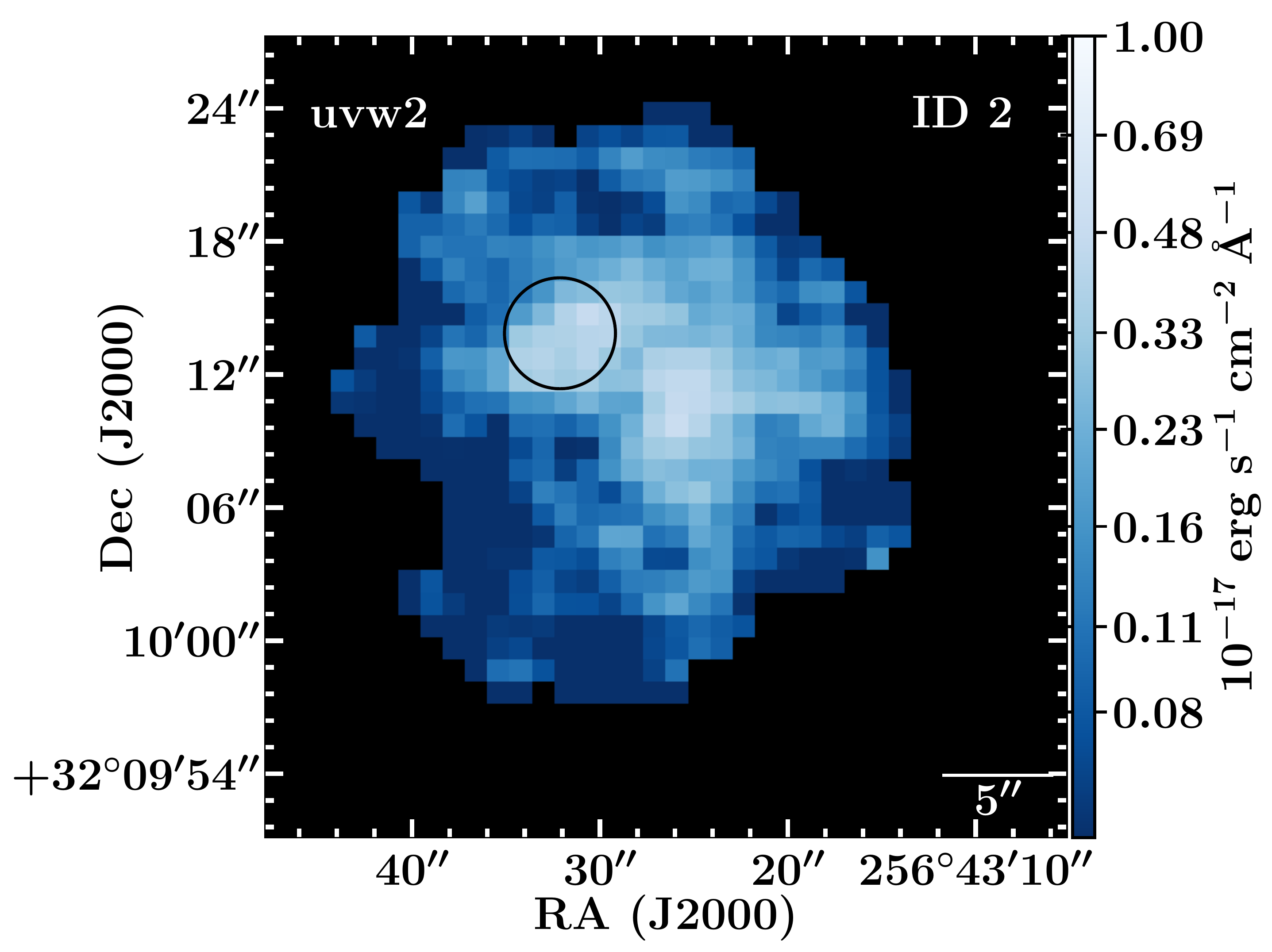}\\
\includegraphics[width=0.3\textwidth]{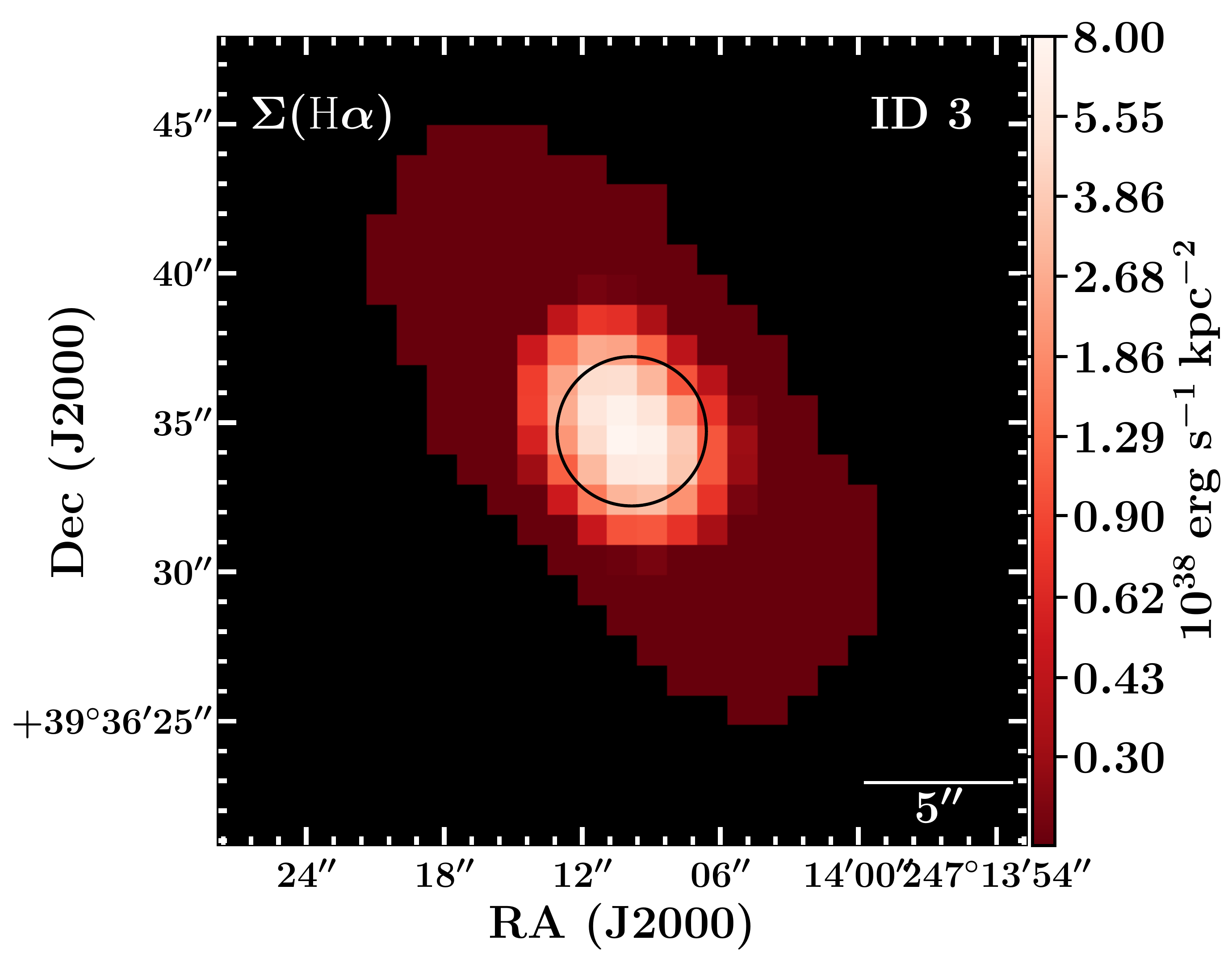}
\includegraphics[width=0.3\textwidth]{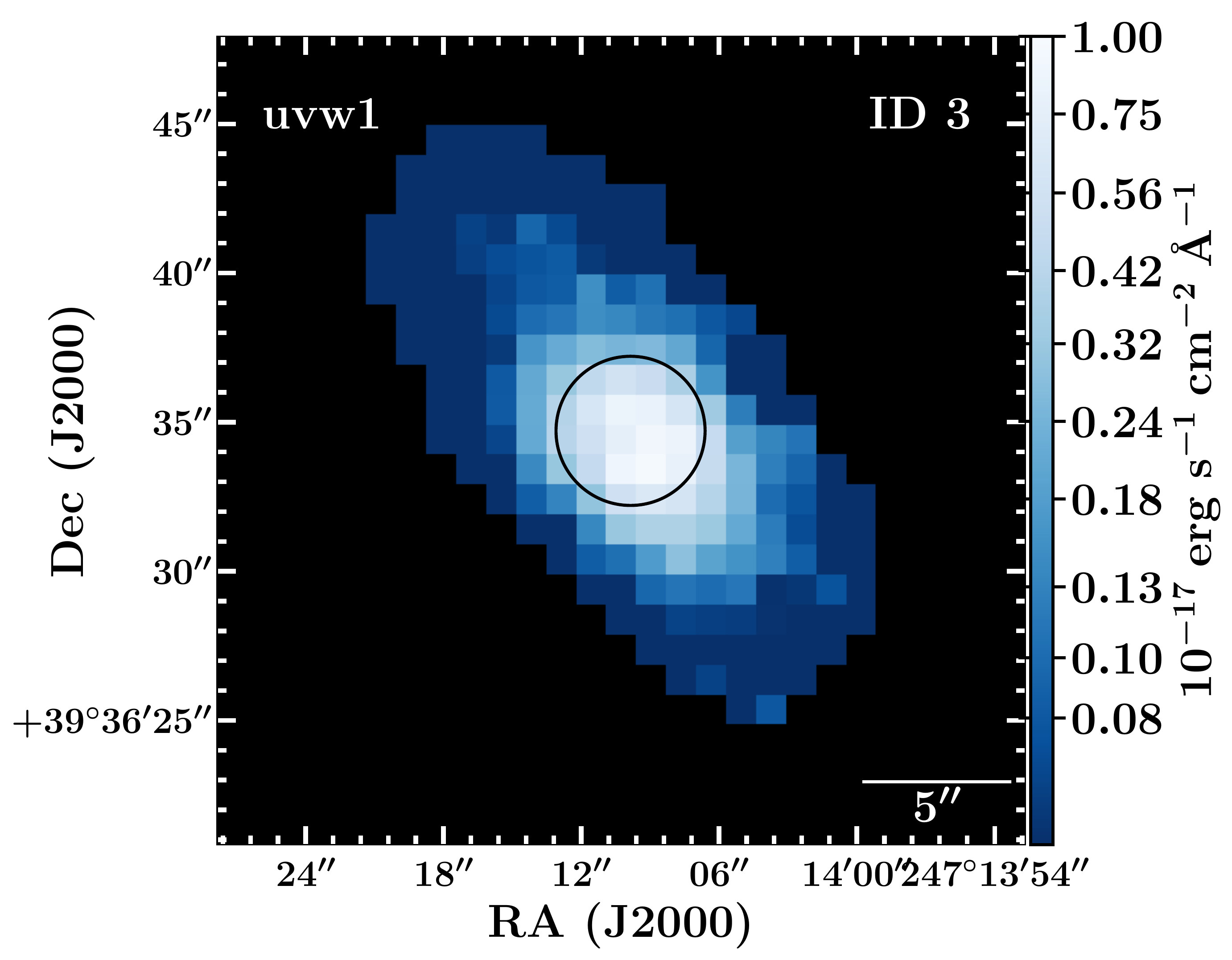}
\includegraphics[width=0.3\textwidth]{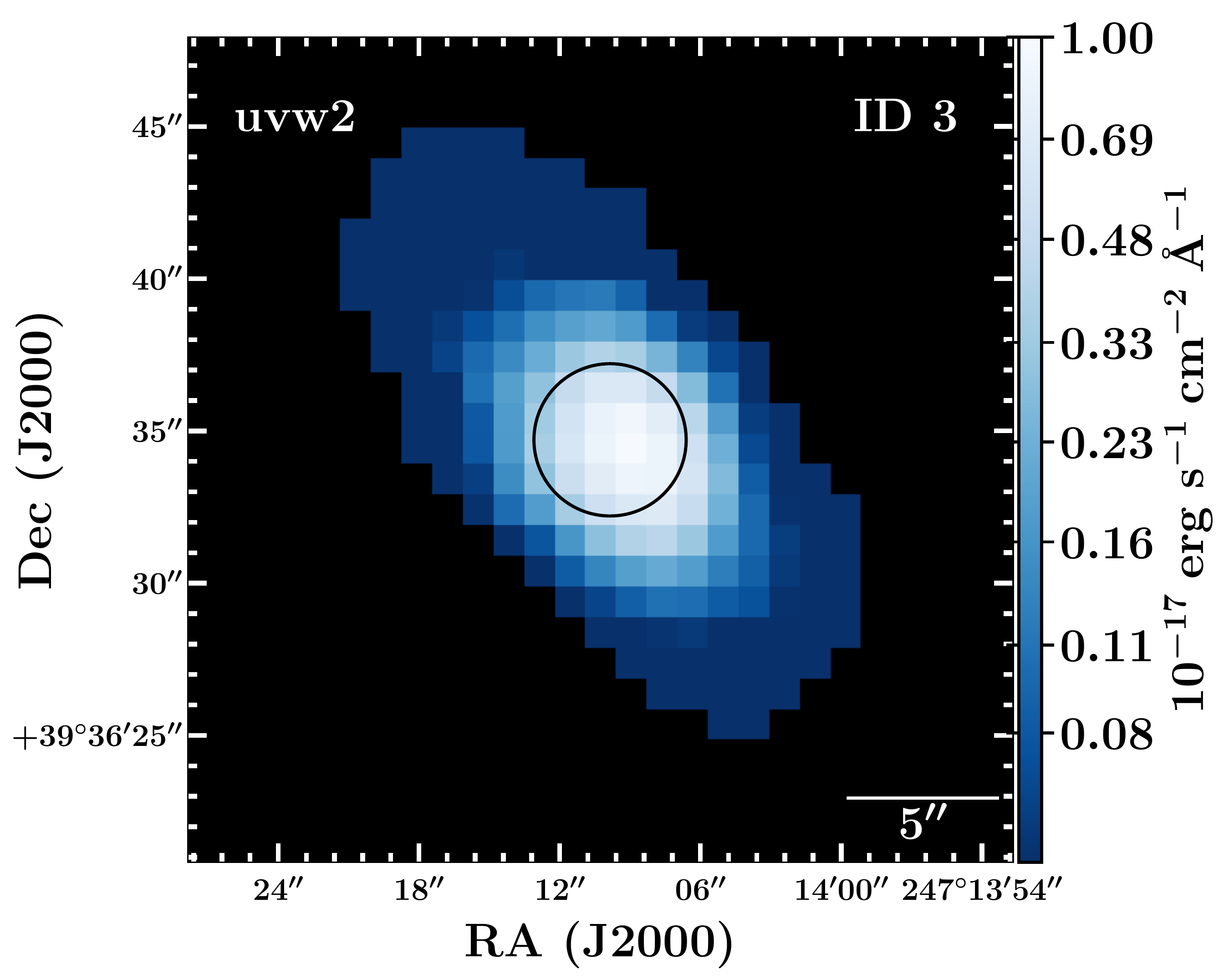}\\
\includegraphics[width=0.3\textwidth]{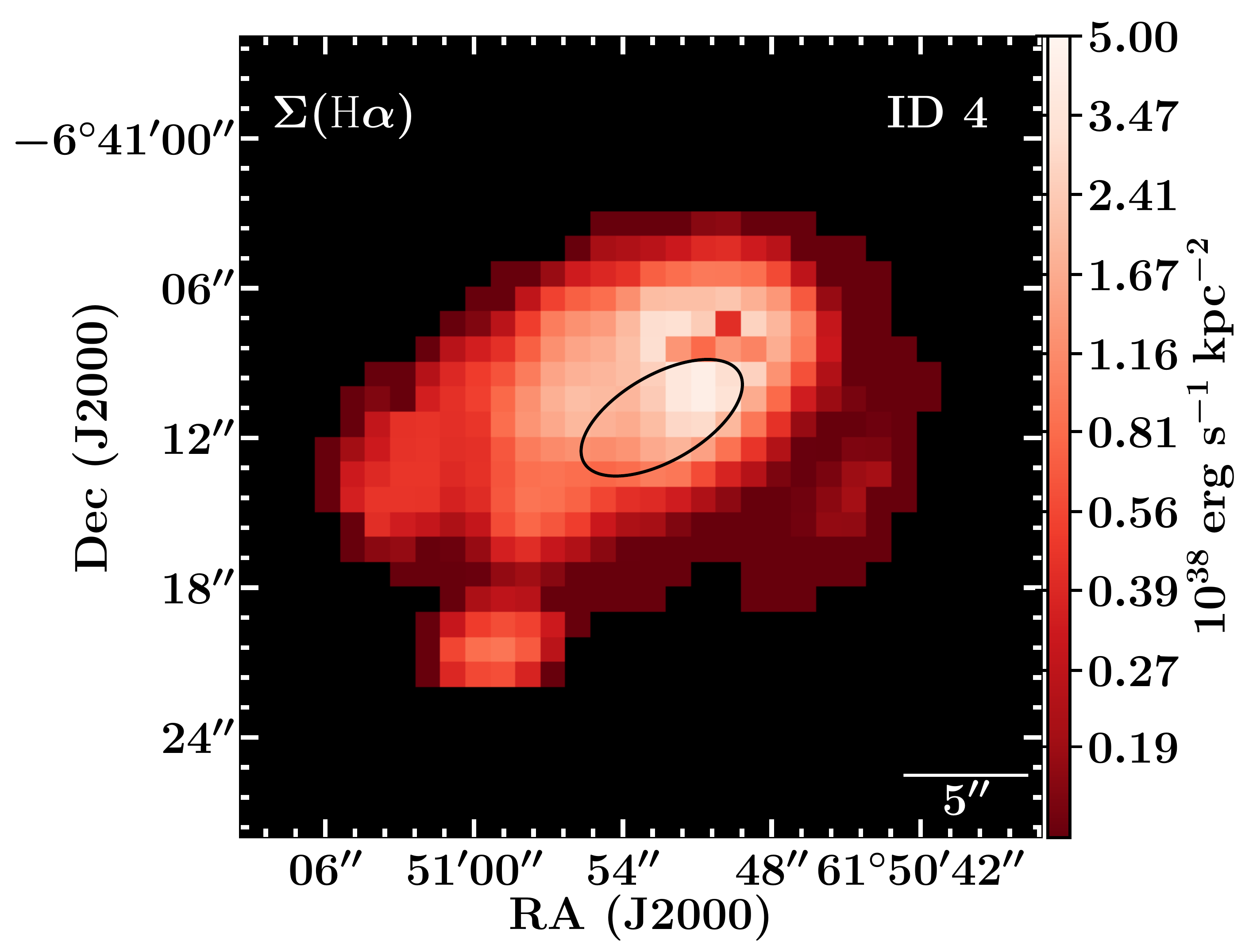}
\includegraphics[width=0.3\textwidth]{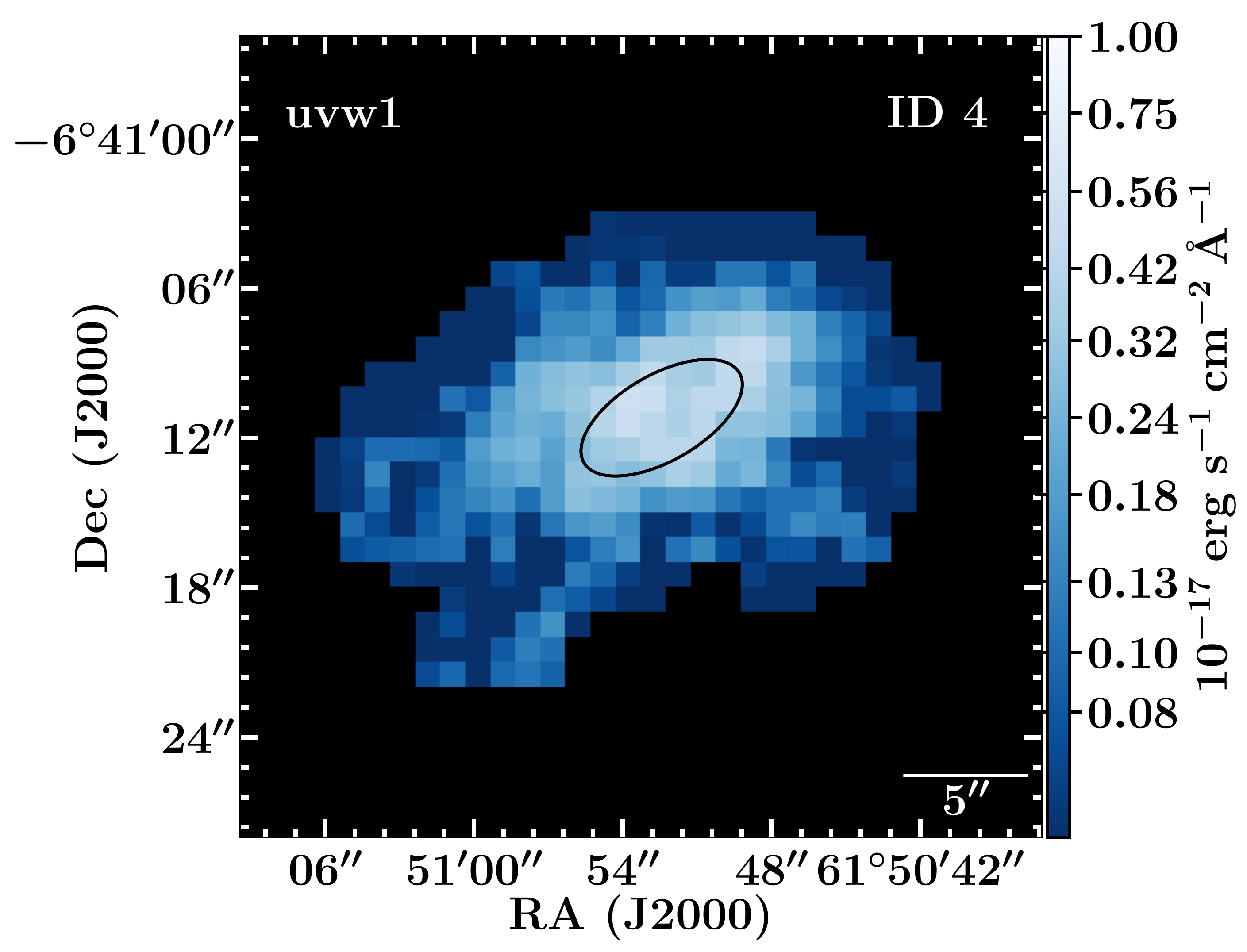}
\includegraphics[width=0.3\textwidth]{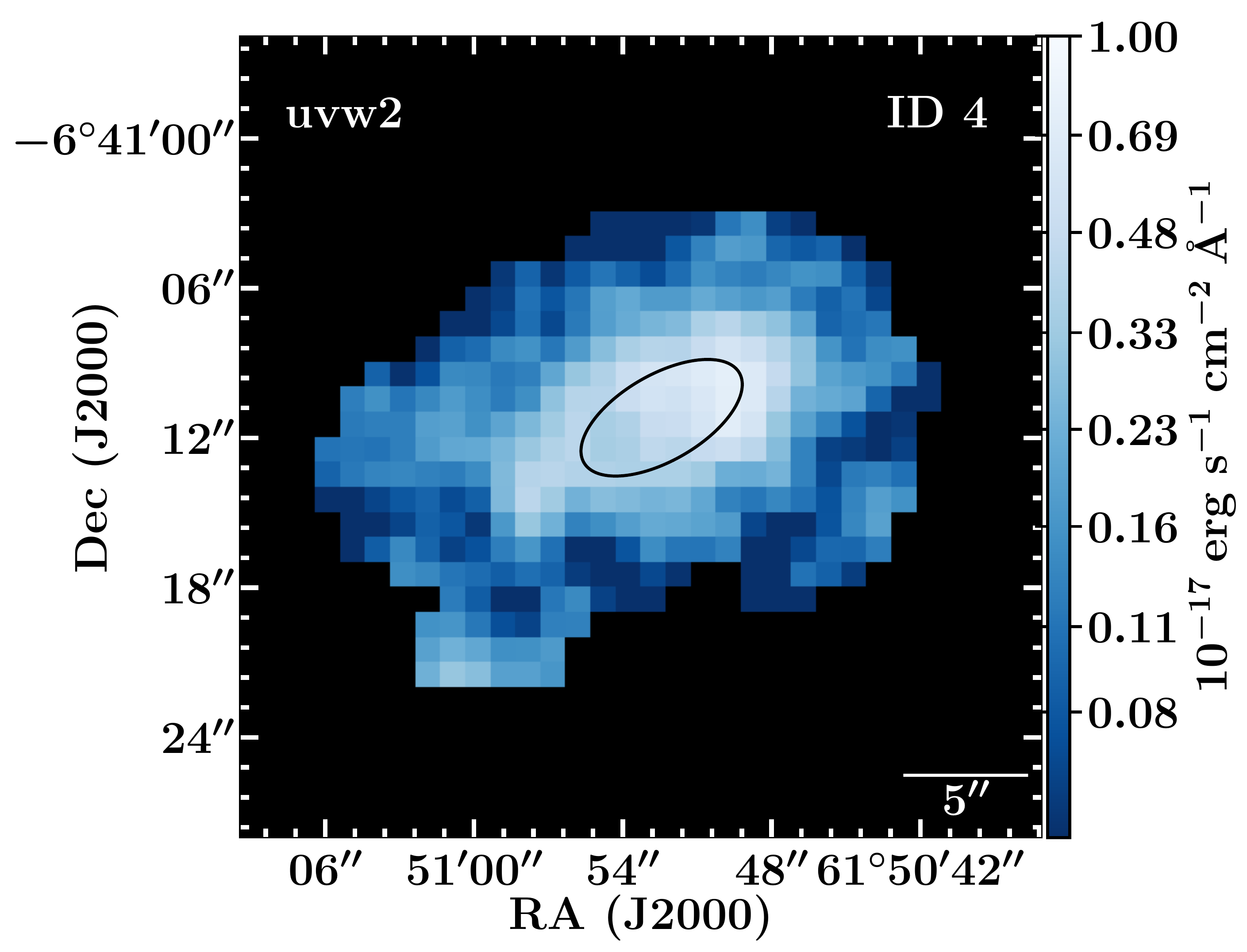}\\
\includegraphics[width=0.3\textwidth]{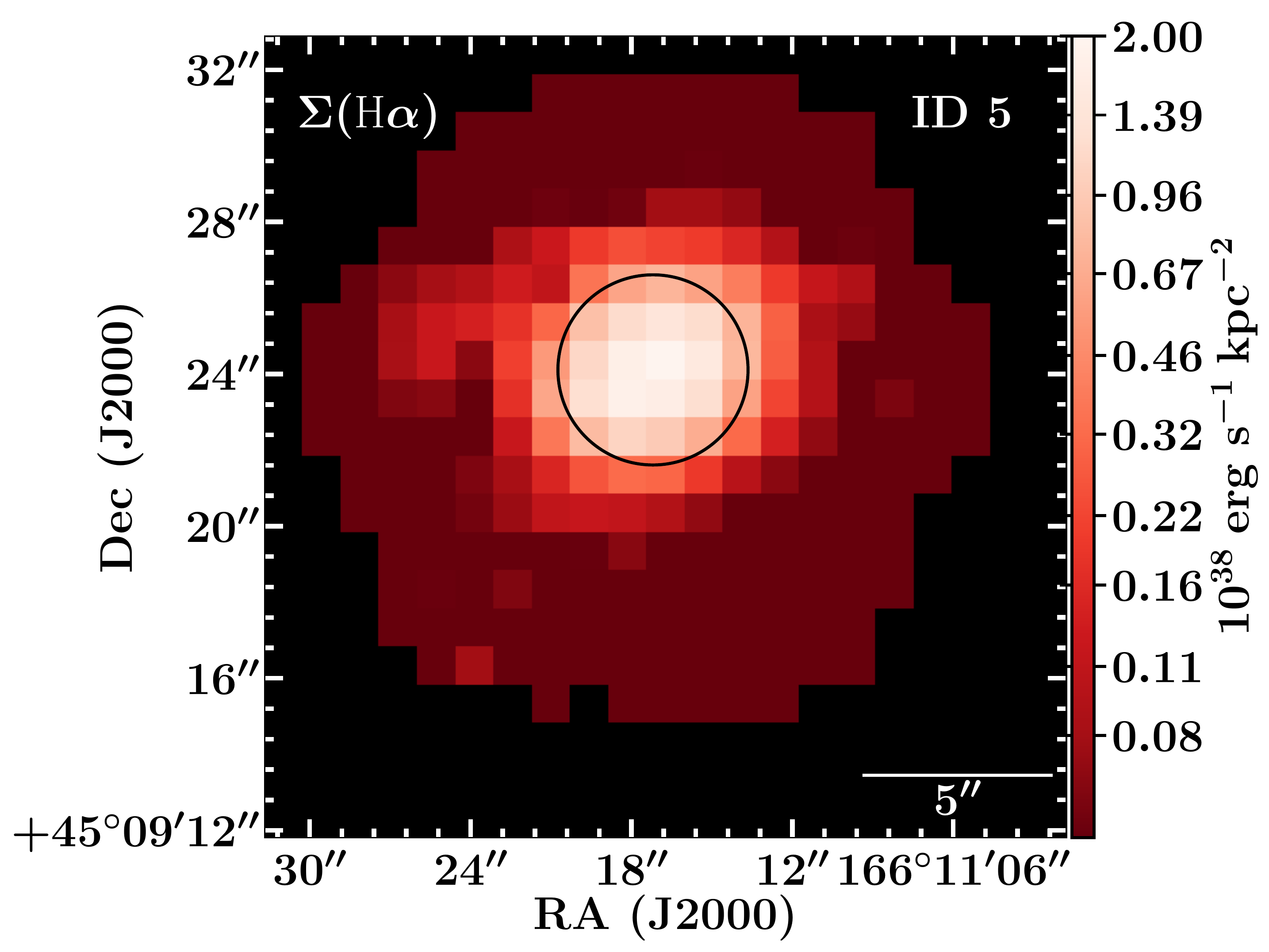}
\includegraphics[width=0.3\textwidth]{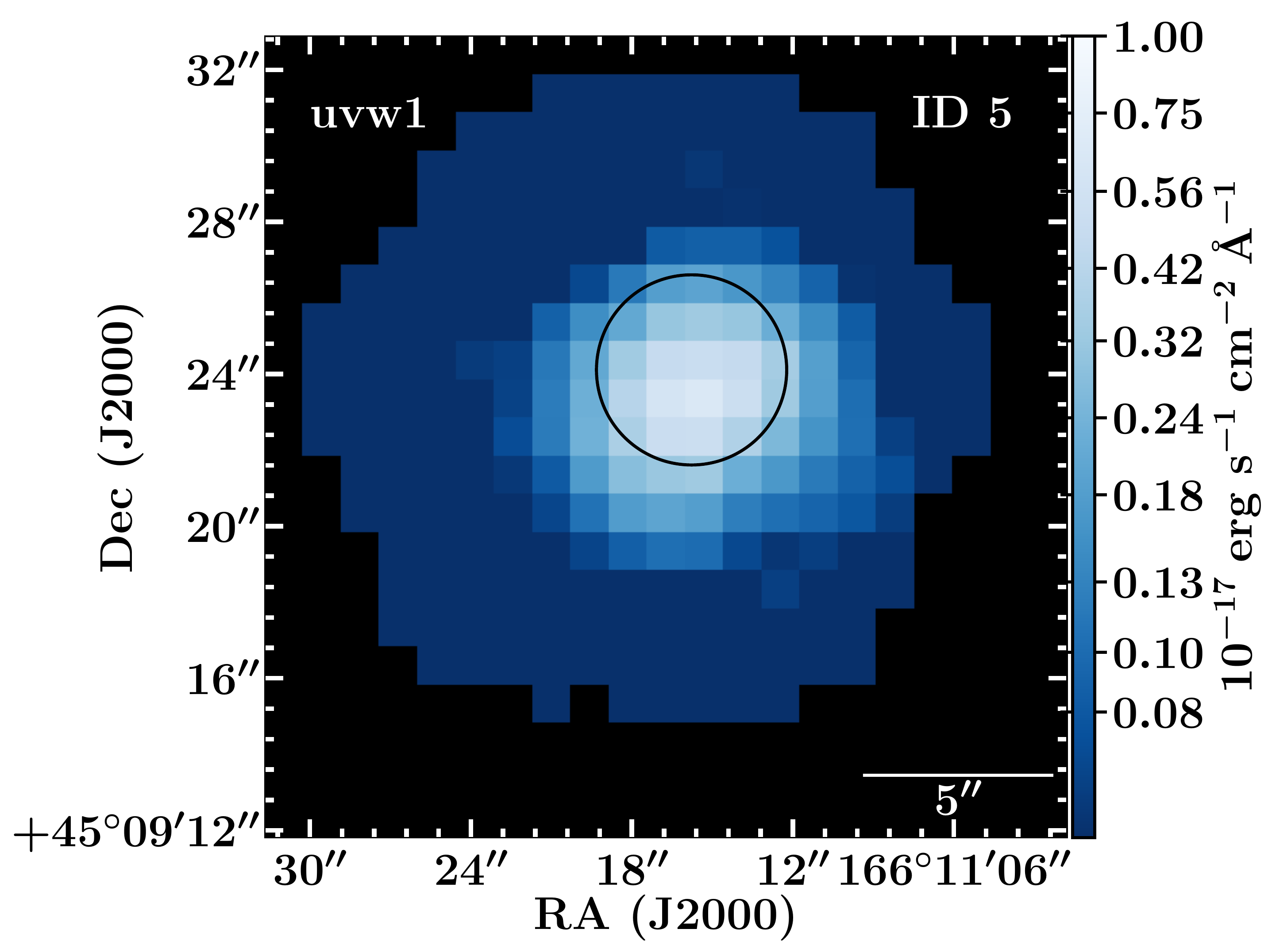}
\includegraphics[width=0.3\textwidth]{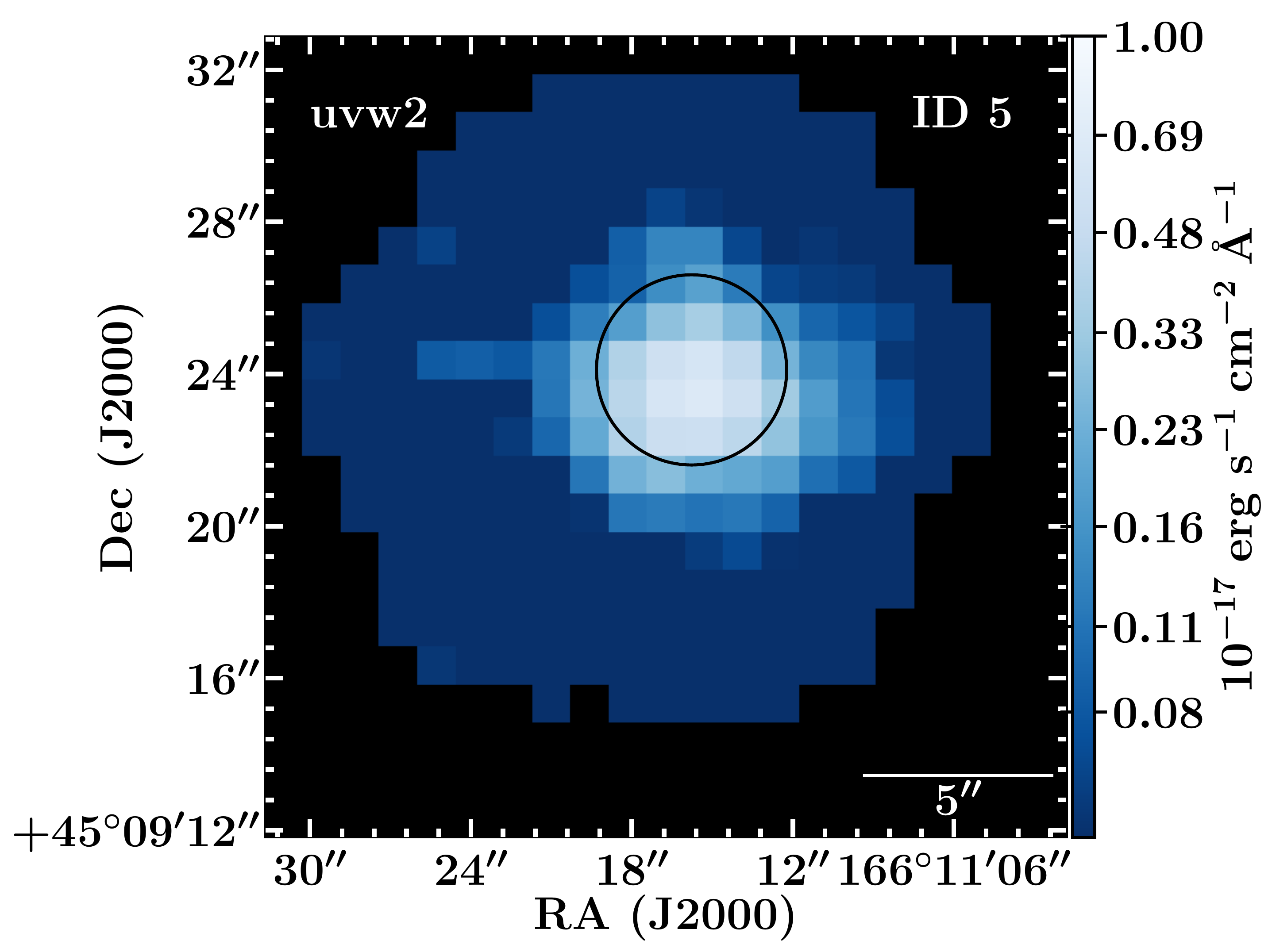}
\end{center}
\clearpage
\begin{center}
\includegraphics[width=0.3\textwidth]{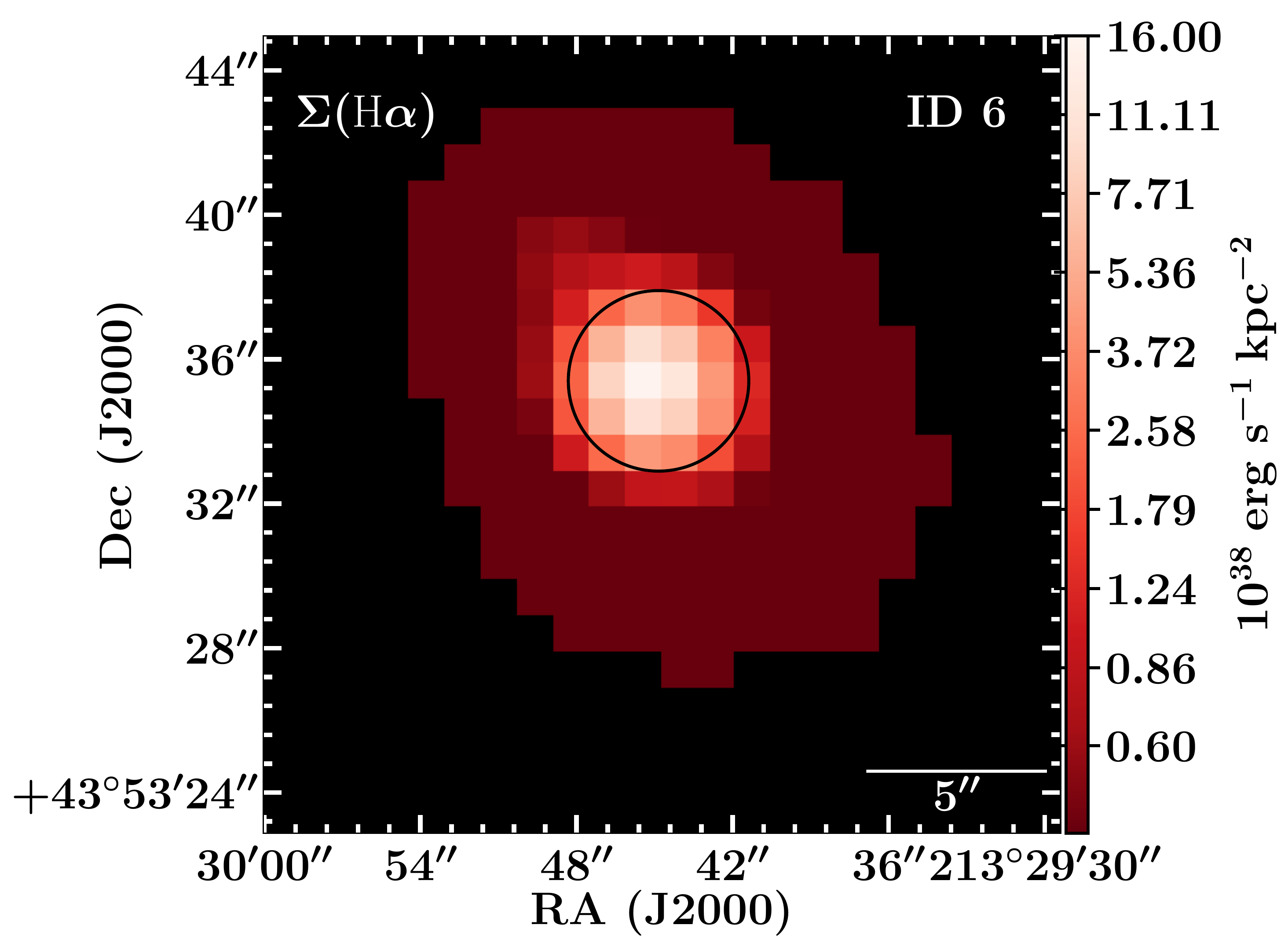}
\includegraphics[width=0.3\textwidth]{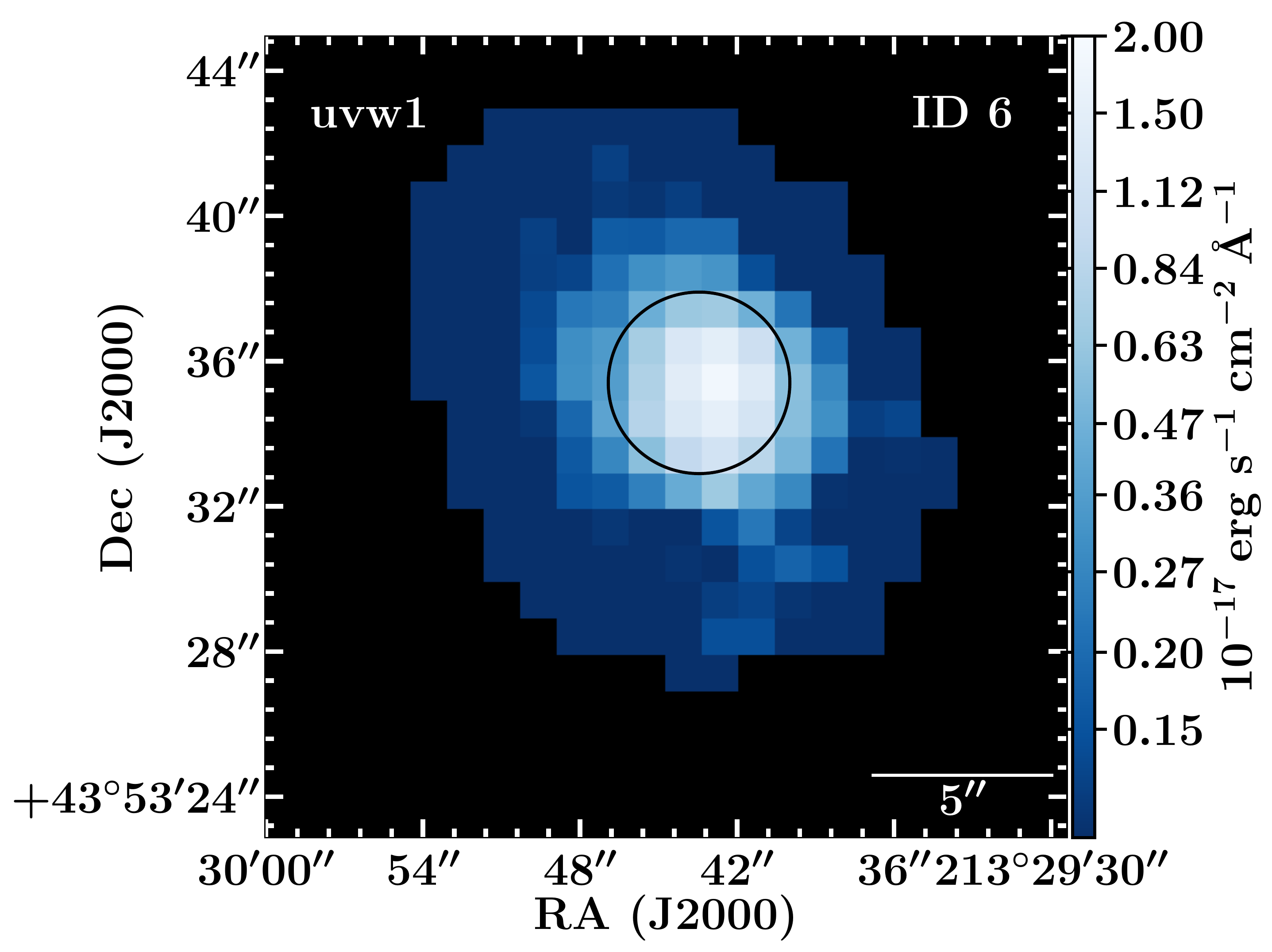}
\includegraphics[width=0.3\textwidth]{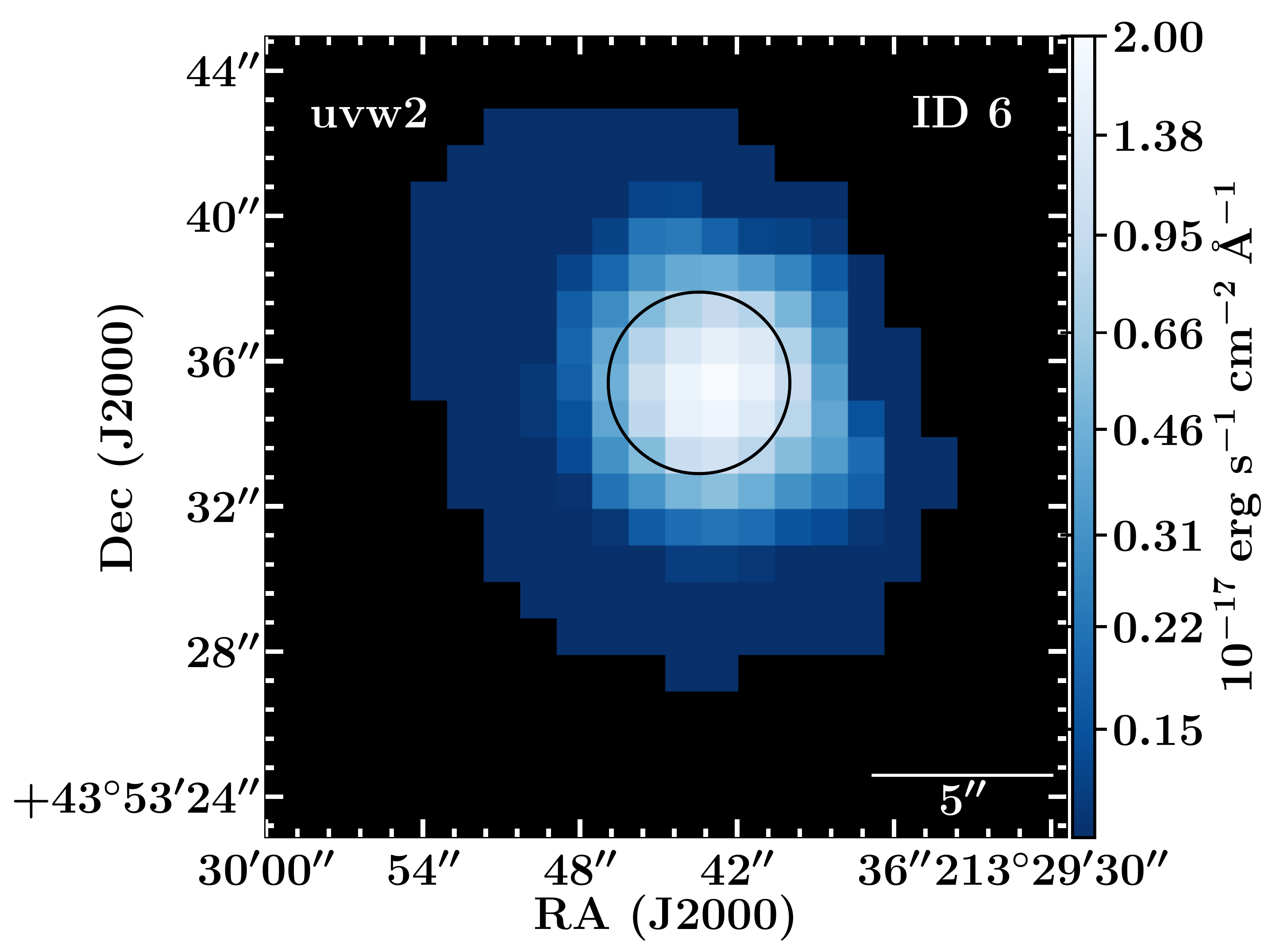}\\
\includegraphics[width=0.3\textwidth]{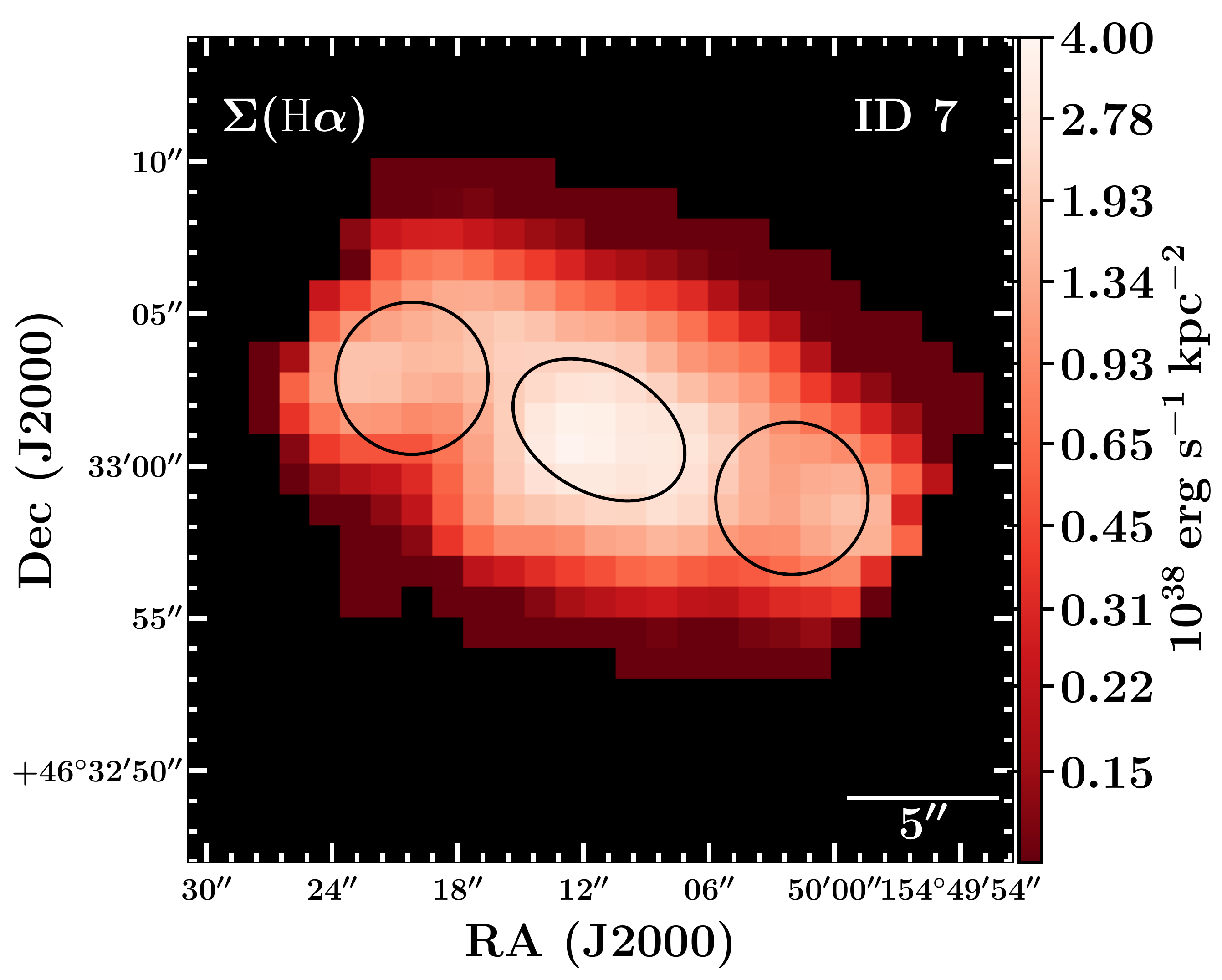}
\includegraphics[width=0.3\textwidth]{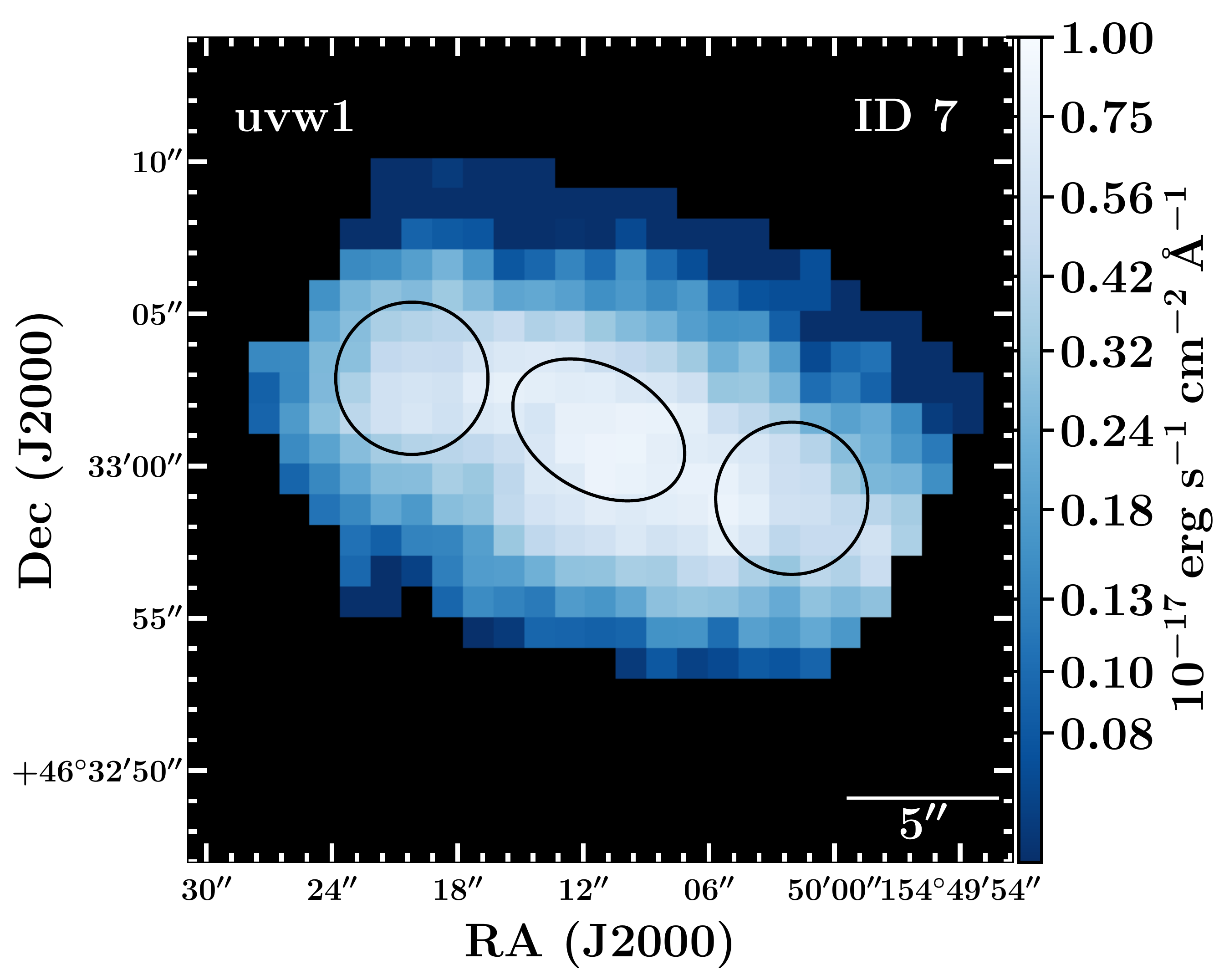}
\includegraphics[width=0.3\textwidth]{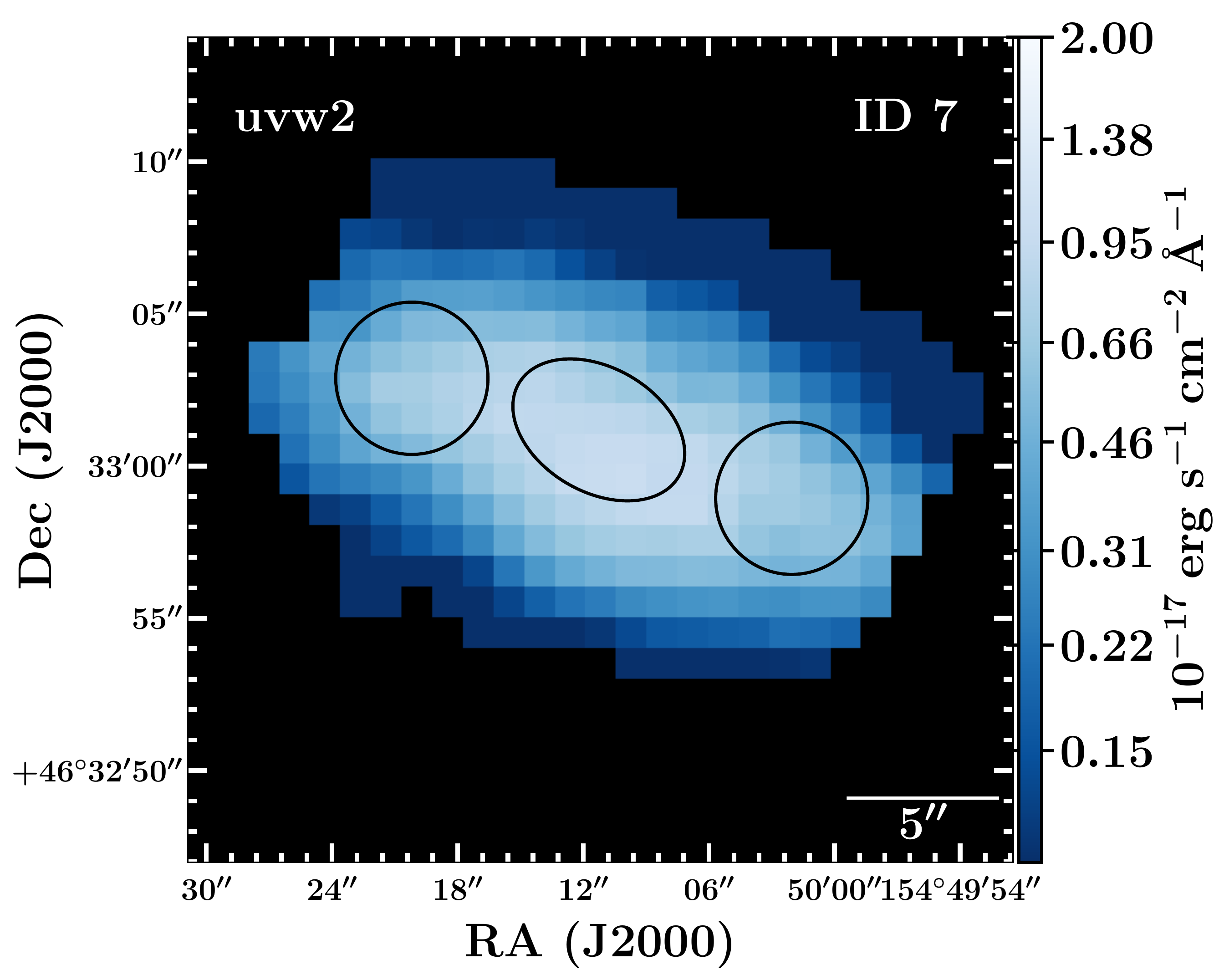}\\
\includegraphics[width=0.3\textwidth]{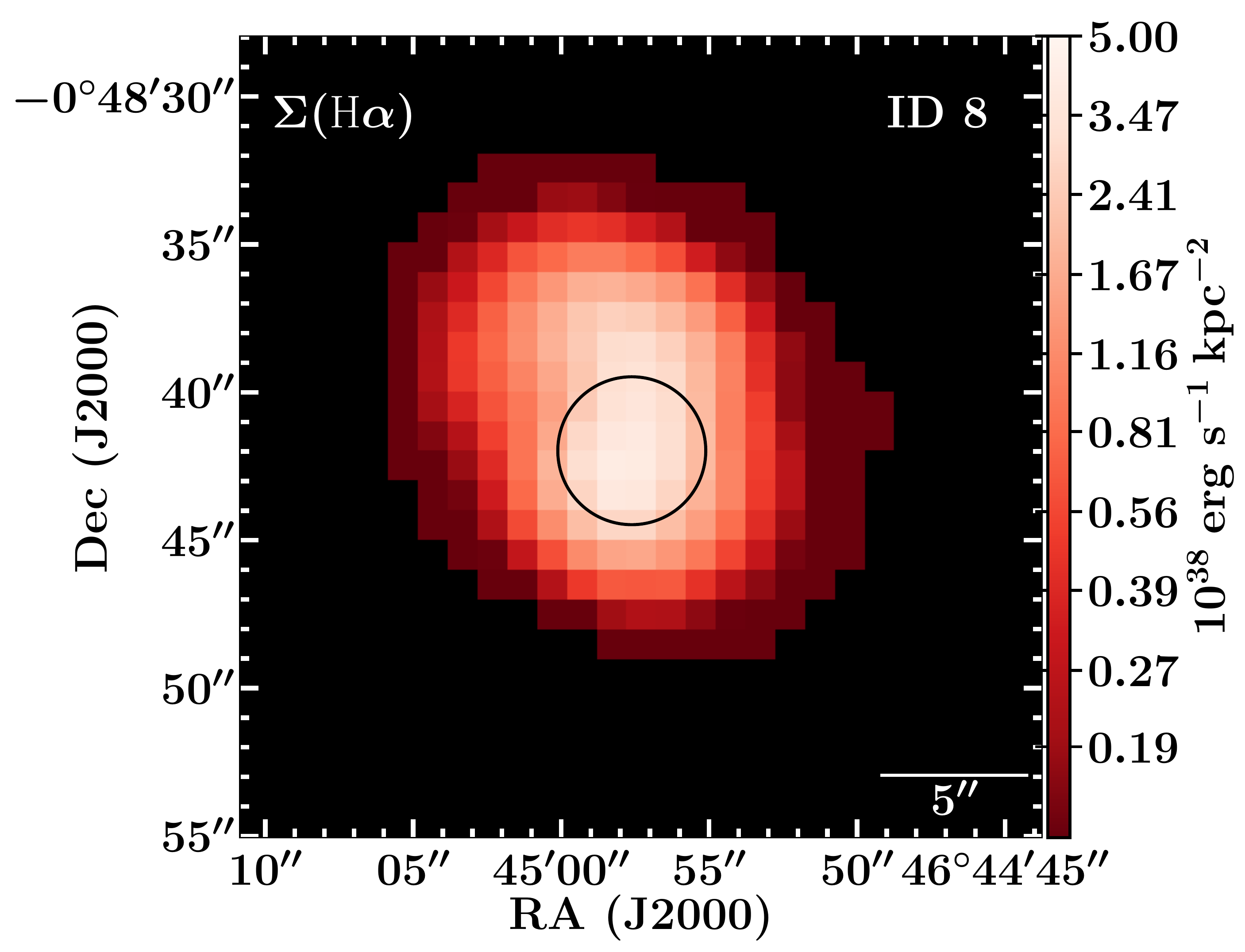}
\includegraphics[width=0.3\textwidth]{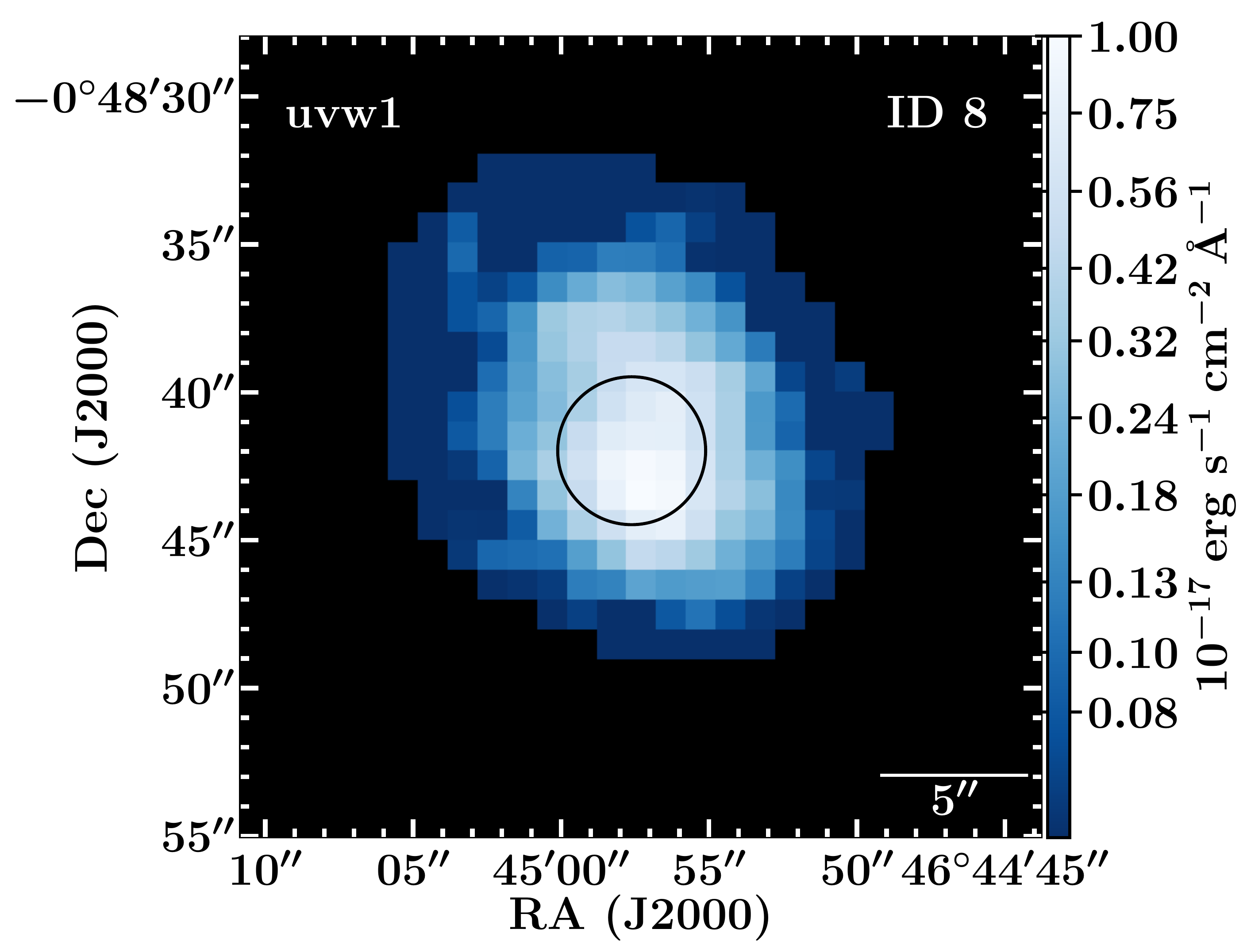}
\includegraphics[width=0.3\textwidth]{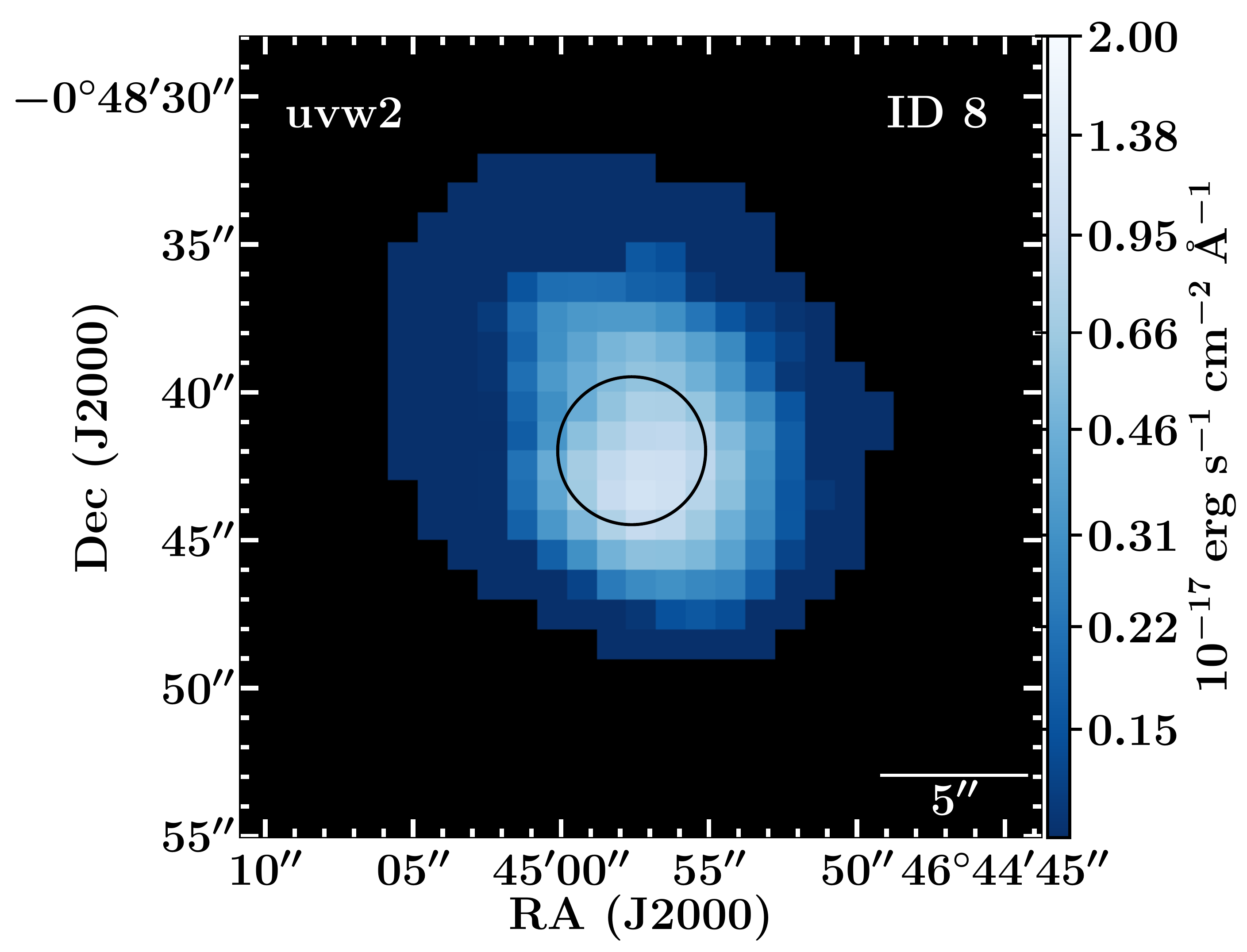}\\
\includegraphics[width=0.3\textwidth]{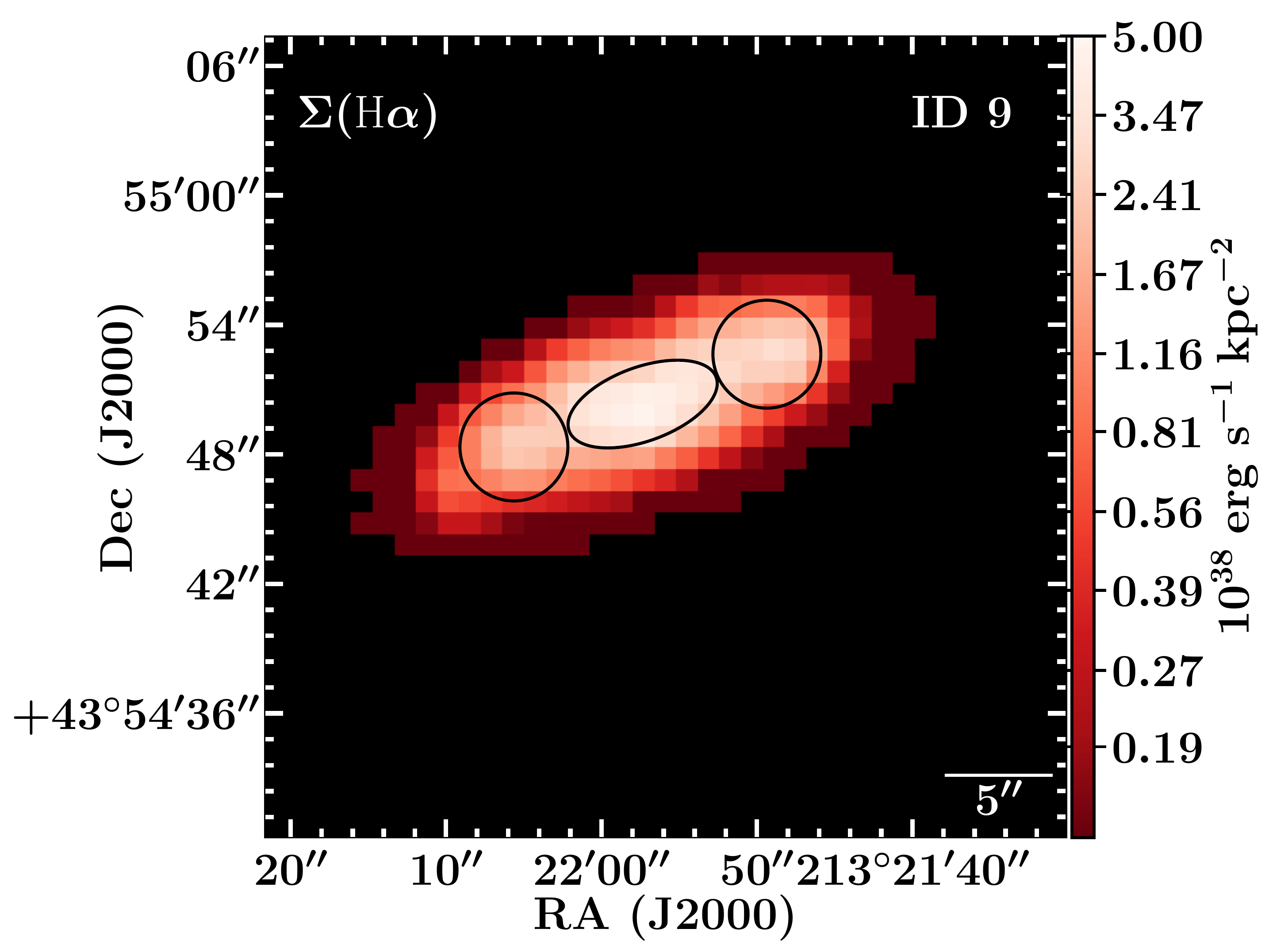}
\includegraphics[width=0.3\textwidth]{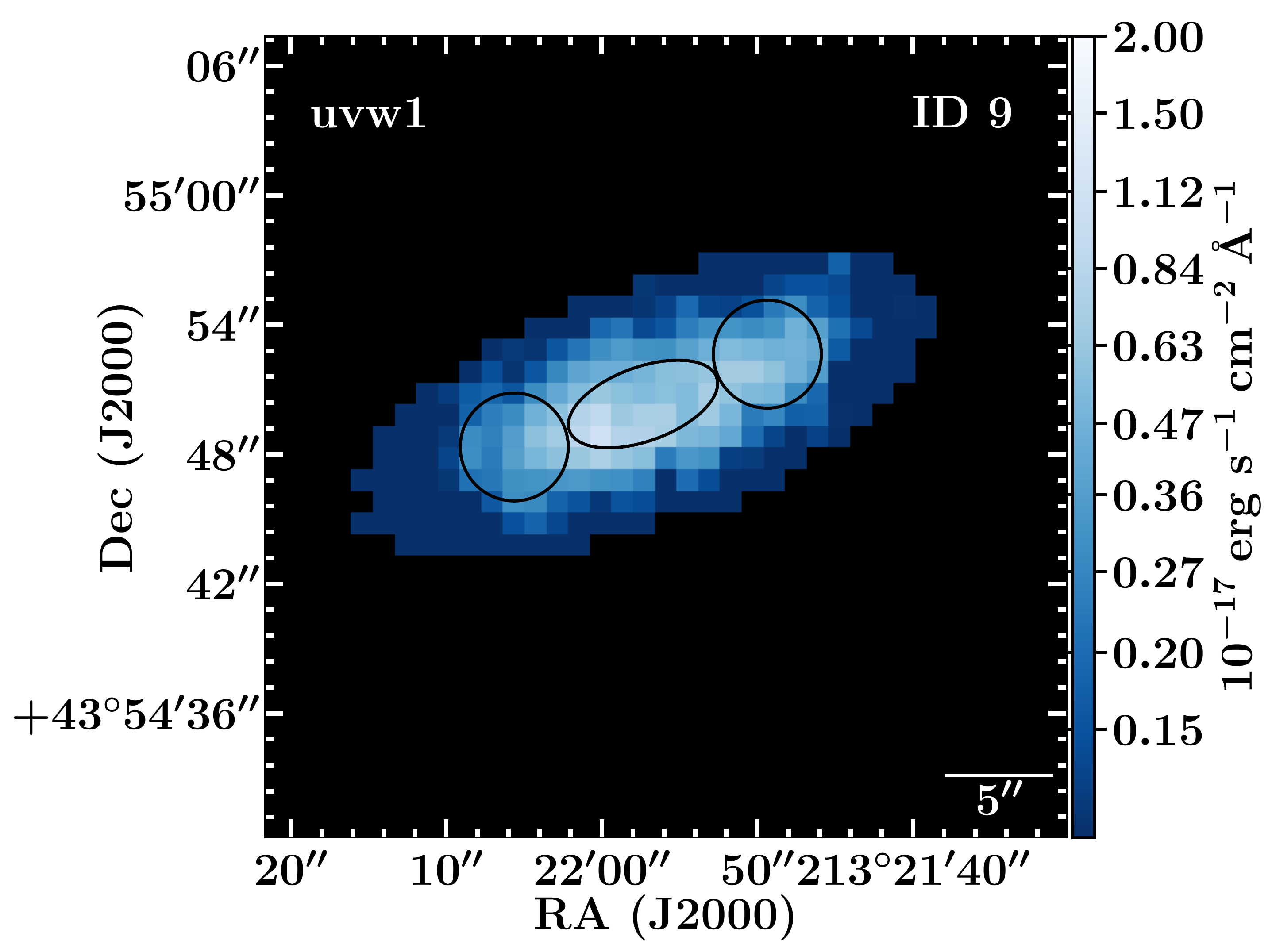}
\includegraphics[width=0.3\textwidth]{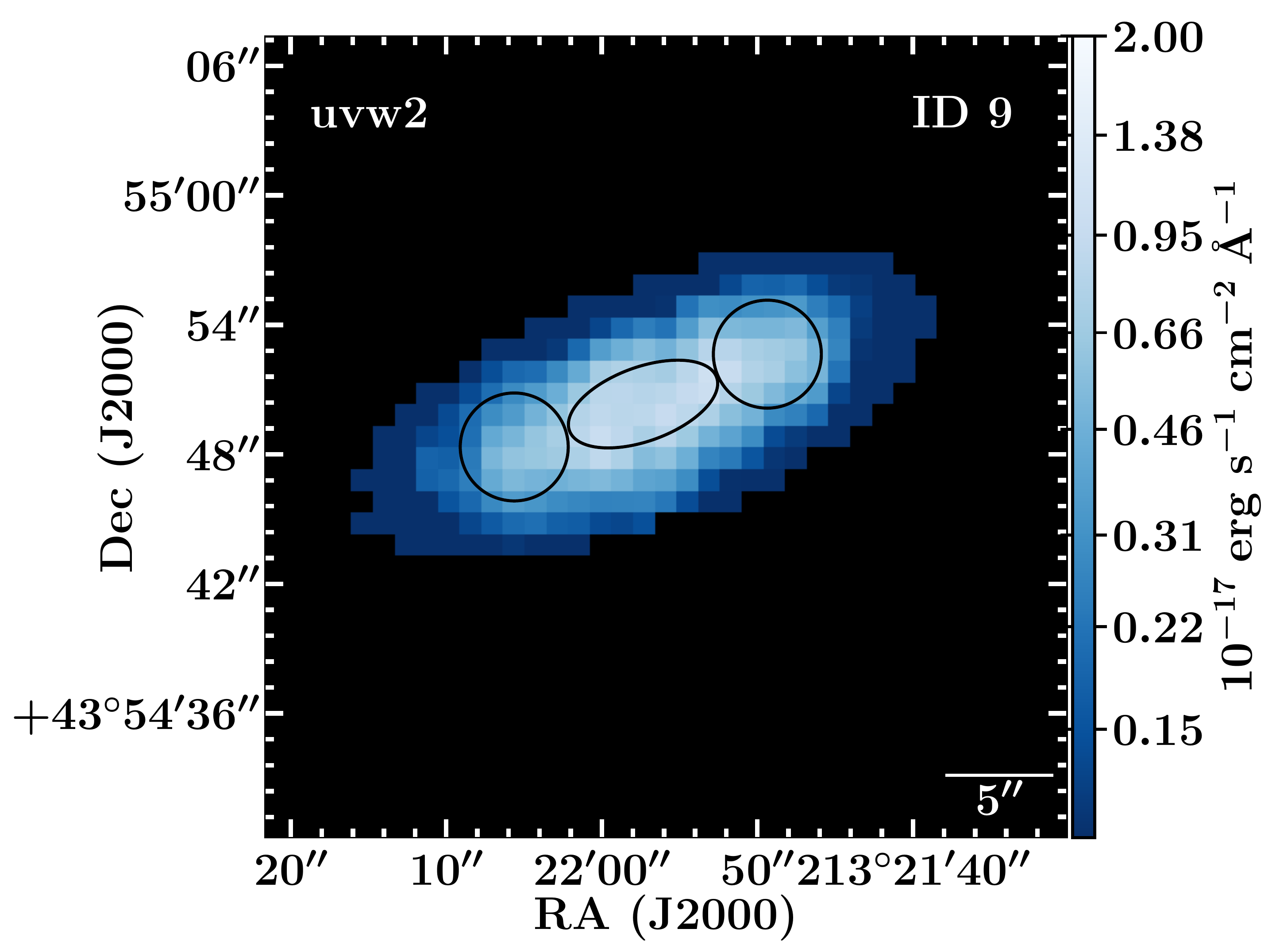}\\
\includegraphics[width=0.3\textwidth]{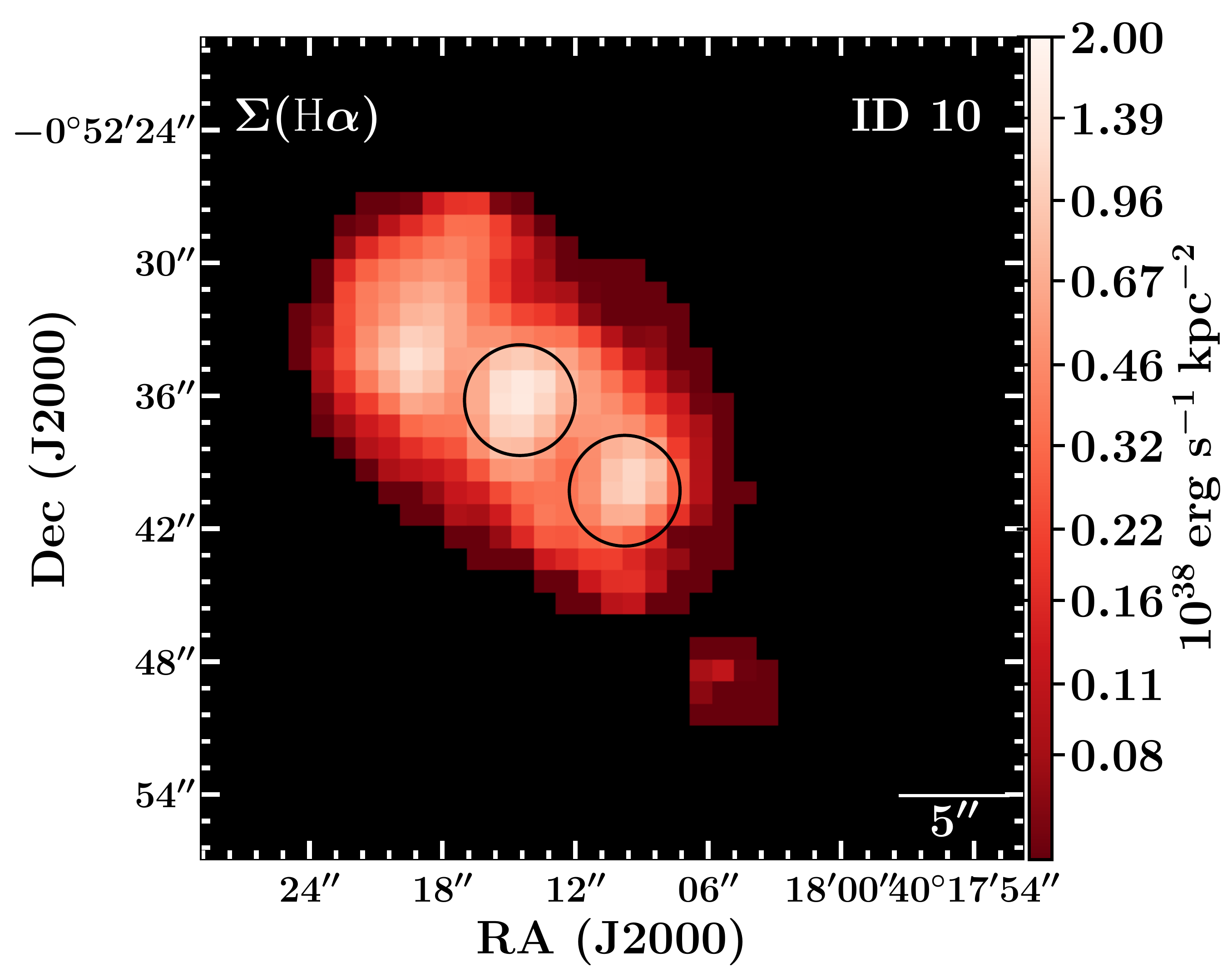}
\includegraphics[width=0.3\textwidth]{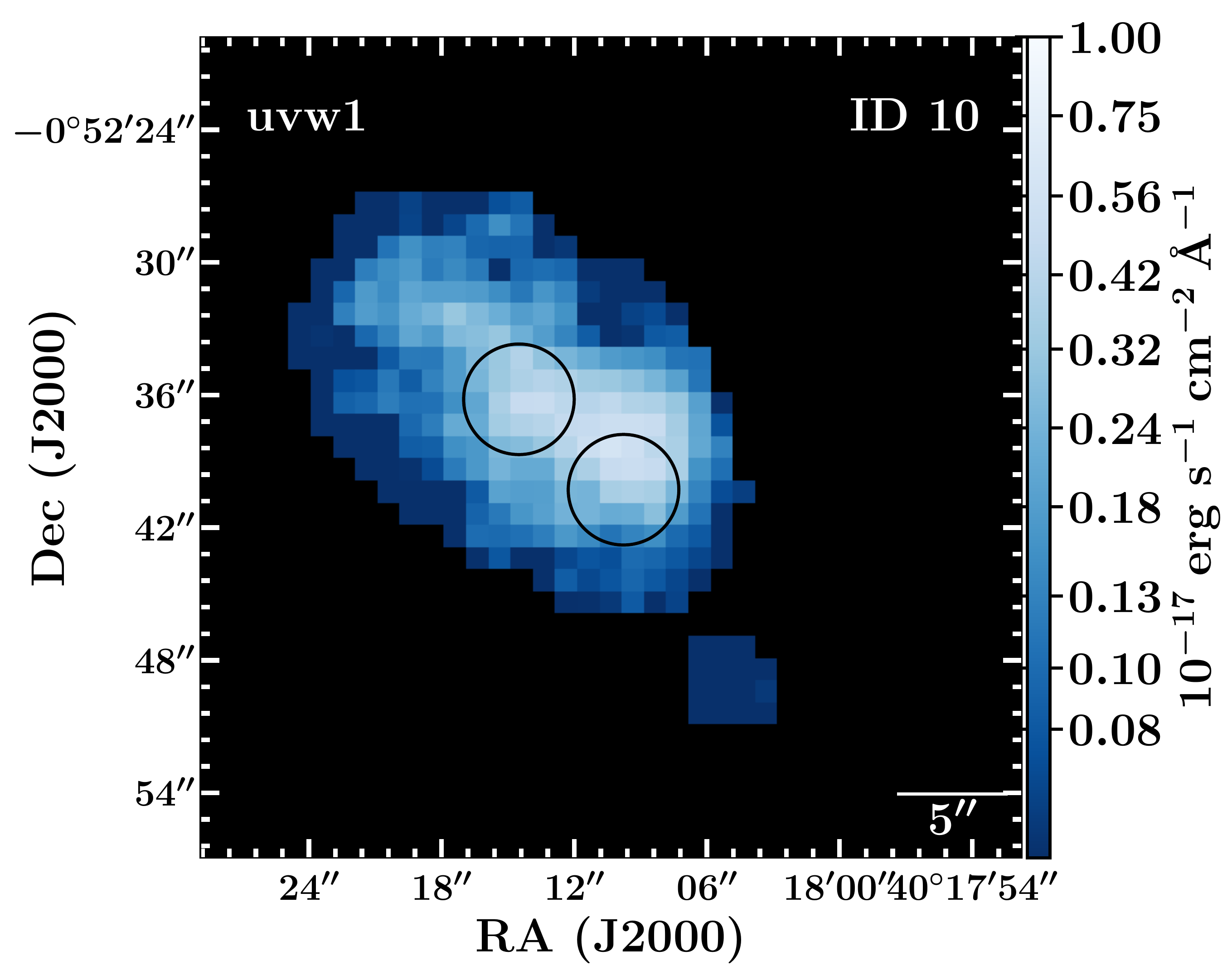}
\includegraphics[width=0.3\textwidth]{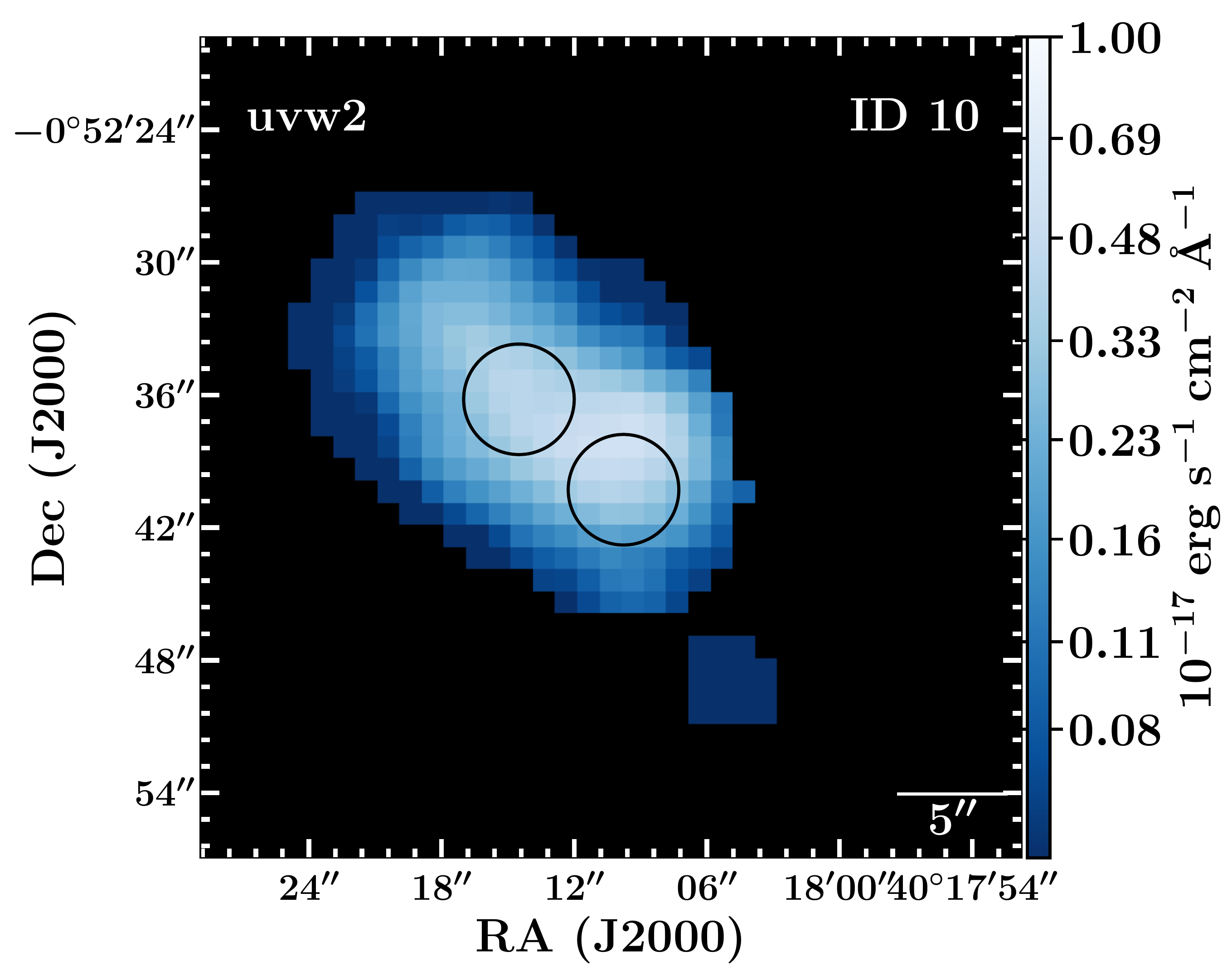}\\
\includegraphics[width=0.3\textwidth]{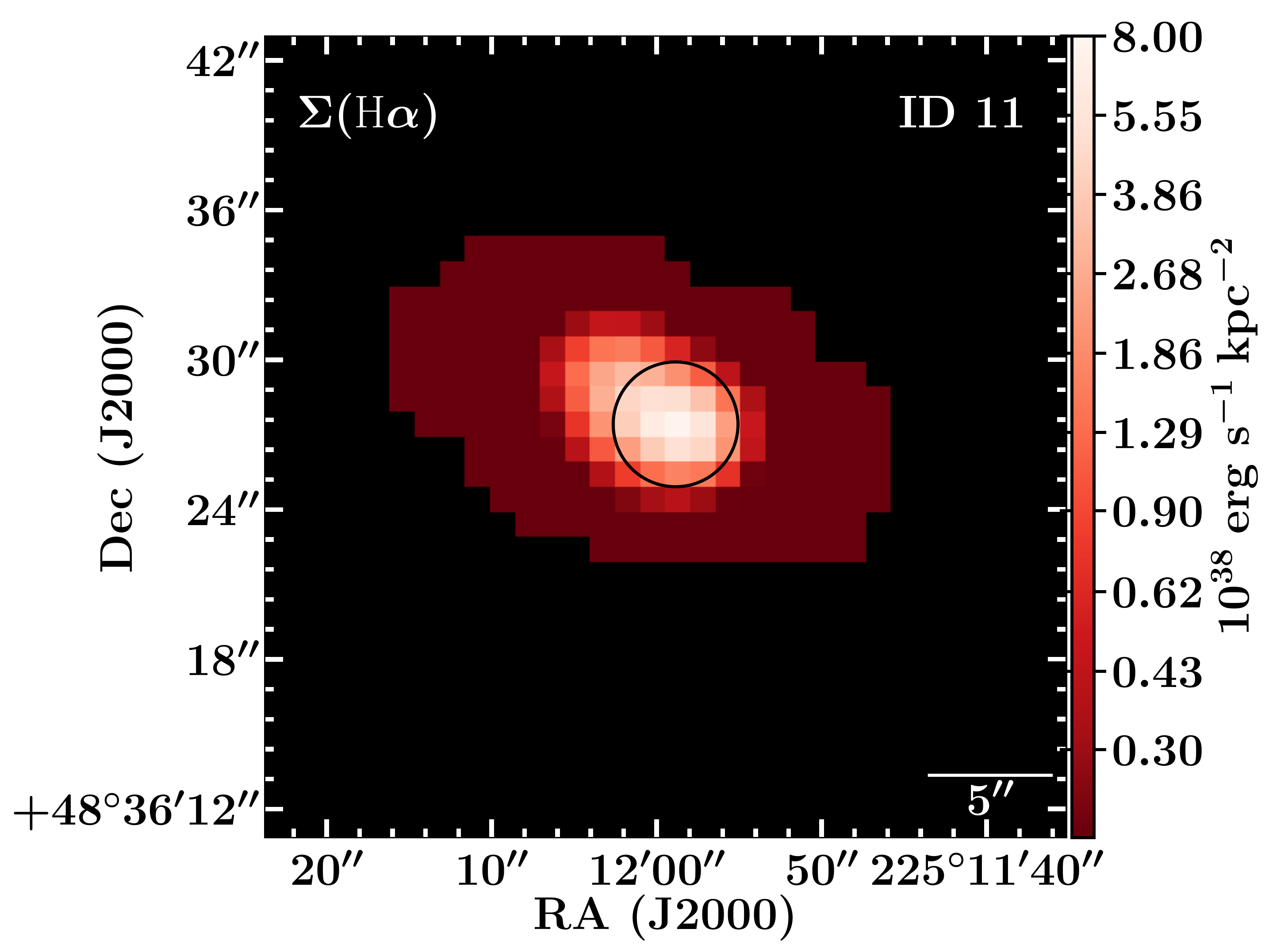}
\includegraphics[width=0.3\textwidth]{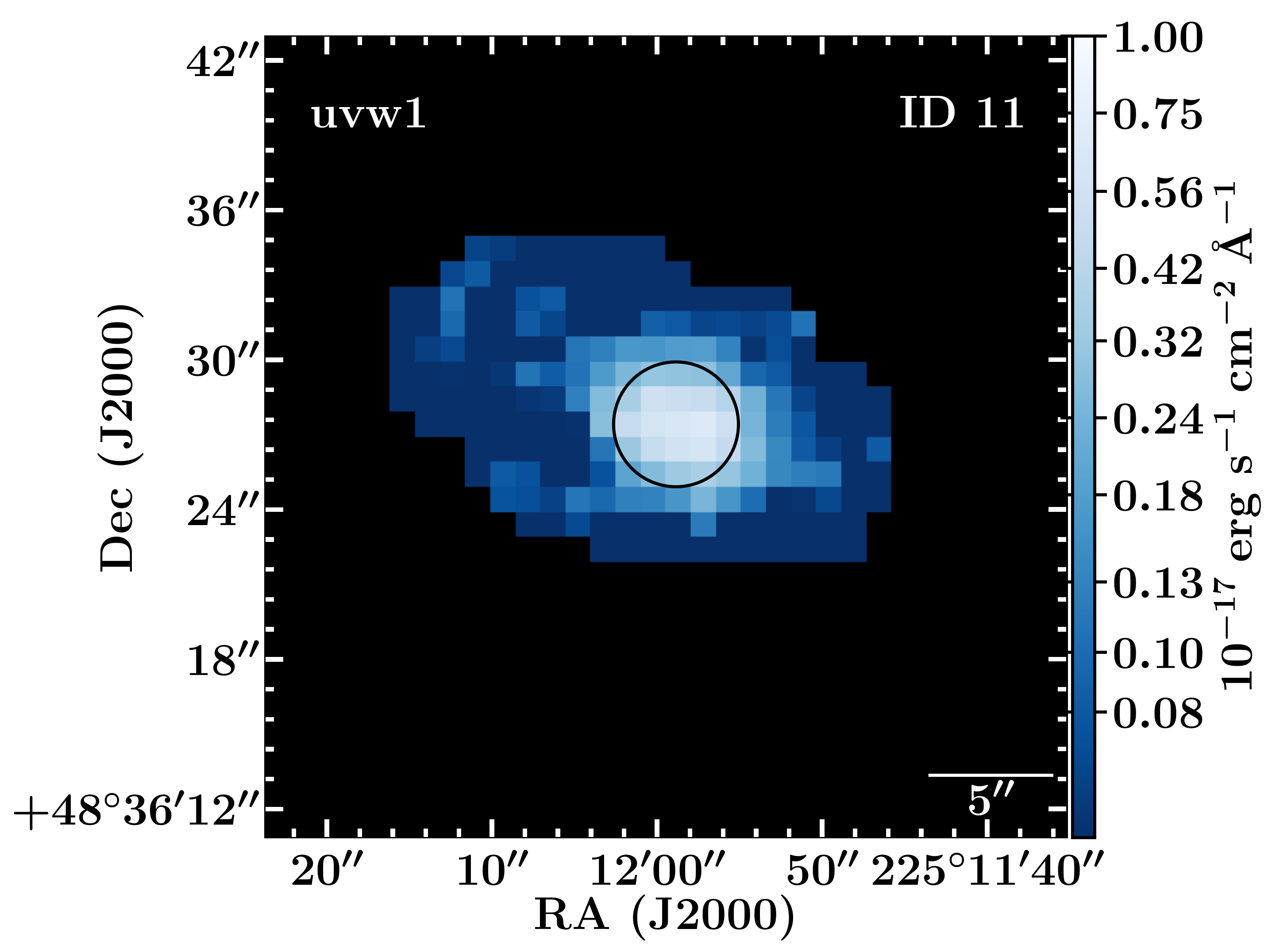}
\includegraphics[width=0.3\textwidth]{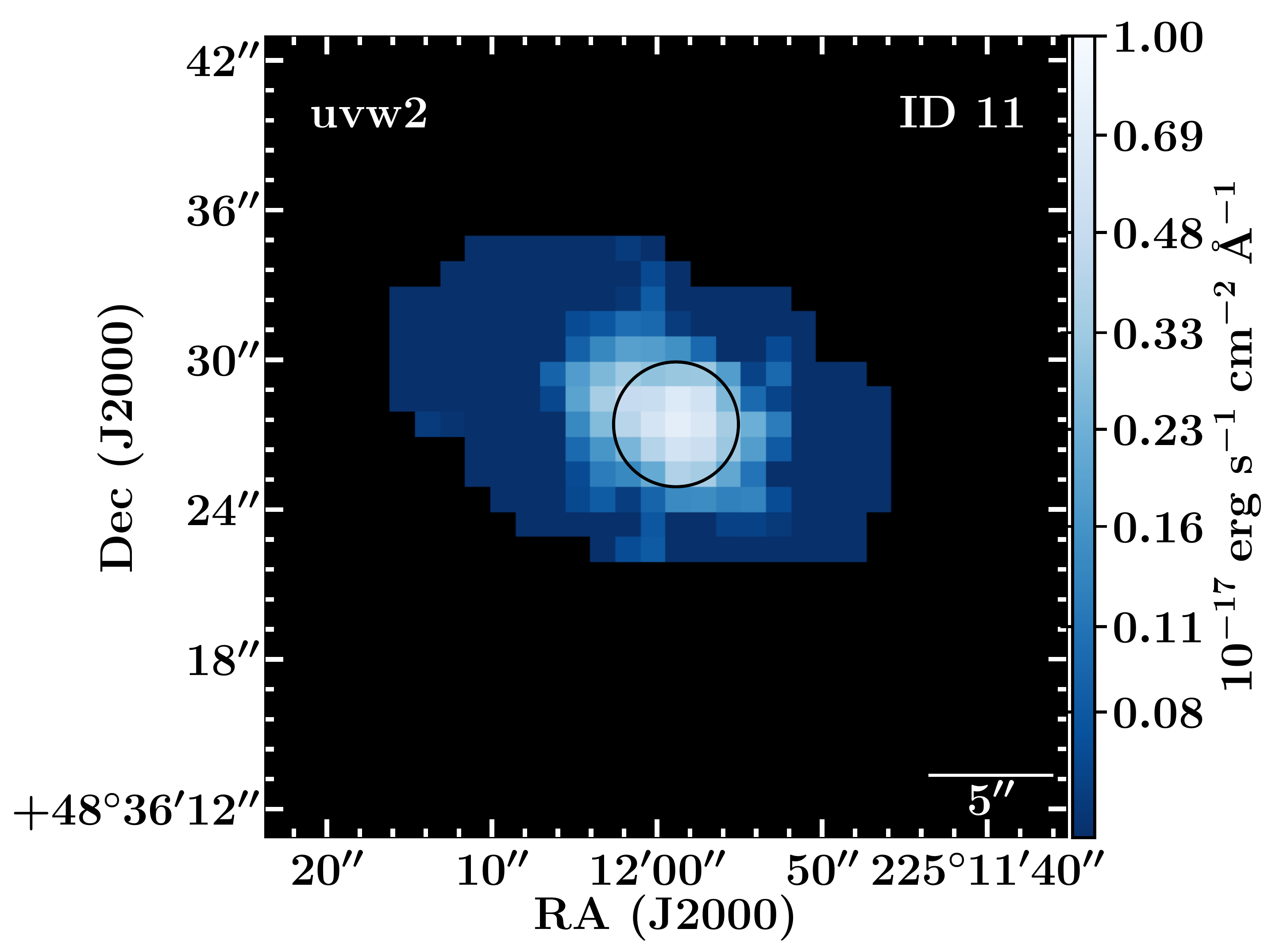}\\
\includegraphics[width=0.3\textwidth]{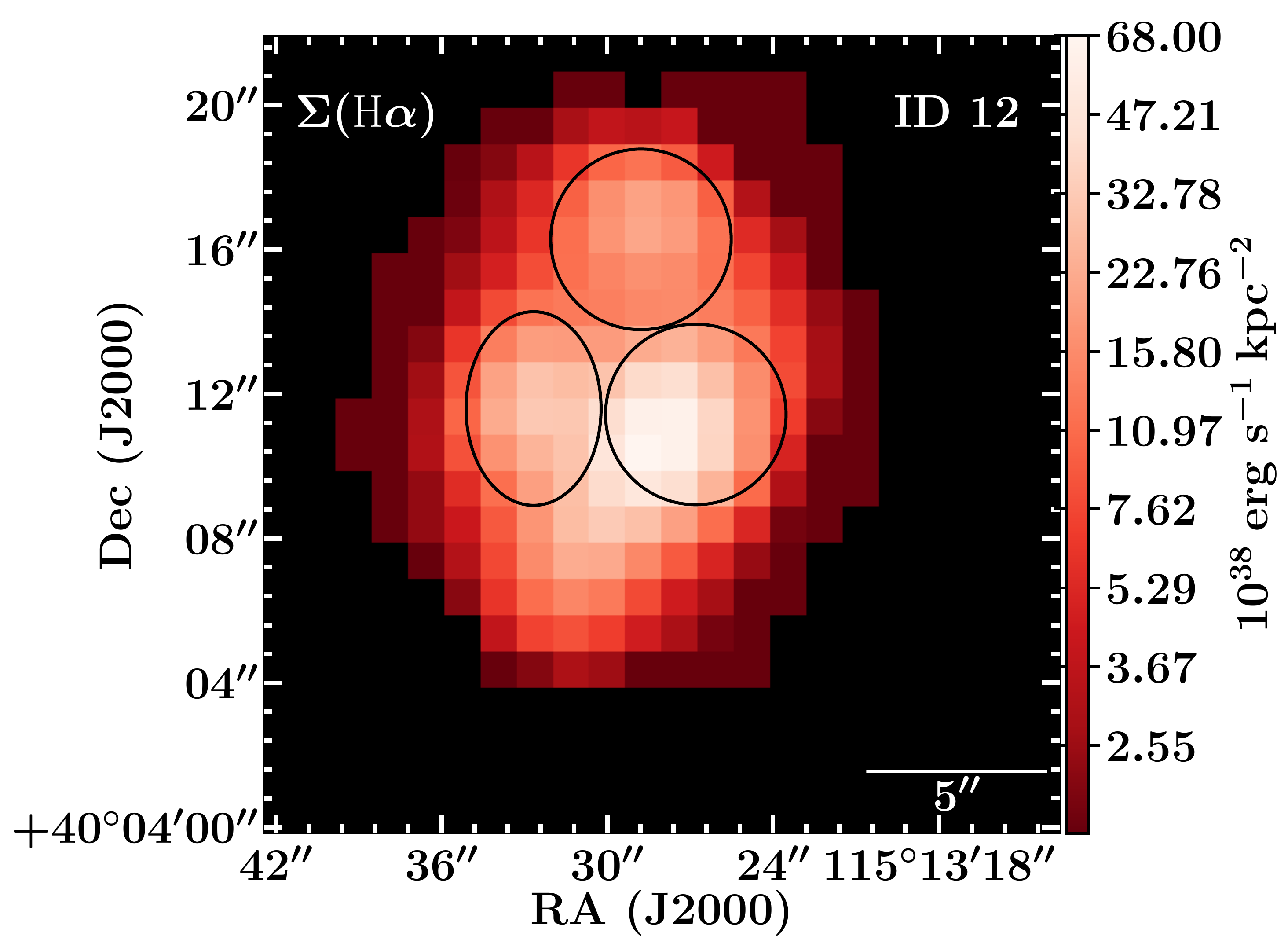}
\includegraphics[width=0.3\textwidth]{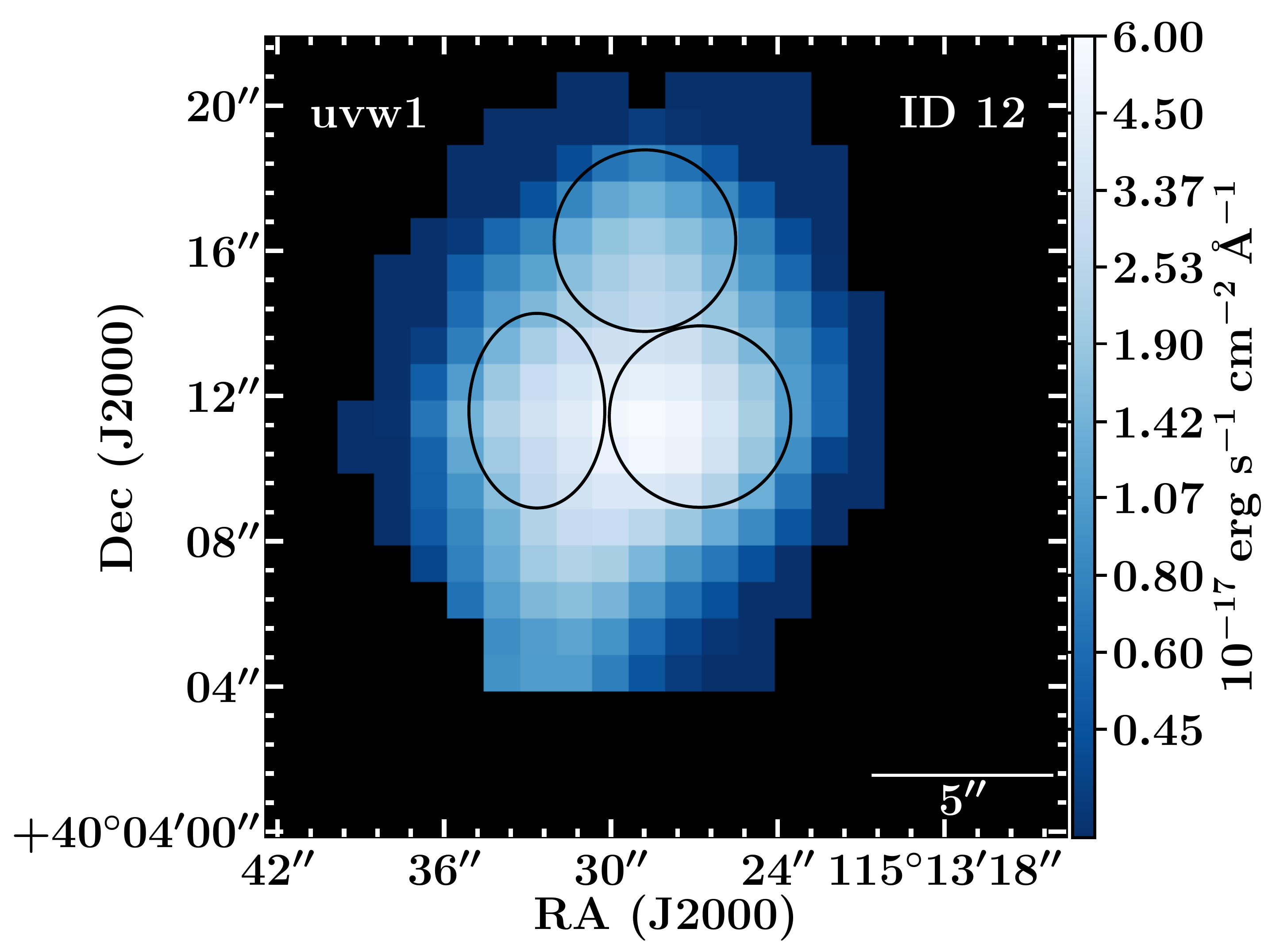}
\includegraphics[width=0.3\textwidth]{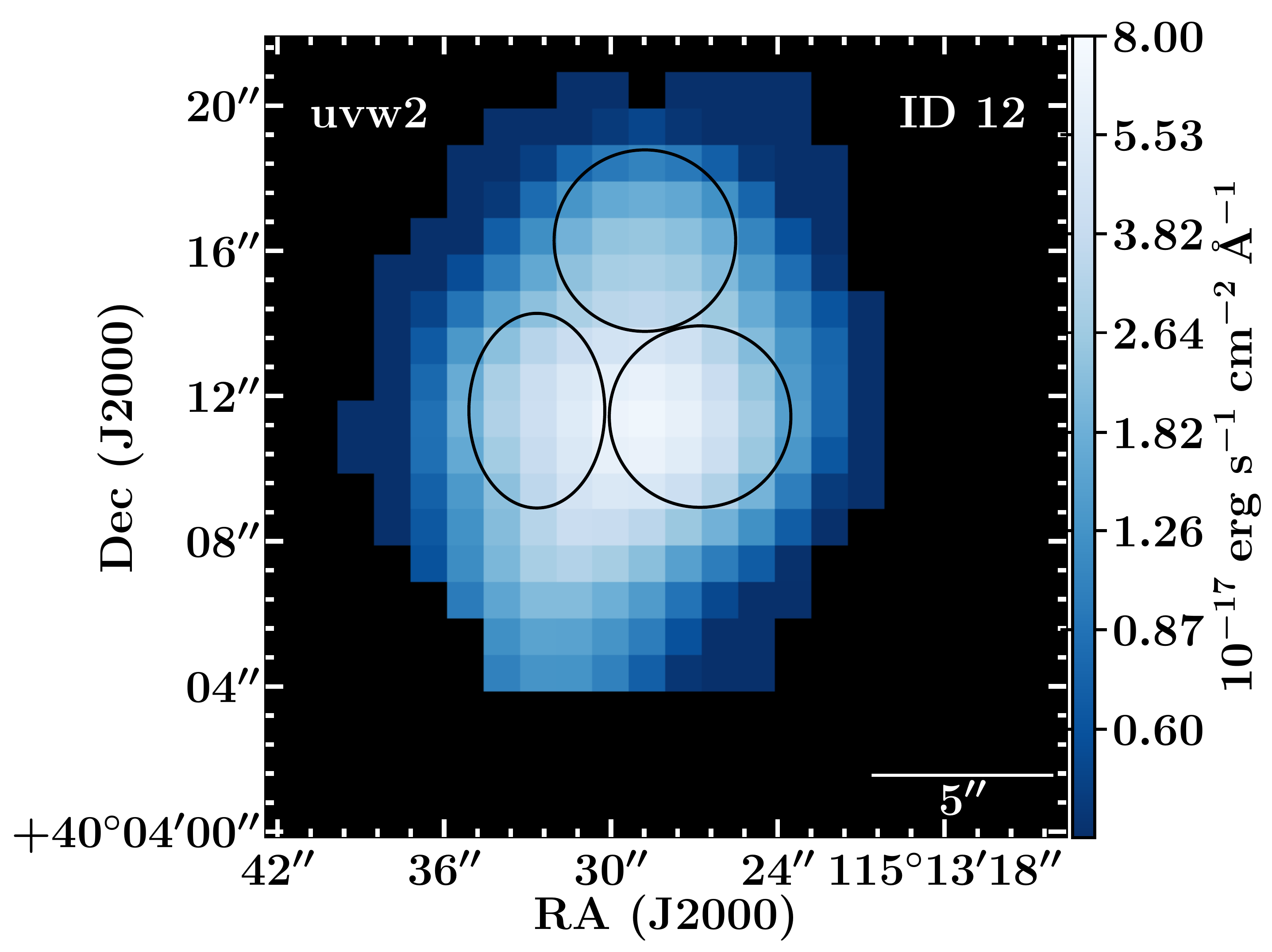}\\
\includegraphics[width=0.3\textwidth]{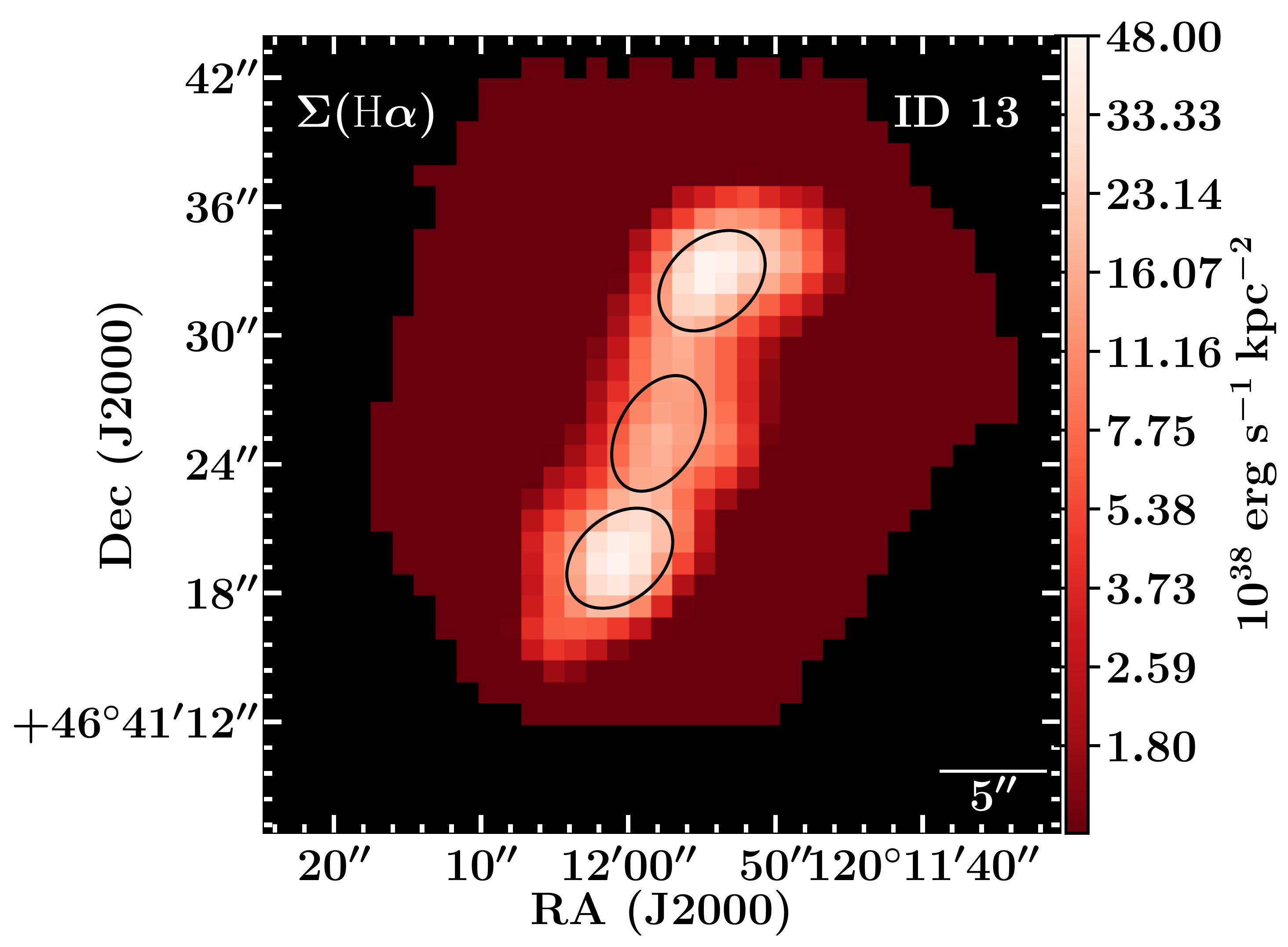}
\includegraphics[width=0.3\textwidth]{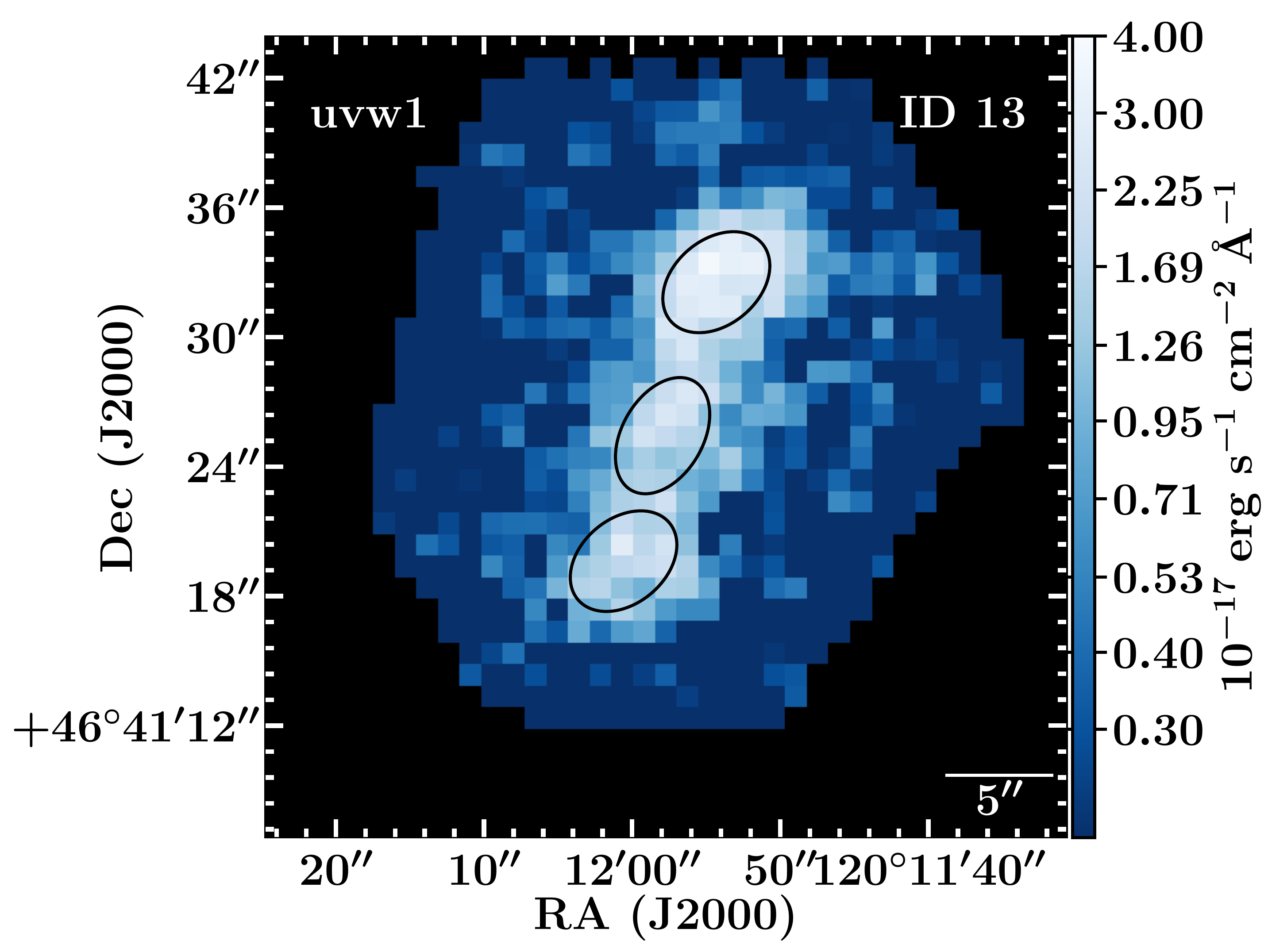}
\includegraphics[width=0.3\textwidth]{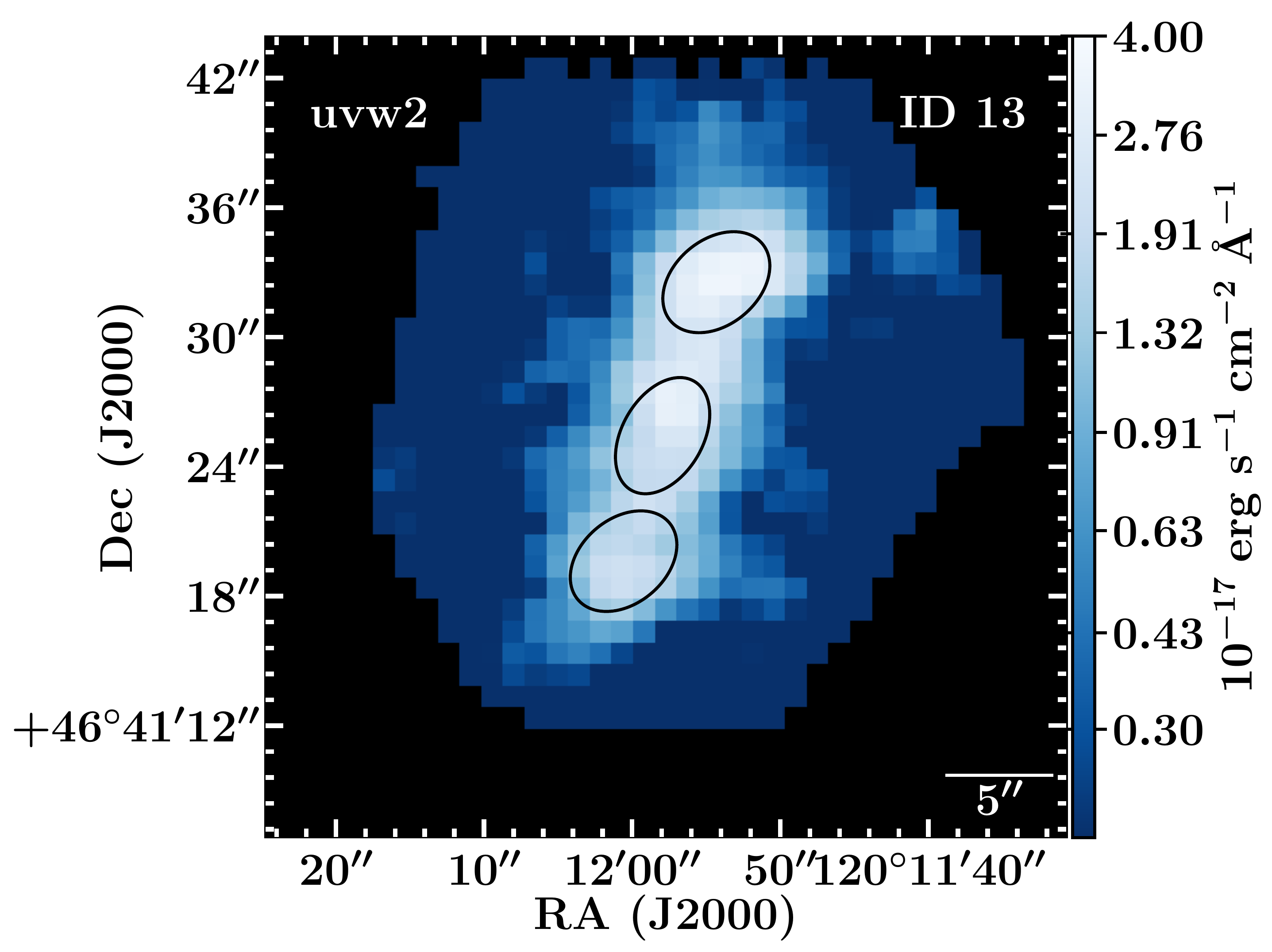}\\
\includegraphics[width=0.3\textwidth]{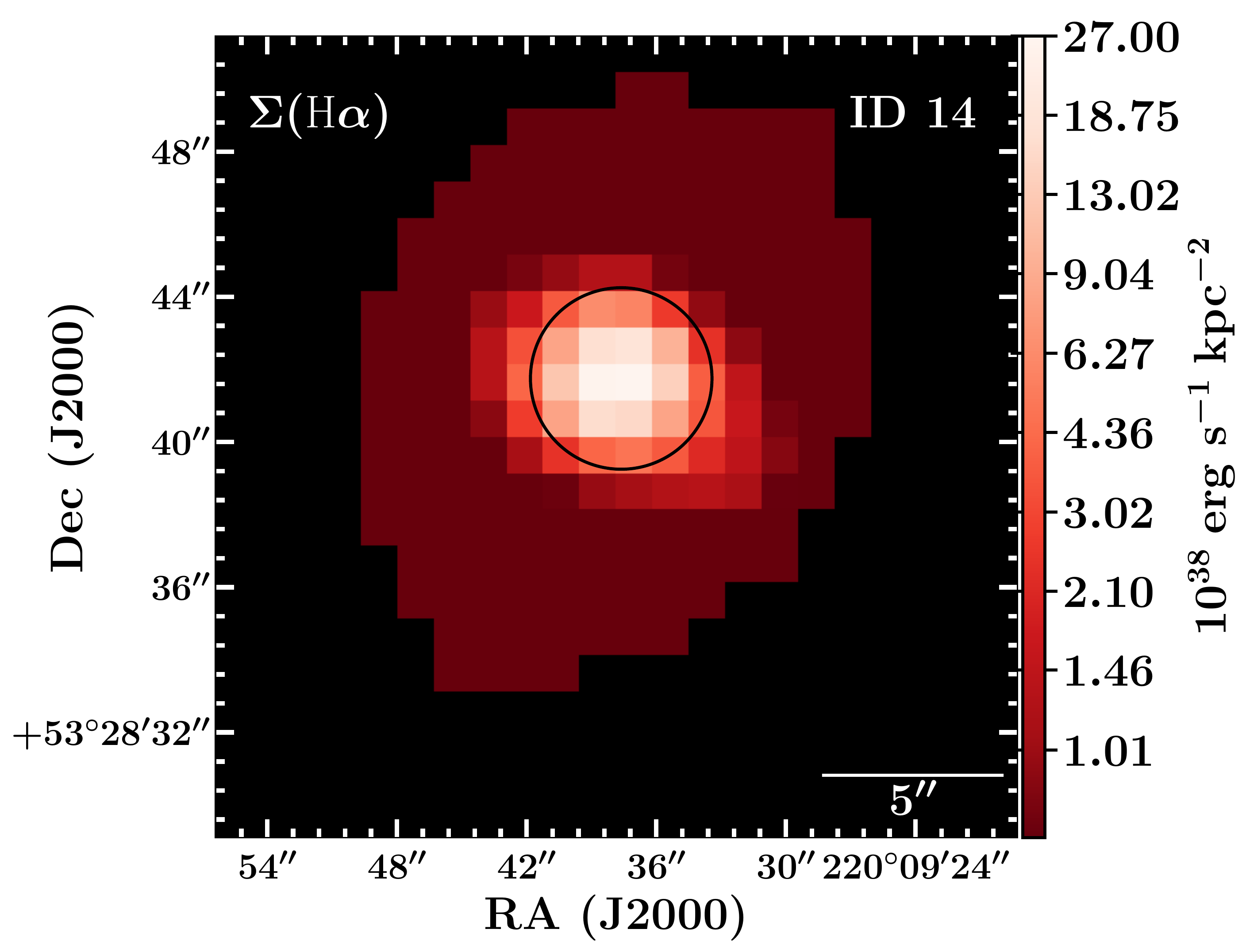}
\includegraphics[width=0.3\textwidth]{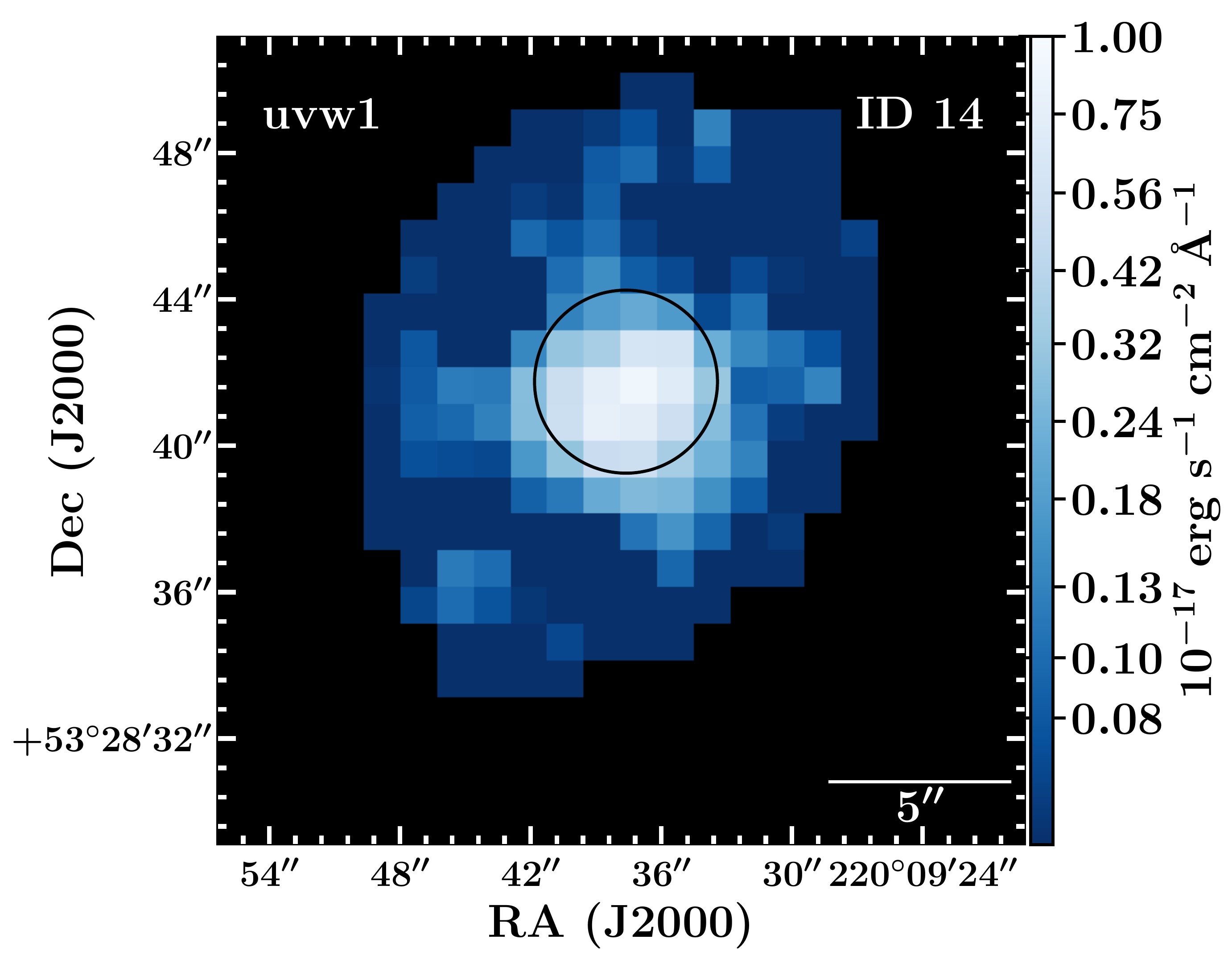}
\includegraphics[width=0.3\textwidth]{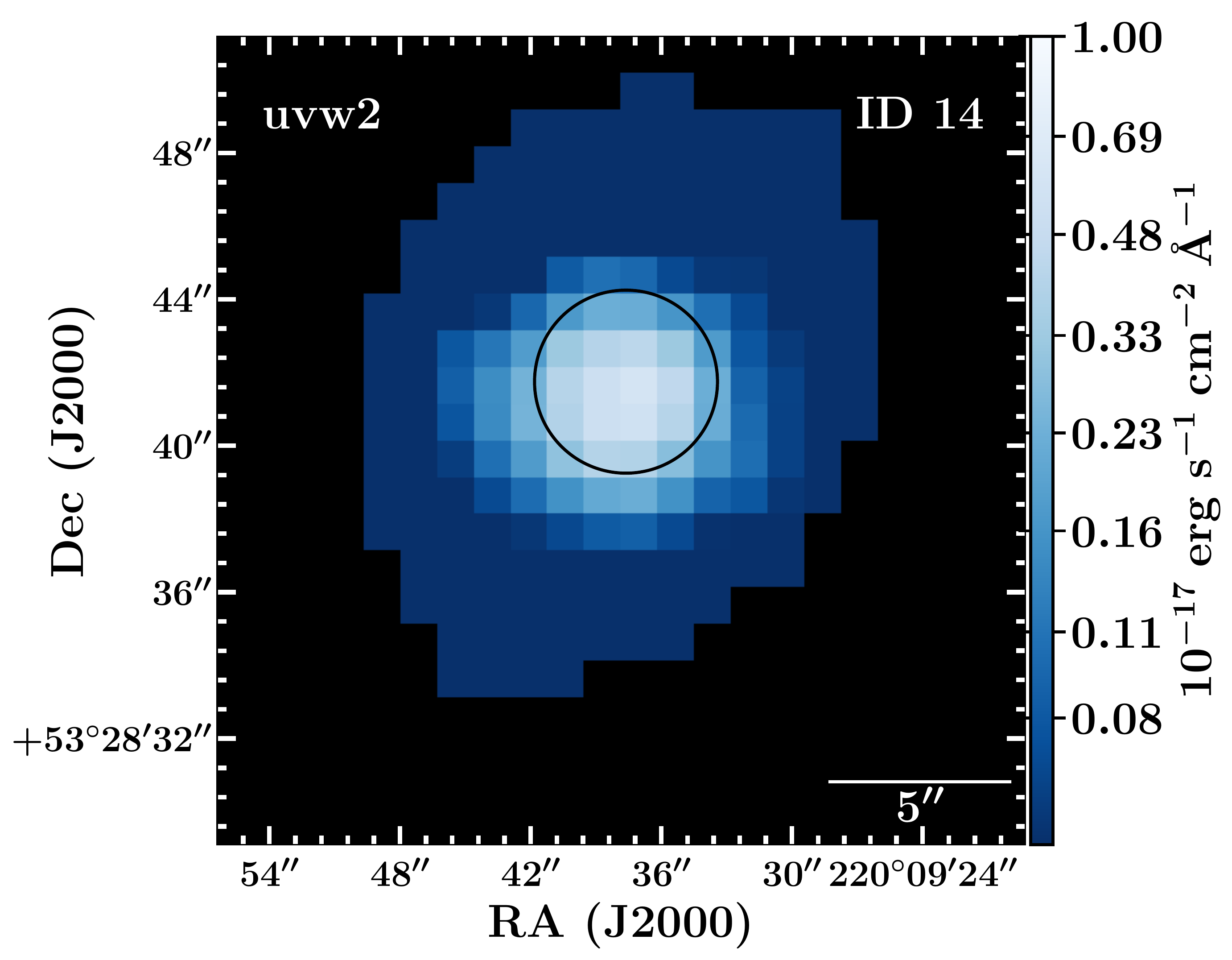}\\
\includegraphics[width=0.3\textwidth]{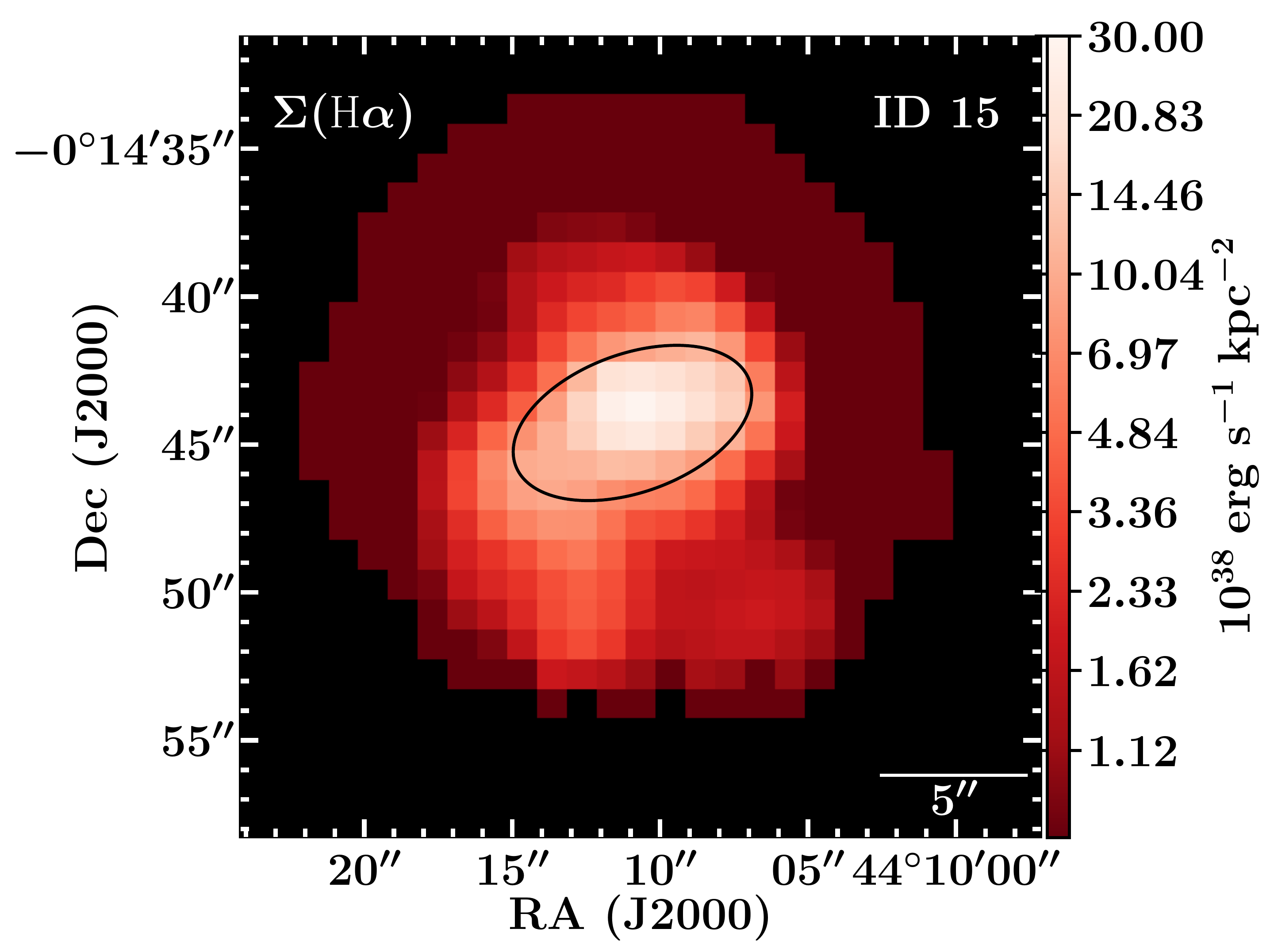}
\includegraphics[width=0.3\textwidth]{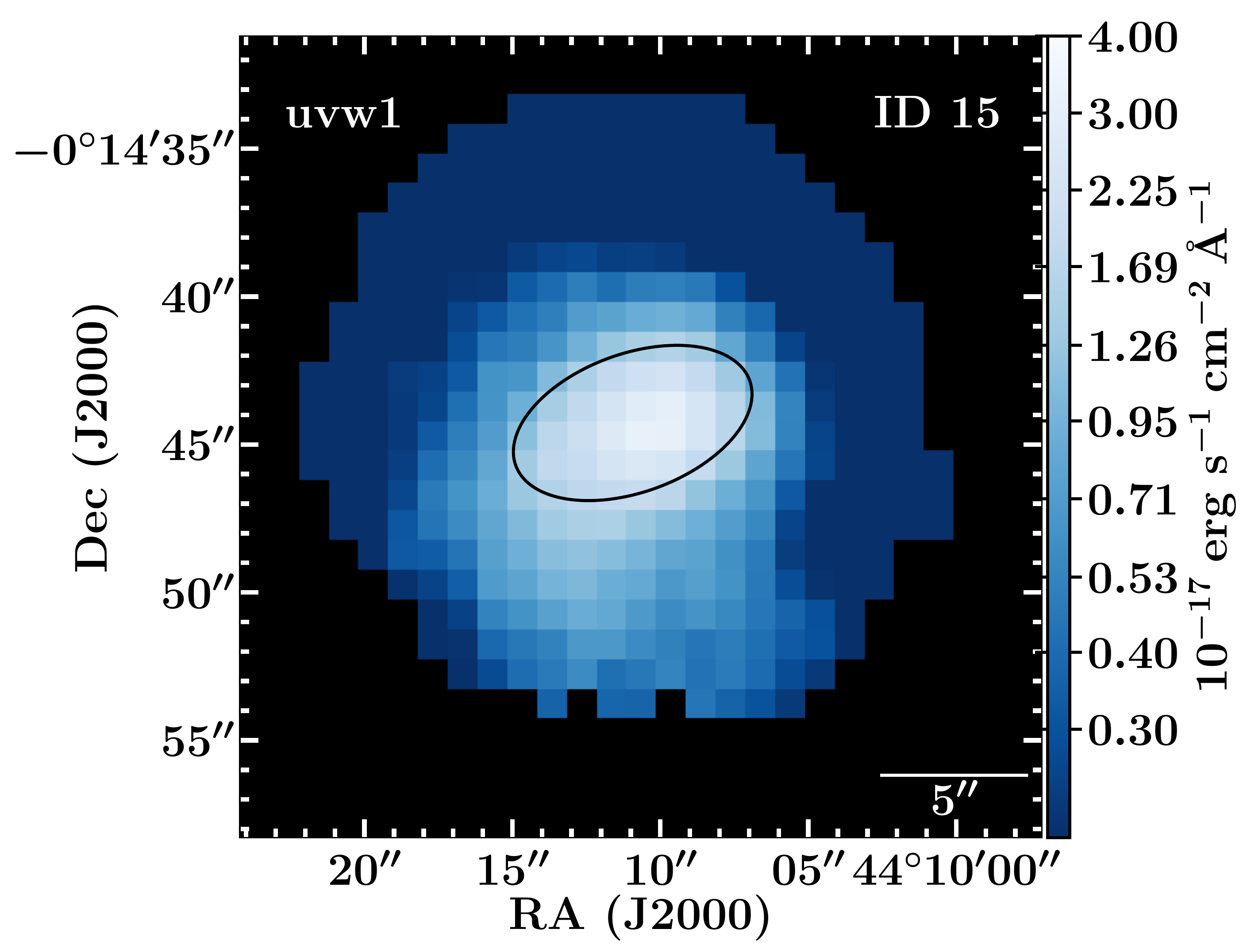}
\includegraphics[width=0.3\textwidth]{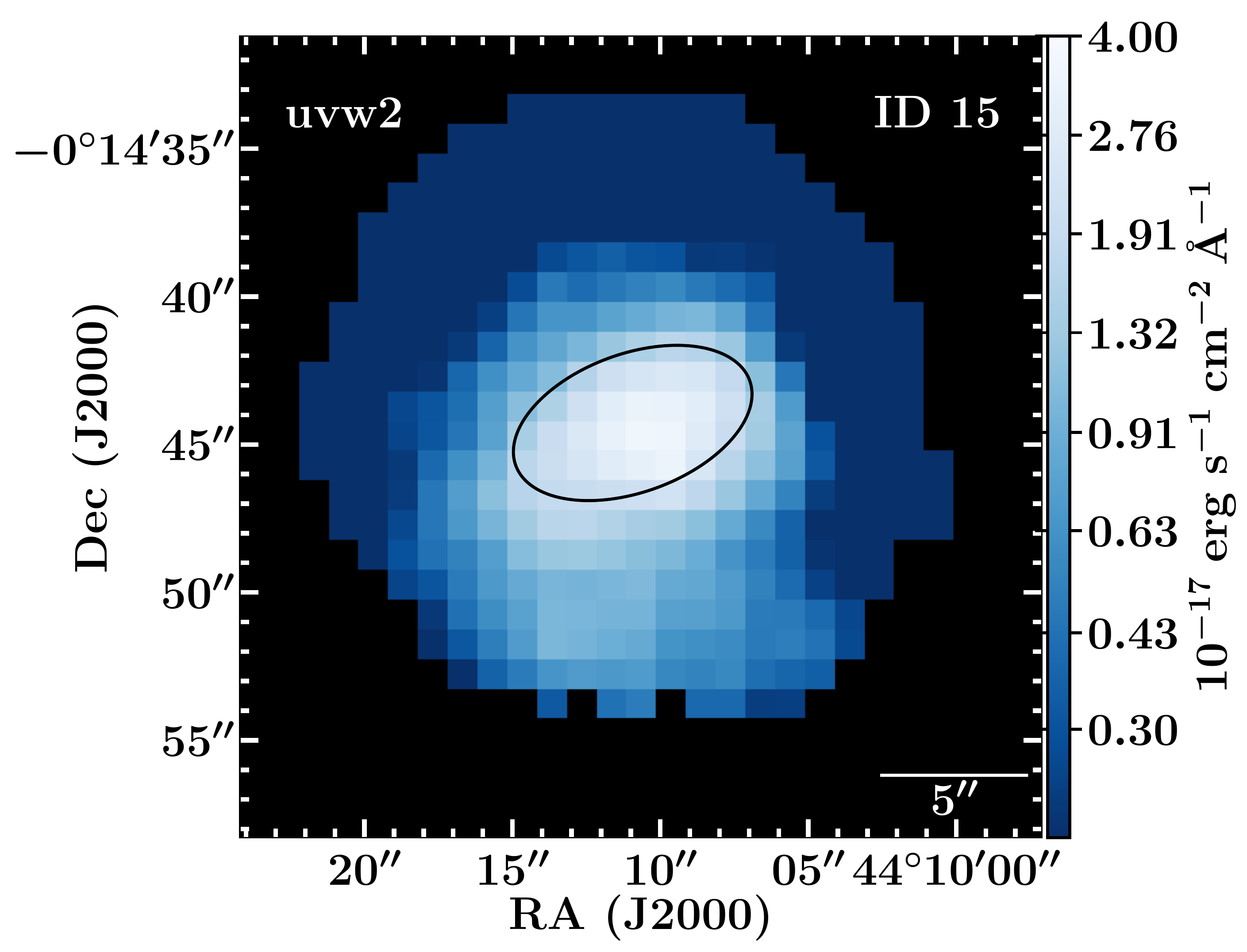}\\
\includegraphics[width=0.3\textwidth]{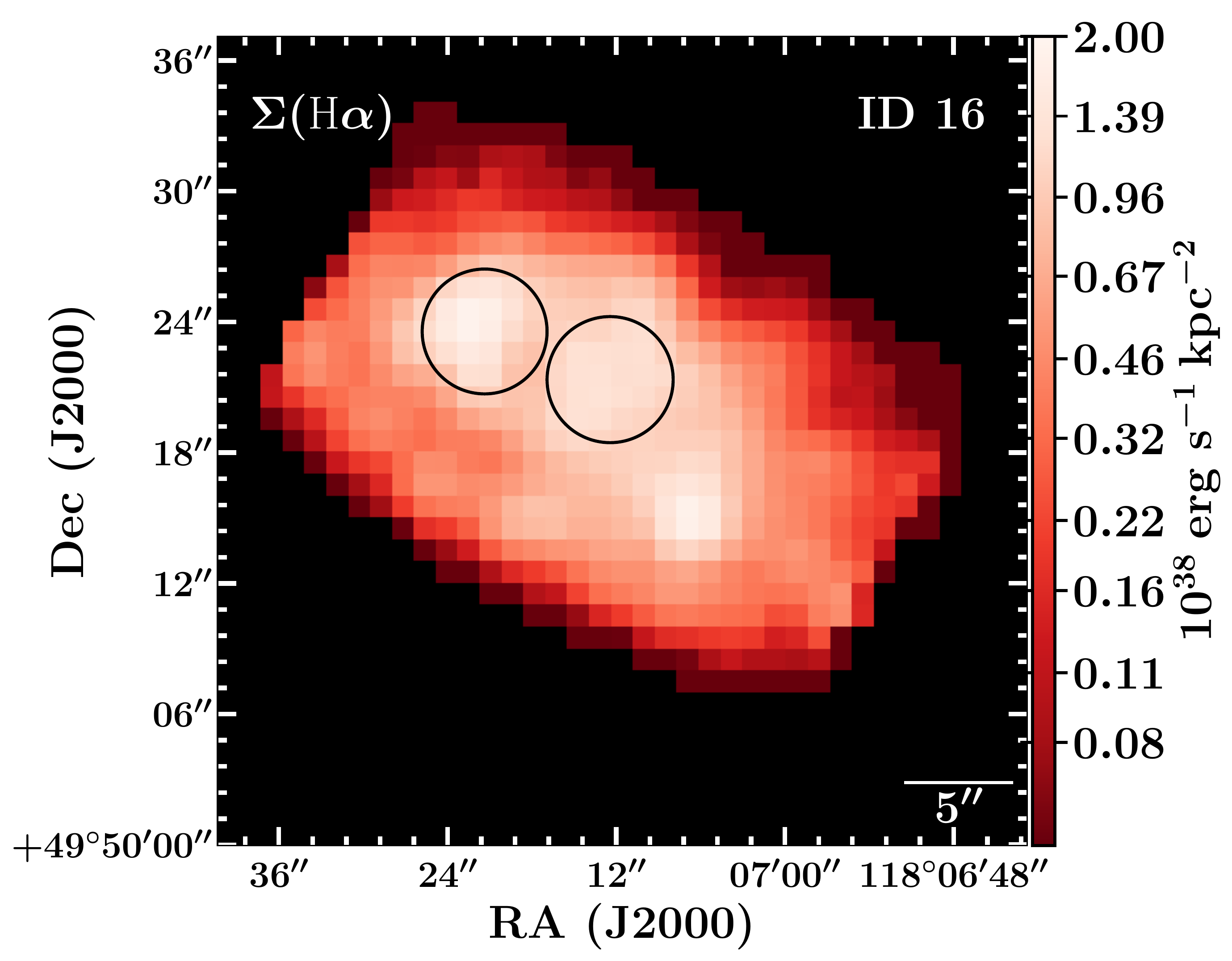}
\includegraphics[width=0.3\textwidth]{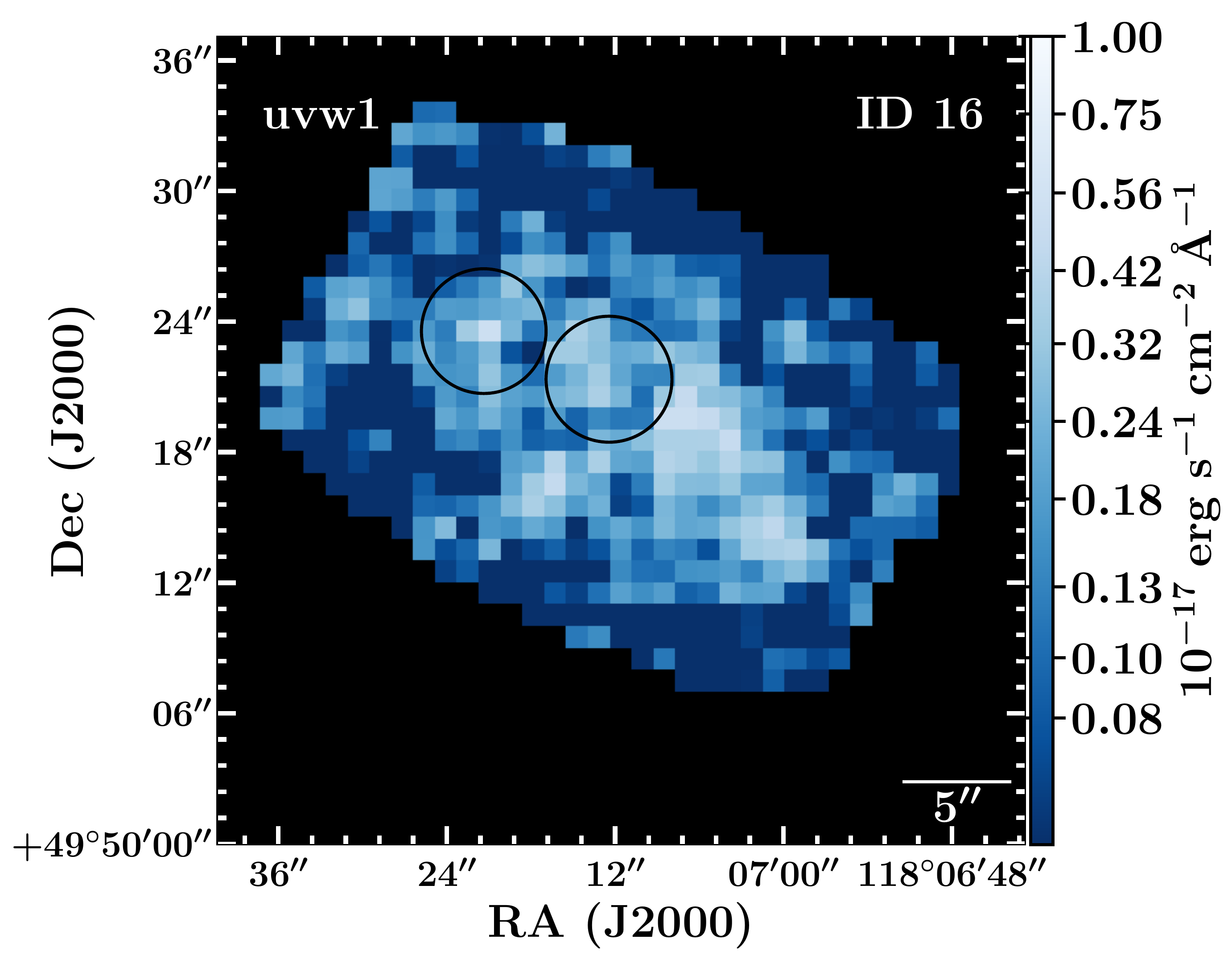}
\includegraphics[width=0.3\textwidth]{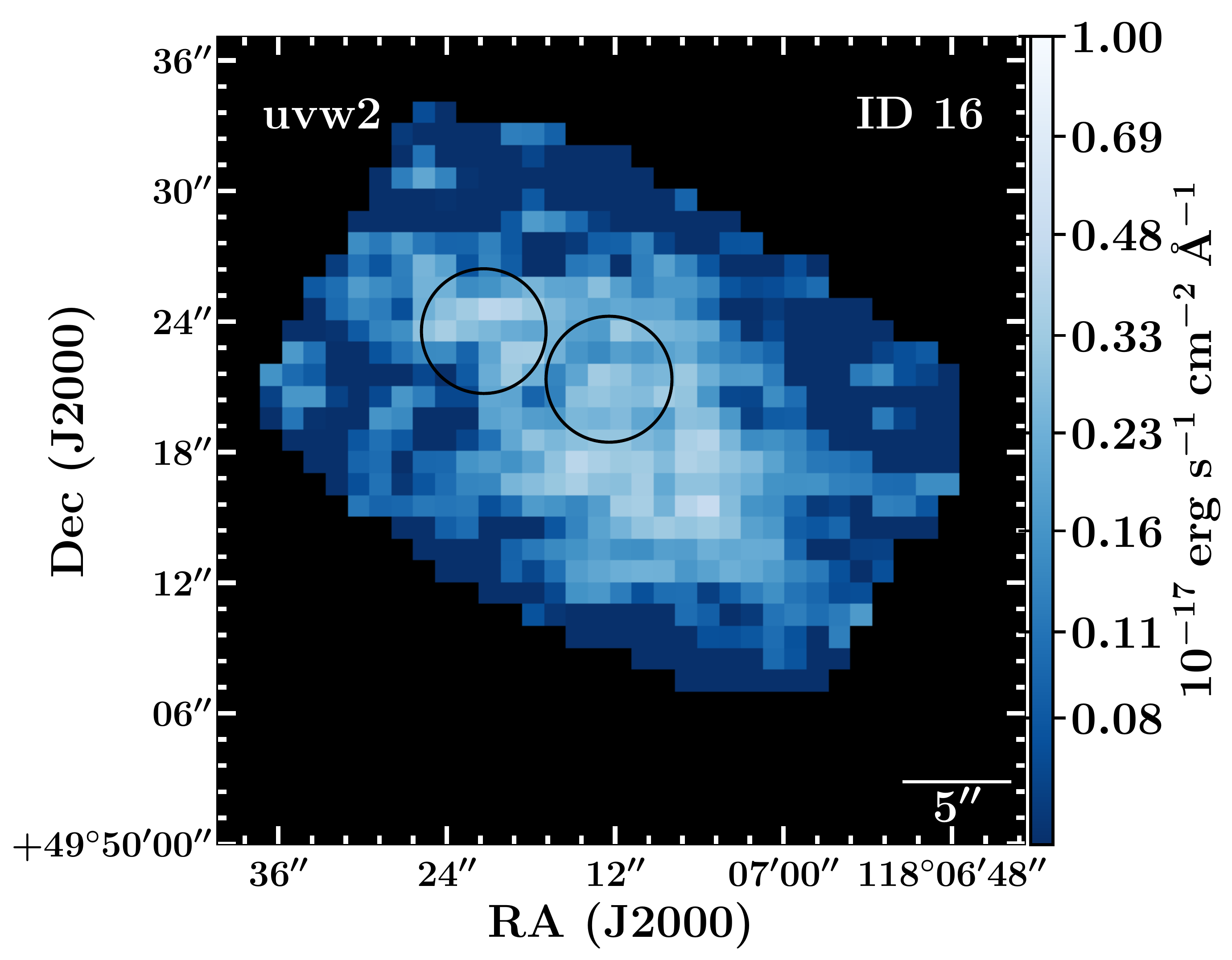}\\
\includegraphics[width=0.3\textwidth]{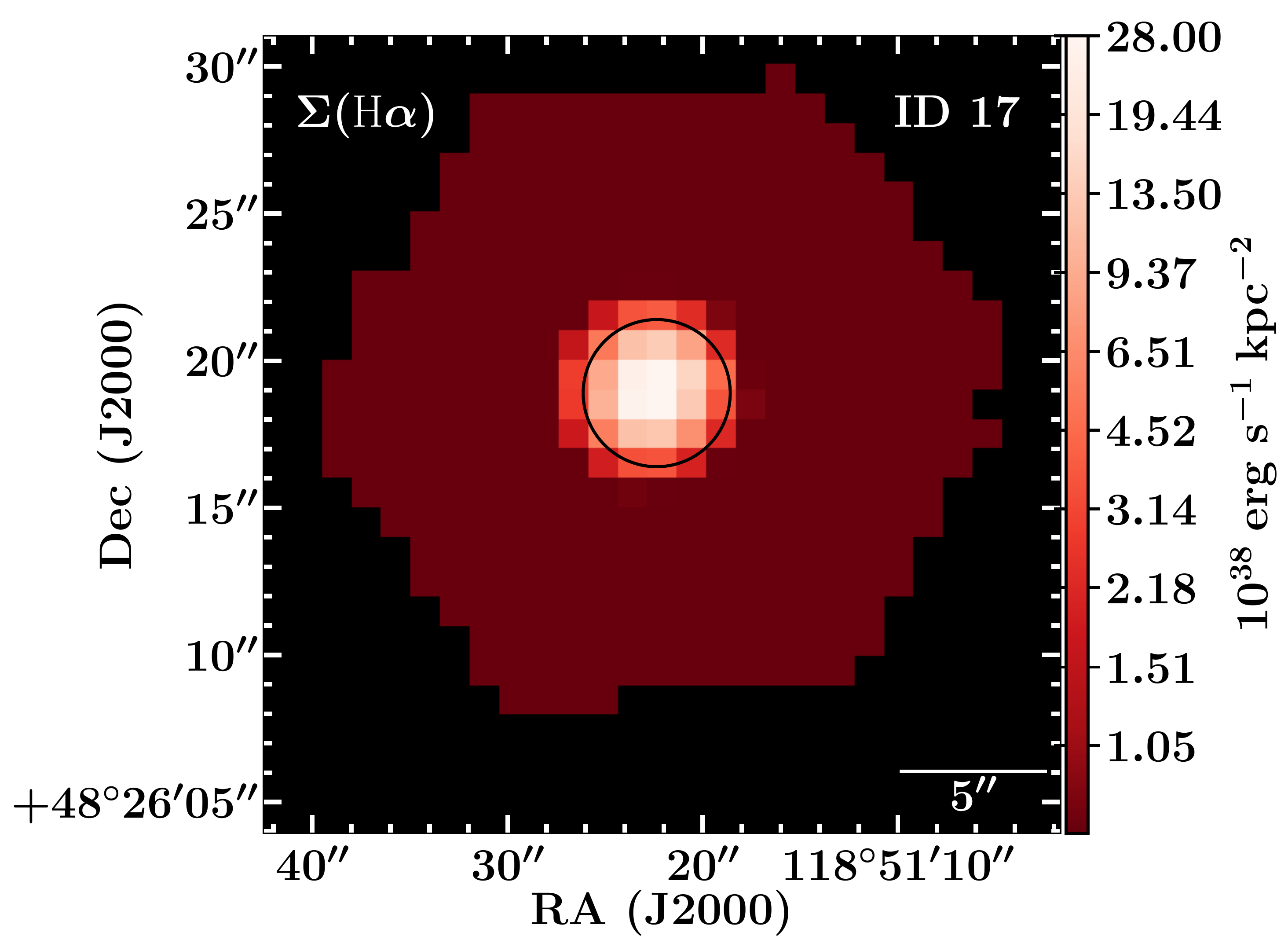}
\includegraphics[width=0.3\textwidth]{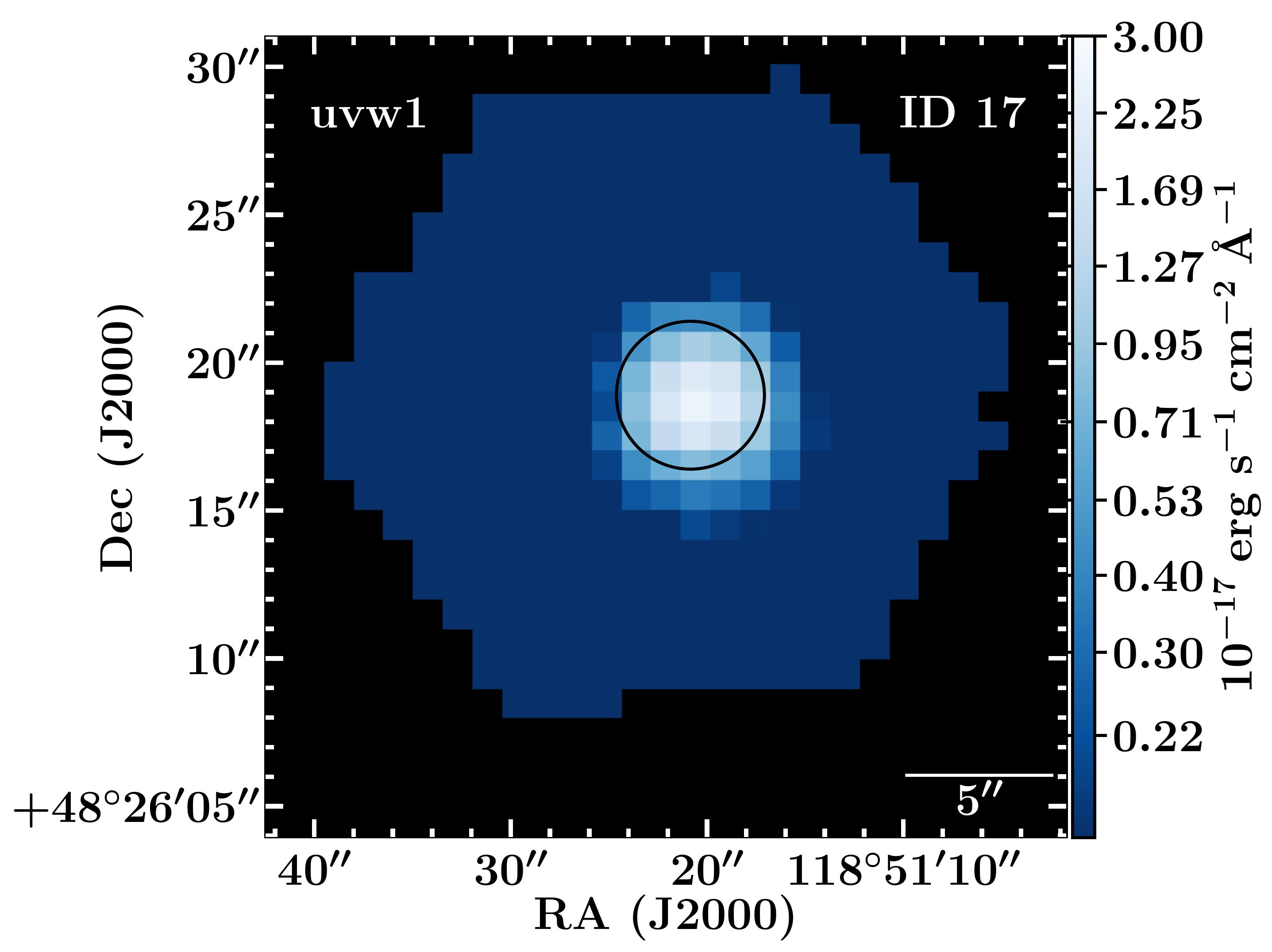}
\includegraphics[width=0.3\textwidth]{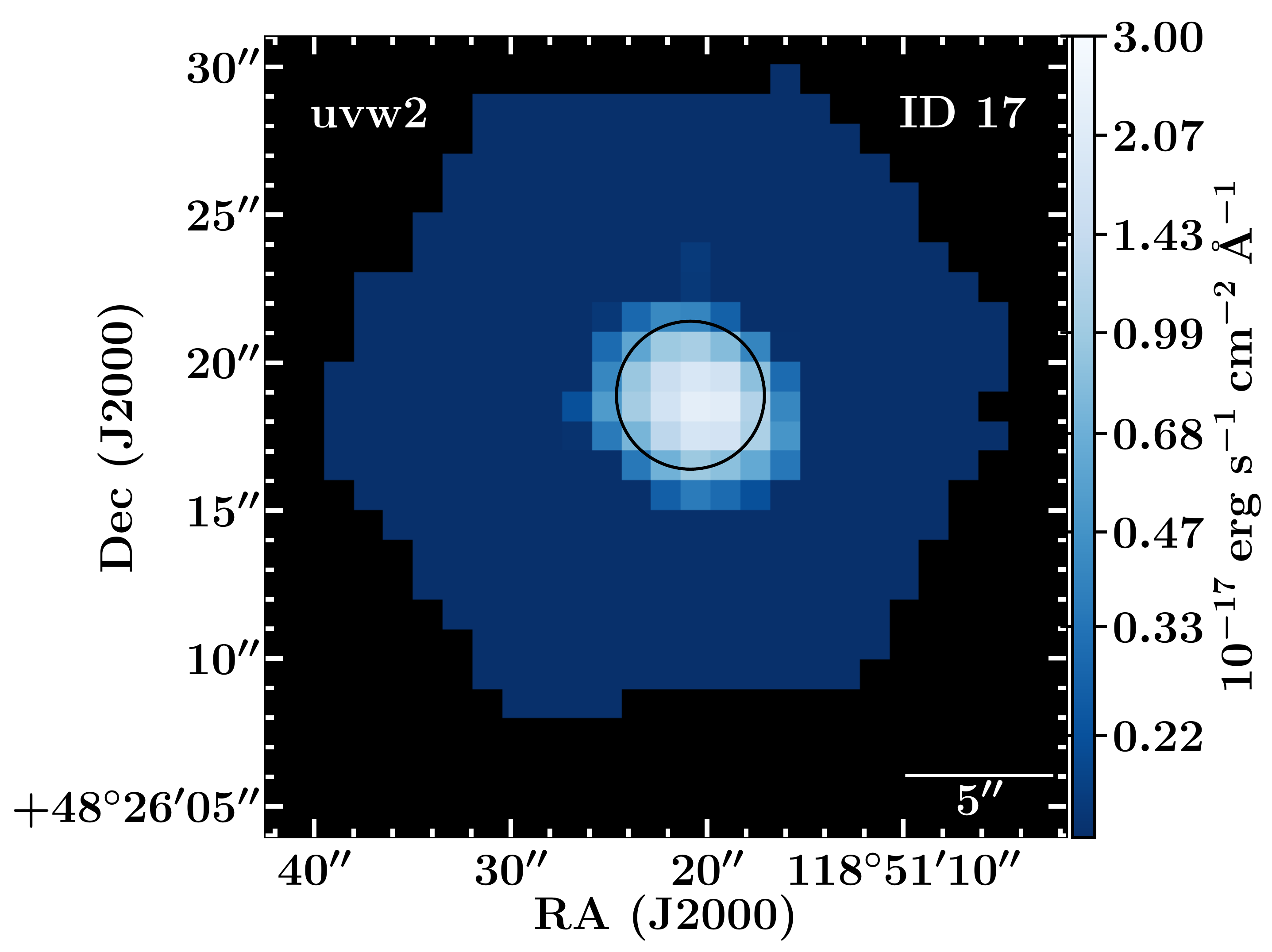}\\
\includegraphics[width=0.3\textwidth]{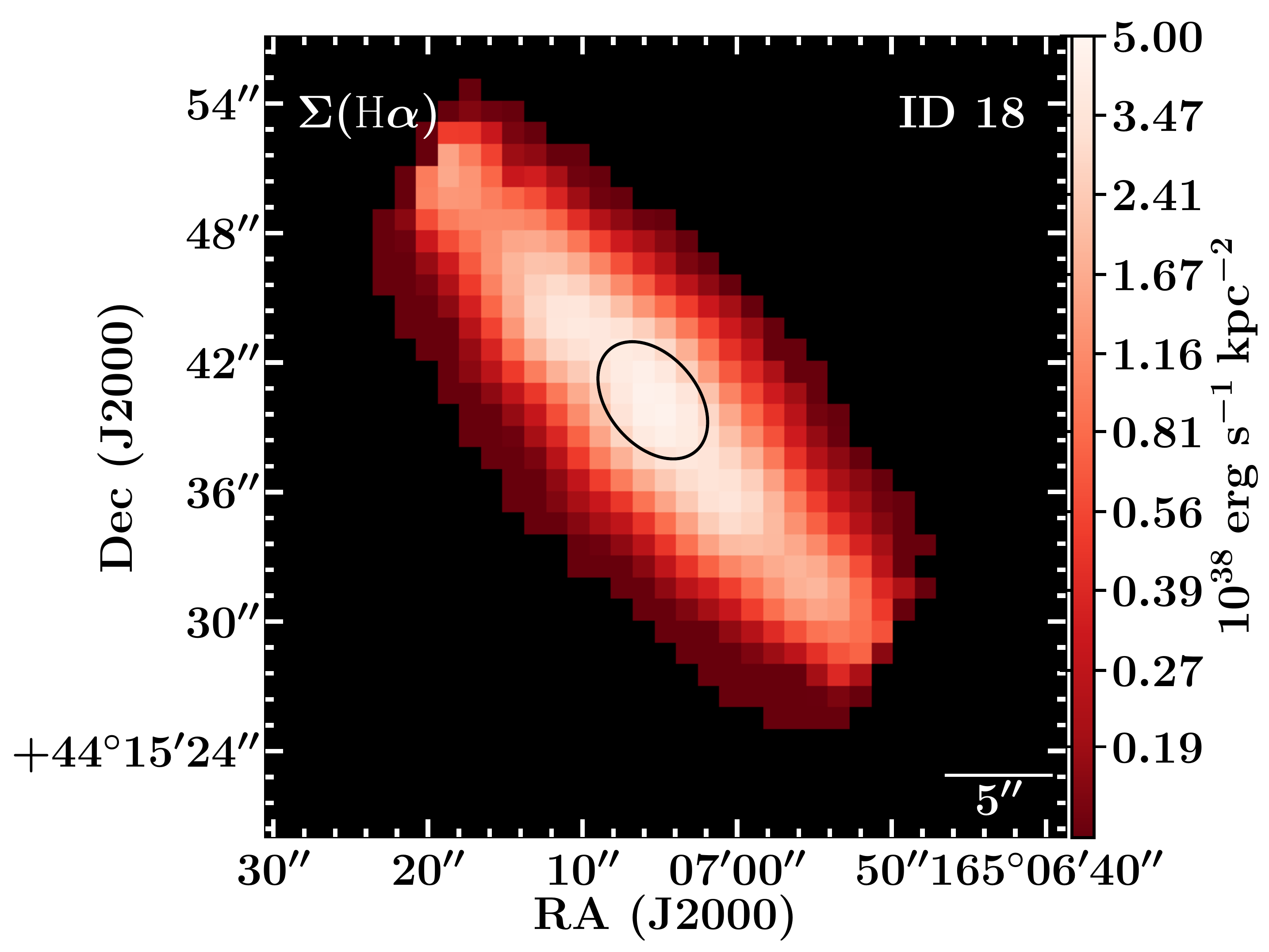}
\includegraphics[width=0.3\textwidth]{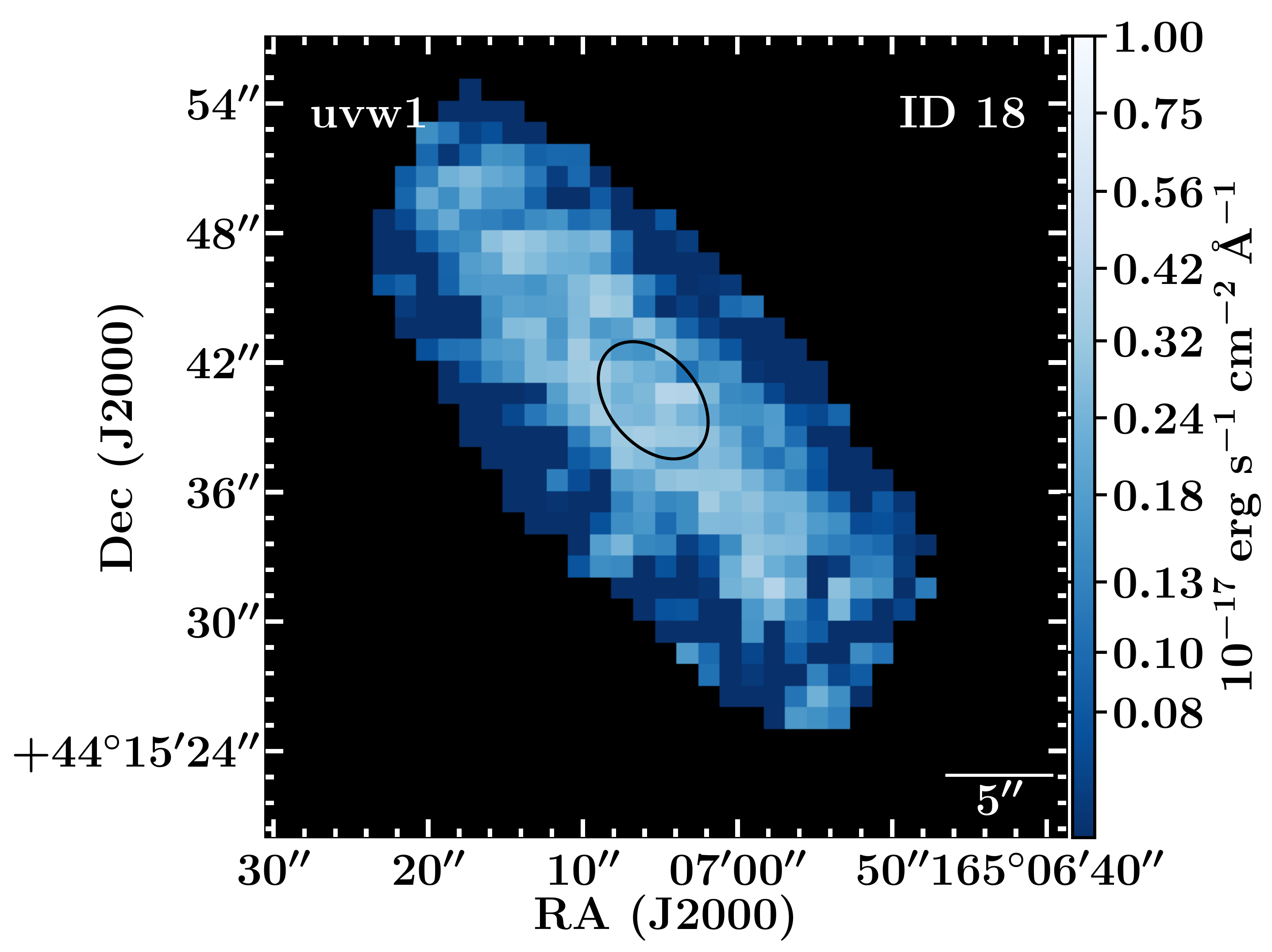}
\includegraphics[width=0.3\textwidth]{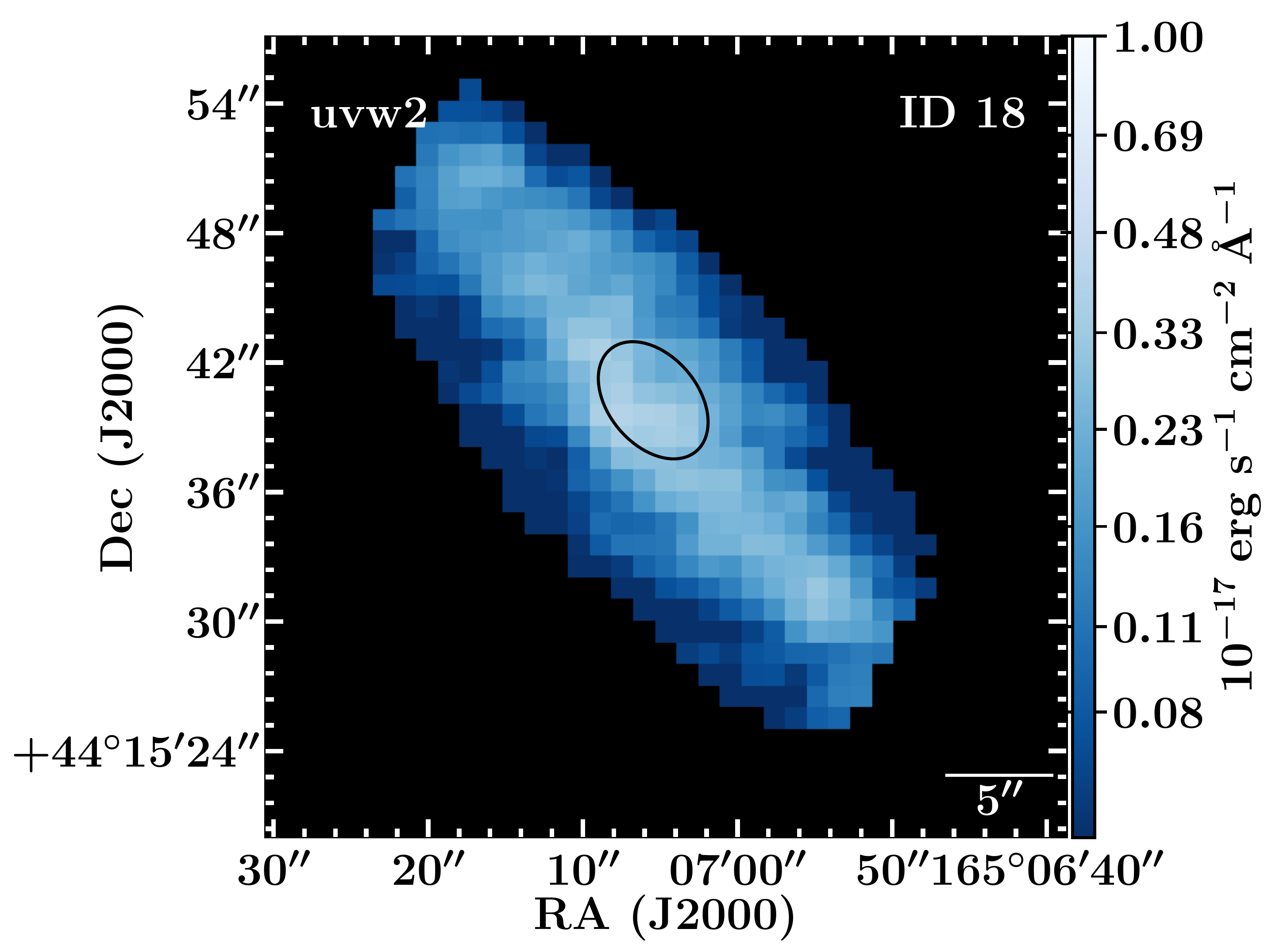}\\
\includegraphics[width=0.3\textwidth]{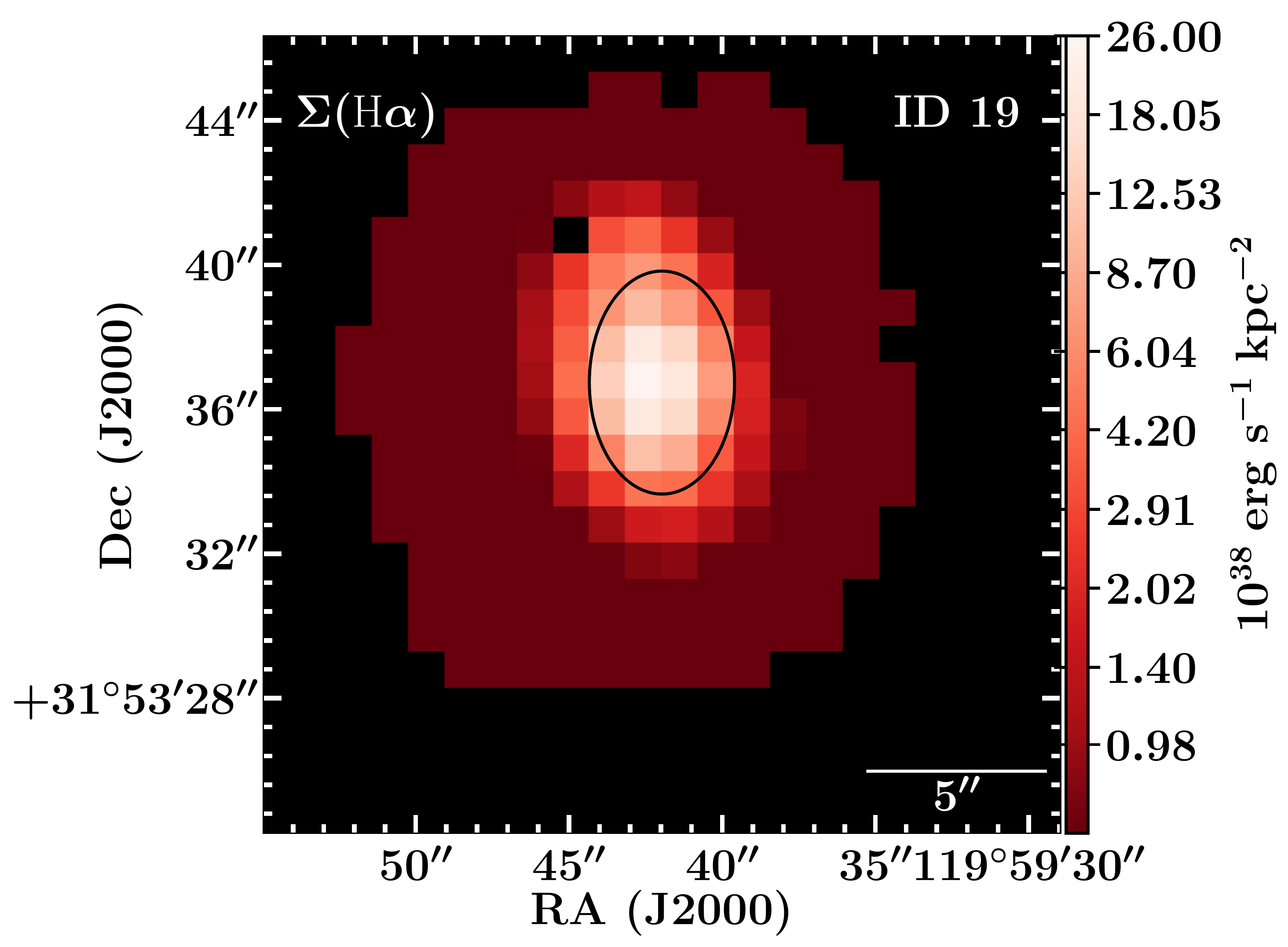}
\includegraphics[width=0.3\textwidth]{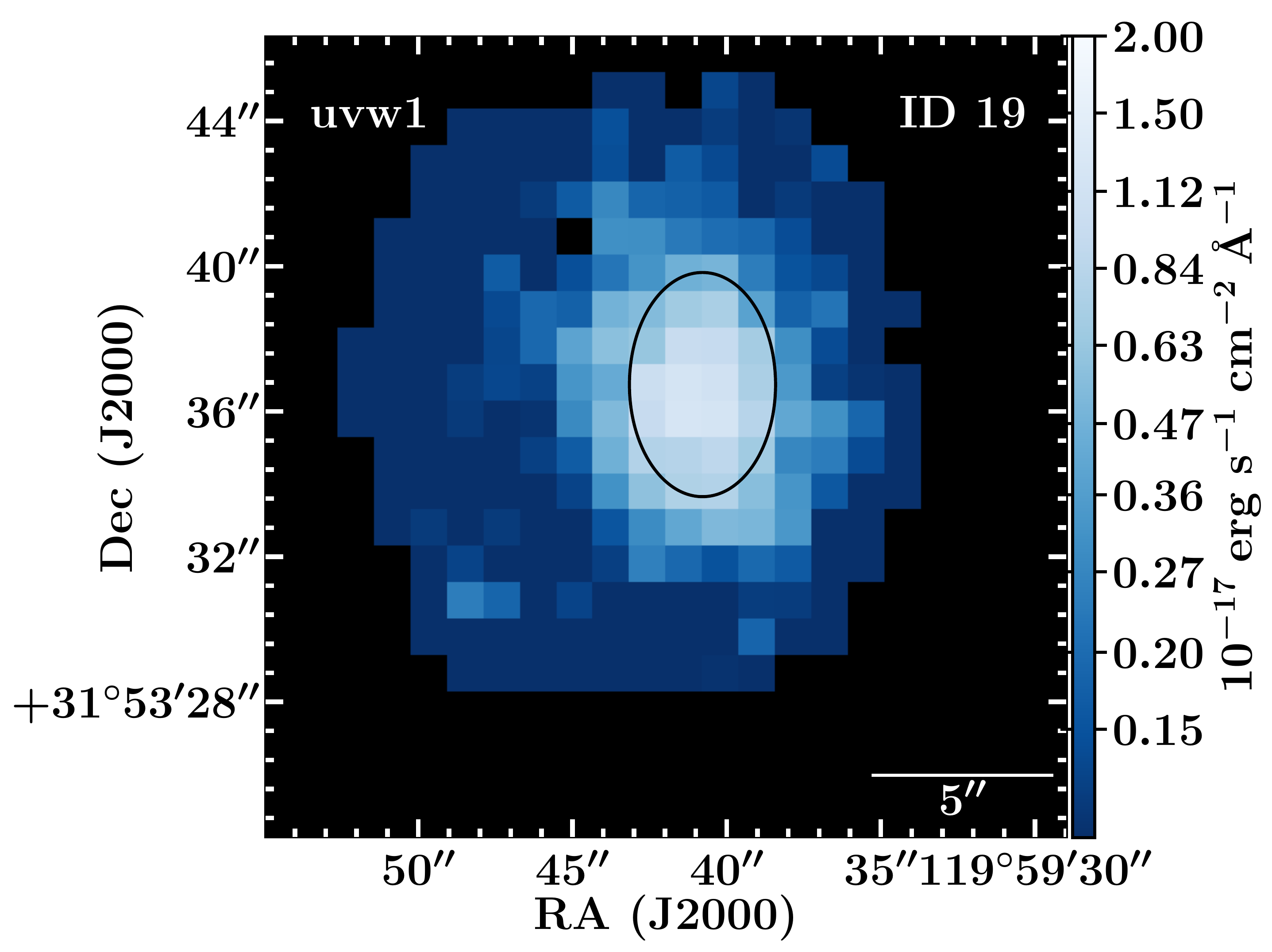}
\includegraphics[width=0.3\textwidth]{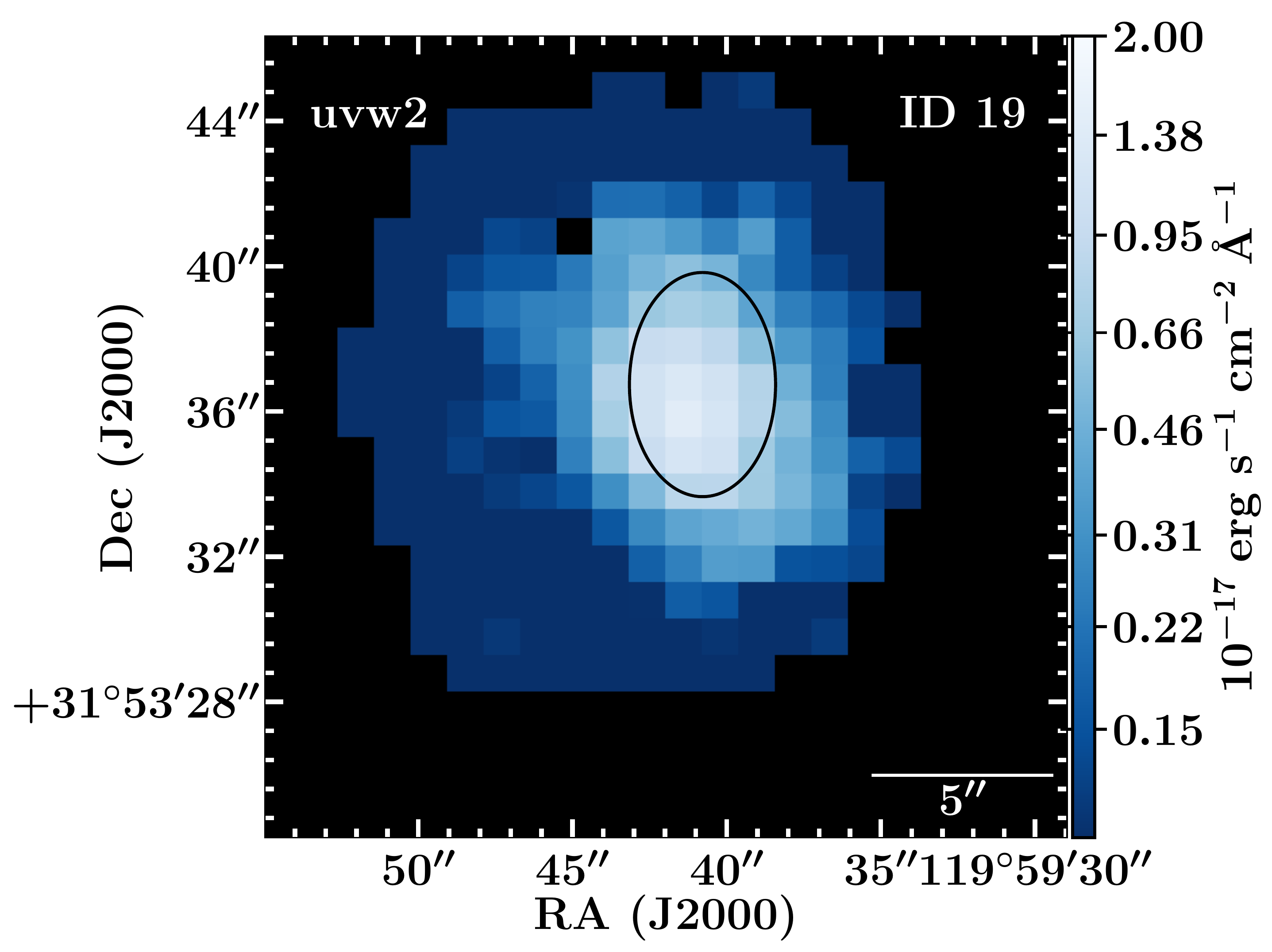}\\
\includegraphics[width=0.3\textwidth]{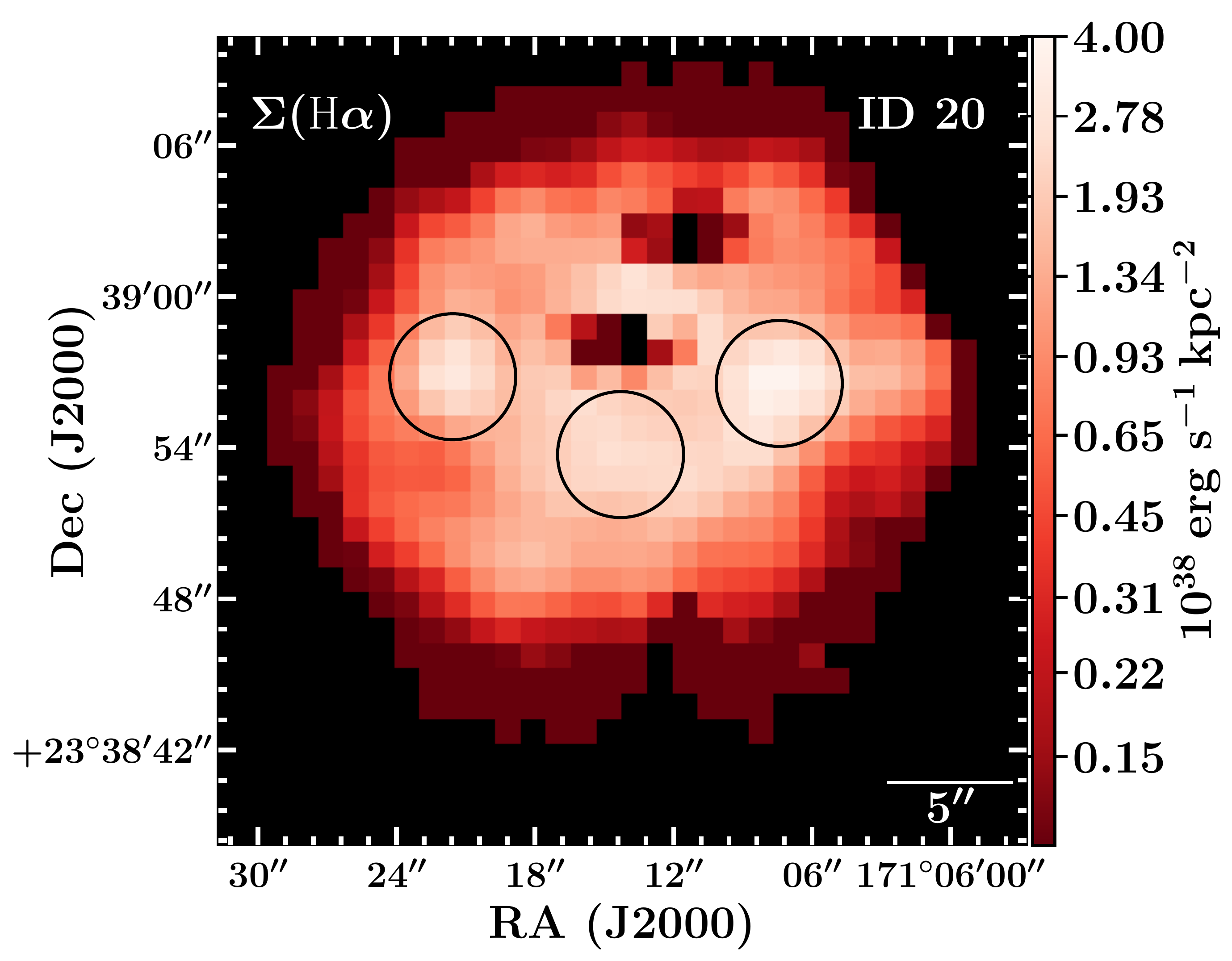}
\includegraphics[width=0.3\textwidth]{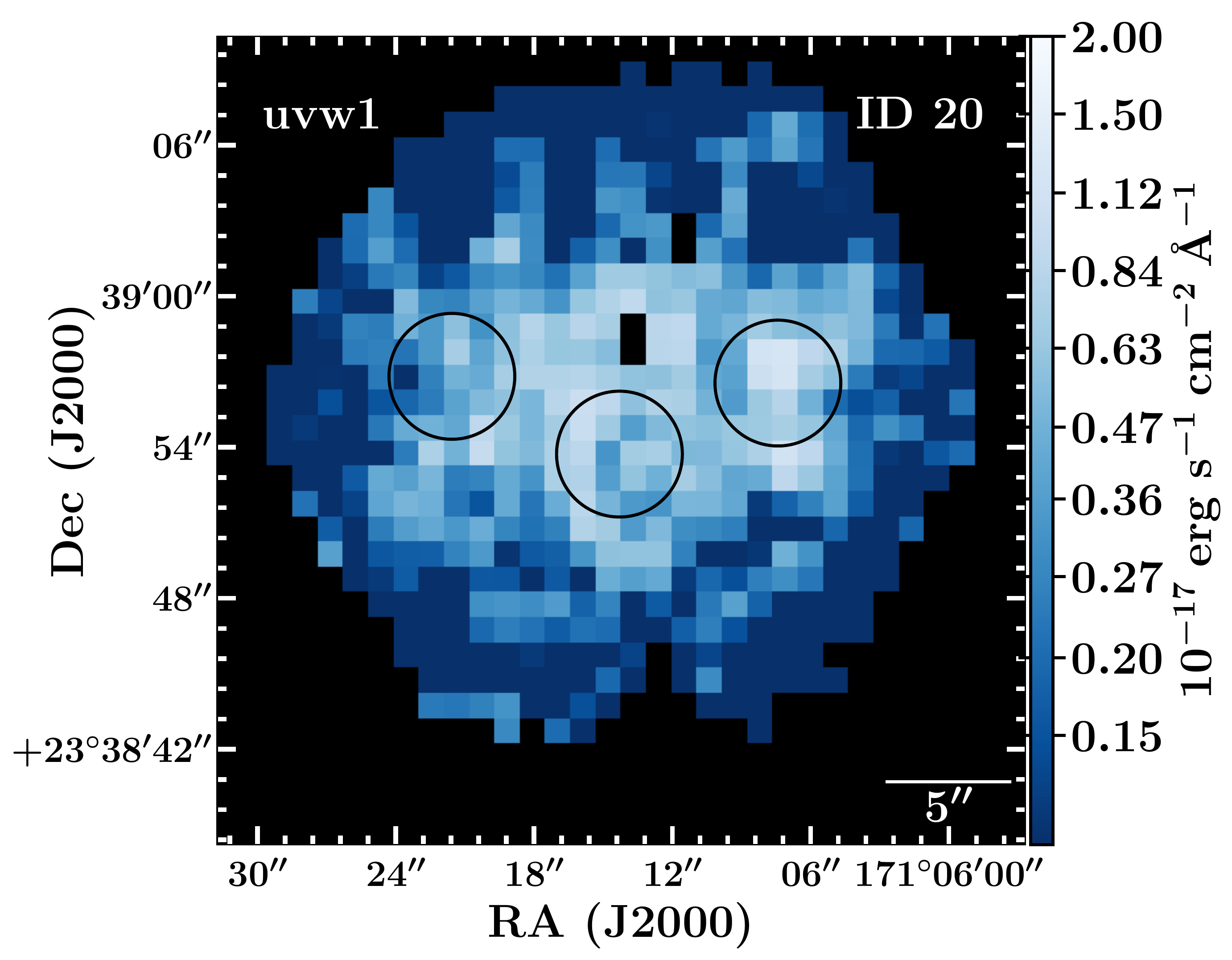}
\includegraphics[width=0.3\textwidth]{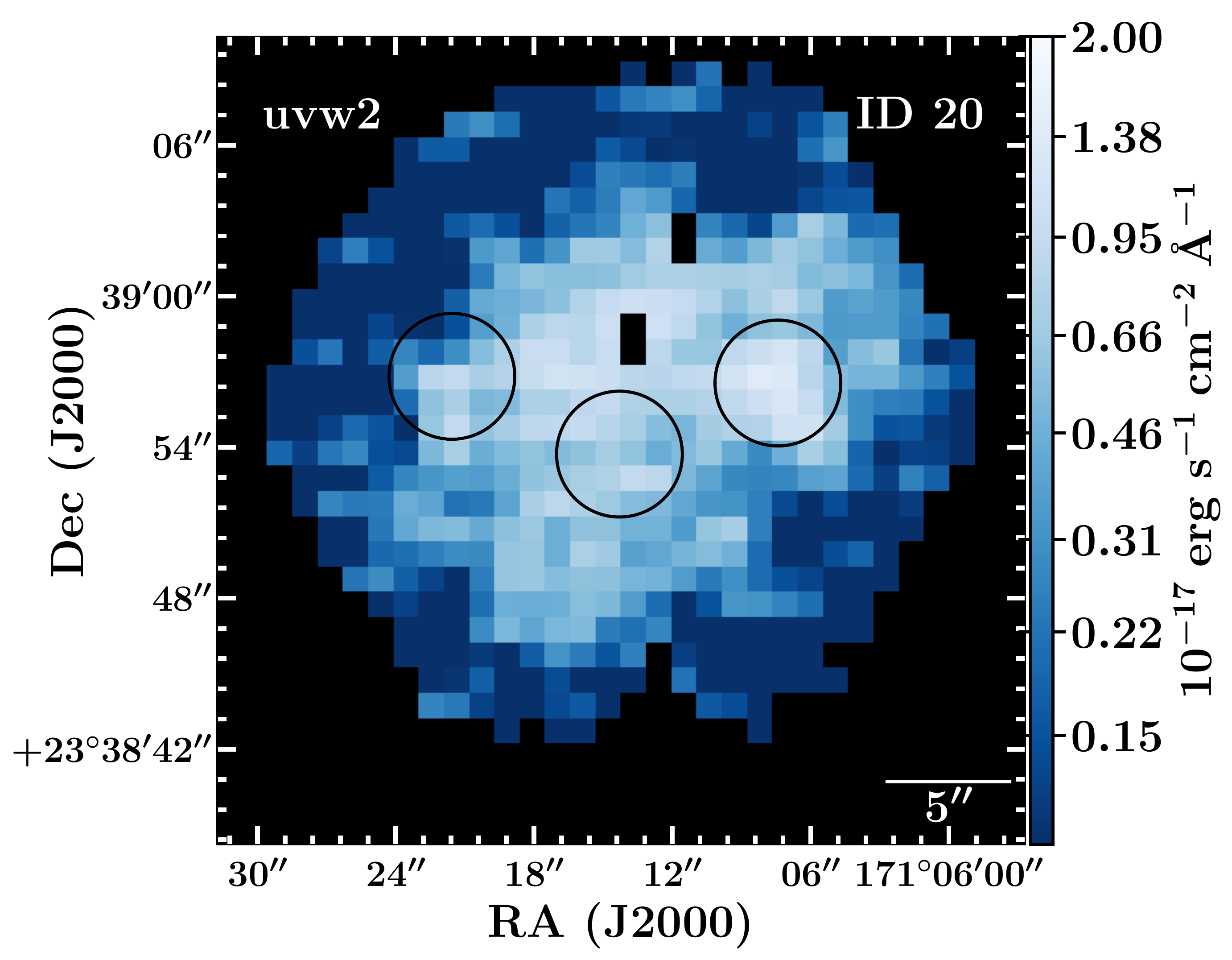}\\
\includegraphics[width=0.3\textwidth]{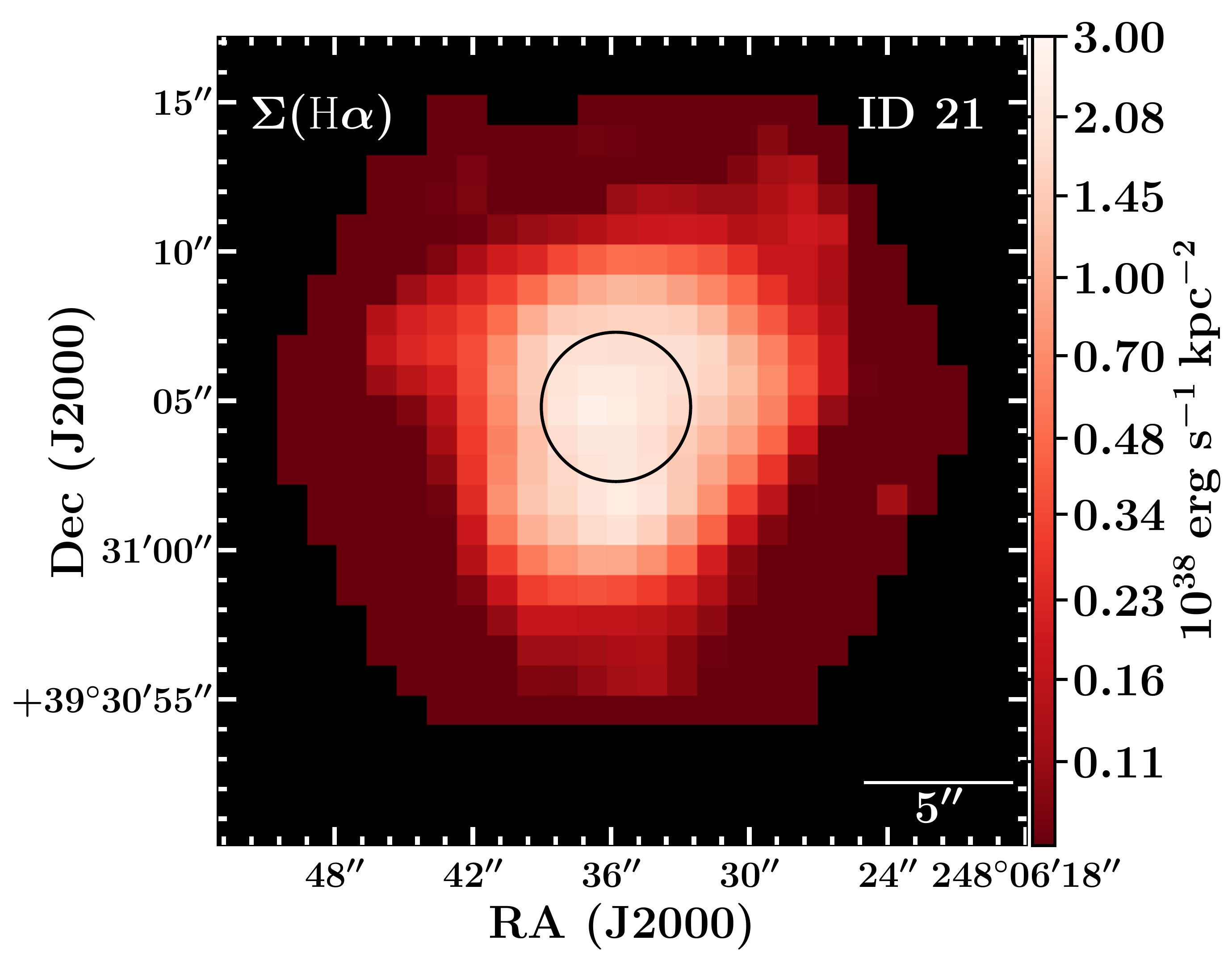}
\includegraphics[width=0.3\textwidth]{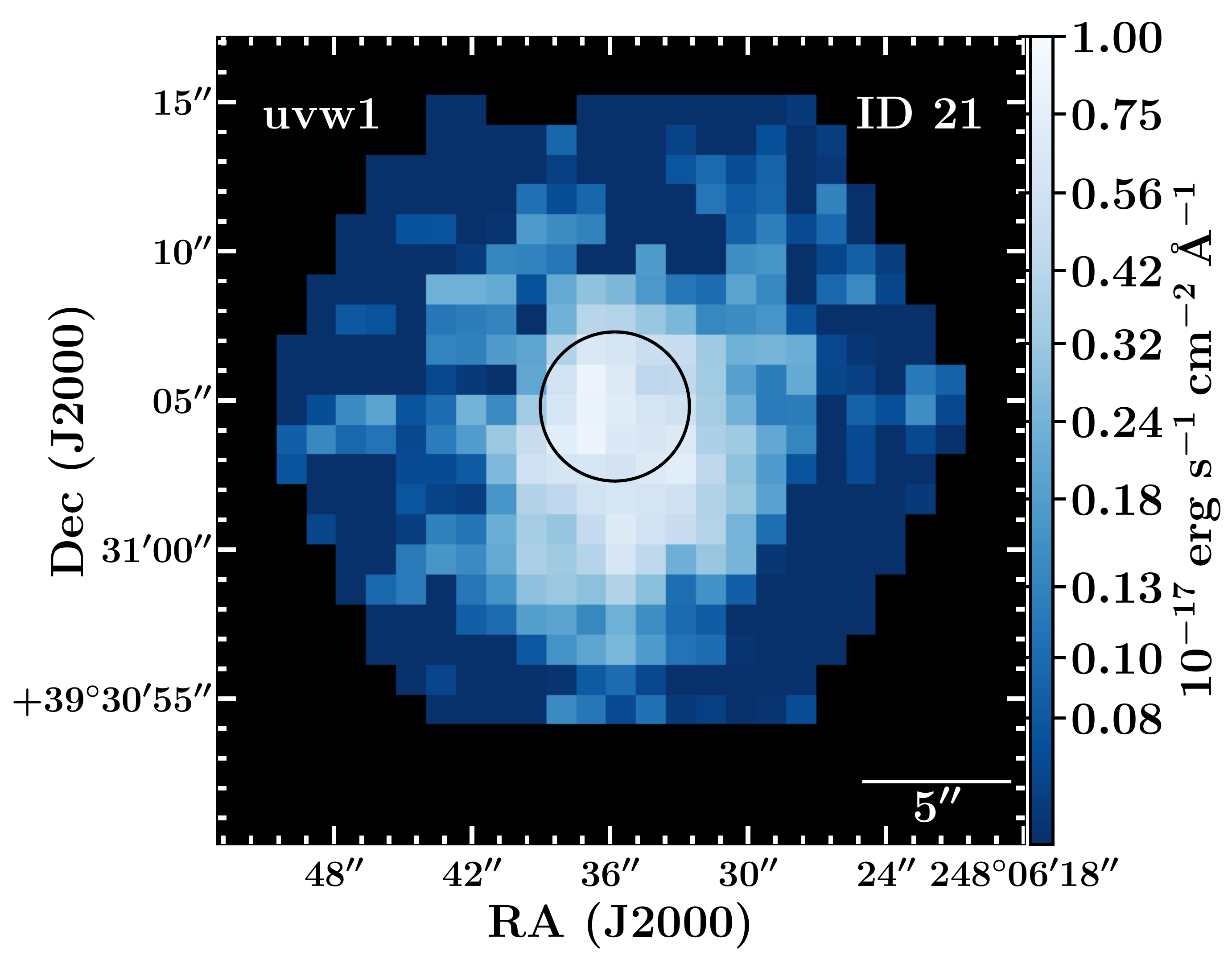}
\includegraphics[width=0.3\textwidth]{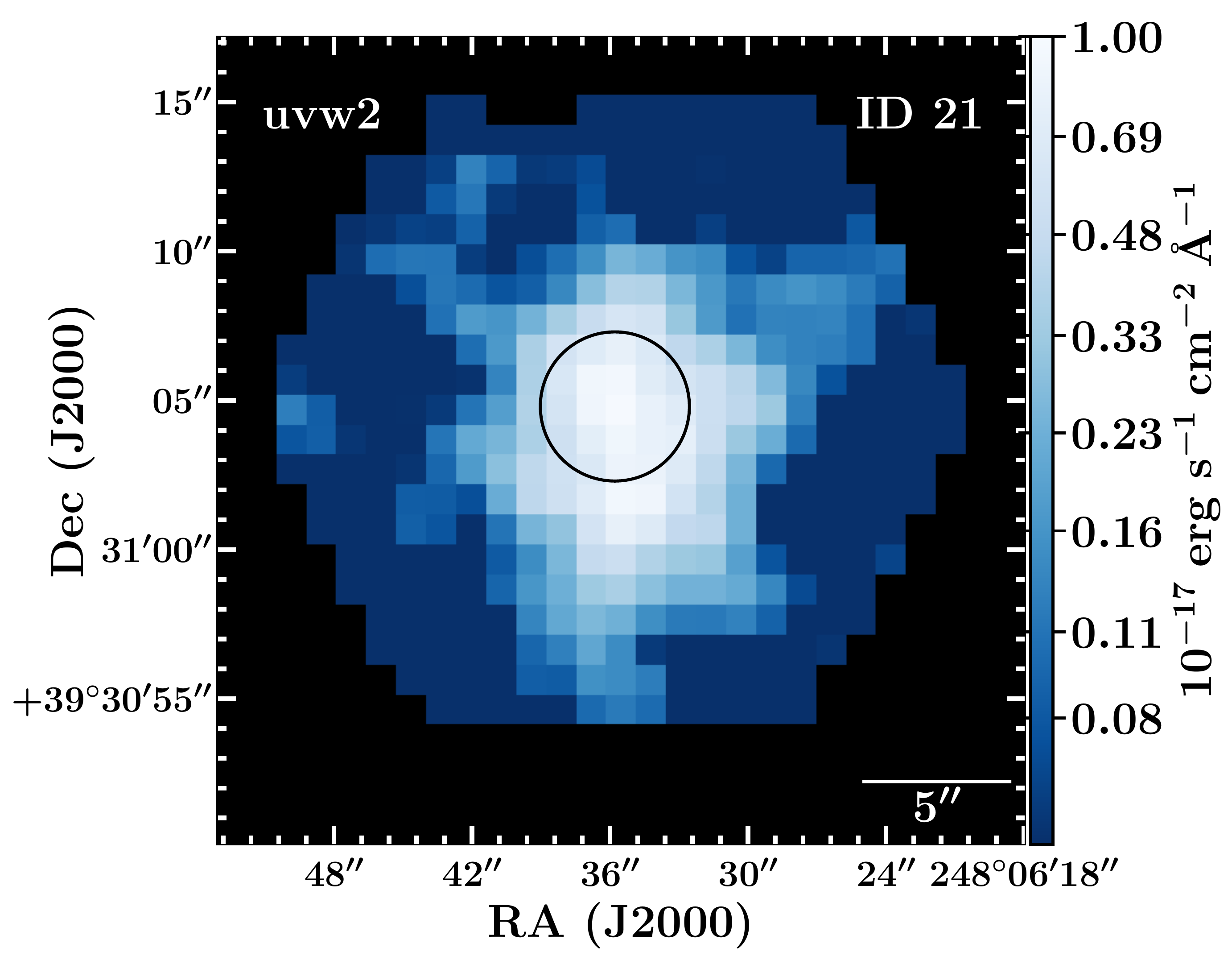}\\
\includegraphics[width=0.3\textwidth]{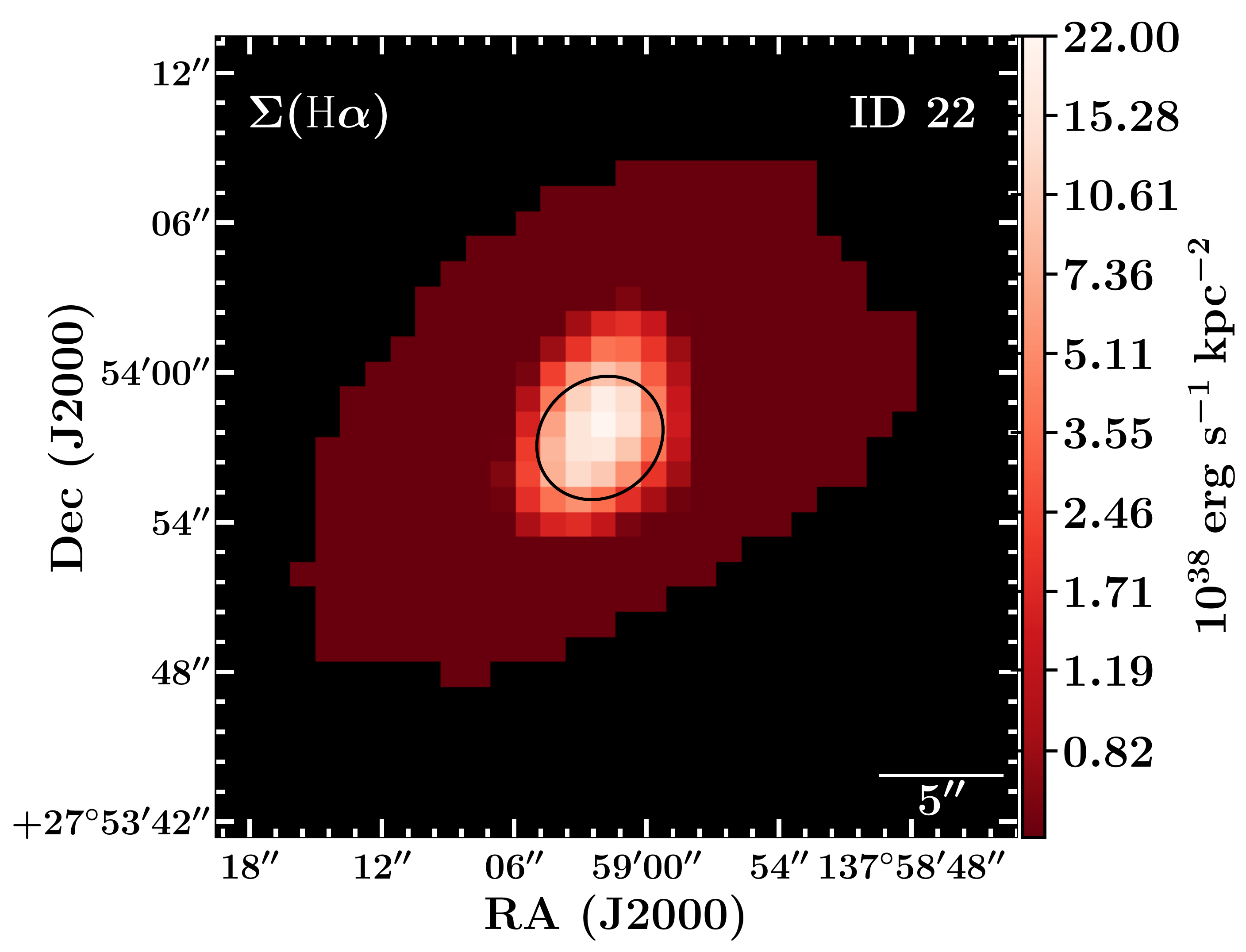}
\includegraphics[width=0.3\textwidth]{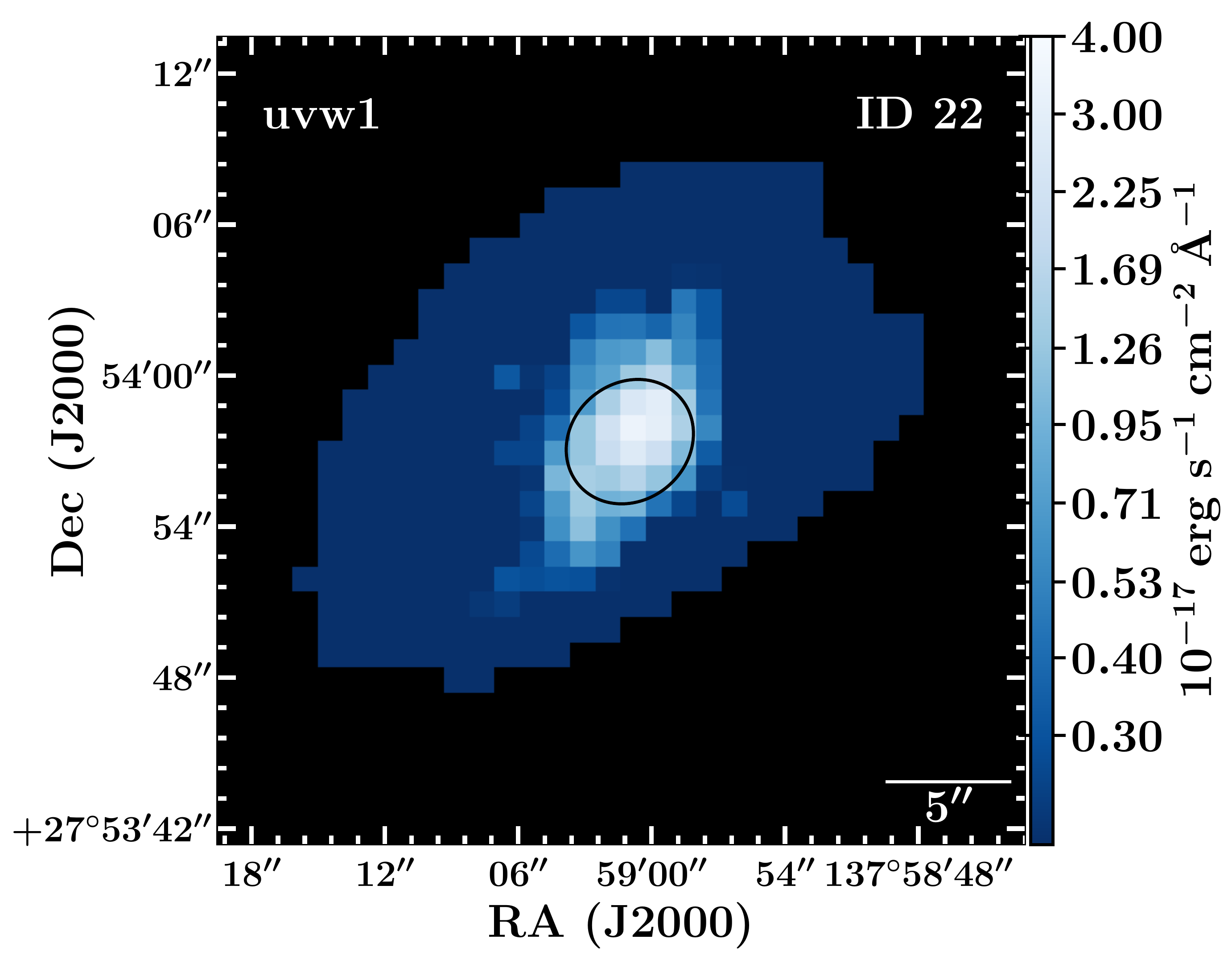}
\includegraphics[width=0.3\textwidth]{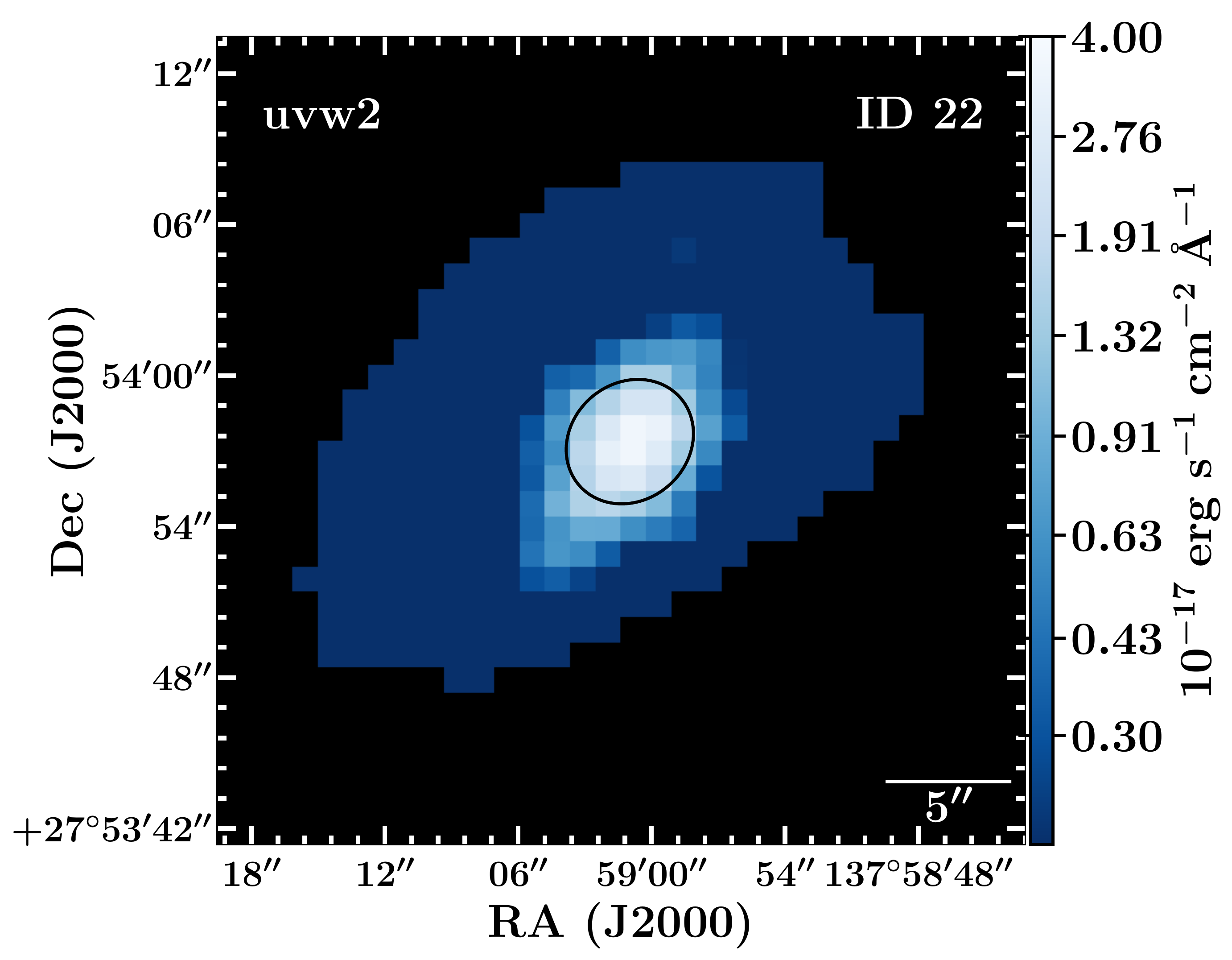}\\
\includegraphics[width=0.3\textwidth]{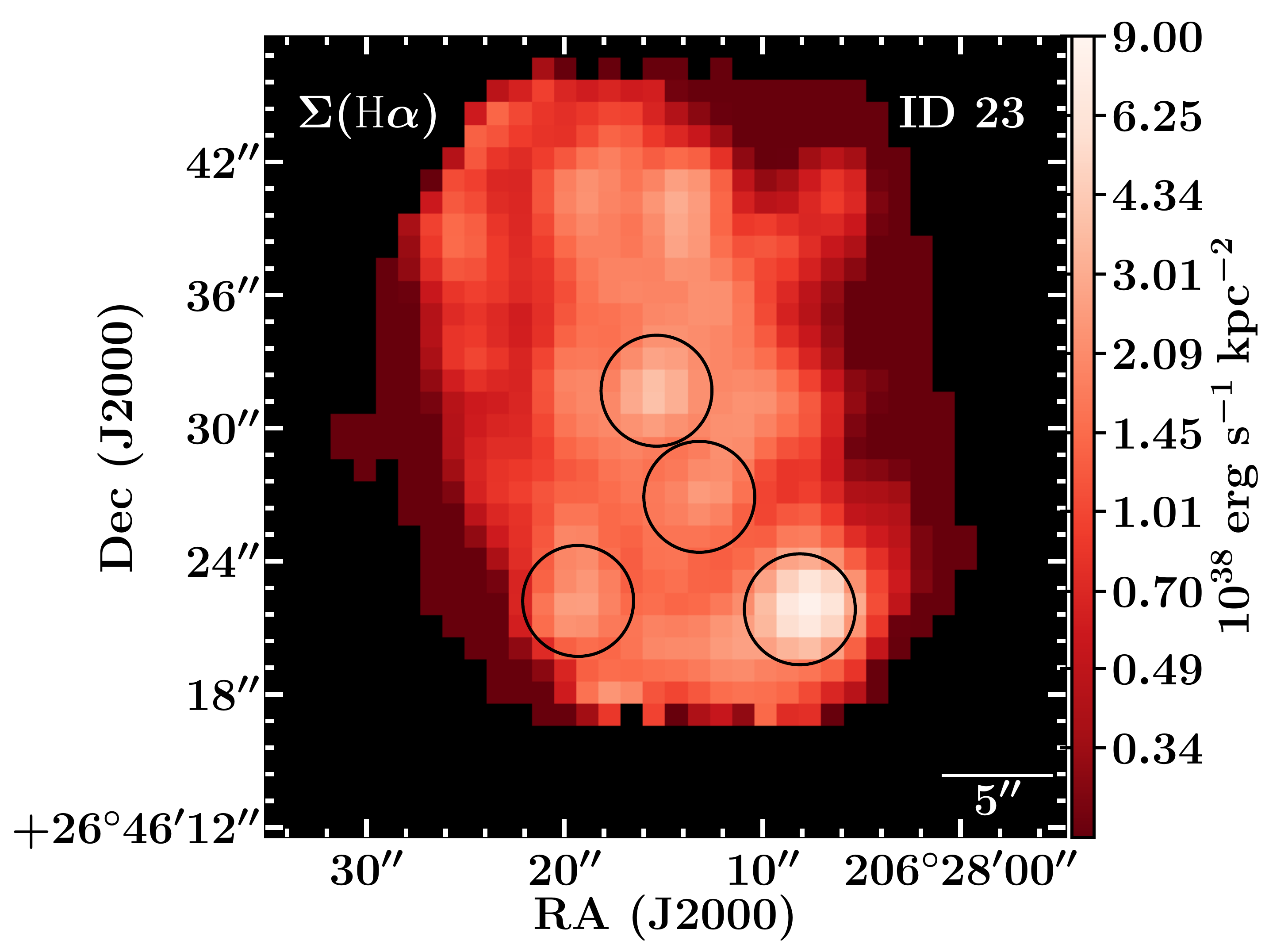}
\includegraphics[width=0.3\textwidth]{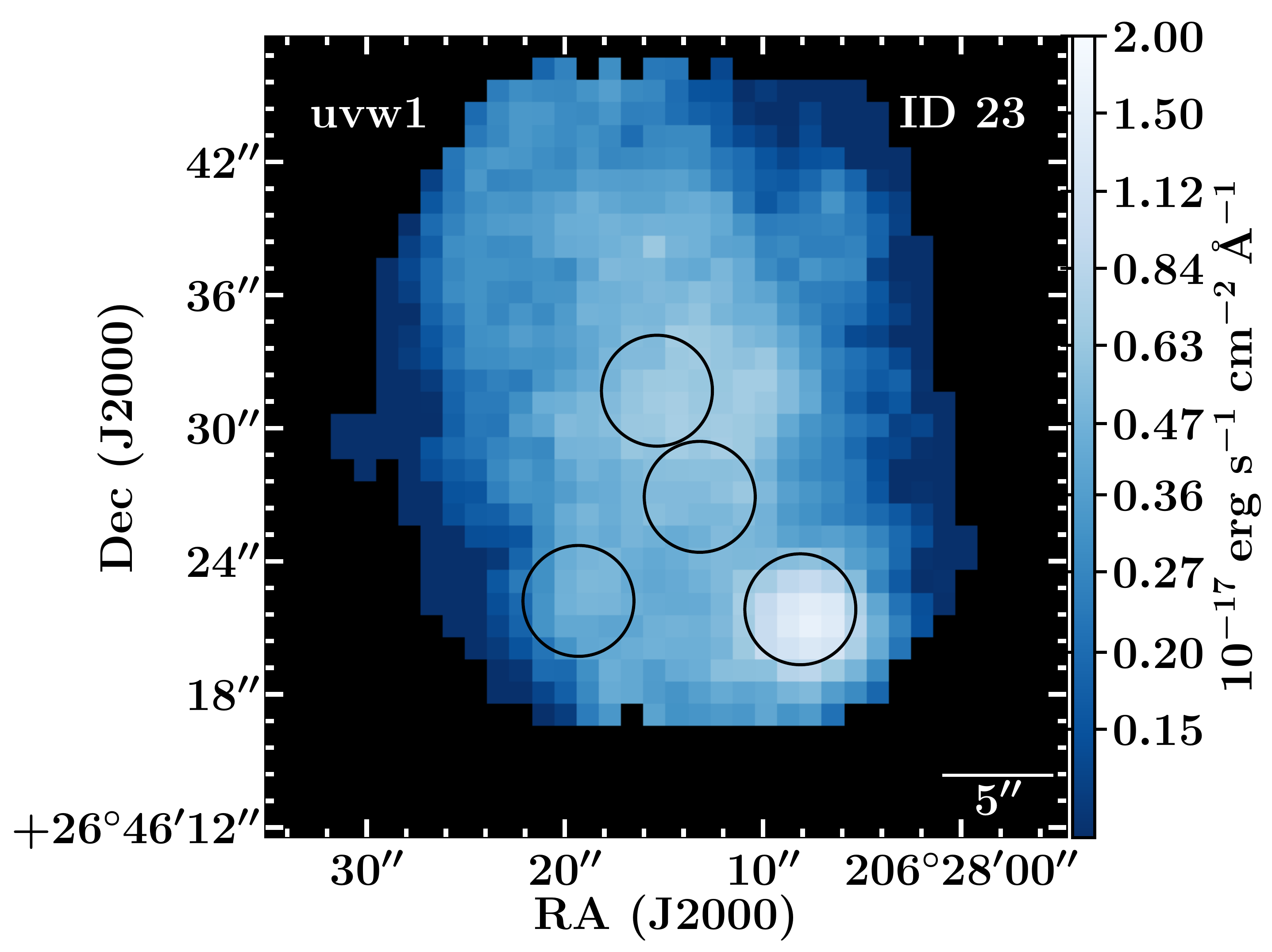}
\includegraphics[width=0.3\textwidth]{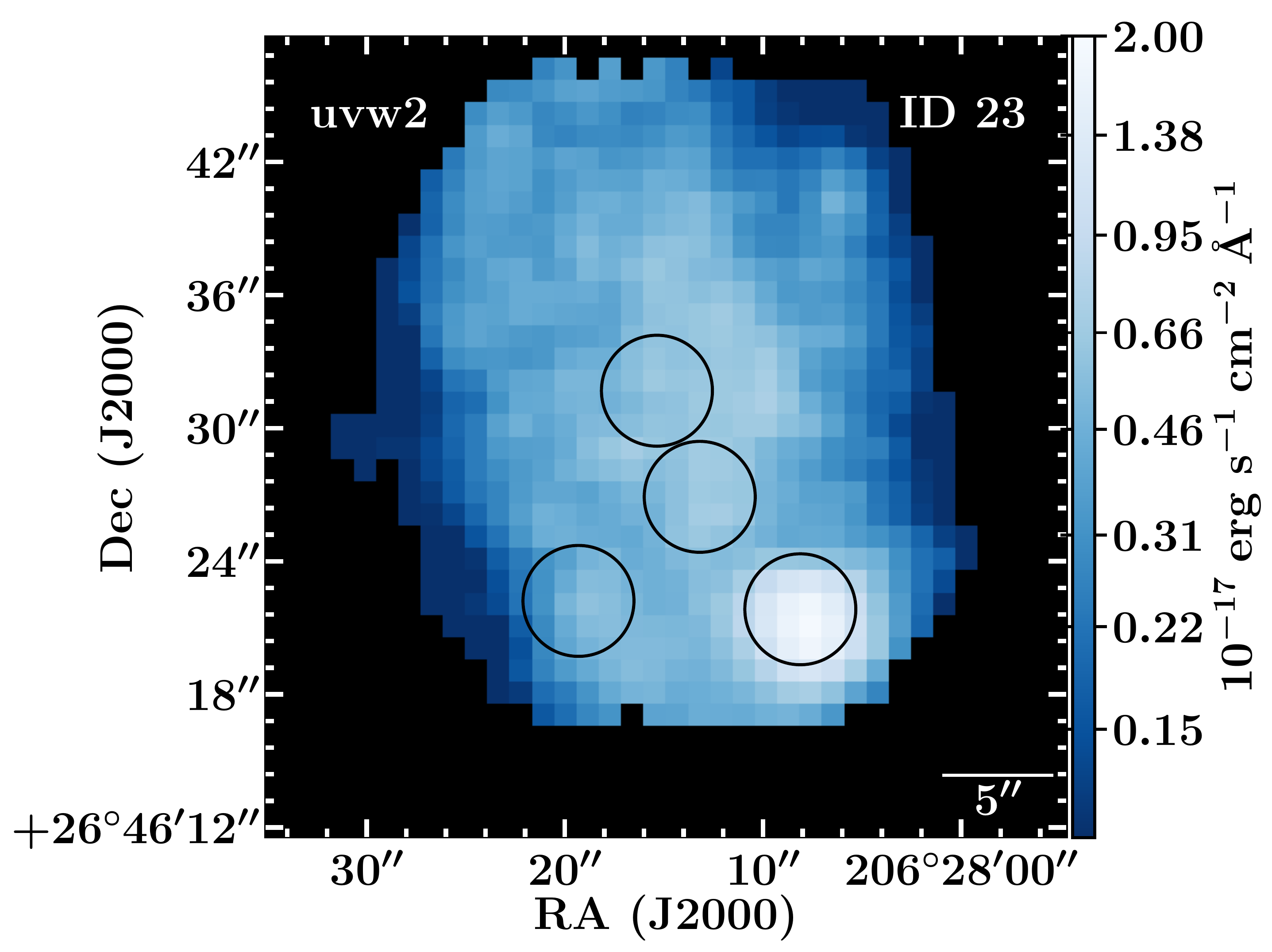}\\
\includegraphics[width=0.3\textwidth]{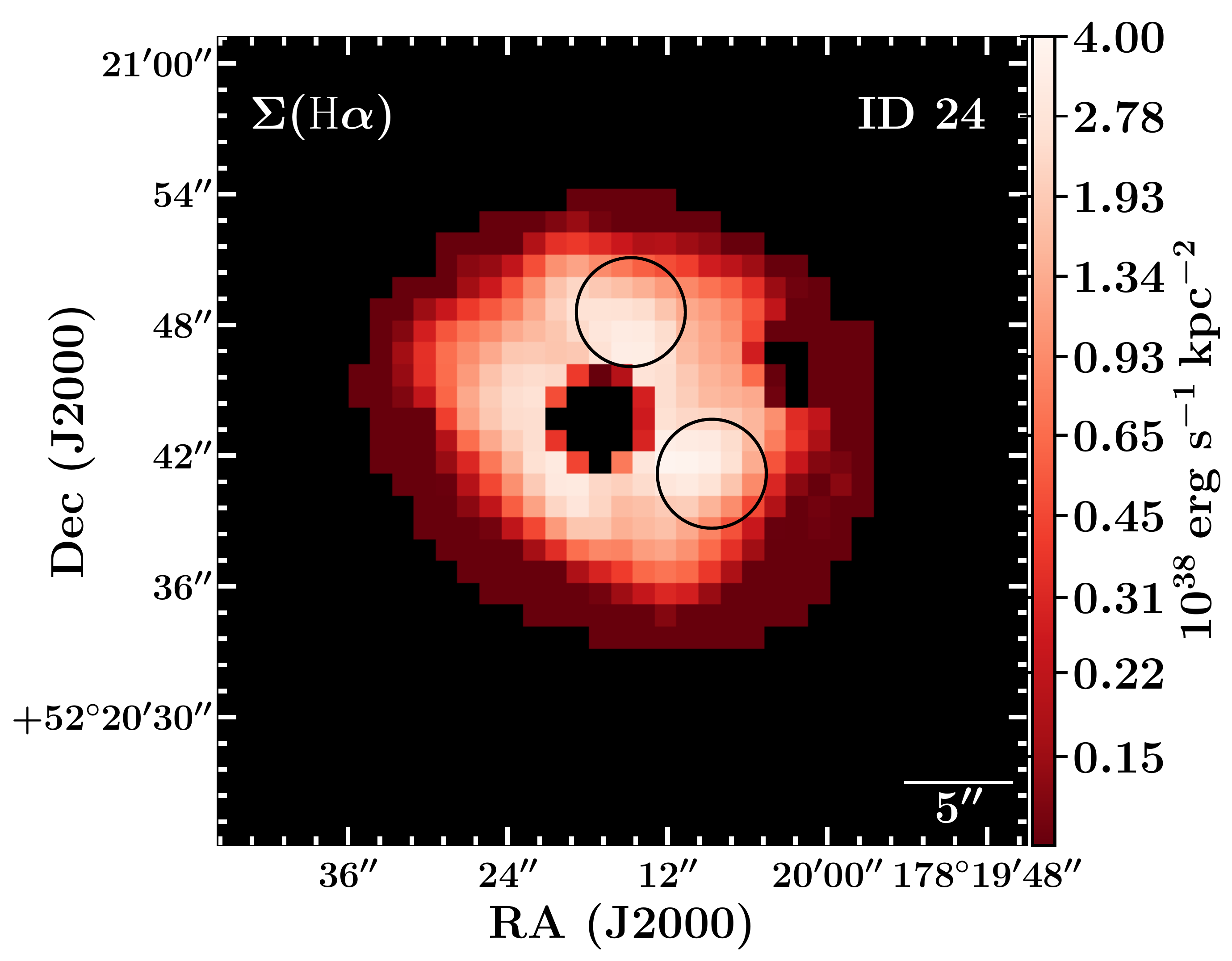}
\includegraphics[width=0.3\textwidth]{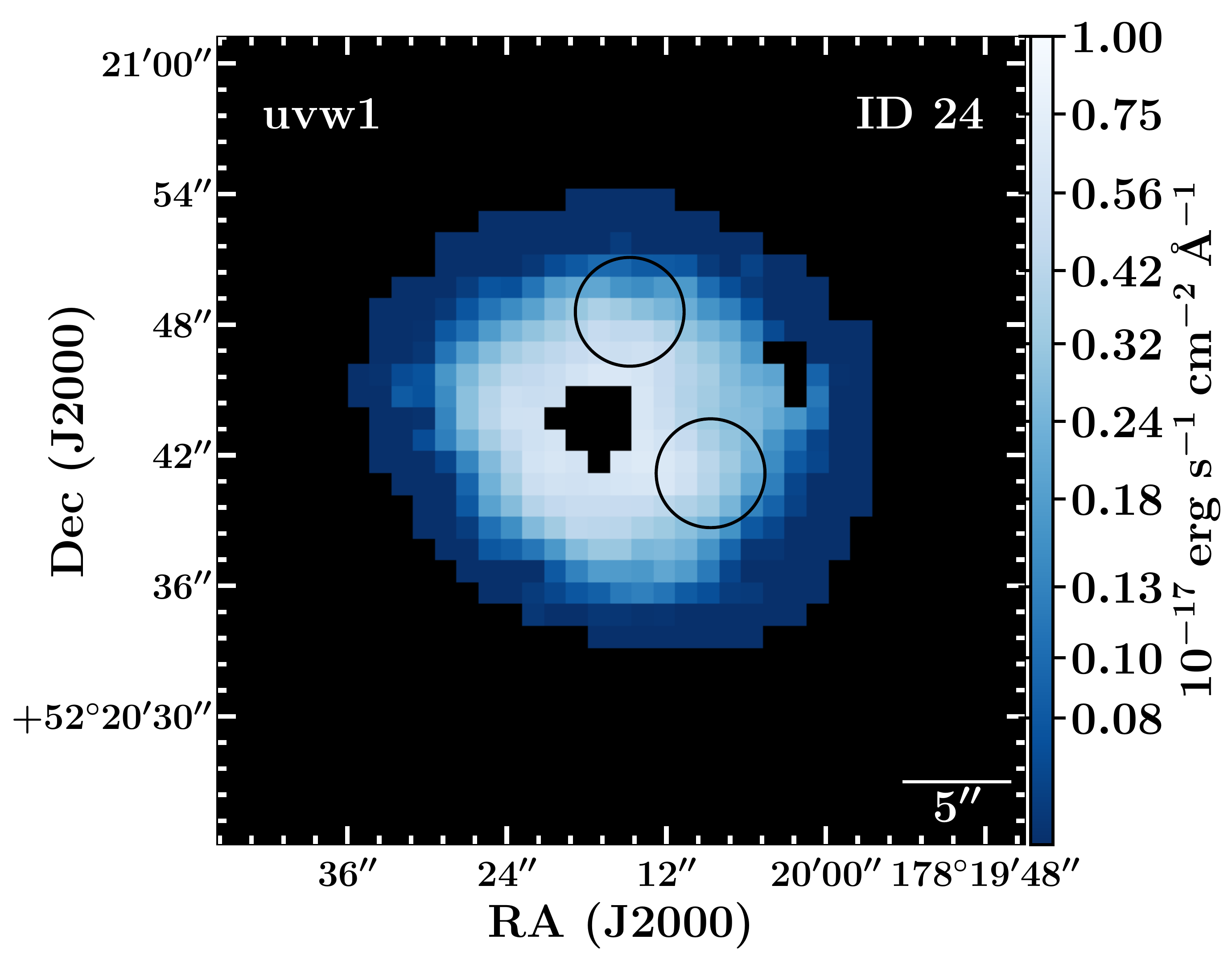}
\includegraphics[width=0.3\textwidth]{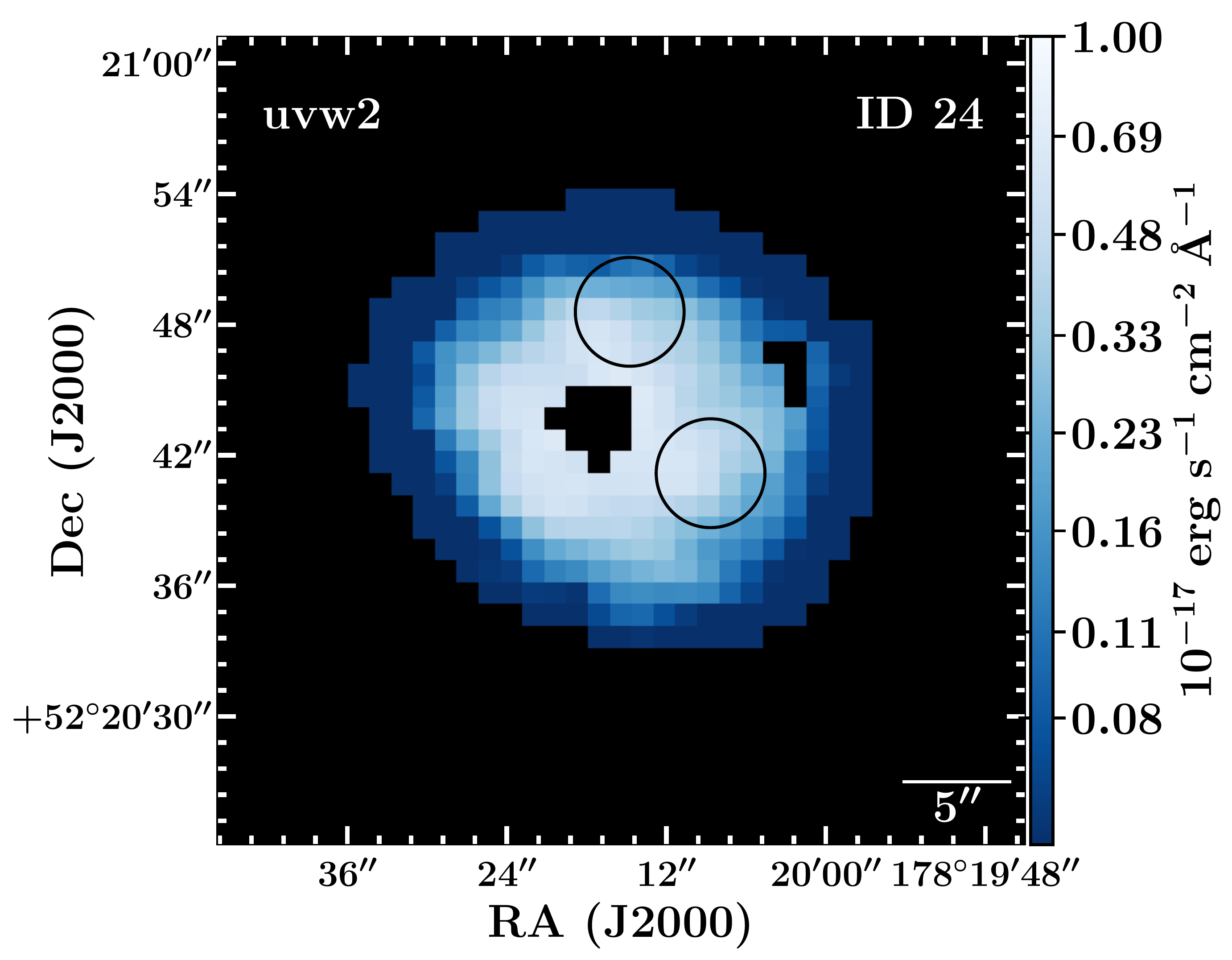}\\
\includegraphics[width=0.3\textwidth]{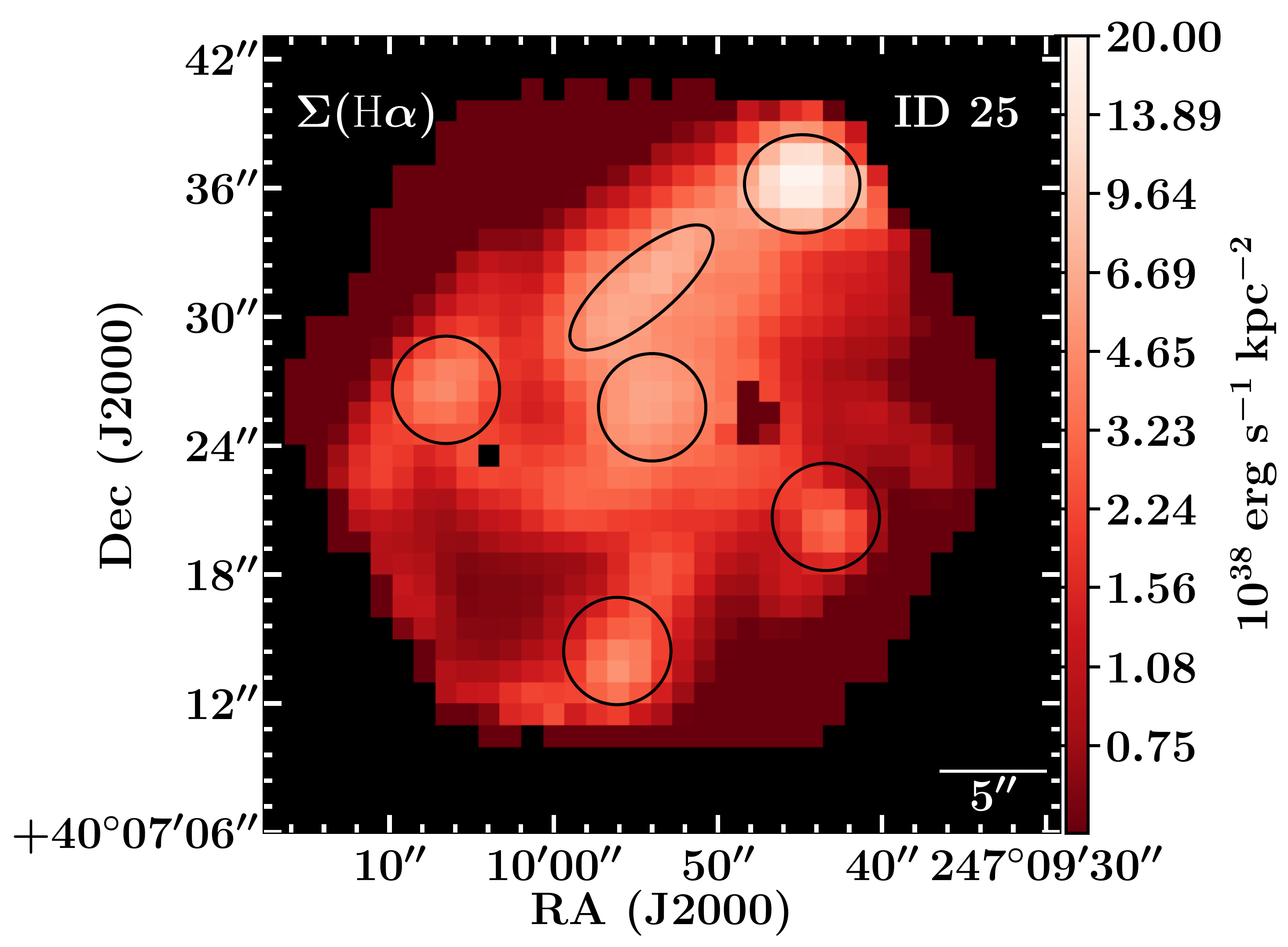}
\includegraphics[width=0.3\textwidth]{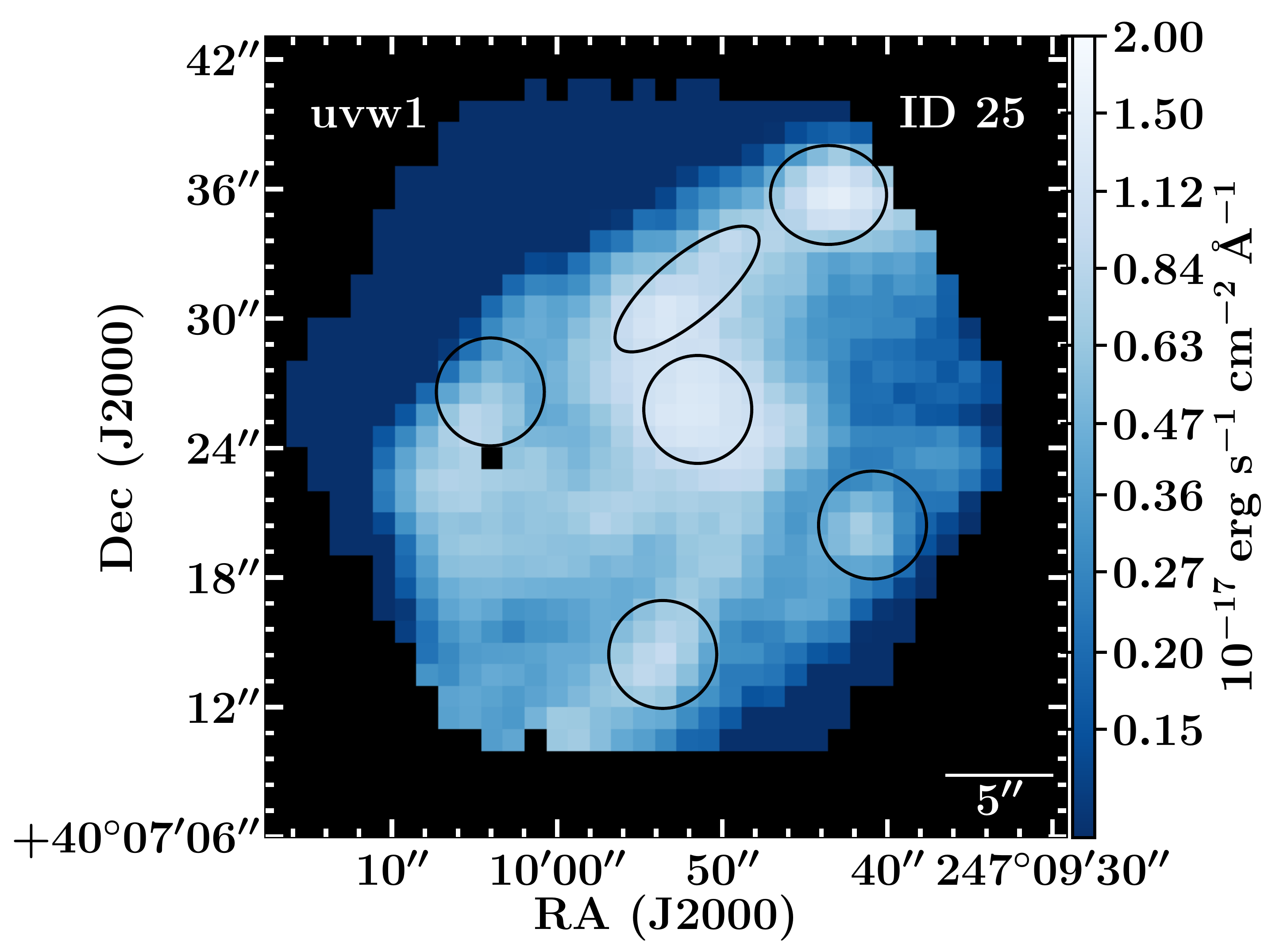}
\includegraphics[width=0.3\textwidth]{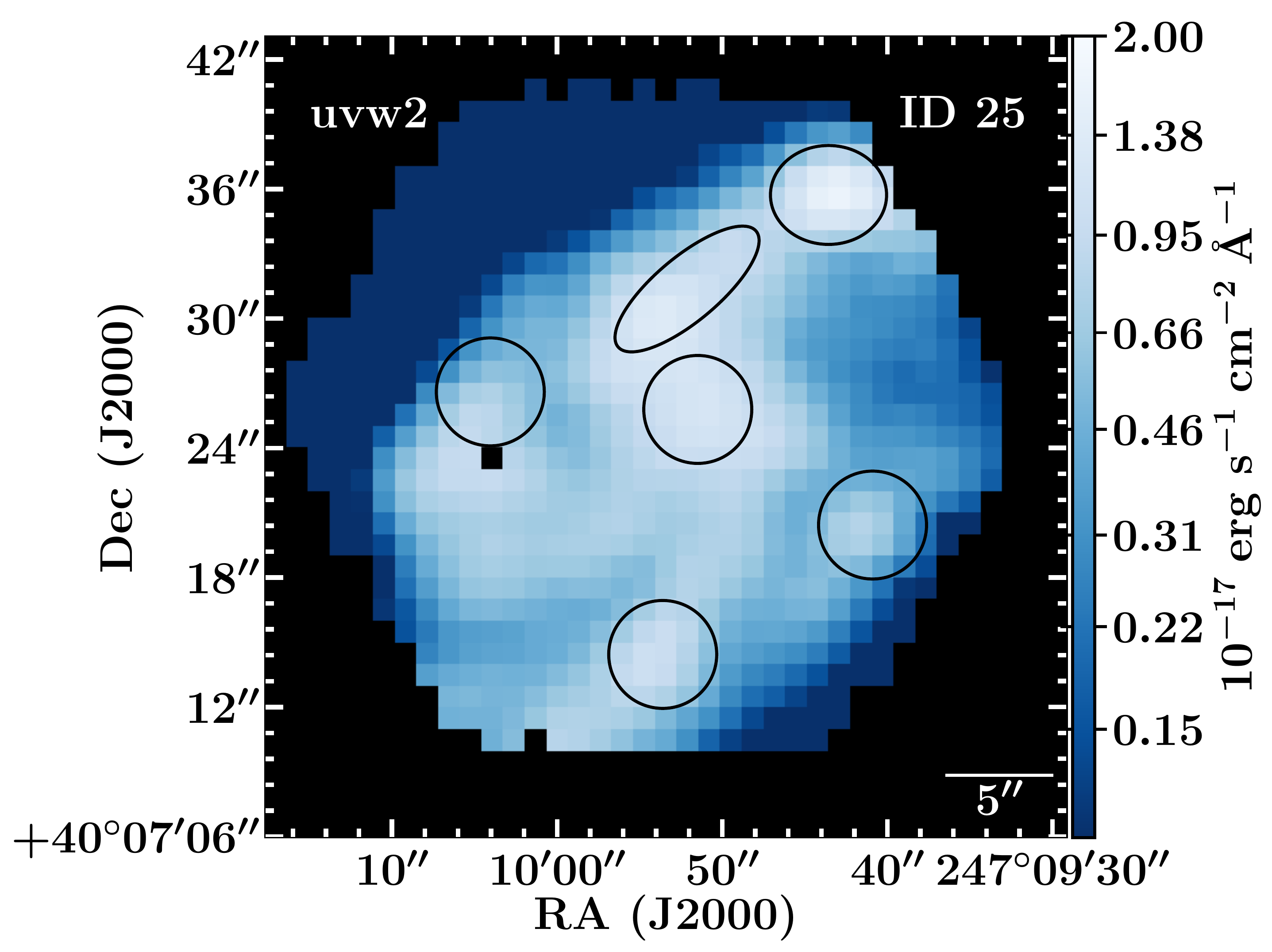}\\
\includegraphics[width=0.3\textwidth]{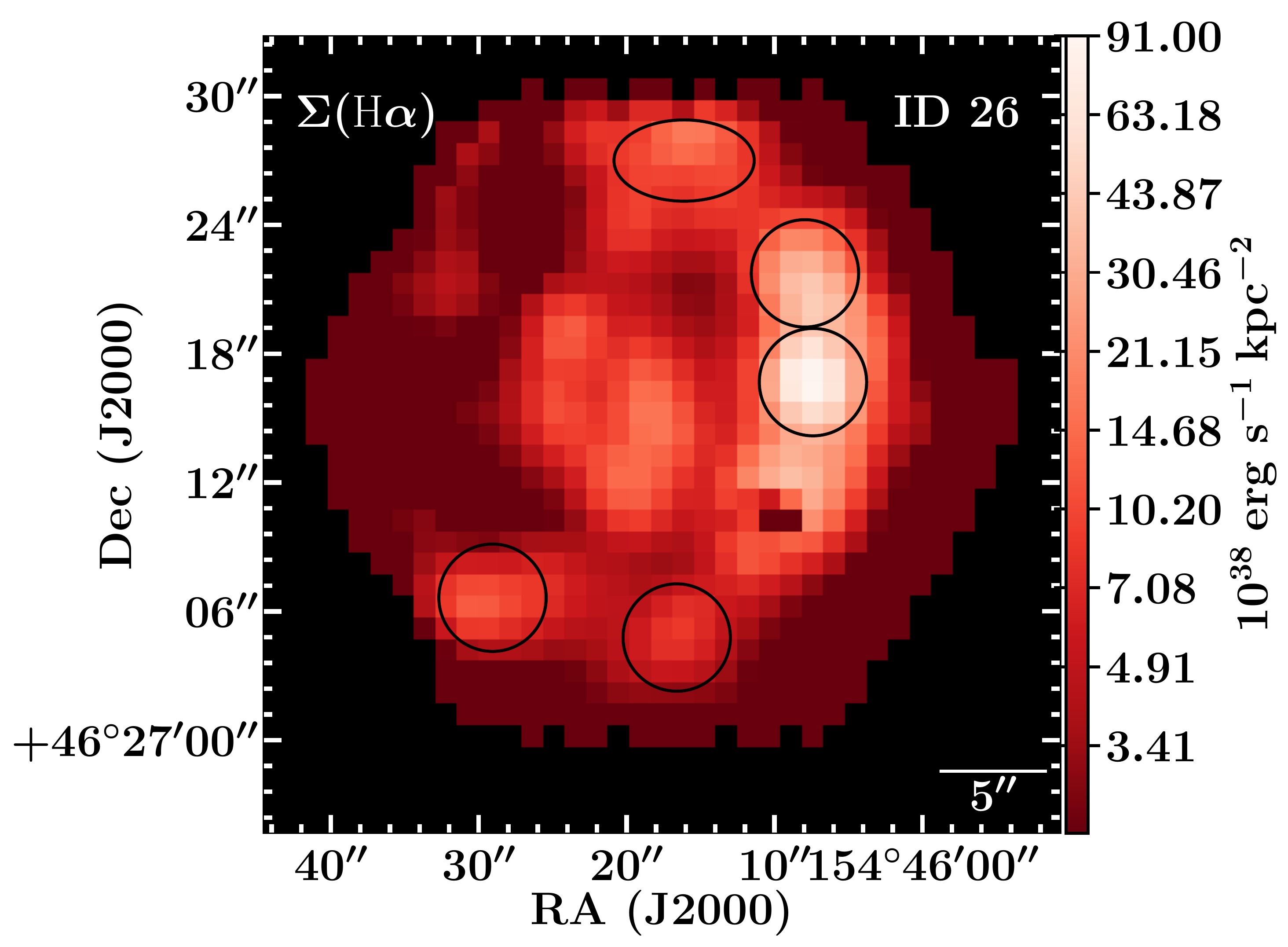}
\includegraphics[width=0.3\textwidth]{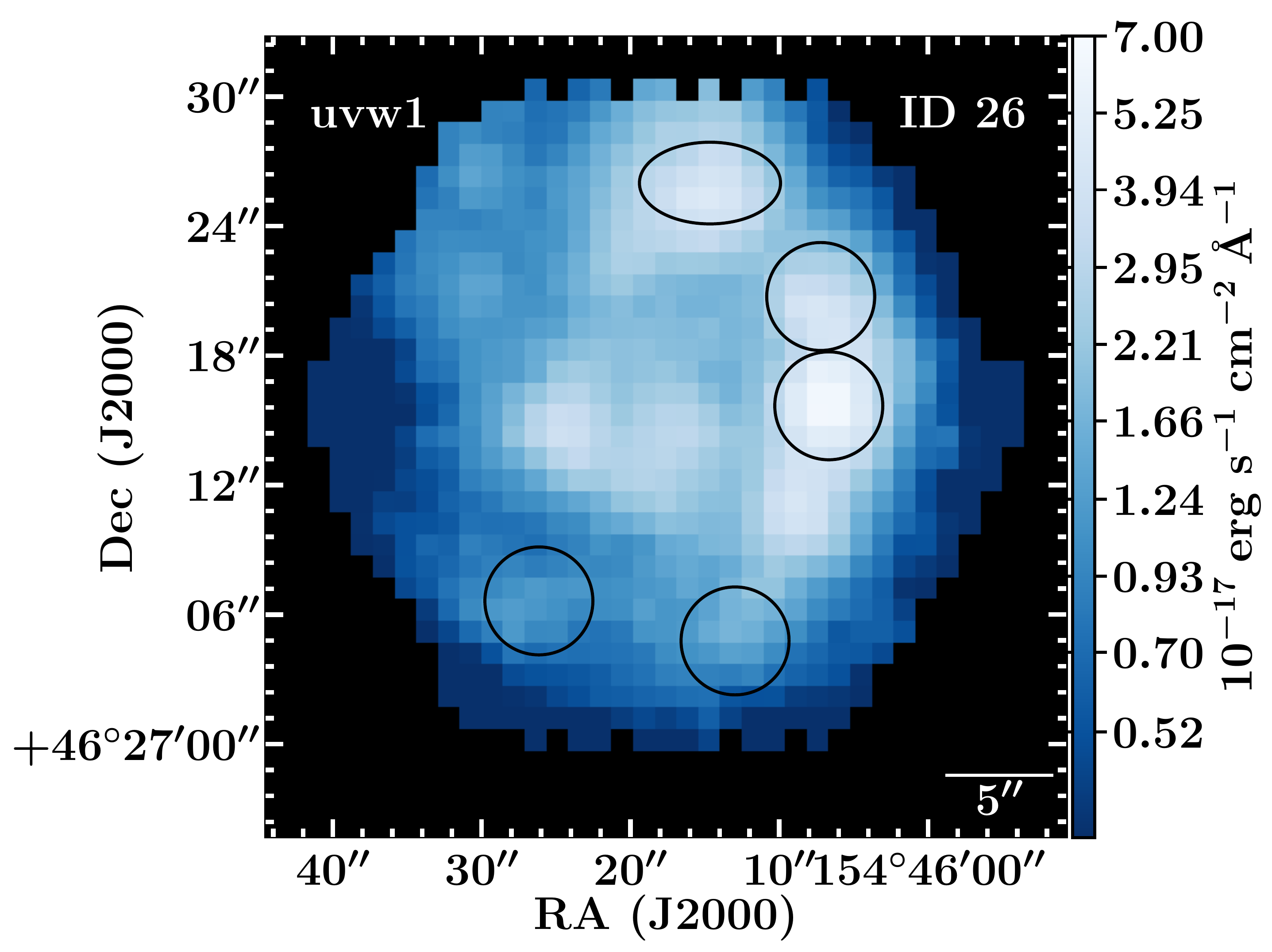}
\includegraphics[width=0.3\textwidth]{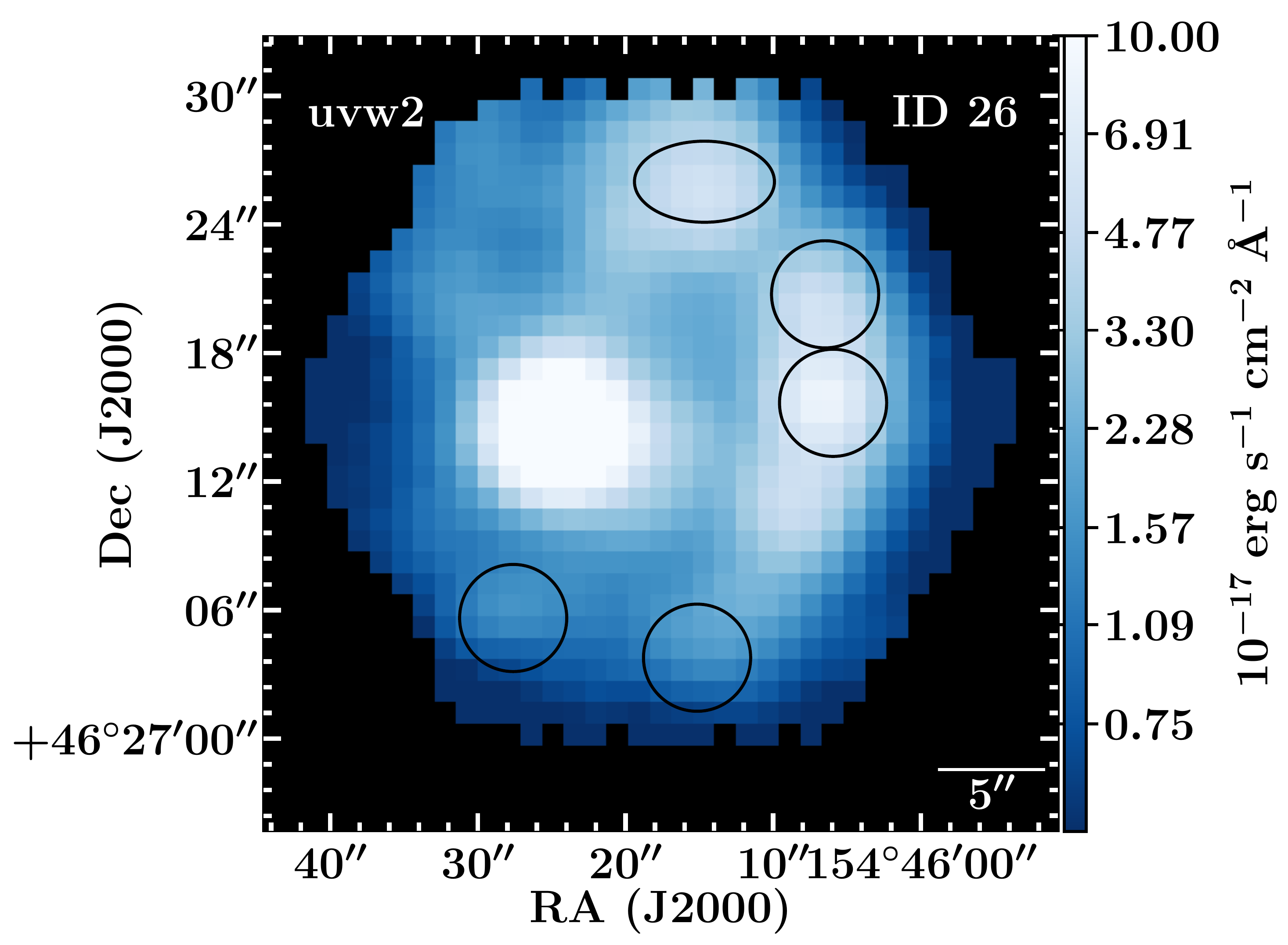}\\
\includegraphics[width=0.3\textwidth]{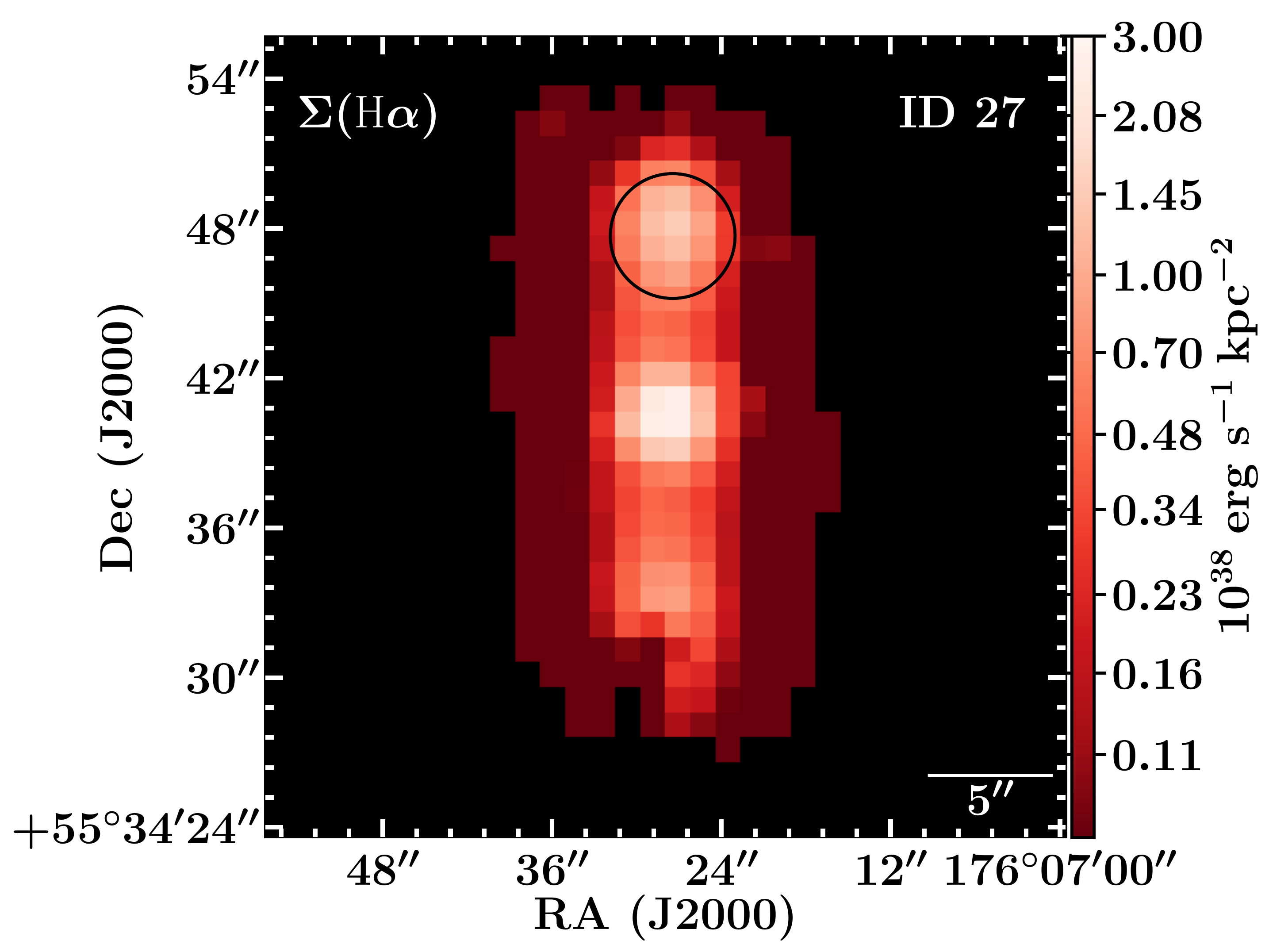}
\includegraphics[width=0.3\textwidth]{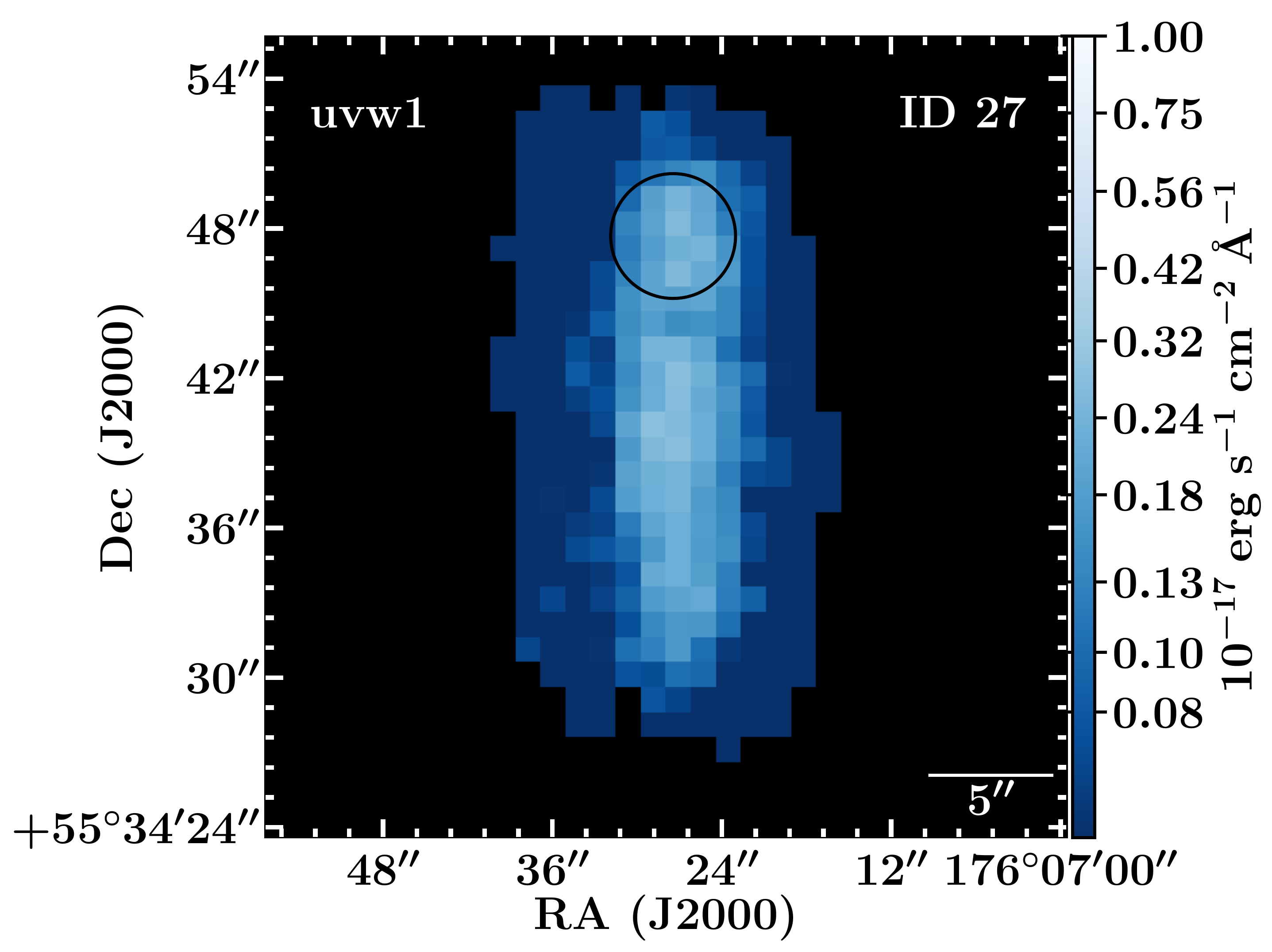}
\includegraphics[width=0.3\textwidth]{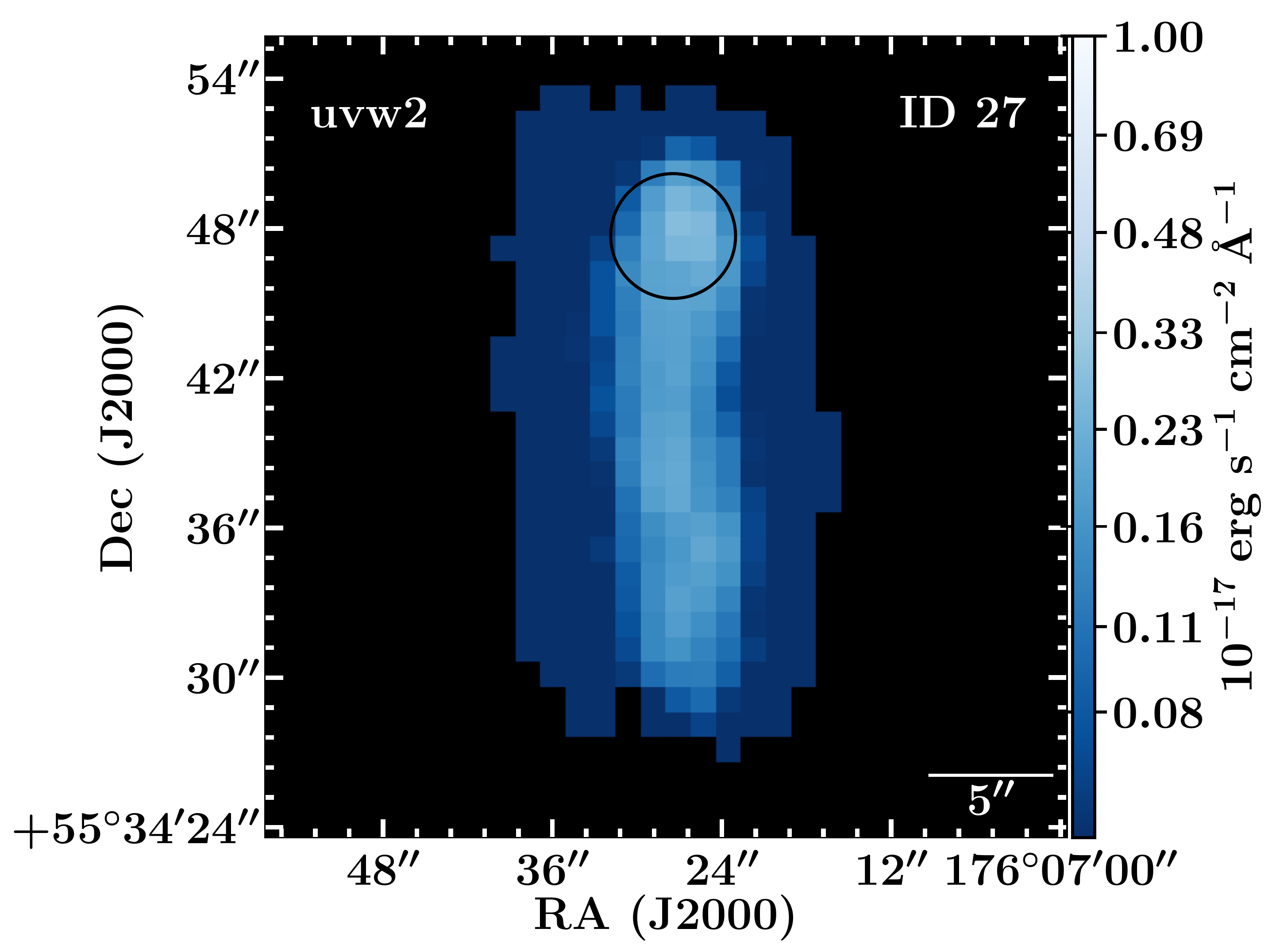}\\
\includegraphics[width=0.3\textwidth]{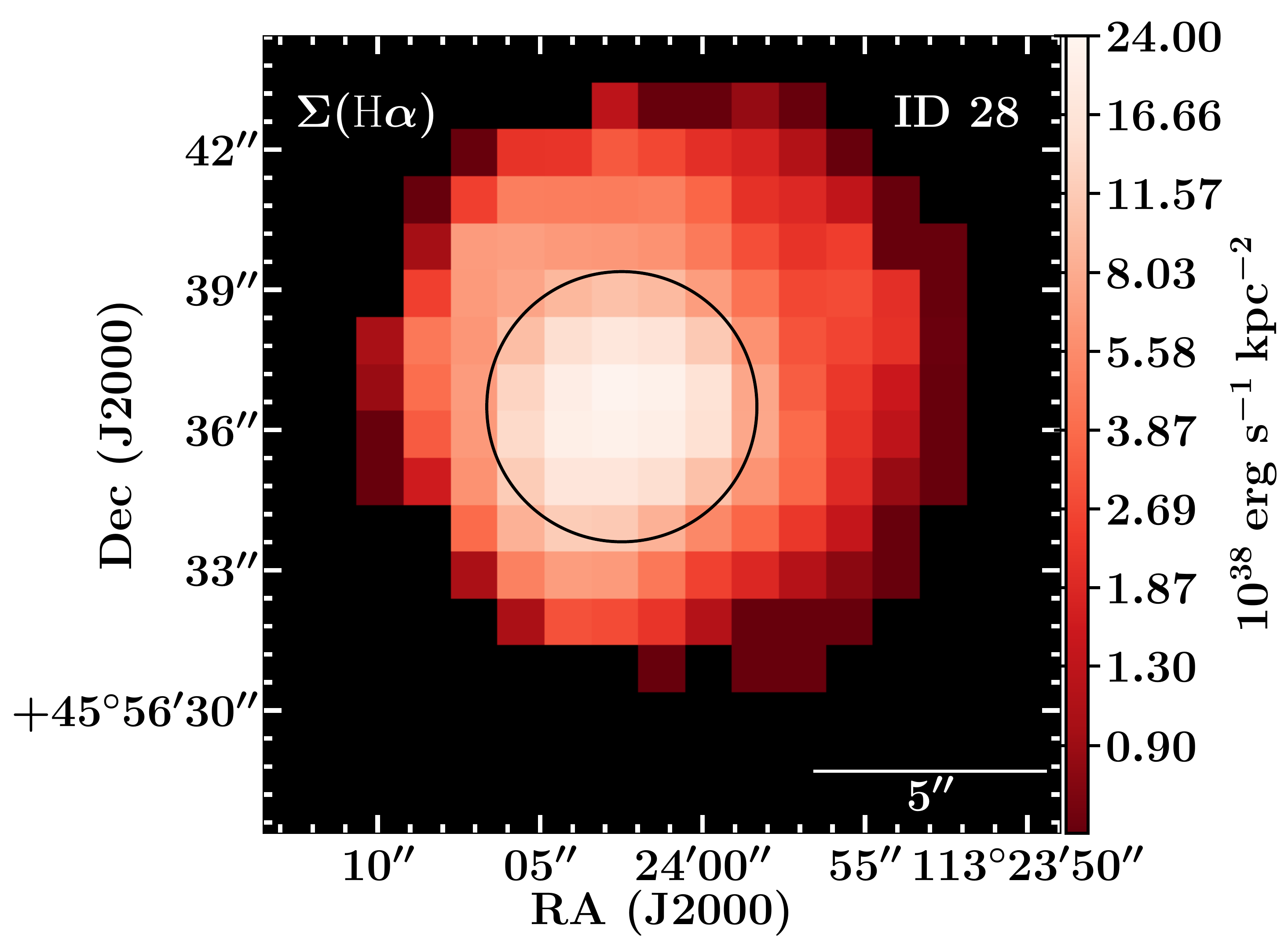}
\includegraphics[width=0.3\textwidth]{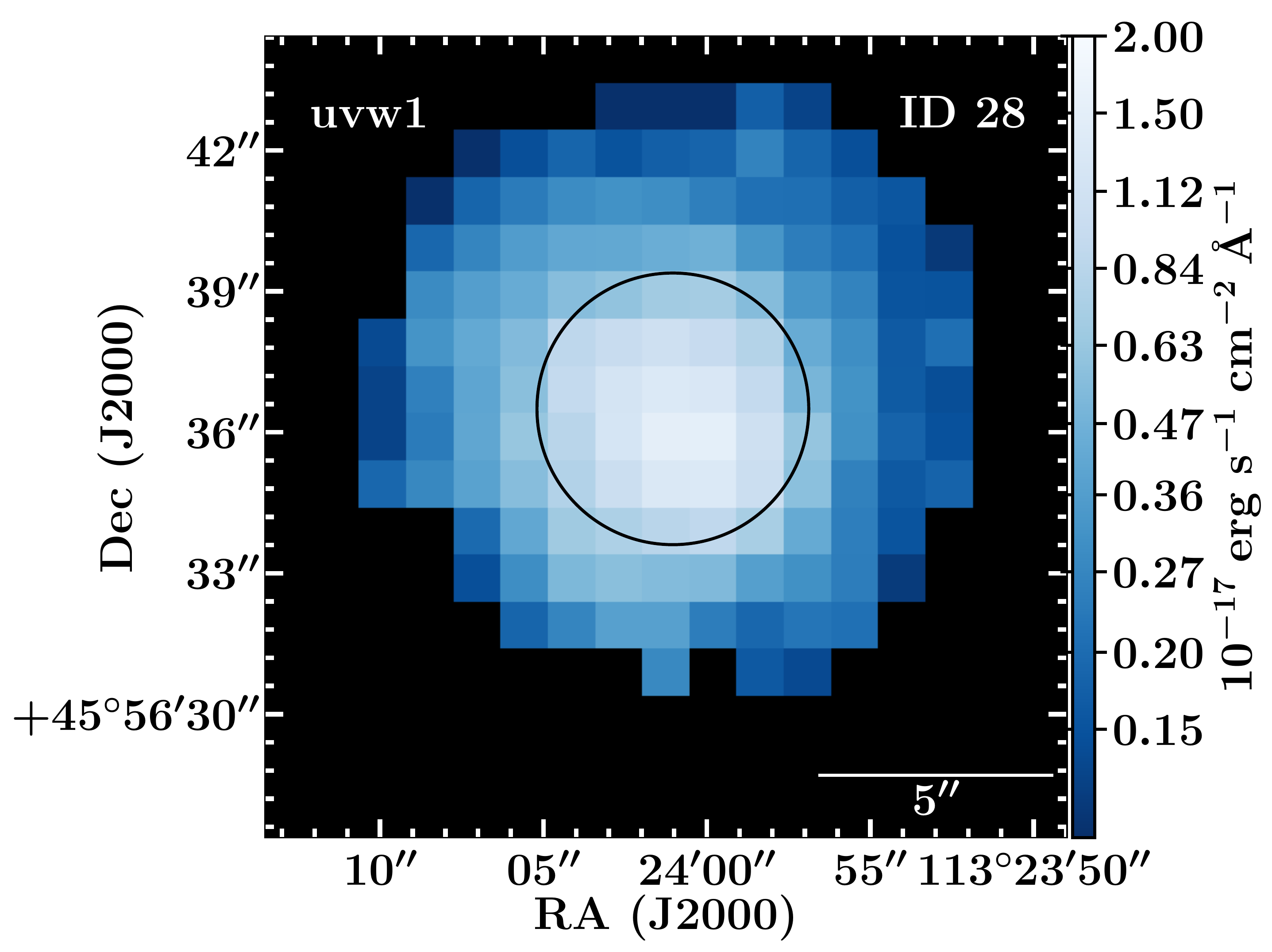}
\includegraphics[width=0.3\textwidth]{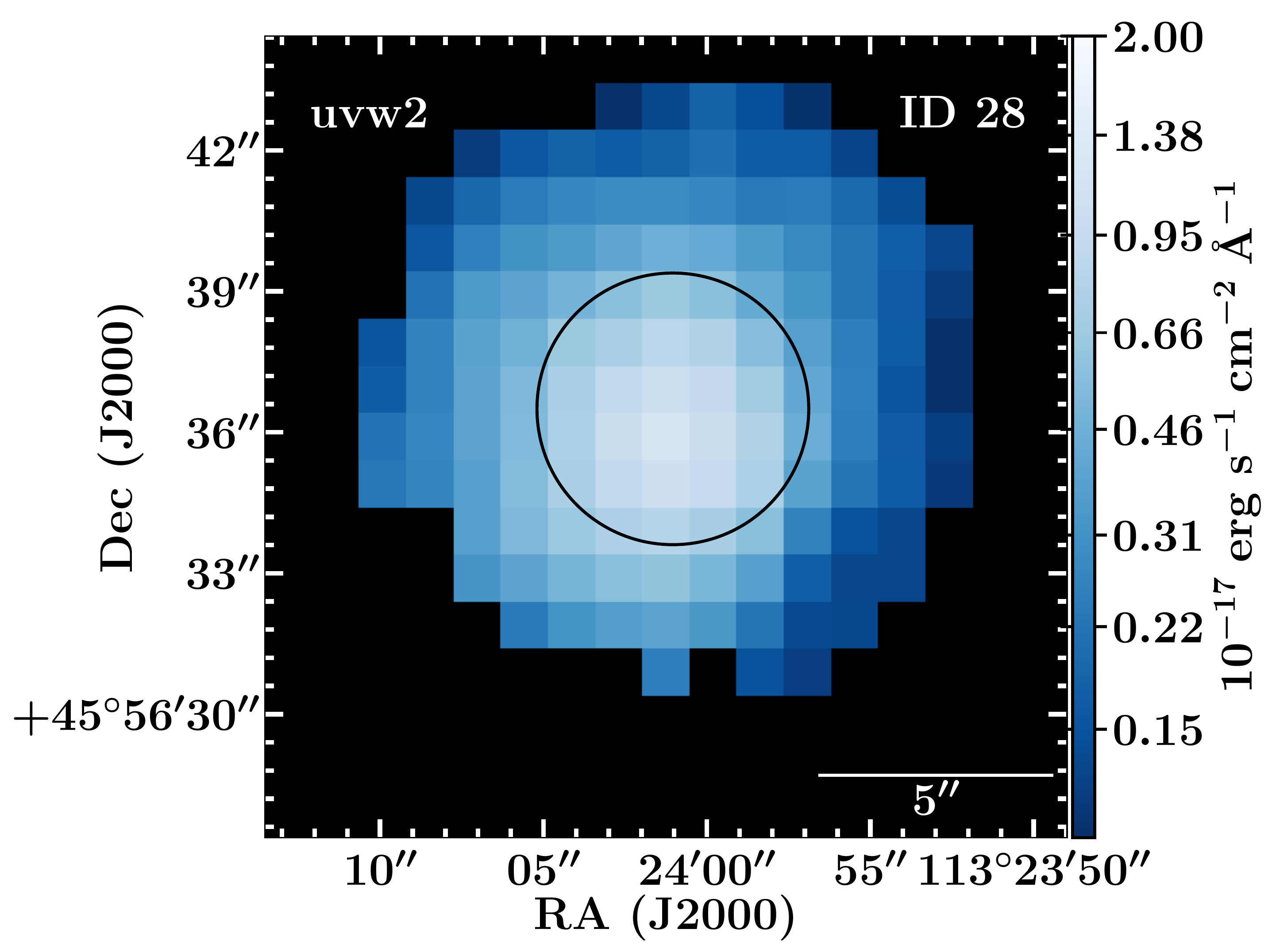}\\
\includegraphics[width=0.3\textwidth]{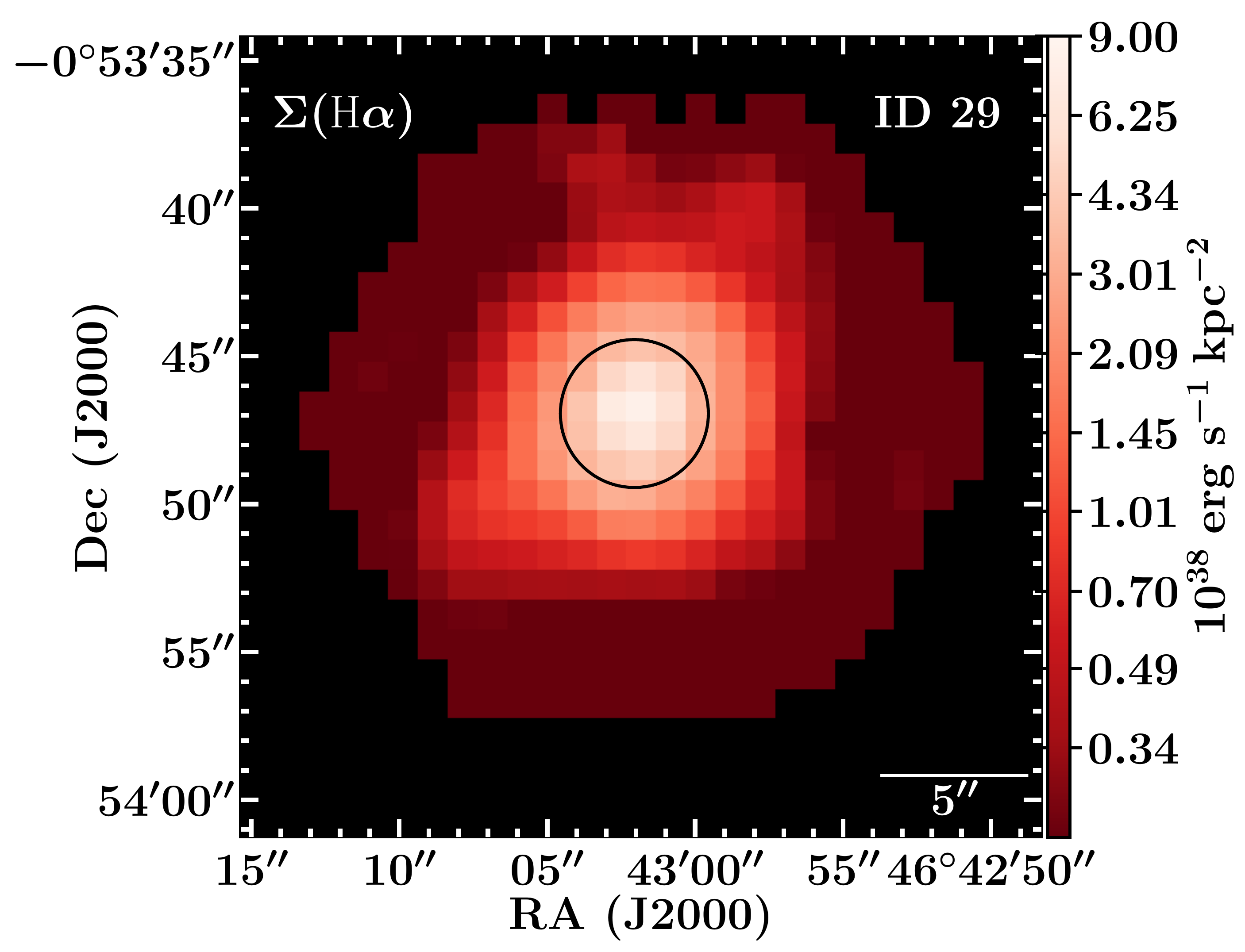}
\includegraphics[width=0.3\textwidth]{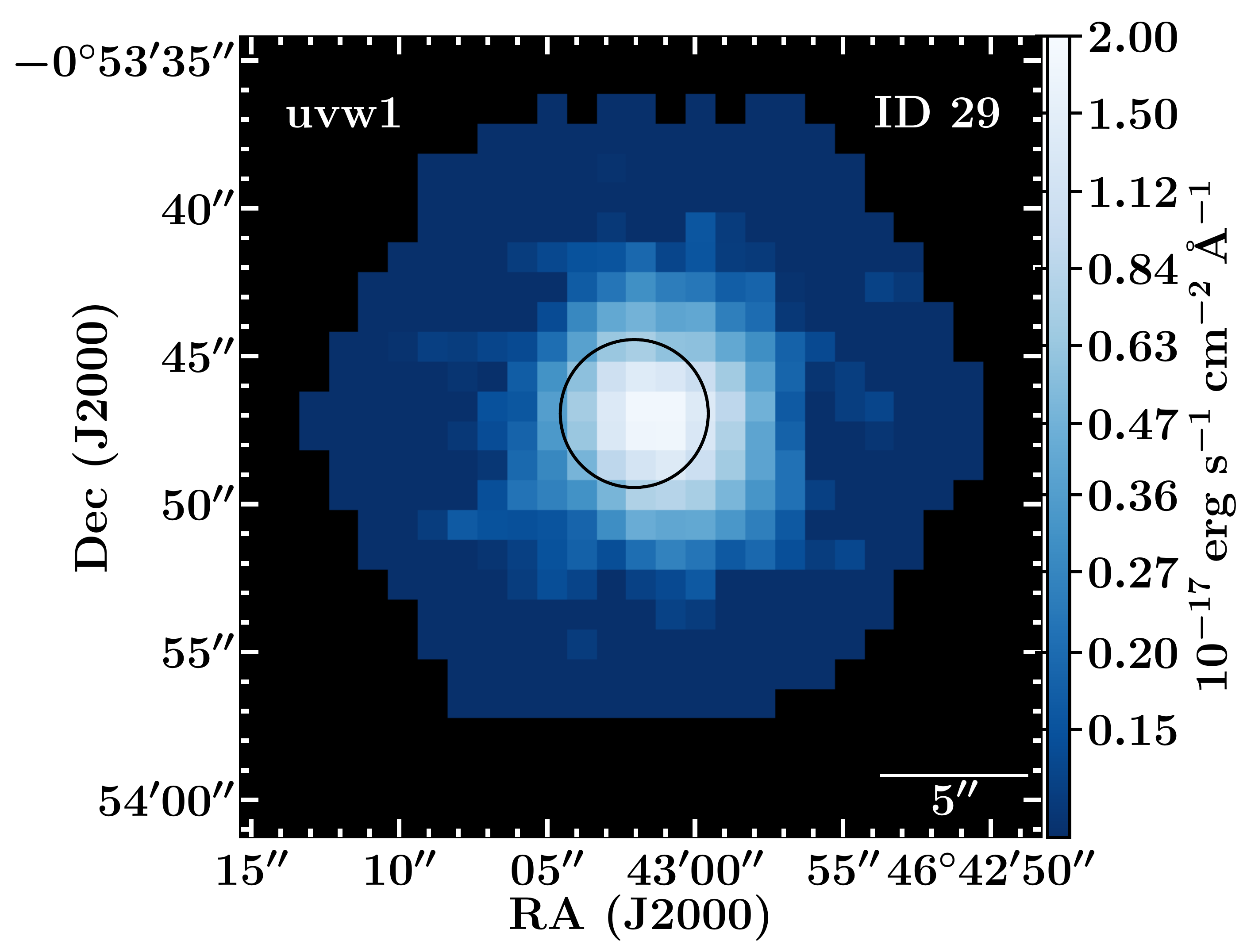}
\includegraphics[width=0.3\textwidth]{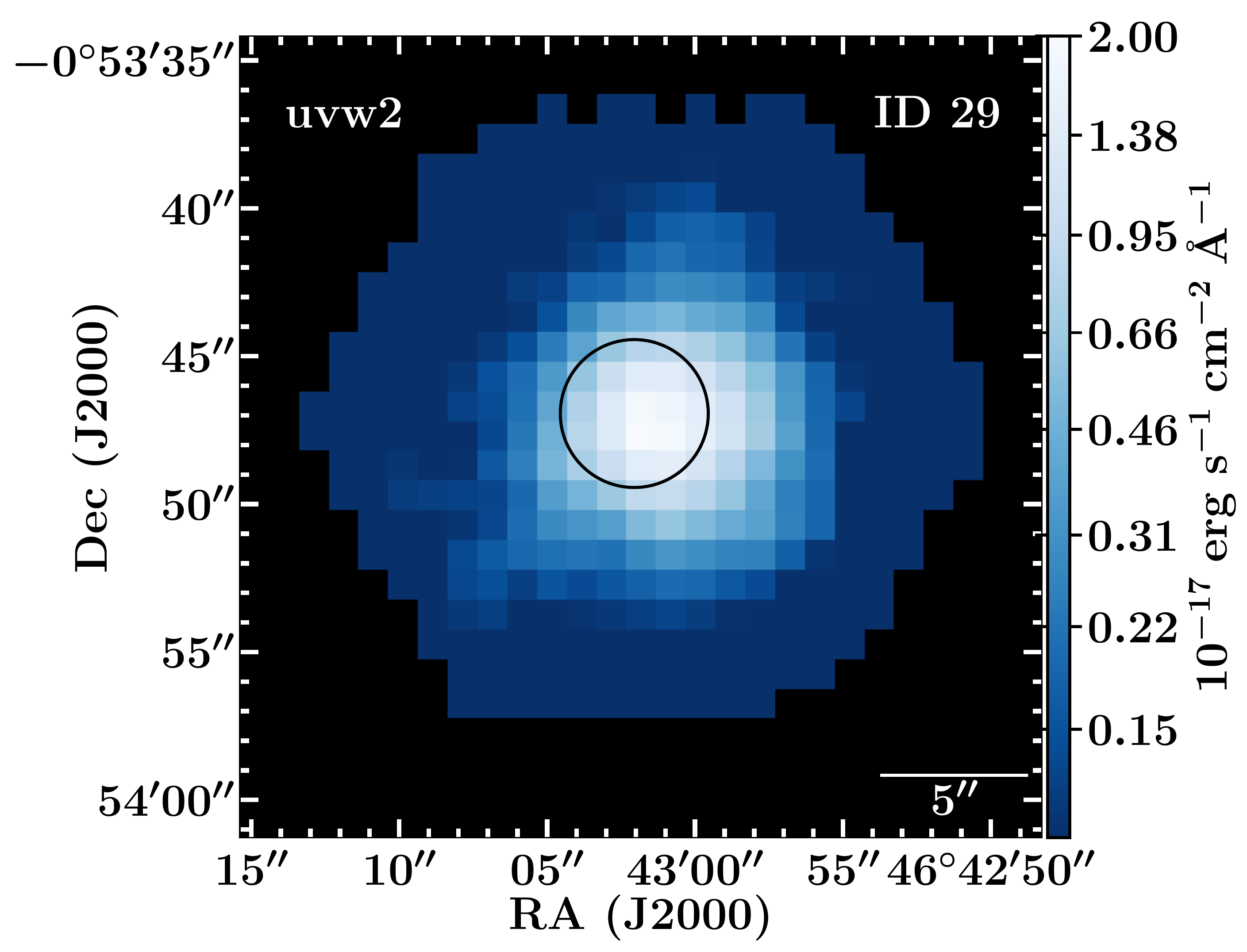}
\end{center}

\begin{table*}
\renewcommand{\thefootnote}{\textrm{\alph{footnote}}}
	\centering
		\setlength{\tabcolsep}{8pt}
	\caption{Aperture Definitions for Star Forming Regions\label{table:aper}}
	\begin{tabular}{cccccc} % four columns, alignment for each
		\hline
		\hline
		{Region} & {R.A.} & {Dec.} & {$a$} & {$b/a$} & {$\theta$}\\
 {I.D.} & {(deg)} & {(deg)}& {(arcsec)} & { } & {(deg)}\\
 		\hline
{$1.1$} & {$318.02666$} & {$11.34723$} & {$3.01130$} & {$1$} & {$\ldots$}\\
{$1.2$} & {$318.02644$} & {$11.34511$} & {$4.32769$} & {$0.34522$} & {$50$}\\
{$1.3$} & {$318.02580$} & {$11.34366$} & {$4.32769$} & {$0.34522$} & {$50$}\\
{$2.1$} & {$256.72559$} & {$32.17052$} & {$2.5$} & {$1$} & {$\ldots$}\\
{$3.1$} & {$247.23607$} & {$39.60964$} & {$2.5$} & {$1$} & {$\ldots$}\\
{$4.1$} & {$61.84818$} & {$-6.68616$} & {$3.59562$} & {$0.47922$} & {$150$}\\
{$5.1$} & {$166.18811$} & {$45.15670$} & {$2.5$} & {$1$} & {$\ldots$}\\
{$6.1$} & {$213.49579$} & {$43.89317$} & {$2.5$} & {$1$} & {$\ldots$}\\
{$7.1$} & {$154.83646$} & {$46.55033$} & {$3.03785$} & {$0.67213$} & {$150$}\\
{$7.2$} & {$154.83390$} & {$46.54971$} & {$2.5$} & {$1$} & {$\ldots$}\\
{$7.3$} & {$154.83894$} & {$46.55080$} & {$2.5$} & {$1$} & {$\ldots$}\\
{$8.1$} & {$46.74934$} & {$-0.81166$} & {$2.5$} & {$1$} & {$\ldots$}\\
{$9.1$} & {$213.36593$} & {$43.91398$} & {$3.62550$} & {$0.47253$} & {$20$}\\
{$9.2$} & {$213.36371$} & {$43.91462$} & {$2.5$} & {$1$} & {$\ldots$}\\
{$9.3$} & {$213.36823$} & {$43.91343$} & {$2.5$} & {$1$} & {$\ldots$}\\
{$10.1$} & {$40.30403$} & {$-0.87672$} & {$2.5$} & {$1$} & {$\ldots$}\\
{$10.2$} & {$40.30271$} & {$-0.87786$} & {$2.5$} & {$1$} & {$\ldots$}\\
{$11.1$} & {$225.19968$} & {$48.60762$} & {$2.5$} & {$1$} & {$\ldots$}\\
{$12.1$} & {$115.22574$} & {$40.06989$} & {$2.68115$} & {$0.69767$} & {$90$}\\
{$12.2$} & {$115.22466$} & {$40.07119$} & {$2.5$} & {$1$} & {$\ldots$}\\
{$12.3$} & {$115.22411$} & {$40.06984$} & {$2.5$} & {$1$} & {$\ldots$}\\
{$13.1$} & {$120.19842$} & {$46.69237$} & {$2.77694$} & {$0.71108$} & {$40$}\\
{$13.2$} & {$120.19943$} & {$46.69040$} & {$2.91361$} & {$0.64391$} & {$60$}\\
{$13.3$} & {$120.20016$} & {$46.68878$} & {$2.76588$} & {$0.71108$} & {$40$}\\
{$14.1$} & {$220.16045$} & {$53.47826$} & {$2.5$} & {$1$} & {$\ldots$}\\
{$15.1$} & {$44.16970$} & {$-0.24563$} & {$4.20687$} & {$0.55542$} & {$20$}\\
{$16.1$} & {$118.12012$} & {$49.83926$} & {$2.89310$} & {$1$} & {$\ldots$}\\
{$16.2$} & {$118.12260$} & {$49.83988$} & {$2.86452$} & {$1$} & {$\ldots$}\\
{$17.1$} & {$118.85621$} & {$48.43858$} & {$2.5$} & {$1$} & {$\ldots$}\\
{$18.1$} & {$165.11818$} & {$44.26118$} & {$3.08765$} & {$0.67097$} & {$130$}\\
{$19.1$} & {$119.99499$} & {$31.89354$} & {$3.08765$} & {$0.65161$} & {$90$}\\
{$20.1$} & {$171.10206$} & {$23.64904$} & {$2.5$} & {$1$} & {$\ldots$}\\
{$20.2$} & {$171.10397$} & {$23.64826$} & {$2.5$} & {$1$} & {$\ldots$}\\
{$20.3$} & {$171.10599$} & {$23.64916$} & {$2.5$} & {$1$} & {$\ldots$}\\
{$21.1$} & {$248.10994$} & {$39.51800$} & {$2.5$} & {$1$} & {$\ldots$}\\
{$22.1$} & {$137.98392$} & {$27.89927$} & {$2.65936$} & {$0.87641$} & {$40$}\\
{$23.1$} & {$206.47093$} & {$26.77547$} & {$2.5$} & {$1$} & {$\ldots$}\\
{$23.2$} & {$206.47033$} & {$26.77414$} & {$2.5$} & {$1$} & {$\ldots$}\\
{$23.3$} & {$206.47203$} & {$26.77284$} & {$2.5$} & {$1$} & {$\ldots$}\\
{$23.4$} & {$206.46895$} & {$26.77273$} & {$2.5$} & {$1$} & {$\ldots$}\\
{$24.1$} & {$178.33743$} & {$52.34683$} & {$2.5$} & {$1$} & {$\ldots$}\\
{$24.2$} & {$178.33574$} & {$52.34477$} & {$2.5$} & {$1$} & {$\ldots$}\\
{$25.1$} & {$247.16206$} & {$40.12241$} & {$2.5$} & {$1$} & {$\ldots$}\\
{$25.2$} & {$247.16246$} & {$40.12662$} & {$2.69824$} & {$0.84901$} & {$0$}\\
{$25.3$} & {$247.16518$} & {$40.12538$} & {$4.17331$} & {$0.35561$} & {$40$}\\
{$25.4$} & {$247.16500$} & {$40.12383$} & {$2.5$} & {$1$} & {$\ldots$}\\
{$25.5$} & {$247.16849$} & {$40.12406$} & {$2.5$} & {$1$} & {$\ldots$}\\
{$25.6$} & {$247.16559$} & {$40.12068$} & {$2.5$} & {$1$} & {$\ldots$}\\
{$26.1$} & {$154.77128$} & {$46.45133$} & {$2.5$} & {$1$} & {$\ldots$}\\
{$26.2$} & {$154.76872$} & {$46.45463$} & {$2.5$} & {$1$} & {$\ldots$}\\
{$26.3$} & {$154.76887$} & {$46.45604$} & {$2.5$} & {$1$} & {$\ldots$}\\
{$26.4$} & {$154.77140$} & {$46.45750$} & {$3.26693$} & {$0.57927$} & {$0$}\\
{$26.5$} & {$154.77514$} & {$46.45156$} & {$2.5$} & {$1$} & {$\ldots$}\\
{$27.1$} & {$176.12429$} & {$55.57992$} & {$2.5$} & {$1$} & {$\ldots$}\\
{$28.1$} & {$113.40069$} & {$45.94347$} & {$2.89125$} & {$1$} & {$\ldots$}\\
{$29.1$} & {$46.71724$} & {$-0.89637$} & {$2.5$} & {$1$} & {$\ldots$}\\
\hline
	\end{tabular}
\end{table*}

\label{lastpage}
\end{document}